\documentclass[preprint]{aastex62}
\usepackage{xspace}
\usepackage{booktabs}
\usepackage{longtable}
\usepackage{morefloats}
\usepackage{blindtext}
\usepackage{placeins}
\usepackage{gensymb}
\usepackage{makecell}
\usepackage{natbib}

\newcommand{\Fermi}{\emph{Fermi}\xspace}
\newcommand{\lat}{\emph{Fermi}-LAT\xspace}
\newcommand{\Swift}{\emph{Swift}\xspace}

\newcommand{\msun}{M$_{\odot}$}

\newcommand{\fov}{FoV\xspace}
\renewcommand{\deg}{\,$^{\circ}$}

\newcommand{\be}{\begin{equation}}
\newcommand{\ee}{\end{equation}}
\newcommand{\ba}{\begin{eqnarray}}
\newcommand{\ea}{\end{eqnarray}}

\newcommand{\ltsima} {$\; \buildrel < \over \sim \;$}
\newcommand{\gtsima} {$\; \buildrel > \over \sim \;$}
\newcommand{\lta} {\lower.5ex\hbox{\ltsima}}
\newcommand{\gta} {\lower.5ex\hbox{\gtsima}}

\newcommand{\de}{$^\circ$\xspace}

\newcommand{\fcat}{1FLGC\xspace}
\newcommand{\tcat}{2FLGC\xspace}
\newcommand{\gcat}{FGGC\xspace}
\newcommand{\tn}{$T_{\rm GBM,90}$\xspace}
\newcommand{\tf}{$T_{\rm GBM,05}$\xspace}
\newcommand{\tnf}{$T_{\rm GBM,95}$\xspace}
\newcommand{\tz}{$T_{\rm \,LAT,0}$\xspace}
\newcommand{\tone}{$T_{\rm \,LAT,1}$\xspace}
\newcommand{\toz}{$T_{\rm \,LAT,100}$\xspace}
\newcommand{\tlln}{$T_{\rm LLE,90}$\xspace}
\newcommand{\tllf}{$T_{\rm LLE,05}$\xspace}
\newcommand{\tllnf}{$T_{\rm LLE,95}$\xspace}

\newcommand{\nor}[1]{\textcolor{red}{ #1}}

\newcommand{\ngrb}{186\xspace}
\newcommand{\nshort}{17\xspace}
\newcommand{\nlong}{169\xspace}
\newcommand{\nredshift}{34\xspace}
\newcommand{\nredshiftl}{33\xspace}
 
\newcommand{\nlike}{169\xspace}
\newcommand{\nlikel}{155\xspace}
\newcommand{\nlikes}{14\xspace}

\newcommand{\nagl}{115\xspace}
\newcommand{\nags}{11\xspace}
\newcommand{\nlle}{91\xspace}
\newcommand{\nllel}{85\xspace}
\newcommand{\nlles}{6\xspace}
\newcommand{\nlleonly}{17\xspace}

\newcommand{\agbpl}{12\xspace}
\newcommand{\agspls}{1\xspace}
\newcommand{\agbpls}{1\xspace}
\newcommand{\agspll}{77\xspace}
\newcommand{\agbpll}{11\xspace}
\newcommand{\agfitl}{86\xspace} 
\newcommand{\agfits}{2\xspace}

\setcitestyle{notesep={ }}

\received{\today}
\revised{\today}
\accepted{\today}
\submitjournal{ApJ}

\shorttitle{The 2$^{nd}$ \lat GRB catalog}
\shortauthors{The \lat collaboration}
\begin{document}
\title{A decade of Gamma-Ray Bursts observed by \lat: \\ The 2$^{nd}$ GRB catalog}
\correspondingauthor{Magnus Axelsson, Elisabetta Bissaldi, Nicola Omodei, Giacomo Vianello}
\email{magnusa@astro.su.se} \email{elisabetta.bissaldi@ba.infn.it} \email{nicola.omodei@stanford.edu} \email{giacomov@slac.stanford.edu}
%
%
%
%
\author{M.~Ajello}
\affiliation{Department of Physics and Astronomy, Clemson University, Kinard Lab of Physics, Clemson, SC 29634-0978, USA}
\author{M.~Arimoto}
\affiliation{Faculty of Mathematics and Physics, Institute of Science and Engineering, Kanazawa University, Kakuma, Kanazawa, Ishikawa 920-1192}
\author{M.~Axelsson}
\affiliation{Department of Physics and Oskar Klein Center for Cosmoparticle Physics, Stockholm University, 106 91 Stockholm, Sweden}
\affiliation{Department of Physics, KTH Royal Institute of Technology, AlbaNova, SE-106 91 Stockholm, Sweden}
\author{L.~Baldini}
\affiliation{Universit\`a di Pisa and Istituto Nazionale di Fisica Nucleare, Sezione di Pisa I-56127 Pisa, Italy}
\author{G.~Barbiellini}
\affiliation{Istituto Nazionale di Fisica Nucleare, Sezione di Trieste, I-34127 Trieste, Italy}
\affiliation{Dipartimento di Fisica, Universit\`a di Trieste, I-34127 Trieste, Italy}
\author{D.~Bastieri}
\affiliation{Istituto Nazionale di Fisica Nucleare, Sezione di Padova, I-35131 Padova, Italy}
\affiliation{Dipartimento di Fisica e Astronomia ``G. Galilei'', Universit\`a di Padova, I-35131 Padova, Italy}
\author{R.~Bellazzini}
\affiliation{Istituto Nazionale di Fisica Nucleare, Sezione di Pisa, I-56127 Pisa, Italy}
\author{P.~N.~Bhat}
\affiliation{Center for Space Plasma and Aeronomic Research (CSPAR), University of Alabama in Huntsville, Huntsville, AL 35899, USA}
\author{E.~Bissaldi}
\affiliation{Dipartimento di Fisica ``M. Merlin" dell'Universit\`a e del Politecnico di Bari, I-70126 Bari, Italy}
\affiliation{Istituto Nazionale di Fisica Nucleare, Sezione di Bari, I-70126 Bari, Italy}
\author{R.~D.~Blandford}
\affiliation{W. W. Hansen Experimental Physics Laboratory, Kavli Institute for Particle Astrophysics and Cosmology, Department of Physics and SLAC National Accelerator Laboratory, Stanford University, Stanford, CA 94305, USA}
% ...
\begin{abstract}
The Large Area Telescope (LAT) aboard the \Fermi spacecraft routinely observes high-energy emission from gamma-ray bursts (GRBs). Here we present the second catalog of LAT-detected GRBs, covering the first 10 years of operations, from 2008 August 4 to 2018 August 4.  A total of \ngrb GRBs are found; of these, \nlle show emission in the range 30--100\,MeV (\nlleonly of which are seen only in this band) and \nlike are detected above 100\,MeV. Most of these sources were discovered by other instruments (\Fermi/GBM, \Swift/BAT, AGILE, INTEGRAL) or reported by the Interplanetary Network (IPN); the LAT has independently triggered on 4 GRBs.

This catalog presents the results for all \ngrb GRBs. We study onset, duration and temporal properties of each GRB, as well as spectral characteristics in the 100\,MeV--100\,GeV energy range. Particular attention is given to the photons with highest energy. Compared with the first LAT GRB catalog, our rate of detection is significantly improved. The results generally confirm the main findings of the first catalog: the LAT primarily detects the brightest GBM bursts, and the high-energy emission shows delayed onset as well as longer duration.  
However, in this work we find delays exceeding 1\,ks, and several GRBs with durations over 10\,ks. Furthermore, the larger number of LAT detections shows that these GRBs cover not only the high-fluence range of GBM-detected GRBs, but also samples lower fluences. In addition, the greater number of detected GRBs with redshift estimates allows us to study their properties in both the observer and rest frames. Comparison of the observational results with theoretical predictions reveals that no model is currently able to explain all results, highlighting the role of LAT observations in driving theoretical models.
\end{abstract}
\keywords{editorials, notices --- 
miscellaneous --- catalogs --- surveys}
\vspace{1cm}
\tableofcontents
\section{Introduction} \label{sec:intro}
Observations by the \emph{Fermi Gamma-Ray Space Telescope} in the ten years since it was placed into orbit on 2008 June 11 have allowed for the opportunity to study the broadband properties of gamma-ray bursts (GRBs) over an unprecedented energy range.  Its two scientific instruments, the Large Area Telescope \citep[LAT;][]{2009ApJ...697.1071A}, and the Gamma-Ray Burst Monitor \citep[GBM;][]{2009ApJ...702..791M}, provide the combined capability of probing emission from GRBs over seven decades in energy.  These ground-breaking observations have helped to characterize the highest energy emission from these events \citep{GRB080825C, LAT081024B, GRB090510:ApJ, LAT_090217, GRB090902B:Fermi, GRB090926A}, furthered our understanding of the emission processes associated with GRBs \citep{Ryde2010, Axelsson+12, Preece2014, Ahlgren2015, Moretti2016, Burgess2016, Arimoto2016}, helped place constraints on the Lorentz invariance of the speed of light \citep{GRB080916C:Science, 2009Natur.462..331A, Shao2010, Nemiroff2012, Vasileiou2013}, the gamma-ray opacity of the Universe \citep{2010ApJ...723.1082A,Desai2017}, and motivated revisions in our basic understanding of collision-less relativistic shock physics \citep{2014Sci...343...42A}.

The first \lat GRB catalog was published in 2013 \citep[][; hereafter \fcat]{2013ApJS..209...11A}. It is a compilation of the 35 GRBs, 30 long-duration ($>$ 2 s) and 5 short-duration ($<$ 2 s) GRBs, detected in the period August 2008 through July 2011 (the first GRB included is GRB\,080825C and the last GRB\,110731A). Of these GRBs, 28 were detected with standard analysis techniques at energies above 100\,MeV, while 7 GRBs were detected only $<100\,$MeV using the LAT Low-Energy (LLE) technique. It established three main observational characteristics of the high-energy emission:
\begin{itemize}
\item Additional spectral components: Many of the bright GRBs observed by \lat\ were not well fit with a single Band function \citep{1993ApJ...413..281B}. In particular, an additional power-law component was required to account for the high-energy data of four bursts.
\item Delayed onset: The emission above 100\,MeV was systematically delayed compared with the emission seen at lower energies, in the keV--MeV energy range. Delays of up to 40\,s have been detected, with a few seconds being a typical value.
\item Extended duration: The emission above 100 MeV also systematically lasted longer than the keV--MeV prompt emission. The flux generally followed a power-law decay with time, $F \propto t^{-\alpha}$, with $\alpha$ close to $-1$.
\end{itemize} 

The \fcat also left some open questions regarding GRB properties at high energy, which we plan to address in the \tcat:
\begin{itemize}
\item Hyper-fluent GRBs: Four GRBs of the \fcat hinted at the possibility of a different class of high-energy fluence, significantly greater than the average fluence of the other GRBs. Because of the small sample, the hyper-fluent class of GRBs was not significant and its confirmation was left for subsequent observations.
\item Late-time temporal decay index: The distribution of the late-time temporal decay index in the \fcat was clustered around $-1$, supporting the hypothesis of an adiabatically expanding fireball as a common origin of the extended emission. Although this scenario could explain all observations, only 9 GRBs had enough data to allow the decay index to be  determined.
\item Breaks in the late time light curve: Three GRBs in the \fcat showed breaks in the temporal decay at late times, similar to the breaks observed in the X-ray light curves. Exploration of this feature was limited by the small sample and the relatively low significance of the breaks.
\end{itemize}

Since the publication of the \fcat, a number of improvements have been made with regard to LAT data processing. The two major changes concern the development of a new event analysis. Since launch, the LAT event classes have undergone a number of versions (or ``passes''), and the latest ``Pass8'' analysis constitutes a major improvement on the previous versions. In addition, a new detection algorithm has been developed running in parallel over a range of different time scales. Taken together, this has increased the detection efficiency by over 60\% \citep{2015arXiv150203122V}, and in particular allow the detection of fainter high-energy GRB counterparts.

This paper presents the second catalog of GRBs detected by the \lat\ (\tcat), covering a 10-year period, from August 2008 to August 2018. During this time the GBM triggered on 2357 GRBs, approximately half of which were in the field of view (FoV) of the LAT at the time of trigger. LAT counterparts are searched for in ground processing, following external triggers provided by the GBM, as well as by other instruments. The LAT instrument is also capable of detecting GRBs through an onboard trigger search algorithm. This is a very rare occurrence, and has so far only happened 4 times (see Sect.\ref{sec:results}). In addition, continual blind searches are performed as part of the standard ground processing, to look for untriggered events. These efforts are further described in \citet{2016ApJ...823L...2A} and \citet{2018ApJ...861...85A}. 

In the \tcat, we have performed an entirely new, standardized analysis to look for LAT counterparts to all GRB triggers reported during the first decade of \Fermi operations. In Section \ref{sec:data} we describe the data used in this study, giving first a short description of the \Fermi instruments, followed by the description of data cuts and sample selection. This is followed in Section \ref{sec:analysis} by a detailed description of the analysis methods used for detection and localization. We also present the methodology that we followed to characterize the temporal and spectral properties of the detected GRBs. The \tcat is focused on the LAT-only properties of the bursts in our sample, and we do not perform any joint spectral analysis of LAT and GBM data. In Sections \ref{sec:results} and \ref{sec:discuss} we present and discuss our results. Finally, we examine the theoretical implications of our results in Sect.~\ref{sec:interpret}.
\section{Data Preparation}
\label{sec:data}
The \tcat presents analysis done with data from the LAT, i.e. covering energies above 30 MeV for GRBs detected through LLE, and from 100 MeV to 300 GeV for GRBs detected with the standard analysis. However, the GBM provides the vast majority of the GRB triggers and is an integral part of the GRB observations made by \Fermi. We therefore begin with a brief description of both instruments. This is followed by a more thorough description of the LAT data used for the analysis. Finally, we present the selection of GRB triggers which formed the seed of the catalog. 
\subsection{Instrument overview}
\label{sec:instruments}
The LAT is a pair production telescope sensitive to $\gamma$ rays in the energy range from $\sim30$ MeV to more than 300 GeV. The instrument and its on-orbit calibrations are described in detail in \citet{2009ApJ...697.1071A} and \citet{LATperform}. The telescope consists of a 4$\times$4 array of identical towers, each including a tracker of 18 x--y silicon strip detector planes interleaved with tungsten foils, followed by an 8.6 radiation length imaging calorimeter made with CsI(Tl) scintillation crystals with a hodoscopic layout. This array is surrounded by a segmented anti-coincidence detector made of 89 plastic scintillator tiles which identifies and rejects charged particle background events with an efficiency above 99.97\% \citep{2012ApJS..203....4A}. 

Whether or not an event is observable by the LAT is primarily defined by two angles: the angle $\zeta$ with respect to the spacecraft zenith, and the viewing angle $\theta$ from the LAT boresight. The LAT performance -- including the dependence of the effective area on energy and $\theta$ -- is presented on the official \lat\ performance web-page\footnote{\url{http://www.slac.stanford.edu/exp/glast/groups/canda/lat_Performance.htm}}. In the analysis performed in this catalog, we do not make any explicit cuts on the angle $\theta$; however, the exposure will drop very quickly for $\theta$ greater than $\sim75^{\circ}$. The wide FoV ($\sim$2.4 sr at 1 GeV) of the LAT, its high observing efficiency (obtained by keeping the FoV on the sky with scanning observations), its broad energy range, its large effective area, its low dead time per event ($\sim27\,\mu$s), its efficient background rejection, and its good angular resolution (the 68\% containment angle of the point spread function is $0.8^\circ$ at 1 GeV) are vastly improved in comparison with those of previous instruments such as EGRET \citep{1999ApJS..123..203E}. As a result, the LAT provides more GRB detections, higher statistics per detection, and more accurate localizations.

The GBM  is composed of twelve sodium iodide (NaI) and two bismuth germanate (BGO) detectors sensitive in the 8 keV--1 MeV and 250 keV--40 MeV energy ranges, respectively. The NaI detectors are arranged in groups of three at each of the four edges of the spacecraft, and the two BGO detectors are placed symmetrically on opposite sides of the spacecraft, resulting in a field of view (FoV) of 9.5\,sr. Triggering and localization are determined from the NaI detectors, while spectroscopy is performed using both the NaI and BGO detectors. The GBM flight software continually monitors the detector rates and triggers when a statistically significant rate increase occurs in two or more NaI detectors. A combination of 28 timescales and energy ranges are currently tested, with the first combination that exceeds the predefined threshold (generally 4.5\,$\sigma$) being considered the triggering timescale. Localization is performed using the relative event rates of detectors with different orientations with respect to the source and is typically accurate to a few degrees (statistical uncertainty). An additional systematic uncertainty has been characterized as a core-plus-tail model, with 90\%\ of GRBs having a 3.7\deg\ \nor{uncertainty} and a small tail suffering a larger than 10\deg\ systematic uncertainty \citep{2015ApJS..216...32C}. The GBM covers roughly four decades in energy and provides a bridge from the low energies (below $\sim 1$ MeV), where most of the GRB emission takes place, to the less studied energy range that is accessible to the LAT.

For GRBs exceeding a preset threshold for peak flux or fluence in the GBM, an autonomous repoint request (ARR) is sent to the spacecraft flight software. If the GBM request is accepted, a special LAT observation mode is initiated. This will keep the GBM flight software location in the LAT FoV for an extended period of time, typically $\sim2.5$ hours, subject to observational constraints. 

On 2018 March 16, one of the solar array drive assemblies on \Fermi  suffered a malfunction. This led to the LAT and GBM being switched off, and normal science operations were only resumed on April 13. However, as of writing one of the solar panels remains in a fixed position, and a modified rocking strategy has been adopted. As a result, the LAT rocks between the northern and southern sky every week, as opposed to every orbit as before. A further impact is that ARRs have been disabled.
\subsection{Sample selection}
\label{sec:sample}
The \tcat presents the results of a search for high-energy counterparts of GRBs that triggered space instruments and have an available public localization. We considered in particular bursts detected by the GBM, the {\it Swift} Burst Alert Telescope \citep[BAT;][]{2005SSRv..120..143B} onboard the {\it Neil Gehrels Swift Observatory} \citep{2004ApJ...611.1005G}, and the {\it International Gamma-Ray Astrophysics Laboratory} Soft Gamma-Ray Imager \citep[INTEGRAL/ISGRI;][]{Lebrun:03} onboard the INTEGRAL Satellite \citep{2003A&A...411L...1W}, or reported by the Interplanetary Network (IPN\footnote{\url{http://www.ssl.berkeley.edu/ipn3/}}). In the 10-year period covered by the \tcat, there were 2357 GBM GRB triggers, 876 {\it Swift}/BAT GRB triggers, 65 INTEGRAL/ISGRI GRB triggers  and 83 events reported by the IPN (a few through private communication by the PI K.~Hurley). We also considered 7 bursts contained in the first catalog of GRBs detected by the {\it Astro-rivelatore Gamma a Immagini Leggero} mini-calorimeter \citep[ AGILE/MCAL;][]{Galli:2013} and not contained in any other list. After accounting for bursts that triggered more than one instrument, we have a total of 3044 independent GRBs. We use the localizations provided by the GBM, unless a better localization is provided by one of the instruments on-board \Swift, namely the BAT, the X--Ray Telescope \citep[XRT;][]{2005SSRv..120..165B}, or the UV--Optical Telescope \citep[UVOT;][]{2005SSRv..120...95R}, or by the IPN. All these latter positions are distributed via the Gamma-Ray Burst Coordinates Network (GCN)\footnote{\url{https://gcn.gsfc.nasa.gov/gcn3_archive.html}}, while the GBM-only localizations are reported in the {\Fermi}-GBM online GRB Catalog\footnote{\url{https://heasarc.gsfc.nasa.gov/W3Browse/fermi/fermigbrst.html}} \citep[hereafter \gcat; see][for more details]{Narayana2016A_223}.
\subsection{LAT data cuts and temporal selection}
For the standard analysis, we use Pass 8 LAT data with energies between 100 MeV and 100 GeV, selecting the time interval around each trigger from 600\,s before to 100\,ks after the GRB trigger time, and defining a standard region of interest (ROI) around the trigger location of 12$^{\circ}$. In order to reduce the contamination of the Earth limb, in some dedicated cases (13 GRBs) we define a smaller ROI with a radius of 8$^{\circ}$. It is worth noting that as a final check we look for high-energy events over a larger energy range (up to 300 GeV), as discussed in Sec.~\ref{Res_HE_Events}.

We then use \texttt{gtmktime} to select only those time intervals when the center of the ROI has a zenith angle $\zeta < 97^{\circ}$ (i.e. every point of the ROI has $\zeta < 105^{\circ}$). For bursts with an initial value of $\zeta > 90^{\circ}$, we increase our selection including all time intervals when the center of the ROI has a zenith angle $\zeta < 102^{\circ}$. This allows to study the prompt emission of those GRBs that started close to the limb of the Earth. The choice of the event class depends on the time scale on which we detect the signal from the GRB and is described in Section~\ref{sec:ltft}.
\subsection{LAT Low Energy data} 
The LAT Low Energy (LLE) technique is an analysis method designed to study bright transient phenomena, such as GRBs and solar flares, in the 30 MeV--1 GeV energy range. The LAT collaboration developed this analysis using a different approach than the one used in the standard photon analysis, which is based on sophisticated classification procedures \citep[a detailed description of the standard analysis can be found in][]{2009ApJ...697.1071A,2012ApJS..203....4A}. The idea behind LLE is to maximize the effective area below $\sim 1$ GeV by relaxing the standard analysis requirement on background rejection. 

The basic LLE selection is based on a few simple requirements on the event topology in the three LAT sub-detectors. First of all, an event passing the LLE selection must have at least one reconstructed track in the tracker and therefore an estimate of the direction of the incoming photon. Secondly, we require that the reconstructed energy of the event be nonzero.

We use the information provided by the flight software in LLE to efficiently select events which are gamma-ray like. 
With the release of Pass 8 data, we have also improved the LLE selection. For events with an incident angle $\theta<40^{\circ}$, we require that no anti-coincidence tiles are in ``veto'' condition (to suppress charged particle contamination), while for angles greater than 40$^{\circ}$, we allow a maximum of 2 tiles in ``veto'' condition, but no tracker hits can be found in proximity of the anti-coincidence hits. This condition helps prevent suppression of large-incident-angle gamma rays due to secondary electrons or positrons interacting with the anti-coincidence detector downstream. 
In order to reduce the number of photons originating from the Earth Limb in our LLE sample we only keep reconstructed events with a zenith angle $\zeta < 90^{\circ}$ or $\zeta < 100^{\circ}$ (depending on the location of the GRB). Finally we explicitly include in the selection a cut on the region of interest, i.e. the position in the sky of the transient source we are observing. In other words, the localization of the source is embedded in the event selection and therefore for a given analysis the LLE data are tailored to a particular location in the sky.
The response of the detector for the LLE class is encoded in a response matrix, which is generated using a dedicated Monte Carlo simulation for each GRB, and is saved in the standard \texttt{HEASARC RMF} File Format\footnote{Described here: \url{http://heasarc.gsfc.nasa.gov/docs/heasarc/caldb/docs/memos/cal_gen_92_002/cal_gen_92_002.html\#Sec:RMF-format}.}. LLE data and the relative response are made available for any transient signal (GRB or Solar Flare) detected with a significance above 4\,$\sigma$ through the {\tt HEASARC FERMILLE} web site\footnote{\url{http://heasarc.gsfc.nasa.gov/W3Browse/fermi/fermille.html}}.
\newpage
\subsection{Low-energy data used for comparisons}
\label{sec:gbmdata}
While we do not perform any joint spectral fitting with GBM data in this work, comparisons with the sample of GBM-detected GRBs are both highly interesting and inevitable in order to characterize our sample. For this purpose we use the official data from the \gcat.

In order to perform the comparisons, we use the standard GRB properties which characterize the GRB emission, in particular the onset time and duration (both calculated by GBM in the 50--300 keV energy range), values of peak fluxes ($F_P$, calculated in 1024-ms and 64-ms intervals for long and short GRBs, respectively), fluences ($F$, calculated over the time interval used in the GBM spectral analysis, which might not always be coincident with the burst duration) and the spectral parameters of the best-fit model derived by the GBM in the 10--1000 keV energy band. The best fit is determined from the four standard spectral models tested in the GBM time--integrated spectral catalog analysis  \citep[see][for more details]{Gruber2014_211}, namely the simple power law ({\it PL}),  the smoothly broken power law ({\it SBPL}), the phenomenological Band function \citep[{\it BAND},][]{1993ApJ...413..281B} and the {\it Comptonized} model ({\it COMP}), which is a subset of the Band function. The latter three models are characterized by a low--energy spectral index $\alpha$, a high--energy spectral index $\beta$, which in the case of the {\it COMP} model goes to minus infinity, and by the energy $E_{\rm \, peak}$, which describes the peak of the ${\nu}F_{\nu}$ distribution.

The classification of GRBs into long and short classes is primarily derived from the low-energy duration as measured by GBM, and follows the standard rule that long and short bursts are longer and shorter than 2\,s, respectively \citep{Kouv93}. For bursts which did not trigger GBM and are not included in the \gcat, we use the durations calculated by the Konus-Wind instrument in the 20 keV - 5 MeV energy range, which have been published in GCN Circulars.
\section{Analysis Methods and Procedures}
\label{sec:analysis}
As presented in Sect.~\ref{sec:intro}, significant improvements have been made to the analysis techniques since the \fcat. The two developments with highest impact are the ``Pass 8'' event analysis and a redesigned detection algorithm. ``Pass 8'' has been rebuilt from the ground up with respect to previous versions (``Pass 6'' was used for the first catalog), resulting in increased sensitivity. It is thoroughly described in \cite{2013arXiv1303.3514A}. In this section, we focus on the improvements to the triggered search and detection algorithms for GRBs used to produce the \tcat.
\subsection{Analysis sequence}
\begin{figure}[t!]
\begin{center}
\includegraphics[width=1.0\columnwidth,trim=10 200 500 200, clip=true]{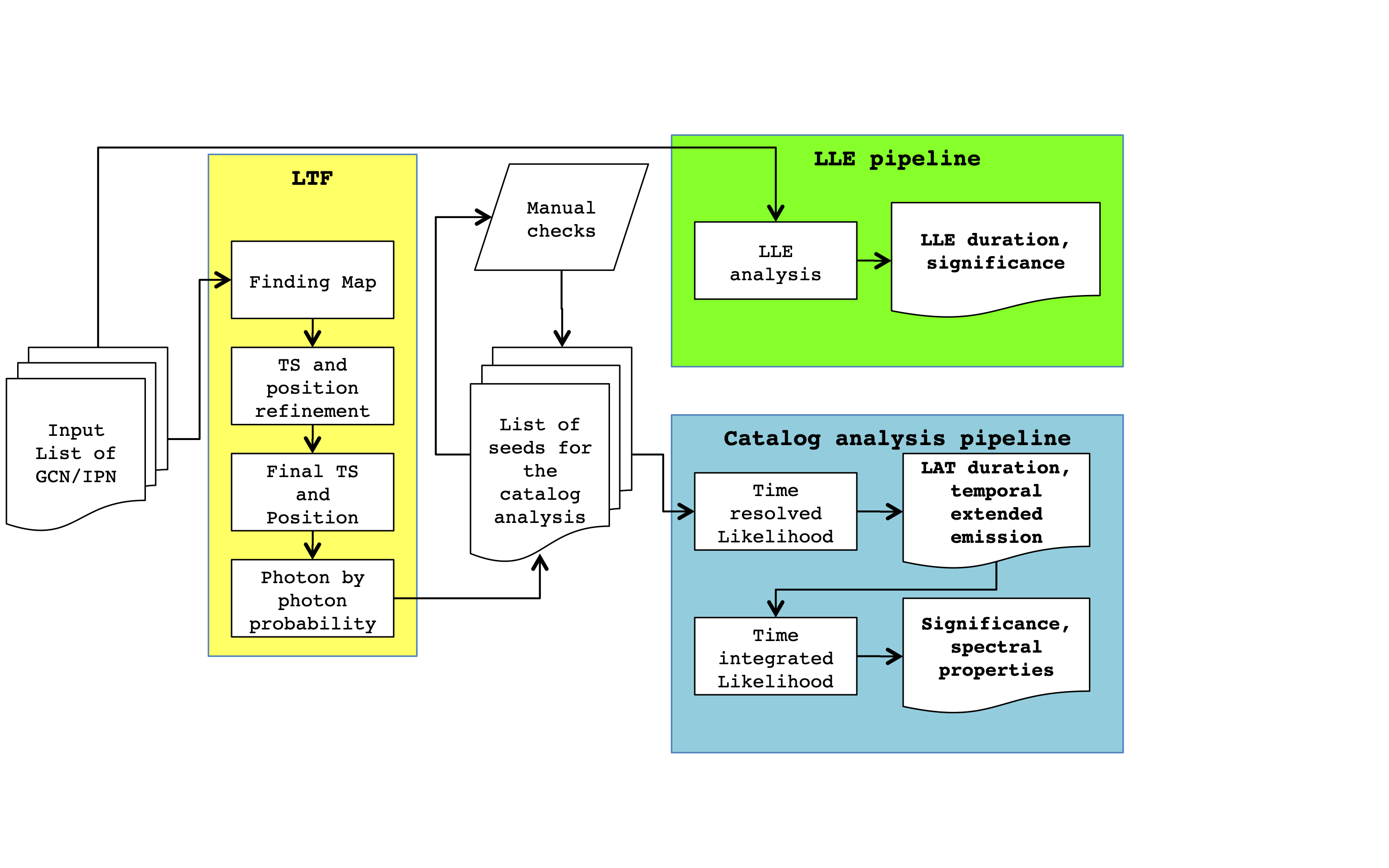}
\caption{Flow diagram representation of the analysis pipeline. The LTF detection algorithm is highlighted in yellow, and the catalog analysis pipeline, which is executed after manually checking the candidate seeds, is highlighted in blue. The LLE analysis pipeline, which is executed on all input triggers, is highlighted in green. All input candidates are analysed by the LTF/LLE pipelines; only those passing the detection criteria are retained. Inspection in the manual checks can also lead to rejection of further cases (see text for details).}
\label{diagram}
\end{center}
\end{figure}

The procedure followed to produce the catalog is summarized in Figure~\ref{diagram}. Each individual step of the analysis is described in detail in the following subsections. 

Each input trigger was first run through a detection algorithm as described in Sect.~\ref{sec:ltft}. All triggers which passed the criteria for detection in any time window were placed in the list of potential candidates for analysis. This list was manually inspected following the procedure in Sect.~\ref{sec:manual}. Only detections retained after this inspection were passed to the analysis pipeline. Here, analyses were performed to determine a number of key properties for each GRB, such as onset, duration, and spectral parameters. The steps of this analysis are described in Sect.~\ref{sec:catanalysis}.

\subsection{The LAT Transient Factory}
\label{sec:ltft}
The LAT Transient Factory (LTF) algorithm was introduced after the publication of the \fcat  and is presented in \citet{2015arXiv150203122V}. LTF has been running continuously ever since, looking in real time for GRB counterparts in LAT data. When compared to the old algorithm on the same dataset as the first catalog, it returned 50\% more GRBs. The adoption of ``Pass 8'' data has led to a further 10\% improvement. 

LTF is based on the application of an unbinned maximum likelihood technique. This analysis starts by selecting all the photons detected by the LAT above 100\,MeV in a circular Region Of Interest (ROI) with radius $r$ and center $(\alpha_{\rm ROI},\, \delta_{\rm ROI})$ in a time window starting at the trigger time $T_0$. Then, the presence of a new point source at a position $(\alpha_{t},\, \delta_{t})$ is tested by using the likelihood-ratio test (LRT). The null hypothesis for the test is represented by a baseline likelihood model including all point sources from the LAT source catalog \citep{2015ApJS..218...23A} with all parameters fixed, as well as the Galactic and isotropic diffuse emission templates \footnote{{\tt gll\_iem\_v06.fits} and {\tt iso\_P8R3\_SOURCE\_V2.txt}; \url{https://fermi.gsfc.nasa.gov/ssc/data/access/lat/BackgroundModels.html}} provided by the \lat\ collaboration \citep{2016ApJS..223...26A} with the normalizations left free to vary. The alternative hypothesis is represented by the baseline model plus the new point source (``test source''), modeled with a power-law spectrum with free index and normalization. The LRT uses as test statistic (TS) twice the logarithm of the maximum of the likelihood function for the alternative hypothesis ($L_1$) divided by the maximum of the likelihood function for the null hypothesis ($L_0$):
\begin{equation}
{\rm TS}=2\left(\log{L_1}-\log{L_0}\right).
\end{equation}    
Detailed instructions on how to perform an unbinned likelihood analysis using the \Fermi Science Tools can be found on the \Fermi website\footnote{\url{https://fermi.gsfc.nasa.gov/ssc/data/analysis/scitools/likelihood_tutorial.html}}.

For each trigger, a search is performed on five time windows starting at the trigger time $T_0$ and ending respectively 10, 100, 500, 4000 and 10000 seconds after $T_0$. This selection slightly differs from the standard LTF real-time analysis, which consists of 10 searches running in parallel over time intervals logarithmically spaced from the trigger time to 10 ks after that, as stated in \citet{2015arXiv150203122V}. 
For the 10 and 100\,s time windows, we use the Pass 8 {\tt P8R2\_TRANSIENT020E\_V6} event class and the corresponding response functions; for the longer time windows, the event class {\tt P8R2\_TRANSIENT010E\_V6} is used. In each time window the LTF starts from the input coordinates and trigger time as measured by the triggering instrument (see section \ref{sec:sample}) and performs the following steps:
\begin{enumerate}
\item {\bf Finding map}: we consider an ROI with radius $r$ centered on the input position, and a square grid of side $\delta x$ inscribed in the ROI with a spacing of $0.8^{\circ}$. The size of the grid is fixed according to the triggering instrument and its typical localization accuracy (statistical + systematic), as well as the typical size of the LAT PSF. Specifically, $\delta x=32^{\circ}$ for triggers localized by the GBM that are dominated by systematic uncertainties \citep{2015ApJS..216...32C}, $\delta x=1.6^{\circ}$ for \Swift and INTEGRAL triggers, and $\delta x = 5^{\circ}$ for IPN triggers. The radius of the ROI is chosen as $r = \delta x / 2 + 12^{\circ}$ in order to have enough data around each point in the grid for performing an LRT test (see below). In order to reduce the contamination from the Earth Limb - a bright source of $\gamma$-rays - all events with $\zeta<100^{\circ}$ are filtered out. The effect of this selection is taken into account when computing the exposure by the tool {\it gtltcube}. We then use the LRT as described above to test for the presence of a source at each position of the grid having at least 3 photons within 10$^{\circ}$. This latter requirement filters out points without any photon cluster around them, in order to reduce the computational cost. The point $(\alpha_{\rm max}, \delta_{\rm max})$ in the grid providing the maximum of the TS is considered the best guess for the position of the new transient, and marked for further analysis.
\item {\bf TS and position refinement}: we consider an ROI centered on $(\alpha_{\rm max}, \delta_{\rm max})$ with a radius of 8$^{\circ}$, and we perform a LRT as described above considering only the time intervals within the time window when the border of the ROI is at a zenith angle smaller than 105$^{\circ}$ (``good time intervals''). This is a different way of reducing the contamination from the Earth Limb that is more effective than the one used in the previous step, but can only be applied on small ROIs. We then use the tool \textit{gtfindsrc} to search for the maximum of the likelihood under the alternative hypothesis (i.e. when the test source is added to the model), varying the position of the test source and profiling the other free parameters. The position $(\alpha_1, \delta_1)$ yielding the maximum of the likelihood is considered the new putative position for the candidate counterpart.
\item {\bf Final TS and position}: the previous step is repeated using an ROI centered on $(\alpha_1, \delta_1)$ which yields the final TS (TS$_{\rm GRB}$). The tool \textit{gtfindsrc} is run again returning the final estimate of the localization uncertainty.
\item {\bf Photon-by-photon assignment of probability}: we run the tool {\it gtsrcprob} using the final optimized likelihood model under the alternative hypothesis. This tool assigns to each detected photon the probability of belonging to the test source, i.e., to the candidate counterpart. We then measure the number of photons $N_{p > 0.9}$ having a probability larger than 90\% of belonging to the candidate counterpart.
\end{enumerate}

The final products of LTF are five sets of results, one for each time window. In order to consider a counterpart detected we consider in particular TS$_{GRB}$ and $N_{p > 0.9}$, as explained in the next section.
\subsection{Detection threshold and False Discovery Rate}
A classic result from \citet{chernoff1954} states that under the null hypothesis the TS of a single LRT as applied in LTF is a random variable which is zero half of the time and is distributed as $\chi^2$ with 1 degree of freedom the other half. This result was confirmed by Monte Carlo simulation in \citet{1996ApJ...461..396M}. Under these circumstances the significance of the detection (z-score) is $\sqrt{\rm TS}$, thus a threshold of $TS=25$ corresponds to a detection $> 5\sigma$ for one LRT. 

As described in the previous section, LTF consists of multiple LRT procedures and the trial factor needs to be taken into account. The effective number of trials for one time window is however difficult to determine because the trials are not independent. Furthermore, we also need to account for the number of time windows and for the number of triggers searched. 

To account for the number of triggers searched, we use the procedure proposed by \citet{10.2307/2346101}. It assumes independent trials and it is simple: all the p-values $p_{i}$ for all the $n_{\rm tr}$ searches, with $i=1...n_{\rm tr}$, are sorted in increasing order. We then find $k$ so that $p_{k}$ is the largest p-value where $p_k \leq \left( \frac{k}{n_{\rm tr}} \right) \alpha$, where $\alpha$ is the error probability for one test. All the triggers with $i=1...k$ are considered detected. In practice, we first compute through Monte Carlo simulations the effective number of trials  for one time window $n_{\rm trw}$. The value of $n_{\rm trw}$ is different depending on the instrument that generated the trigger, and it reflects the size of the finding map (see previous section) so it is larger for larger finding maps. We find $n_{\rm trw}=110$ for GBM triggers, $n_{\rm trw} = 12.28$ for IPN triggers and $n_{\rm trw} = 1.25$ for Swift and INTEGRAL triggers. We then compute the post trial p-value for one time window by using the Binomial distribution as $p_{\rm w} = 1 - (1-p_{\rm pr})^{n_{\rm trw}}$, where $p_{\rm pr}$ is the p-value coming from the LRT applied to the current time window. There is also another independent trial factor $n_{\rm et}$ which we consider to be equal to the number of time windows where we effectively searched for a counterpart. Note that this is a slightly conservative approach, as the time windows are not independent and thus $n_{\rm et}$ is in reality a little smaller. The number $n_{\rm et}$ can vary from zero to 5 depending on how many time windows had an exposure larger than zero after our data cuts. For example, if the trigger was never in the field of view or it was always at a zenith angle larger than our cut during a time window, this will not constitute a search and it will not contribute to $n_{\rm et}$.  We can now compute the p-value for a GRB corrected for both the spatial and the time trials as $p = 1 - (1-p_{\rm w,min})^{n_{\rm et}}$, where $p_{\rm w,min}$ is the minimum $p_{\rm w}$ corresponding to the maximum final TS$_{\rm GRB}$ found by LTF among the time scales searched. We then apply the \citet{10.2307/2346101} procedure using these p-values and correct for the number of triggers searched as explained above.

We further apply the quality cut $n_{\rm p > 0.9} \ge 3$ to the list of detections, i.e., we require at least 3 photons with a probability larger than 90\% of belonging to the GRB. This neutralizes the effect of isolated high-energy photons ($\gtrsim 10$\,GeV) within the search region that tend to return high TS values during the unbinned analysis but also very hard spectra, with photon indexes close to 0. Moreover, 3 photons are required in order to have both the normalization and the photon index free during the likelihood maximization under the alternative hypothesis and still have at least 1 degree of freedom.
\subsection{The Bayesian Blocks Burst Detection algorithm for LLE data}
\label{sec:llebb}
In order to detect GRB counterparts in LLE data we use a counting analysis based on the well-known Bayesian Blocks (BB) algorithm of \citet{2013ApJ...764..167S}. The BB algorithm is capable of dividing a time series in intervals of constant rate, opening a new block only when the rate of events changes in a statistically significant way. In particular, we use the unbinned version of the algorithm which presents as the only parameter the probability $p_{0}$ of opening a new block when the rate is constant (false positive). However, before we can apply the BB algorithm, we need to introduce a pre-processing step to account for the time-varying background in LLE data. Otherwise, the BB algorithm will find many blocks following the variation in the event rate due to the variations in the background. 

\begin{figure}[t!]
\centering
\begin{tabular}{cc}
\includegraphics[width=0.48\columnwidth]{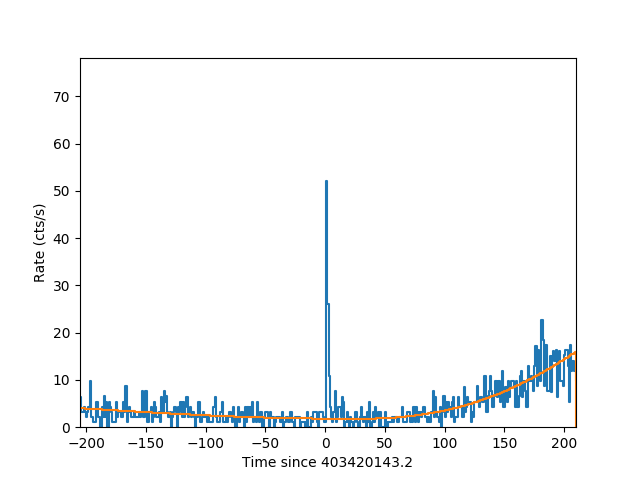} & 
\includegraphics[width=0.48\columnwidth]{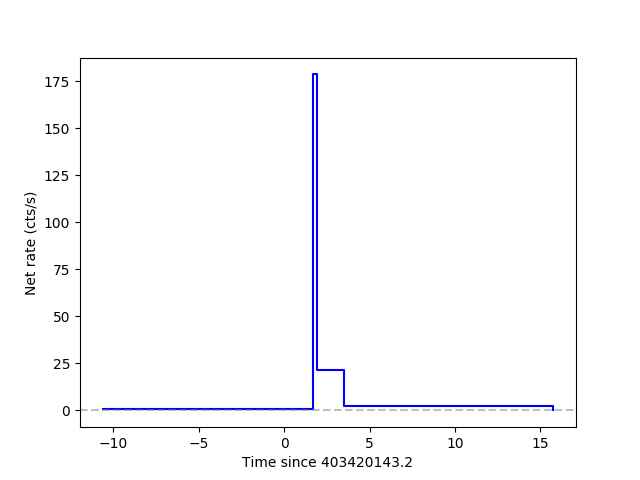}
\end{tabular}
\caption{{\it Left panel}: LLE light curve of GRB\,131014 in the 30 MeV -- 1 GeV energy range. The polynomial fit to the background is marked by a red line. {\it Right panel}: Bayesian Blocks representation of the light curve of GRB\,131014. All light curves are centered around the GBM trigger time $T_{\,0}=403420143.2$ (in MET).}
\label{bb}
\end{figure}

We start by fitting a polynomial function to the data in two off-pulse time windows, one before and one after the trigger window, as shown in the left plot of Figure~\ref{bb} for GRB\,131014A. The trigger window is defined on the basis of the duration measured by the GBM in the 50--300 keV energy range. Then, we exploit the fact that a non-uniform Poisson process with expected value $\lambda(t)$ can be converted to a uniform one by transforming the time reference system to:
\begin{equation}
t^{\prime} = \int_{0}^{t}~\lambda(t) dt.
\end{equation}
In our case $\lambda(t)$ is the expected number of events of the time-varying background as modeled by the polynomial. We then transpose the time series of LLE events to the $t^{\prime}$ time reference system and then apply the standard BB algorithm. If more than one block is found in the trigger window it means that there is a rate change on top of what is predicted by the background model. We then deem the transient detected and we transform back to the original time reference system to yield the time interval of the detection. For all searches we use $p_{0} = 10^{-3}$. The analysis is illustrated in the right plot of Figure~\ref{bb}, which shows the final BB representation of the light curve for GRB\,131014A.
\subsection{Final manual checks}
\label{sec:manual}
The normalization of the Galactic and diffuse templates are allowed to vary in the analysis, but the parameters for the point sources are kept fixed to the values found in the updated \lat\ point source list\footnote{\url{https://fermi.gsfc.nasa.gov/ssc/data/access/lat/fl8y/}}. This means that the algorithm can potentially pick up non-GRB sources, such as AGN flares. Moreover, GRBs detected far from zenith might also suffer from Earth Limb contamination. Therefore, a final manual check of all potential detections was also performed. The list of detections derived by the LTF pipeline previously described was divided into random subsets of bursts which were assigned for analysis to the members of the \lat\ GRB team. Each putative event was independently cross checked by two people, whose task was to either confirm or reject the detection. In case of agreement, the classification was seen as final; otherwise the case was reviewed by a third person.

The manual checks included a series of tasks to be carried out. First, the LTF results were evaluated in each of the 5 temporal intervals, taking into account (1) the number of photons with probability $>90\%$; (2) the distance to the nearest known source; (3) the localization error; (4) the spectral index; and (5) the final TS value. We identified many ``simple" cases, in which both the number of detected photons was high and the final TS was above 80 in several time intervals, and no other known sources were present in the ROI. These candidates were marked as confirmed with no further inspection. 

Intermediate cases that needed deeper investigations included (1) cases where the final TS in all time intervals was close to the threshold; (2) cases where only 3-4 high-energy photons were detected; (3) cases where a bright source (an AGN, Solar Flares, etc.) was at an angle $< 1^{\circ}$ from the GRB candidate in the ROI; (4) cases where a high TS value was obtained only by integrating over the longest timescale (from 0 to 10000\,s, see Section \ref{sec:ltft}). In order to check for other active sources in the ROI around the time of the GRB trigger, we looked for flaring blazars using the publicly available FAVA tool\footnote{\url{https://fermi.gsfc.nasa.gov/ssc/data/access/lat/FAVA/}} \citep{2017ApJ...846...34A, 2013arXiv1303.4054C}, and for solar activity we checked the Solar Monitor public pages\footnote{\url{http://www.solarmonitor.org}}.
In case of particularly uncertain candidates, we performed an ad-hoc likelihood analysis, similar to the one performed by the LTF pipeline, but running on dedicated time intervals which might differ from the catalog ones.

Through these manual checks, $\sim$15--20\%\ of the examined cases were rejected as not connected to a GRB. The remaining events were processed in the dedicated analysis pipeline.
\subsection{Catalog Analysis description}
\label{sec:catanalysis}

In this section we describe the analysis steps we performed on each GRB of the final sample. The idea is to perform an automated analysis, which is implemented in a series of \texttt{python} scripts that are used to control the various steps. The analysis is based on \texttt{ScienceTools v11r05p03}, available for download at the \Fermi Science Support Center\footnote{\url{https://fermi.gsfc.nasa.gov/ssc/data/analysis/}}.
\\
\subsubsection{Time-integrated likelihood analysis}
\label{sec:timeintegrated}
We perform an unbinned likelihood analysis in four different time intervals. The ``GBM" time interval represents the GRB duration as given by \tn reported in the \gcat. \tn is the interval during which the instrument measures from $5\%$ to $95\%$ of the total GRB flux in the 50--300 keV energy range (i.e., from \tf to \tnf).  The ``LTF'' interval corresponds to the time interval in the LAT Transient Factory analysis where the highest TS was found. The ``LAT'' interval encompasses the signal detected by the LAT, as defined in \ref{sec:time-resolved}. The ``EXT'' interval is defined as the time interval including LAT emission (if any) after the \tnf. 
Table \ref{tab_intervals_fit} summarizes the definition of these intervals.
\begin{deluxetable}{lll}[t!]
\tablecolumns{3}
\tablewidth{0pt}
\tablecaption{Definition of the four time intervals (with start and stop times) used in the time-integrated spectral analysis
\label{tab_intervals_fit}}
\tablehead{\colhead{Name}&\colhead{Interval}&\colhead{Description}}
\startdata
{\bf GBM} & $\left[T_{\rm GBM,05},\,T_{\rm GBM,95}\right]$ & GRB duration measured by GBM in the 50--300 keV energy range\\
{\bf LTF}   & $\left[T_0,\,T^{\rm LTF}_{\rm max} \right]$ & Time interval showing the highest TS value as calculated by the LAT transient \\
& &  factory, starting from the GRB trigger time \\
{\bf LAT} & $\left[T_{\rm \,LAT,0},\, T_{\rm \,LAT,1} \right]$ &  GRB duration measured by LAT by performing a time-resolved likelihood analysis \\
& & in the 100 MeV - 100 GeV energy range \\
{\bf EXT} & $\left[T_{\rm GBM,95},\, T_{\rm \,LAT,1}\right]$ & Interval from end of GBM to end of LAT duration. \\
\enddata
\end{deluxetable}
\subsubsection{Time-Resolved likelihood analysis}
\label{sec:time-resolved}
In order to perform time-resolved likelihood analysis we have developed an algorithm for adaptively binning the LAT events. Starting from the result of the analysis in the ``LTF'' time window, we apply \texttt{gtsrcprob} to calculate the probability of each LAT event to be associated with the GRB source. Starting with pre-selected logarithmically spaced time bins (48 bins from 0.01\,s to 50\,ks after the GBM trigger), we merge them until at least 3 events with probability $> 0.9$ are present in each final bin. In practice, we have 3 degrees of freedom (N$_{dof}$): 2 associated with the power law describing the GRB, and 1 with the normalization of the isotropic diffuse component The normalization of the Galactic model has been fixed to its nominal value (1). We require at least N$_{dof}$ events with probability $> 0.9$ in every bin. In this way, we optimize the duration of the time intervals in order to always have enough photons to perform the fit. Once we have identified the time bins, we perform unbinned likelihood analysis in each bin, calculating the value of flux or, in case of a TS-value $<$10, we calculate the flux upper limit (95\%) by profiling the likelihood function.

\begin{figure}[t!]
\centering
\begin{tabular}{cc}
\includegraphics[width=0.48\columnwidth,trim=0 0 0 0,clip=false]{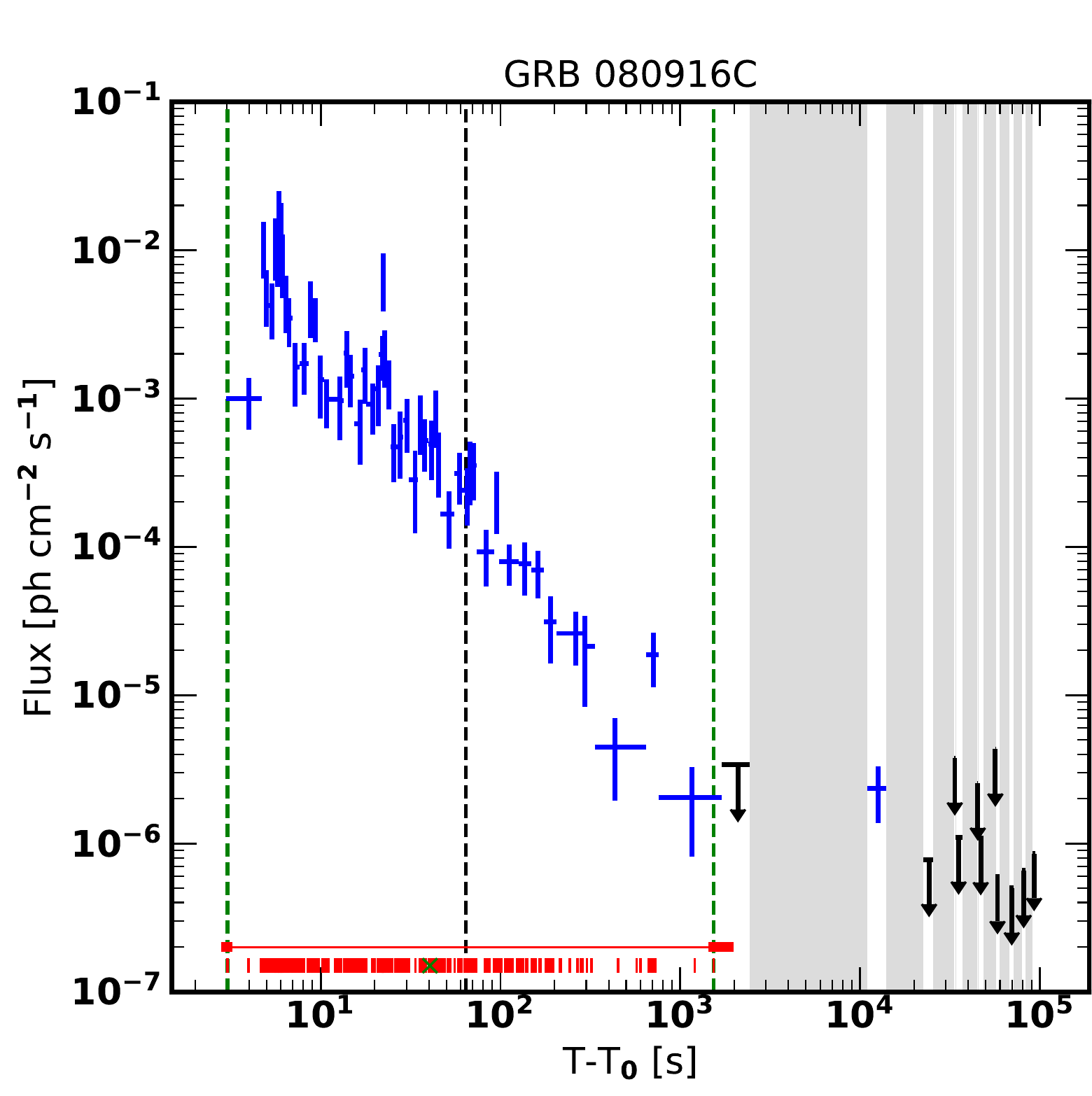} & 
\includegraphics[width=0.48\columnwidth,trim=0 0 0 0,clip=true]{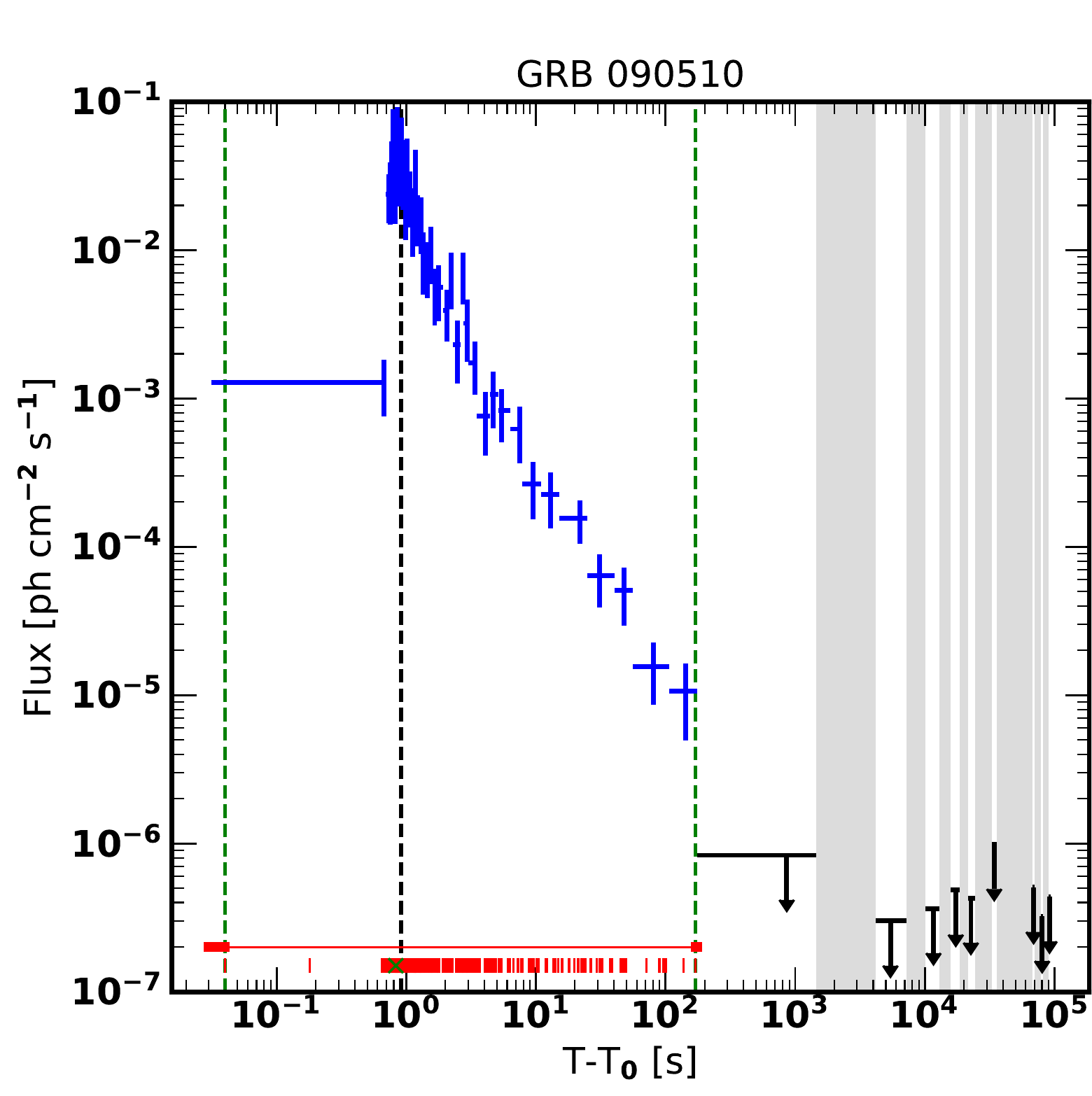}
\end{tabular}
\caption{Temporal extended emission for two bright LAT GRBs, the long GRB~080916C ({\it left panel}), and the short GRB~090510 ({\it right panel}). Blue points show the flux in each time bin, while black arrows mark upper limits. The green vertical dashed lines indicate the first and the last LAT-detected event, while the vertical dashed black line marks the end of the GBM emission (\tnf). Shaded grey areas mark intervals when the GRB is outside the FoV. The red vertical markers at the bottom of each panel indicate the arrival times of the events with probability $>$0.9 to be associated with the GRB, with the green cross being the event with maximum energy. The horizontal red line indicates the estimated duration of the GRB (\toz).}
\label{fig_LC_timeresolved}
\end{figure}

In the \fcat, the duration in the LAT was calculated based on the concept of \tn, i.e., the time during which 90\% of the flux is collected. As the LAT observes each photon individually, this requires the simulation of light curves. In this analysis, we instead use a technique based on the individual photons intrinsic to the LAT. The total duration of the signal in the LAT, defined as \toz, is estimated starting from the results of the time-resolved analysis. The LAT onset time \tz corresponds to the time when the first photon with probability $p > 0.9$ to be associated with the GRB is detected, while \tone corresponds to the last event with $p > 0.9$. \toz of the signal is simply \tone-\tz. These are also the quantities which define the ``LAT'' time interval, as previously discussed (see Table~\ref{tab_intervals_fit}). 

In order to correctly estimate the uncertainty on \tone  ($\delta$\tone) for an event with $n$ detected photons with probability $p > 0.9$, we define $\Delta t_{n-1,n}$ as the time interval between the second to last and the last event. Assuming Poisson statistics, the probability to measure an event between $t$ and $t+dt$ is $P(t,t+dt) = \lambda\,dt$, where $\lambda$ is the rate: in our case $\lambda = 2/\Delta t_{n-1,n}$. Therefore, we conservatively compute the uncertainty as $\delta$\tone$ = 1/\lambda = \Delta t_{n-1,n}/2$. Similarly, considering the first two events with probability $p > 0.9$, we define the uncertainty on \tz as $\delta$\tz$= \Delta t_{1,2}/2$. The error on \toz follows using standard error propagation.

In order to better illustrate this analysis, Figure~\ref{fig_LC_timeresolved} shows two light curves of two bright GRBs, the long GRB\,080916C ({\it left panel}) and the short GRB\,090510 ({\it right panel}). For the long burst, the arrival time of the last event is substantially smaller than the end of the bin of the last detection. This could indicate that in the last bin the GRB is only marginally detected. For most of the other bursts, the arrival time of the last event is very close to the end of the last bin with positive detection, as in the case of GRB\,090510.
\subsubsection{Calculation of energetics}
\label{subsec:energy}
In addition to reporting the flux and fluence of each GRB, for the subset of GRBs with measured redshift $z$ we also calculate their total radiated energy ($E_{\,iso}$). Starting from the measured spectrum of each burst, this is done by using the best-fit model over a specific energy range, and by assuming that the energy emitted by a GRB at the source in the cosmological source frame is isotropically radiated. The isotropic radiated energy is defined by following expression
\begin{equation}
E_{\,iso} = \frac{4\pi \,d_{\,L}^{\,2}}{1 + z} \;\, S(E_1, E_2, z)\,,  
\label{Eiso_sbolo}
\end{equation}
where $d_{\,L}$ is the luminosity distance, and
$S(E_1, E_2, z)$ is the fluence integrated between the minimum energy $E_{\, 1}$ and the maximum energy $E_{\, 2}$. It can be expressed as
\begin{equation}
S(E_1, E_2, z) = T_{\,100}\displaystyle\int_{{E_{\,1}/{(1 + z)}}}^{{E_{\,2}}/{(1 + z)}} E \, N(E)\, dE \, .
\label{sbolo}
\end{equation}
Here, $N(E)$ describes the best-fit spectral model, and $T_{\,100}$ represents the total duration of the burst as defined in the previous section. LAT data are always fit with a simple power-law model in the energy range 100\,MeV to 10\,GeV, i.e.,
\begin{equation}
N(E) = A E^{\Gamma}.
\end{equation}
Finally, assuming a spatially flat universe $\Lambda$CDM model with $\Omega_{\Lambda} = 0.714$, $\Omega_{M} = 0.286$ and $H_0 = 69.6$ km s$^{-1}$ Mpc$^{-1}$ \citep{2014ApJ...794..135B,2016A&A...594A..13P},   the luminosity distance is given by \citep{1972gcpa.book.....W}:
\begin{equation}
d_L(z,\Omega_{\Lambda},\Omega_M) = (1+z)\,\frac{c}{H_0}\int_0^z \frac{dz'}{\sqrt{\Omega_{M}\,(1+z')^3+\Omega_{\Lambda}}}\,.
\label{eq_dL}
\end{equation} 
\subsubsection{Localization}
The LTF algorithm described in Sect.~\ref{sec:ltft} returns the position of the GRB as well as its detection probability. The steps of the procedure include refining the source location, and the position given in the final step is taken as the definitive one; no further optimization is performed.
\subsubsection{LLE light curve and duration}
The Bayesian Blocks Burst Detection algorithm described in Sec.~\ref{sec:llebb} provides a way of binning the data taking into account background fluctuations: blocks are defined only when an intrinsic rate variation above the background is detected, as opposed to an absolute variation. In our analysis we therefore define the onset of the LLE signal (\tllf) as the starting time of the first block above background. Similarly, we define the \tllnf as the ending time of the last block above background. The LLE duration (\tlln) is simply defined as \tllnf-- \tllf. 
\section{Results}
\label{sec:results}
In the following subsections, we examine the main results of our analysis. The focus will be on the properties of the overall population, rather than a presentation of individual GRBs.
\subsection{LAT detections}
\label{sec_LAT_det}
\begin{figure}[t!]
\begin{center}
\begin{tabular}{cc}
\includegraphics[height=0.4\columnwidth]{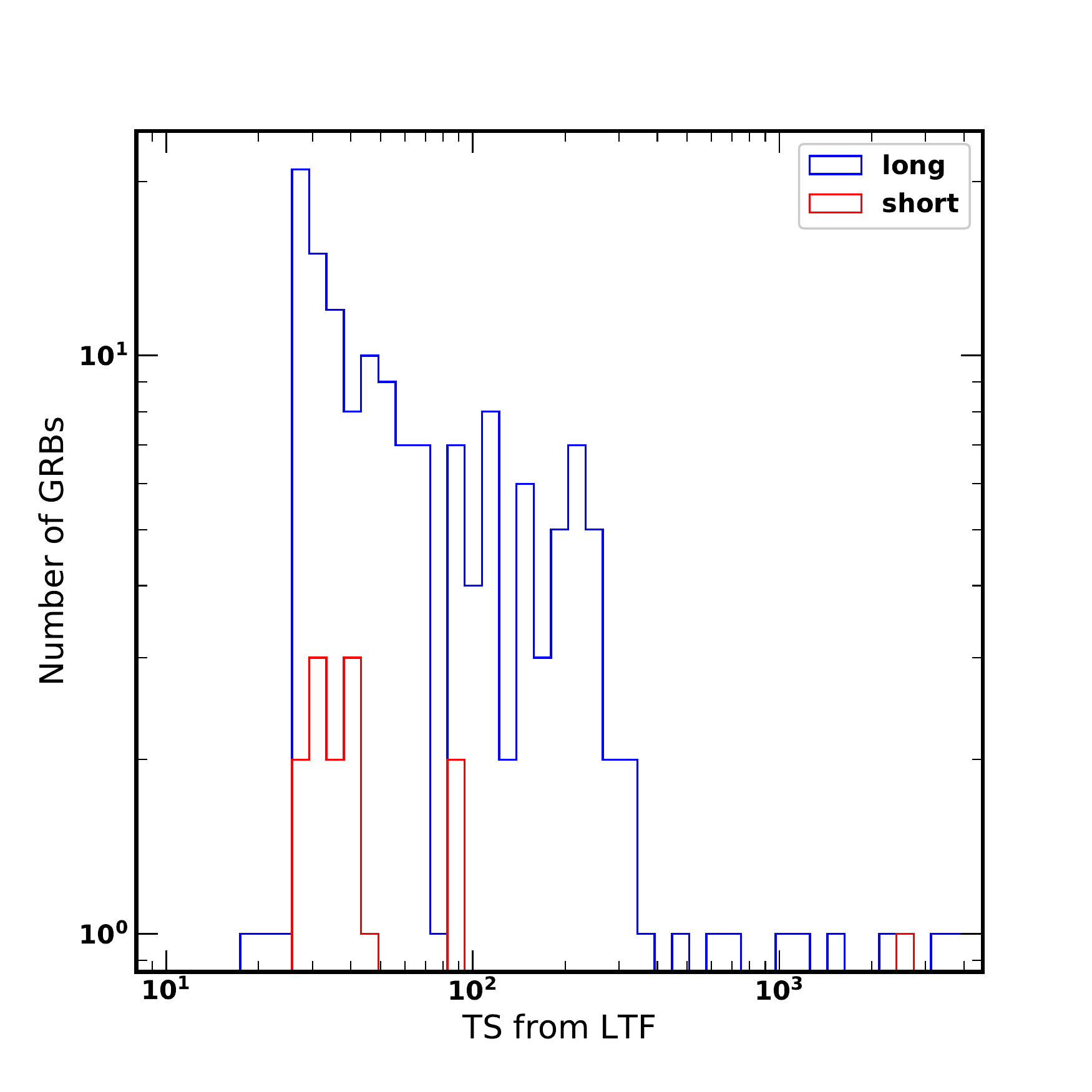} & 
\includegraphics[height=0.4\columnwidth]{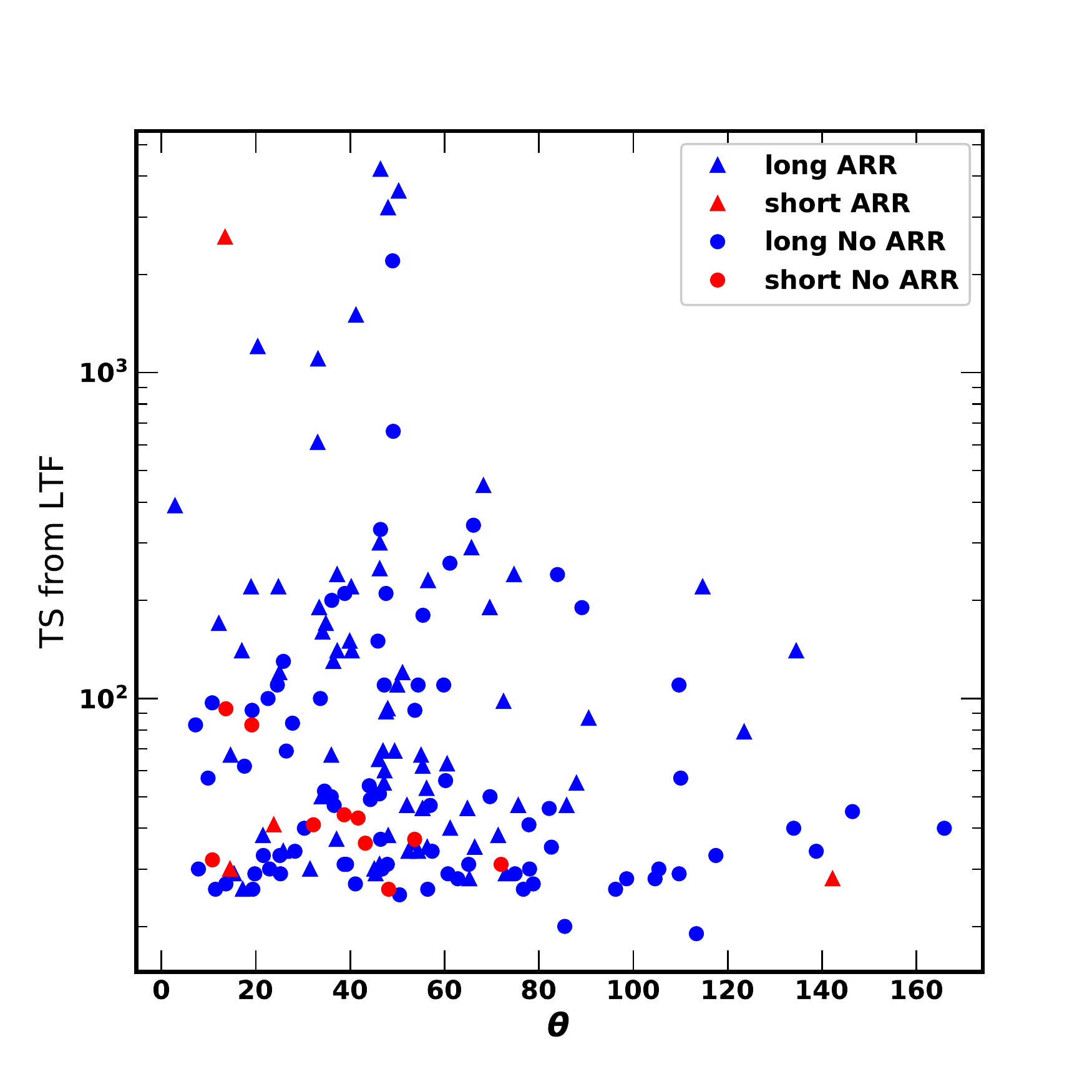}
\end{tabular}
\caption{Left panel: Distribution of the  Test Statistic (TS) for \nlikes short (red) and \nlikel long (blue) GRBs detected by LTF algorithm. Right panel: TS values for long and short bursts as a function of the angle $\theta$ at the trigger time. Bursts which triggered an ARR are marked with a triangle.}
\label{fig_SU_TSINPUT}
\end{center}
\end{figure}

This 10-year catalog comprises \ngrb\ detections, \nshort\ short GRBs (sGRBs) and \nlong\ long GRBs (lGRBs). Adopting the analysis methods described in Sect.~\ref{sec:analysis}, we detect \nlike GRBs with our likelihood analysis above 100 MeV. Of these,  \nlikel are lGRBs and \nlikes are sGRBs. The distribution of the Test Statistic (TS) obtained by the LTF algorithm is shown in the left panel of Figure~\ref{fig_SU_TSINPUT}. The distribution peaks at relatively low values of TS ($\sim30$), and then smoothly falls with increasing TS value. Only a handful of GRBs ($\sim$5\,\%) form a tail at very high TS (above 1000).

Using the LLE technique, \nlle GRBs are found below 100\,MeV. Out of those, \nllel are lGRBs and \nlles are sGRBs. Moreover, \nlleonly of these GRBs (of which 2 sGRBs) are found only with the LLE technique, and are not detected at higher energies with the LAT standard analysis chain.

Of all 3044 triggers in our initial list, we thus detect $\sim 6\%$ at high energies with the LAT. About 18\%\ of the LAT-detected bursts were outside the nominal FoV of the LAT at the time of the trigger. We note that the position in the sky of events which were outside the FoV at trigger time may have entered the FoV at a later time. Moreover, in 10 years 220 triggers initiated an autonomous repoint request of the satellite, a small fraction ($< 10 \%$) of which are caused by other sources, such as solar flares or particle events. 83 of these ARRs successfully resulted in a LAT detection. The distribution of the LTF TS values as a function of $\theta$ at the trigger time is shown in the right panel of Figure~\ref{fig_SU_TSINPUT}. The highest TS values are seen for GRBs with $\theta < 50^{\circ}$.

Furthermore, this catalog includes four GRBs which triggered the LAT directly: one short burst, GRB\,090510, and three long ones, GRB 131108A, GRB\,160509A and GRB\,160821A. This underscores that onboard LAT GRB detections are relatively rare, implying exceptional brightness in high-energy gamma rays. It is worth to note that the very bright GRB\,130427A did not result in a LAT onboard trigger, since the GBM had triggered and issued an ARR on the first emission peak \citep{Preece2014}, which was very bright at low energies but not particularly strong above 100\,MeV.

\begin{figure}[t!]
\begin{center}
\includegraphics[width=0.95\columnwidth]{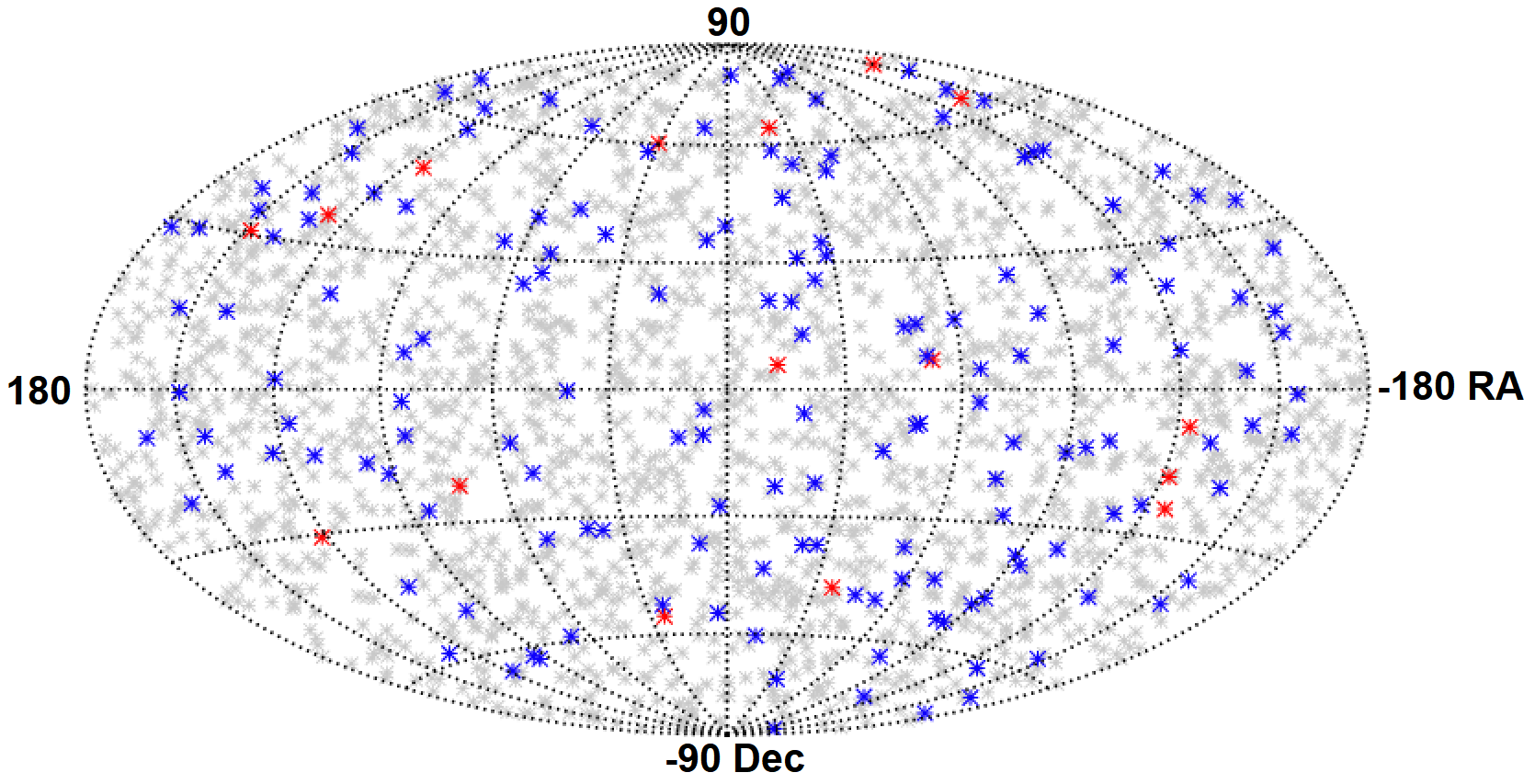}
\caption{Sky distribution of 2357 GBM-triggered GRBs (from 2008 July 14 to 2018 July 31) in equatorial coordinates ({\it grey asterisks}). {\it Blue (red) asterisks} indicate 160 (16) long (short) LAT-detected GRBs included in the \tcat over the same time period.}
\label{fig_GRB_skymap}
\end{center}
\end{figure}

Figure \ref{fig_GRB_skymap} shows the position in equatorial coordinates of 2357 GBM GRB triggers (grey symbols) detected over the 10-year period of the catalog. 160 lGRBs and 16 sGRBs also detected by LAT are marked by blue and red asterisks, respectively.
In fact, out of the \nlike likelihood-detected GRBs, 10 did not trigger the GBM instrument. Of these, two GRBs triggered Swift-BAT, namely GRB\,081203A and GRB\,130907A, while six GRBs were reported by the IPN: GRB\,090427A, GRB\,110518A, GRB\,120911B, GRB\,140825A, the short GRB\,160702A, and GRB\,180526A. While GRB\,120911B did not trigger GBM, it was the only burst to be later found in on-ground analysis of GBM data and announced by \citet{2012GCN.13757....1G}. Furthermore, we report for the first time the detection of GRB\,100213C and GRB 111210B. These triggers were reported via private IPN communication, as stated in Sec.~\ref{sec:sample}. 

A list of all LAT detections is given in Table \ref{tab_GRBs}. For each event, we state the trigger date and time (both in UT and in MET), the final LTF localization with error, the off-axis and zenith angles at trigger time, whether an ARR was issued, the likelihood TS value and LLE significance, the redshift, and the references to the corresponding GCN circulars published by the LAT collaboration. For those events detected only with the LLE technique, we report the best possible localization of the burst as determined by e.g. GBM or \Swift.

In our sample, \nredshift GRBs have a measured redshift (19\%), as compared to 10 (29\%) in \fcat. For comparison, the fraction of \Swift-detected bursts with redshift is $\sim29\%$\footnote{\url{https://swift.gsfc.nasa.gov/archive/grb_table/}}. The smaller fraction of LAT bursts with a measured redshift in the \tcat with respect to the \fcat is not surprising, as $\sim 50$ new GRBs were discovered by our analysis, which have not been previously reported to the community. In addition, the improvements to the analysis techniques enable us to detect fainter GBRs, which are more difficult targets for follow-up observations.

On average, the (90\% containment, statistical only) uncertainty in LAT detections is $0.36^{\circ}$ with a range from $0.04^{\circ}$ to $2.0^{\circ}$. 
In order to assess the LAT location accuracy, we also checked for joint detections by \lat and \Swift, and found that 75 bursts ($\sim$40\%) have a BAT-position, while 67 bursts ($\sim$36\%) have an XRT position. By comparing LAT and \Swift/XRT localizations of the co-detected GRBs, we find that $\sim$70\% of the \Swift localizations are inside the LAT 90\% confidence region. The majority of the remaining XRT positions are only marginally outside the LAT region, indicating that the LAT localization error is slightly underestimated ($\sim 0.1^{\circ}$).
\subsubsection{Comparison with the first LAT GRB catalog}
The changes and improvements in the \tcat mean that the results reported here will differ from those in the \fcat. In the time interval of the \fcat, August 2008 to July 2011 (3 years), we now recover more events: instead of 28 standard likelihood detections we now have 50 detections. Three of these new detections are short GRBs, namely GRB\,081102B, GRB\,090228A and GRB\,110728A. Four of the new detections come from non-GBM triggers.

The \fcat included 21 GRBs also detected with the LLE technique below 100 MeV. During the same period, we now find 25 LLE detections. Four of those -- GRB\,090531B, GRB\,100225A, GRB\,101123A and GRB\,110529A -- are LLE-only bursts as reported also in the \fcat, with the first and the last one being short GRBs. The total number of LLE-only detections is lower with respect to the \fcat, where we retrieved 7 LLE-only bursts. Indeed this is not surprising, since we now detect more events with the likelihood analysis thanks to Pass 8 and to the improved LTF pipeline. 

As a result of the new analysis, we do not include in the current catalog two events which were included in the \fcat: GRB\,091208B and GRB\,110709A. Both GRBs were long, with estimated LAT durations of $\sim40$\,s; however, only 3 photons were detected for each GRB and their detection was marked as marginal. The highest-energy photon in GRB\,091208B was 1.2 GeV, while GRB\,110709A had no detected emission above 500\,MeV. By selecting Pass 8 data and applying the new detection algorithm, the significance of these two detections further decreased, thus resulting in their exclusion from the \tcat.
\subsubsection{LAT detections after July 2011}
We have also cross checked the current catalog with the LAT detections that were publicly announced through GCN circulars in the time period from July 2011 until August 2018. Using the standardized catalog analysis described in \ref{sec:ltft}, we now detect 31 previously unreported GRBs, for which no GCN has been issued. As expected, this is a much smaller relative increase than during the period of the \fcat, since Pass 8 data and the improved detection algorithm have been used since 2015. 

Moreover, we do not retrieve 8 GRBs which have previously been publicly announced by the \lat\ Collaboration, namely GRB 120916A (GCN 13777), GRB\,130206A (GCN 14190), GRB\,131018B (GCN 15357), GRB 140329A (GCN 16047), GRB\,150127A (GCN 17356), GRB\,150724B (GCN 18065), GRB\,161202A (GCN 20229) and GRB\,170810A (GCN 21452). In general, these are all GRBs which at the time of detection were reported with low significance, or with few photons. All these cases were analysed at the time of the GCN writing either on ad-hoc time intervals chosen by the burst advocates or on the 10 real-time LTF temporal windows. These differ from the five fixed time intervals chosen for the catalog analysis presented in Sec.~\ref{sec:ltft}, thus leading to different results. GRB\,130206A and GRB\,150127A were previously reported through GCNs as marginal LLE detections, both with a significance $<3\sigma$, again not matching the current catalog requirements.
\subsection{LAT onset times and duration}
\label{sec:onset_duration}
In the following paragraphs we discuss the temporal properties of the bursts in our sample. As presented in Sect.~\ref{sec:gbmdata}, the classification of GRBs into long and short classes is derived from the low-energy duration as measured by GBM in the 50--300 keV energy band. 
The LAT durations are calculated in the 100\,MeV--10\,GeV energy range. Table \ref{tab_durations} summarizes the various temporal characteristics of the GRBs in our catalog. This includes the values of \tf\, \tnf\ and \tn\ for GBM; \tllf, \tllnf and \tlln for LLE, \tz, \tone and \toz\ for the LAT. Two GRBs, GRB 100213C and GRB 111210B, were reported only by the IPN through private communication: we do not provide any duration information for those. We mark all non-GBM durations in Table \ref{tab_durations} with an asterisk in the \tf column.

\begin{figure}[t!]
\centering
\begin{tabular}{cc}
\includegraphics[width=0.45\columnwidth,trim=0 0 0 0,clip=false]{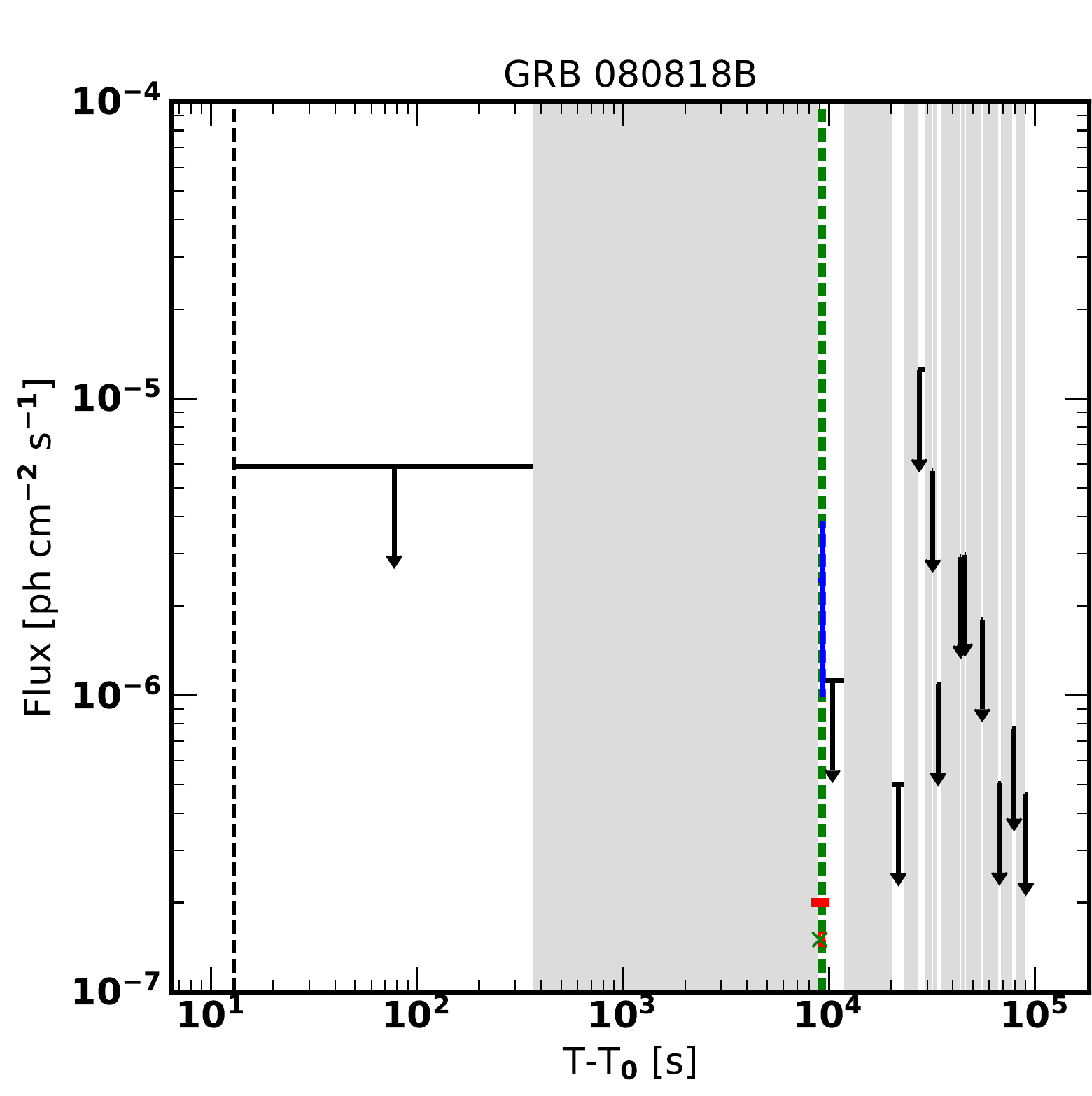} &
\includegraphics[width=0.45\columnwidth,trim=0 0 0 0,clip=true]{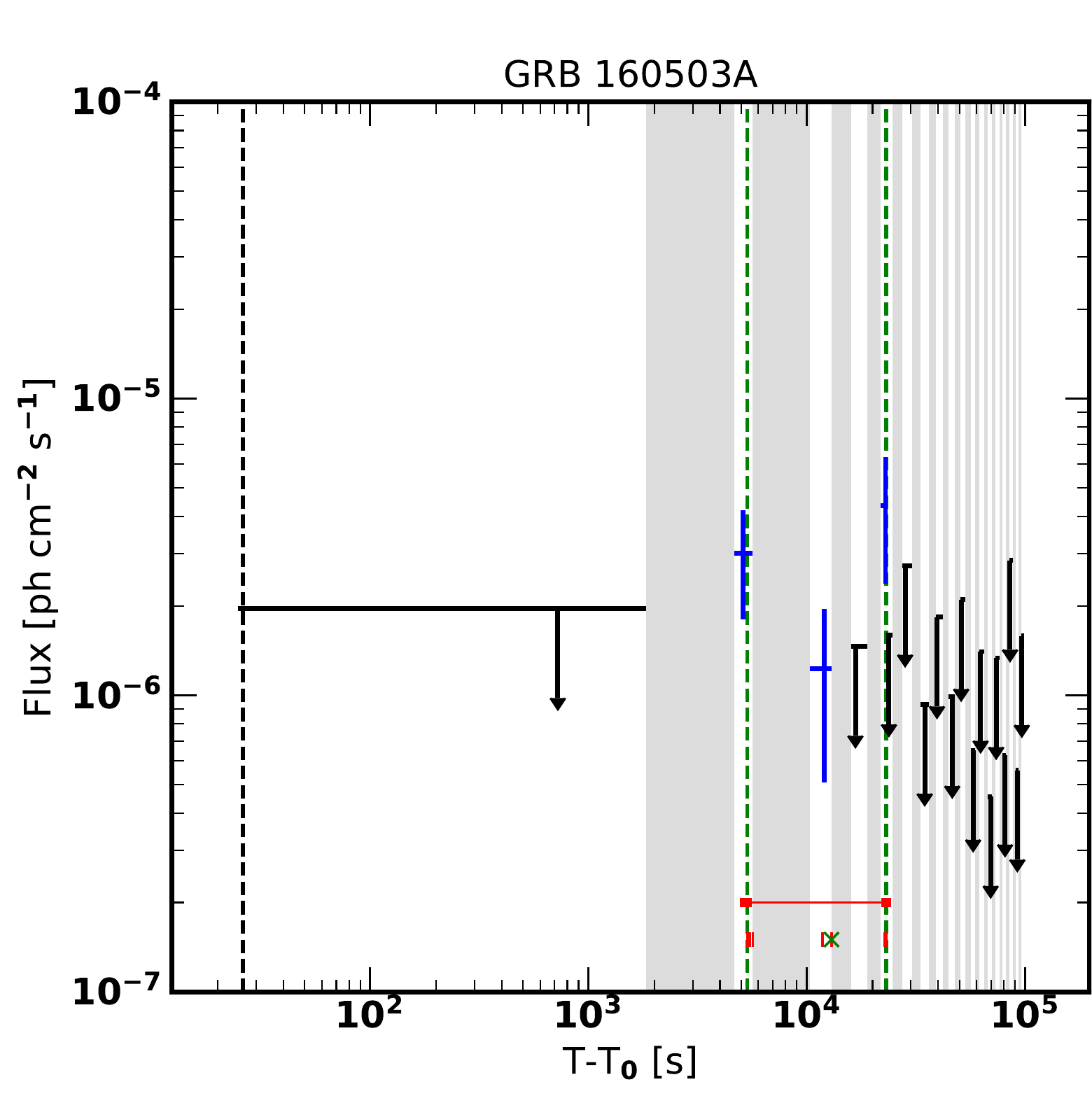}
\end{tabular}
\caption{
Temporal extended emission for two GRBs detected by LAT at very late times, namely GRB~080818B ({\it left panel}), 
and GRB~160503A ({\it right panel}). Markers and colors are the same as in Figure~\ref{fig_LC_timeresolved}.}
\label{fig_Like_T0}
\end{figure}

For some GRBs, the LAT detection of the first photon occurs at very late times. This could be due to high-energy photons not being emitted during the initial phase, but could also be due to observational constraints, where the GRB location is outside the FoV for long intervals. This is illustrated for two GRBs in Figure~\ref{fig_Like_T0}. In both panels, blue points are photon flux measurements, while upper bounds are displayed as black arrows. In the left panel, the first shaded grey area marking when GRB\,080818B was outside the FoV spans almost 10\,ks (\tz$=9.0 \pm 0.6$ ks). The estimated duration of the burst, \toz$=500 \pm 200$\,s, is almost not visible due to the late time of the detection. Similarly, in the right panel, the first detection of GRB\,160503A occurs at 5.3\,ks, again after a period of several ks where the burst was first not detected and then outside the FoV. In this case, the duration was \toz$\sim 18$ ks.

\begin{figure}[t!]
\centering
\begin{tabular}{ccc}
\includegraphics[width=0.32\columnwidth,trim=0 5 40 0,clip=true]{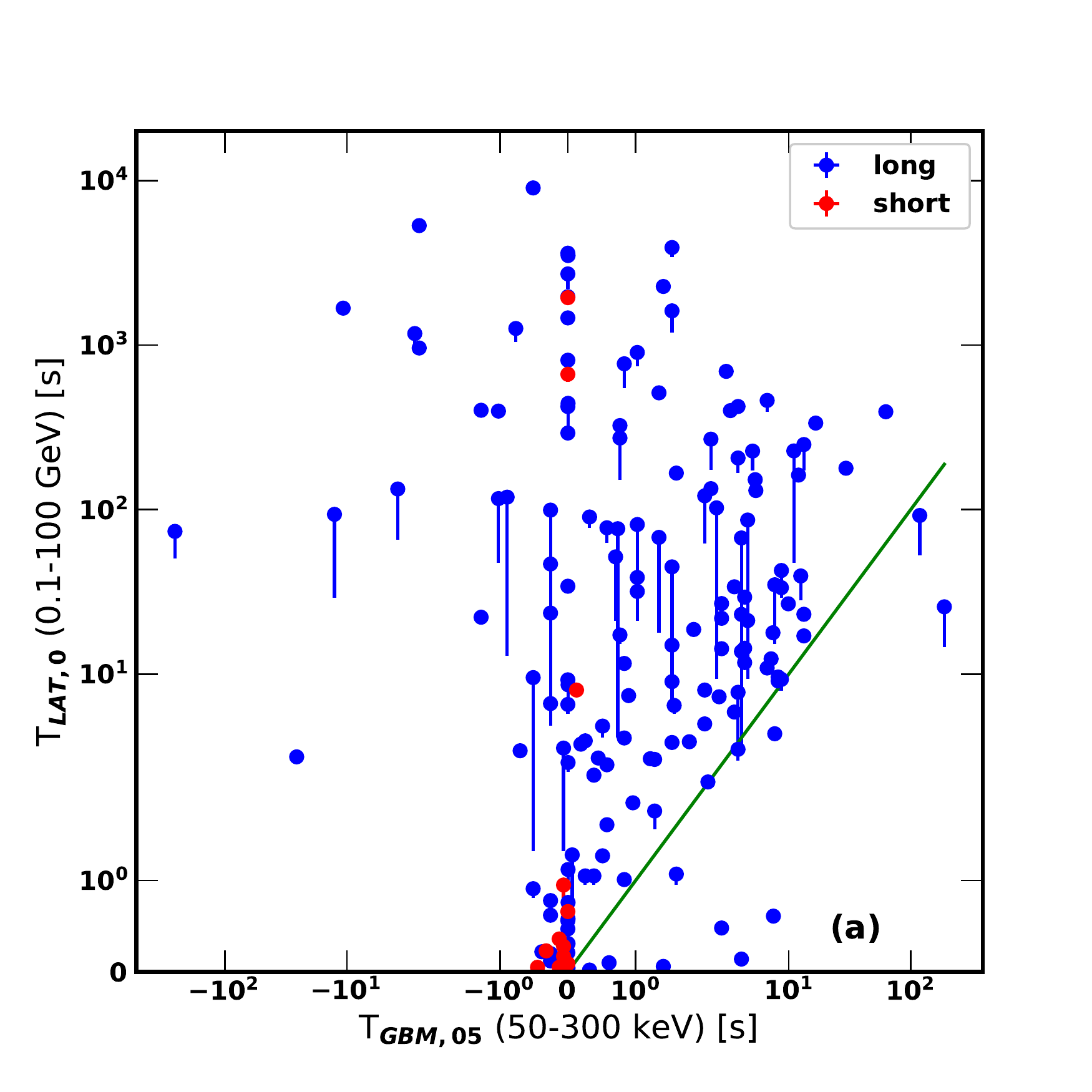} &
\includegraphics[width=0.32\columnwidth,trim=5 5 40 40,clip=true]{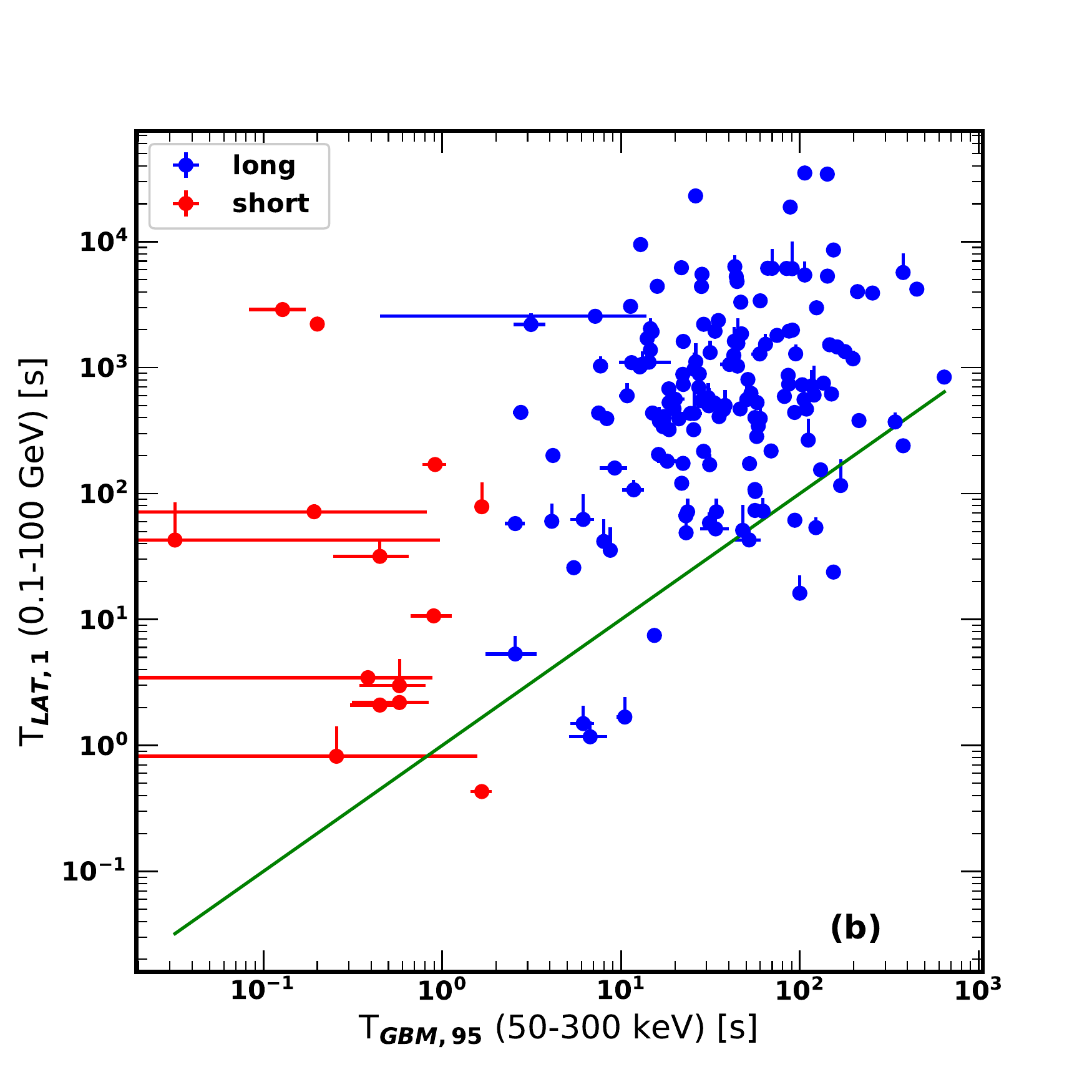} &
\includegraphics[width=0.32\columnwidth,trim=5 5 40 40,clip=true]{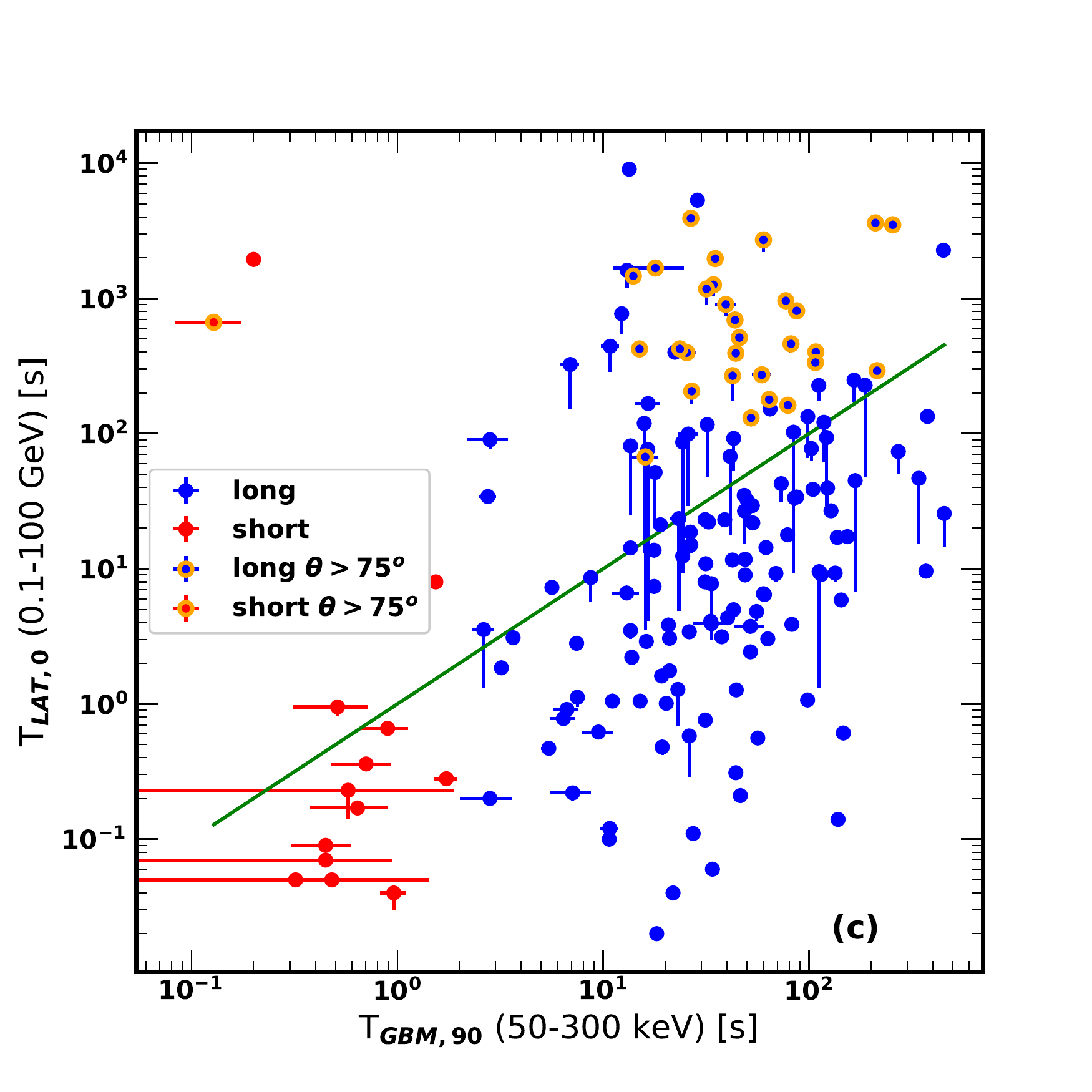}
\end{tabular}
\caption{
\tz (a) and \tone (b) calculated in the 100 MeV--100 GeV energy range vs. the same quantities calculated in the 50--300 keV energy range. Panel (c) shows the onset times (\tz) in the 100 MeV--100 GeV vs. the durations (\tn) in the 50--300 keV energy range. The solid line denotes where values are equal. Blue and red circles represent long and short GRBs, respectively. In panel (c), we additionally mark the GRBs which were outside FoV at trigger time with a thick orange contour.}
\label{fig_GBMT05_LATT05}
\end{figure}

In panel (a) of Figure~\ref{fig_GBMT05_LATT05} we compare the onset times estimated in the LAT energy band (100 MeV--100 GeV) with the ones estimated in the GBM energy band (50--300 keV). A negative \tf value in the low--energy band means that the burst onset occurred before the trigger time. In general, we notice that the high--energy emission starts significantly later with respect to the low--energy one, for both long and short bursts. Burst durations are compared in  panel (b) of the same figure. Here, the end of the signal at high energies (\tone) appears to be significantly later than the one measured in the GBM energy band. Both these characteristics were already reported in the \fcat. Our results confirm and strongly support the claim that when high--energy emission is observed in GRBs, this emission is delayed and lasts longer compared to that in the low--energy band.

\renewcommand{\tabcolsep}{2pt}

In panel (c) of Figure~\ref{fig_GBMT05_LATT05} we show the onset time (\tz) of the high--energy emission versus the burst duration (\tn) in the 50--300 keV range. It is worth noting that the \tz of the majority of GRBs (both long and short ones) occurs before the prompt emission measured by the GBM is over. Events that were outside the nominal LAT FoV ($\sim$75\de) at the time of the GBM trigger are marked with thick orange contours. They comprise the majority of GRBs where the onset of the high-energy emission came after the low-energy emission had faded, indicating that most such events are due to observational bias. This effect will be further investigated below. 

\begin{figure}[t!]
\centering
\includegraphics[width=0.6\columnwidth]{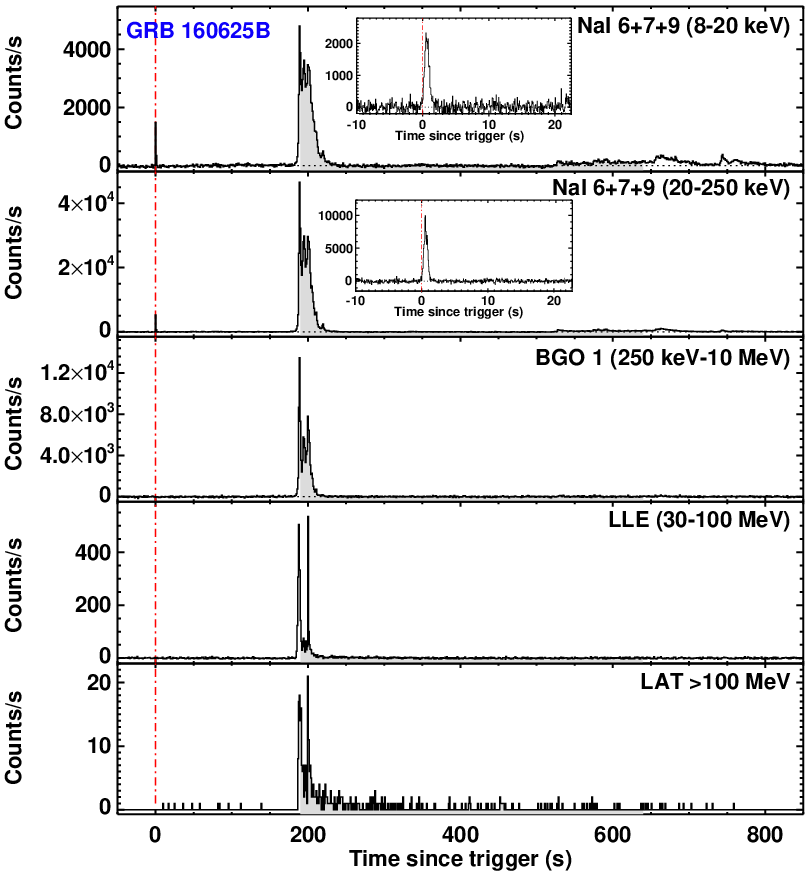}
\caption{
Composite light curve of GRB\,160625B: summed GBM/NaI detectors (first two panels), GBM/BGO (third panel), LLE (fourth panel), and LAT rates above 100 MeV (bottom panel). The grey shaded area indicates the \tn calculated in the 50--300 keV energy range. The dashed red lines marks the GBM trigger time. A zoom around trigger time is shown in two insets for each energy range of the NaI detectors, highlighting the peak detected at low energies which caused the trigger in GBM. This peak does not show any counterpart at higher energies. The association of the low-energy ($<250$ keV) late-time emission (from 500 to 800 s post trigger) with the main GRB emission episode has been cross-checked by the GBM Team (\gcat).}
\label{LC_160625B}
\end{figure}

As shown in panels (a) and (b) of Figure~\ref{fig_GBMT05_LATT05}, there are just a few outliers which have high-energy emission that is not delayed and/or has shorter duration compared to the low-energy band. However, since the procedure to calculate onset times and durations differs between the two energy ranges, we caution that further analysis is needed before strong conclusions can be drawn about individual GRBs. The difference is in most cases less than a few seconds. The most prominent outlier to the right of the line is GRB\,160625B, where the GRB \tf is $\sim 190$\,s, whereas {\tz} is $\sim 25$\,s. However, this burst showed three emission episodes spread over a period of more than $\sim 10$ minutes, as shown in Figure \ref{LC_160625B}. The first one triggered the GBM, a second one three minutes later resulted in a LAT onboard trigger, and then the GBM triggered again 10 minutes after the first trigger. It is thus not surprising that the \tf is much greater than the arrival time of the first LAT photon.

Short GRBs in general have more similar onset times in LAT and GBM. They also exhibit shorter durations in the high-energy range, although they last still significantly (generally more than an order of magnitude) longer than at lower energies. The short GRB\,170127C is the short burst with the longest lasting high-energy duration, more than 2\,ks. 

\begin{figure}[t!]
\centering
\begin{tabular}{cc}
\includegraphics[width=0.45\columnwidth,trim=0 0 0 0,clip=true]{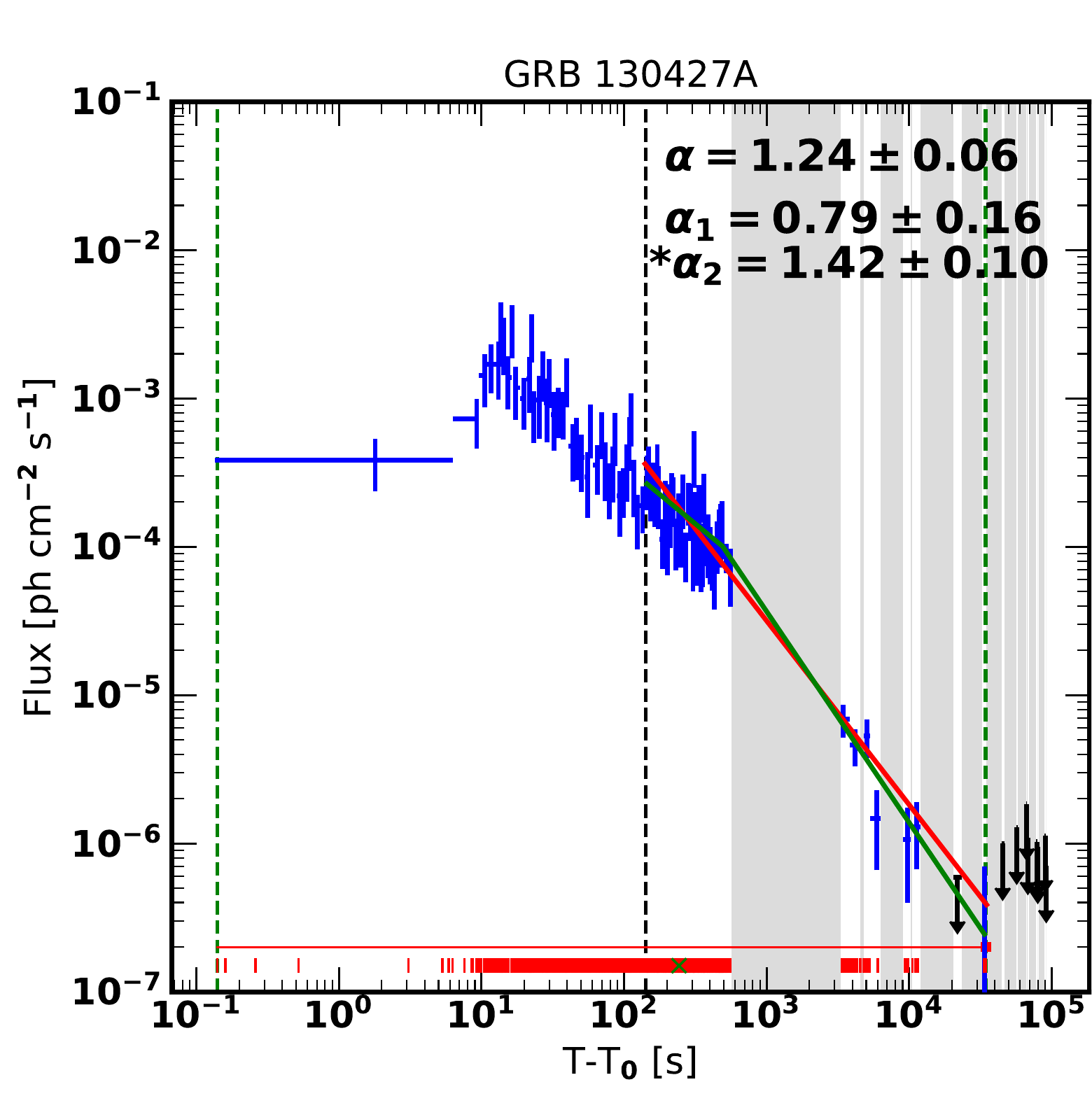} & 
\includegraphics[width=0.45\columnwidth,trim=0 0 0 0,clip=true]{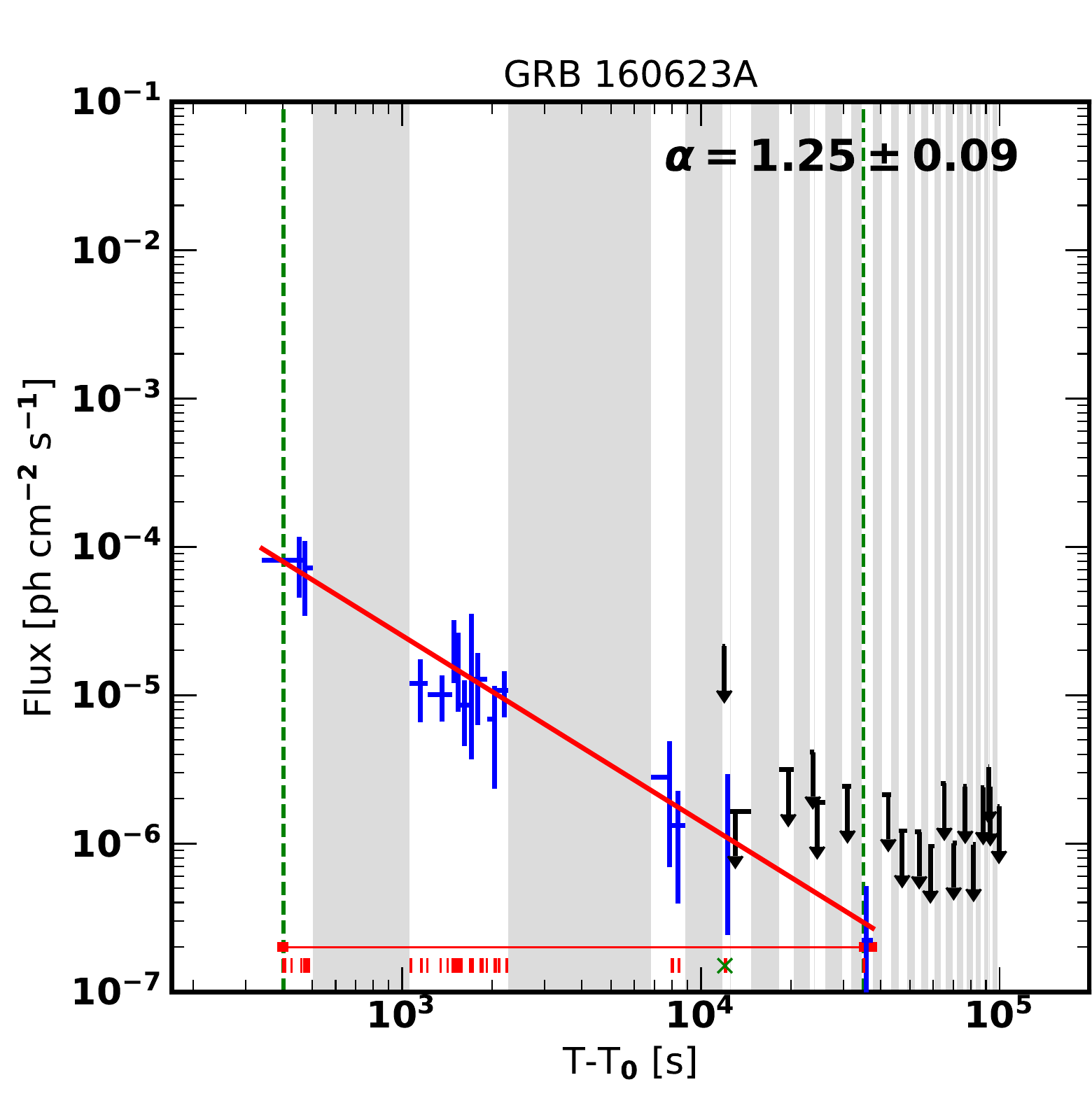}
\end{tabular}
\caption{
Temporal extended emission for the two longest LAT GRBs, GRB\,130427A (left panel), and GRB\,160623A (right panel).
For the bright GRB\,130427A we display the results of the fit for a simple power law model with temporal index $\alpha$ (red line) and with a broken power law with indexes  $\alpha_{\rm 1}$ and $\alpha_{\rm 2}$ (green line). For GRB\,160623A, only the result of the fit with a simple power law is shown.}
\label{fig_Longest_T100}
\end{figure}

In our sample, 16 GRBs have high-energy emission lasting over 5\,ks, and 4 have durations over 10\,ks, namely GRB\,160623A ($\sim 35$ ks), GRB\,130427A ($\sim 34$ ks), GRB\,140810A ($\sim 18$ ks), and GRB\,160503A ($\sim 18$ ks). Figure~\ref{fig_Longest_T100} shows the temporal extended emission
for the two longest bursts, GRB\,130427A in the left panel, and GRB\,160623A in the right panel. In each panel, we also indicate the fit results to the temporal decay, giving the corresponding model parameters in the top right corner. This will be further discussed in Sect.~\ref{sec:res:temporal_decay}.

\begin{figure}[ht!]
\centering\begin{tabular}{cc}
\includegraphics[width=0.45\columnwidth,trim=5 5 40 40,clip=true]{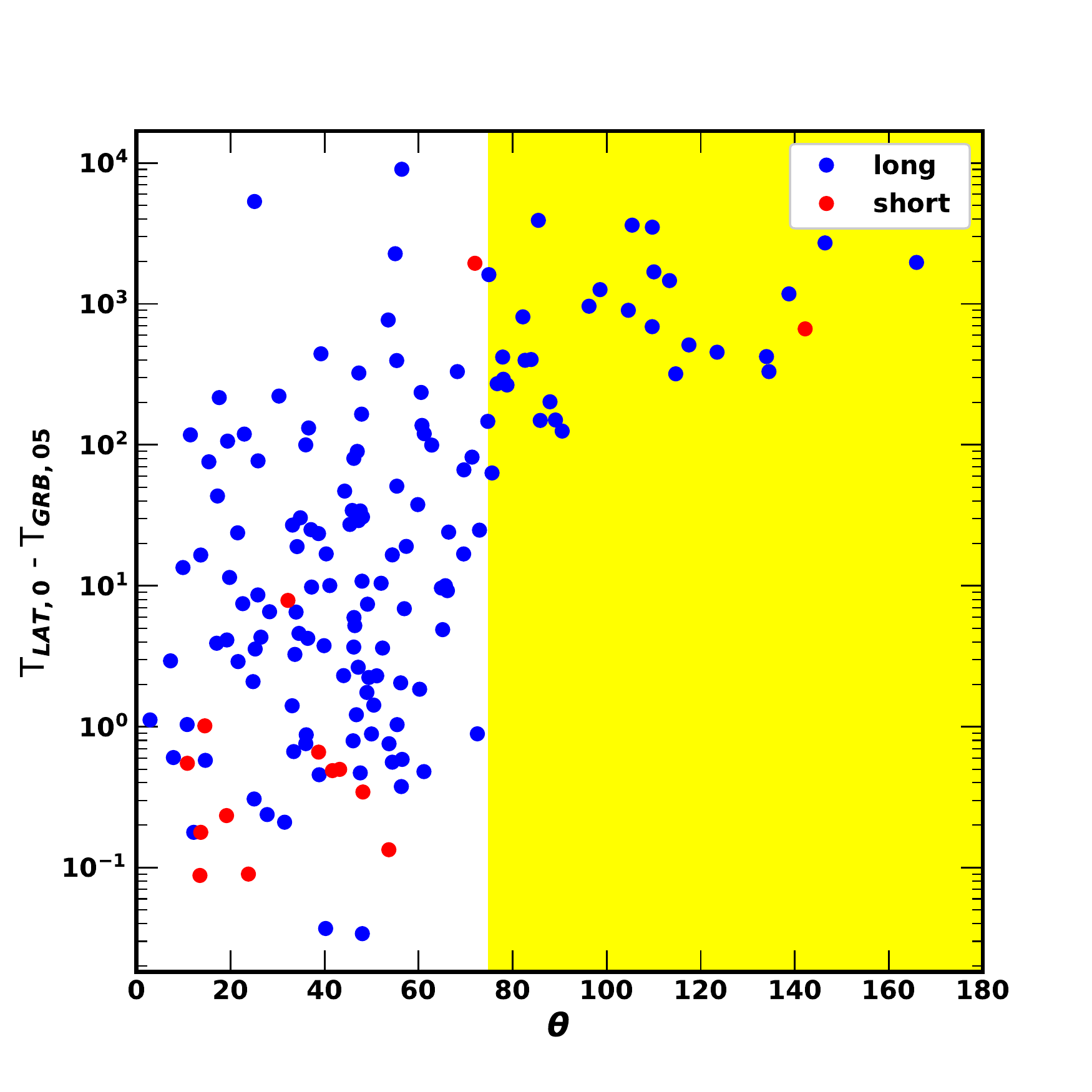} & 
\includegraphics[width=0.49\columnwidth,trim=5 5 0 40,clip=true]{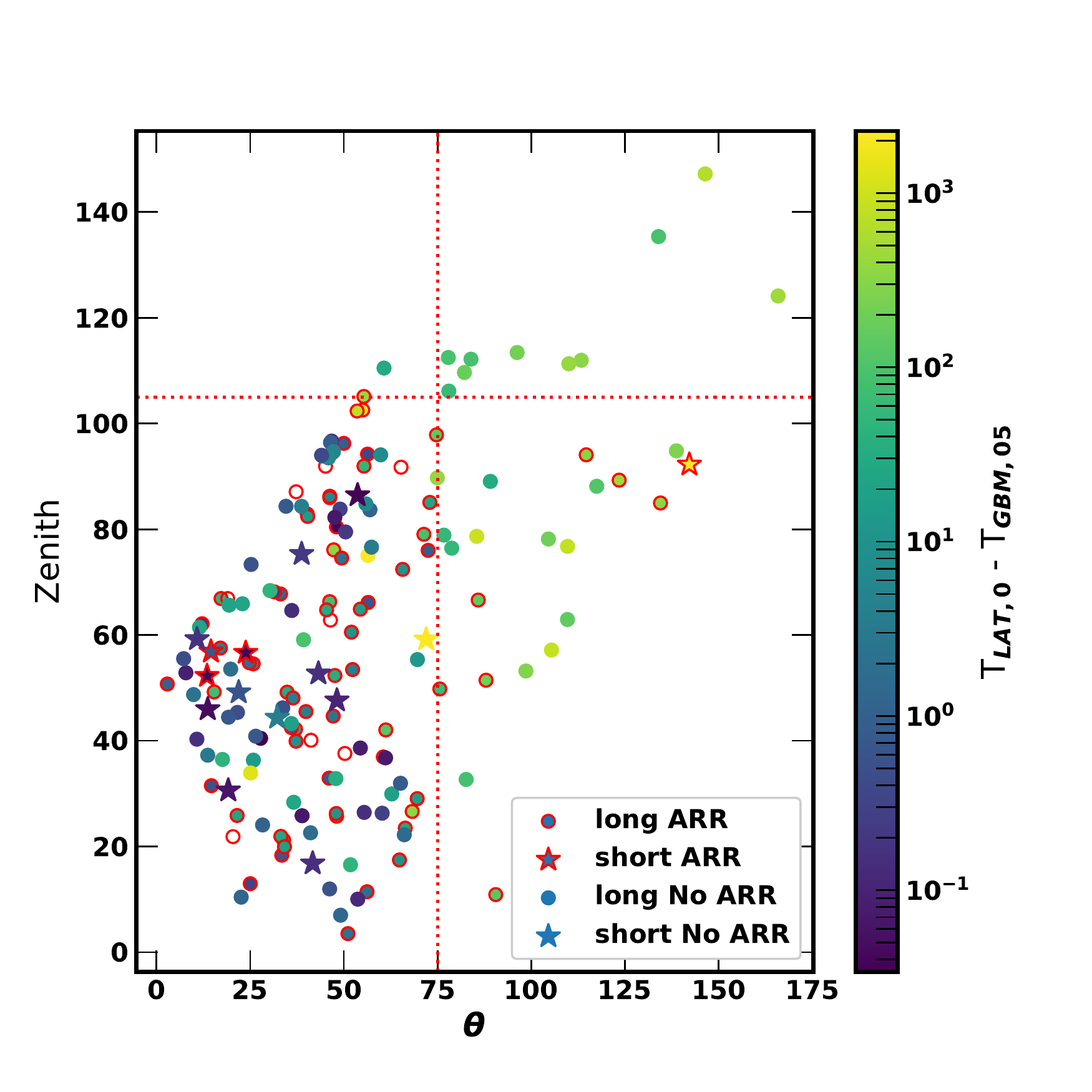}
\end{tabular}
\caption{Left panel: Difference of the \tz and the \tf with respect to the incident angle of the GRB at the time of the trigger. The yellow region highlights the GRBs that were outside the LAT FoV ($\theta \sim 75^{\circ}$) at the time of the GBM trigger. 
Right panel: Zenith angle ($\zeta$) vs incident angle ($\theta$). The symbols are colored as a function of the difference of the \tz and the \tf (left plot, y-axis). Circles mark long GRBs, while stars mark short GRBs. Bursts which triggered an ARR are marked with red contours. The vertical dashed line marks the LAT FoV, while the horizontal line marks the zenith angle cut used in the analysis ($\zeta_{\rm MAX} = 105^{\circ}$). In general, LAT-detected events at $\zeta > 105^{\circ}$ and $\theta > 75^{\circ}$ are seen at very late times or thanks to ARRs.}
\label{fig_THETA}
\end{figure}

As already mentioned, a possible bias in the estimation of the onset time in the LAT is related to the initial position of the GRB at the time of the GBM trigger. For a GRB outside the nominal LAT FoV at trigger time, the first significant detection would happen only when the GRB re-enters the FoV. We further illustrate this effect in Figure~\ref{fig_THETA}. In the left panel the delay of the LAT onset time with respect to the GBM one is plotted as a function of the incident angle of the GRB, while in the right panel we plot the zenith angle as a function of the incident angle. It is evident that all GRBs that were outside the LAT FoV at trigger time have a large delay (\gta100 s) with respect to the GBM trigger, which corresponds to the time needed for the GRB to re-enter the LAT FoV. On the other hand, we also measure significant delays for GRBs that were in the FoV at the time of the GBM trigger, supporting the intrinsic nature of the delay of the high-energy component. 

\begin{figure}[t!]
\centering
\begin{tabular}{cc}
\includegraphics[width=0.62\columnwidth,trim=90 80 90 550,clip=true]{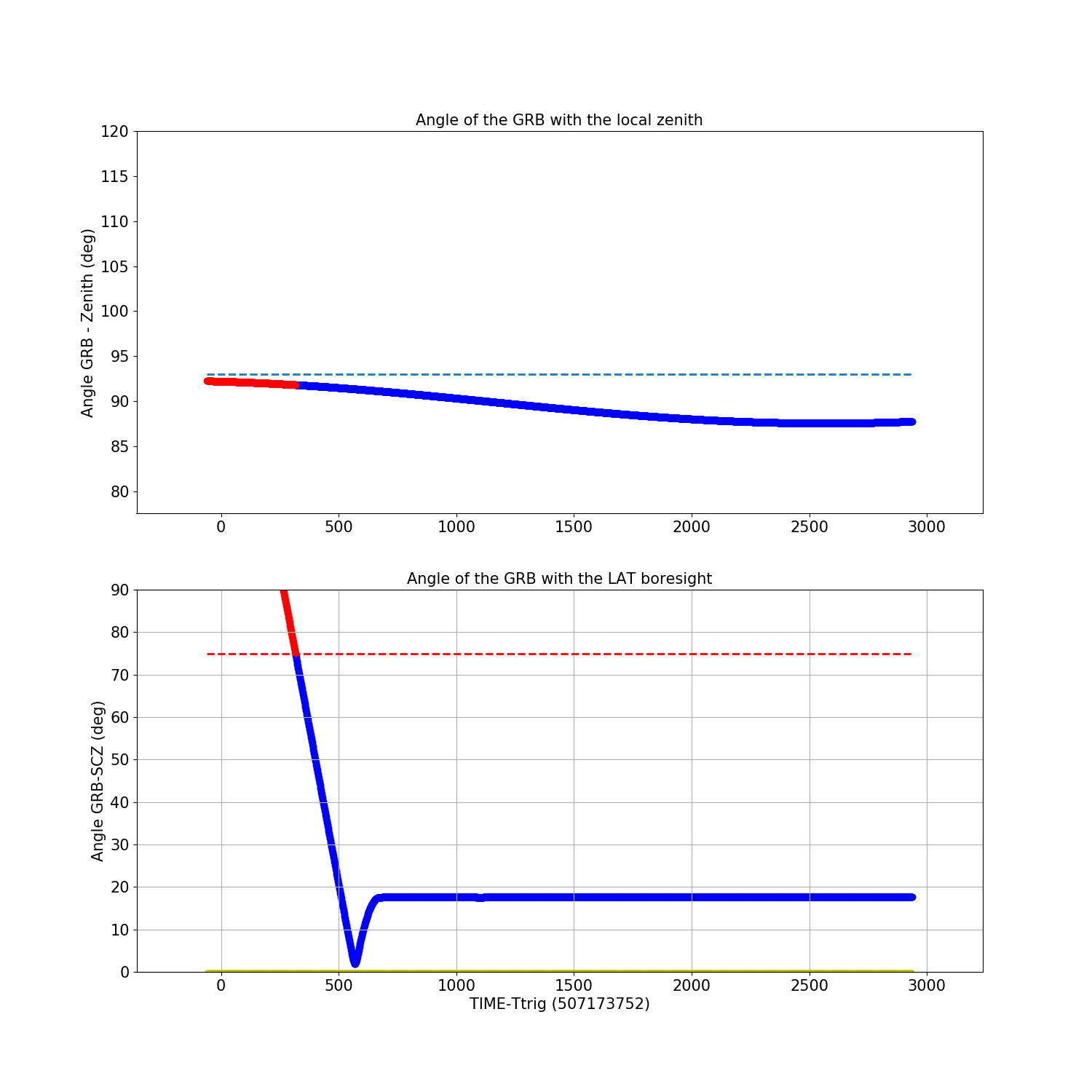} &
\includegraphics[width=0.32\columnwidth,trim=0 0 0 0,clip=true]{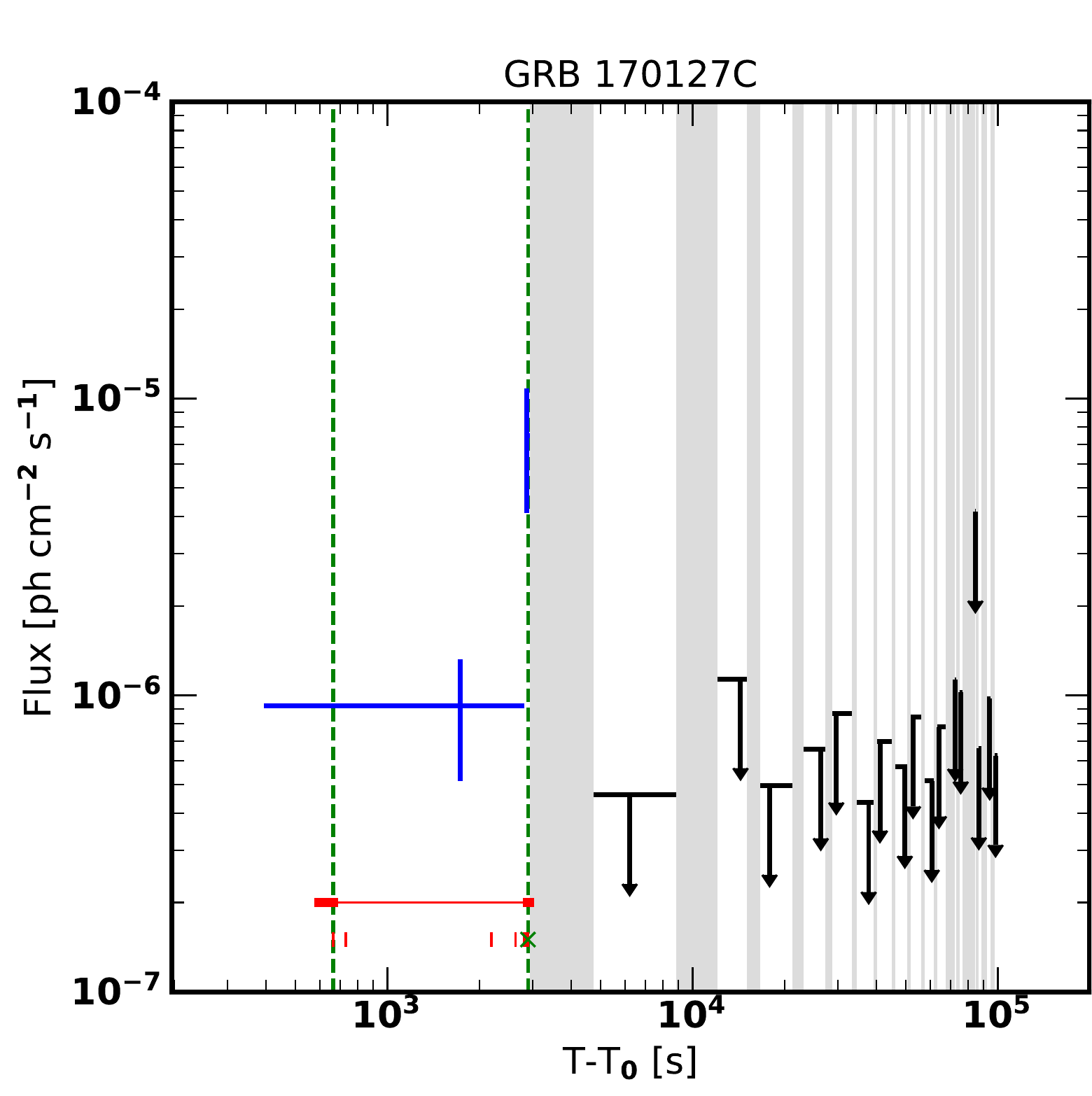}
\end{tabular}
\caption{Left panel: Off axis angle for GRB\,170127C as a function of time since the GBM trigger. The horizontal red line corresponds at $\theta = 75^{\circ}$. Right panel: Temporal extended emission for GRB\,170127C. The blue point is the photon flux measurement with a significance of TS\,$> 10$, while, for lower value of TS, upper bounds are displayed as black arrows. The horizontal red line indicates the estimated duration of the burst, and the thick parts of it correspond to the uncertainty on the \tz and \toz parameters. The red vertical markers indicate the arrival time of each photon with probability $> 90$\% to be associated with the GRB, with the green cross being the event with maximum energy.}
\label{fig_170127067}
\end{figure}

In the right panel of Figure~\ref{fig_THETA}, we highlight GRBs which resulted in an ARR, marking each symbol with a red contour. Most of the bursts for whom an ARR was issued were in the LAT \fov at the time of the trigger, whereas in 7 cases the GRBs were outside the LAT \fov and the detection happened only at later times. To better illustrate this effect, we display the case of the short GRB~170127C in Figure~\ref{fig_170127067}. This burst was at a zenith angle of $\zeta\sim 94^{\circ}$, and at an off-axis angle $\theta\sim 142^{\circ}$ when it triggered the GBM (it is the outlier sGRB seen to the far right in both panels of Figure~\ref{fig_THETA}). The trigger resulted in an ARR and the spacecraft slewed to move the location of the burst close to the center of the field of view (at $\sim 17^{\circ}$). This can be seen in the left panel of Figure~\ref{fig_170127067}, where the blue dots show the evolution of $\theta$ as a function of time after the trigger. In the right panel, we show the photon flux light curve resulting from the time dependent analysis. The first detection is at $\sim 400$\,s, well beyond the end of the GBM signal (\tnf\,$=0.13$\,s).

\FloatBarrier
\begin{figure}[t!]
\centering
\begin{tabular}{cc}
\includegraphics[width=0.45\columnwidth,trim=5 5 40 40,clip=true]{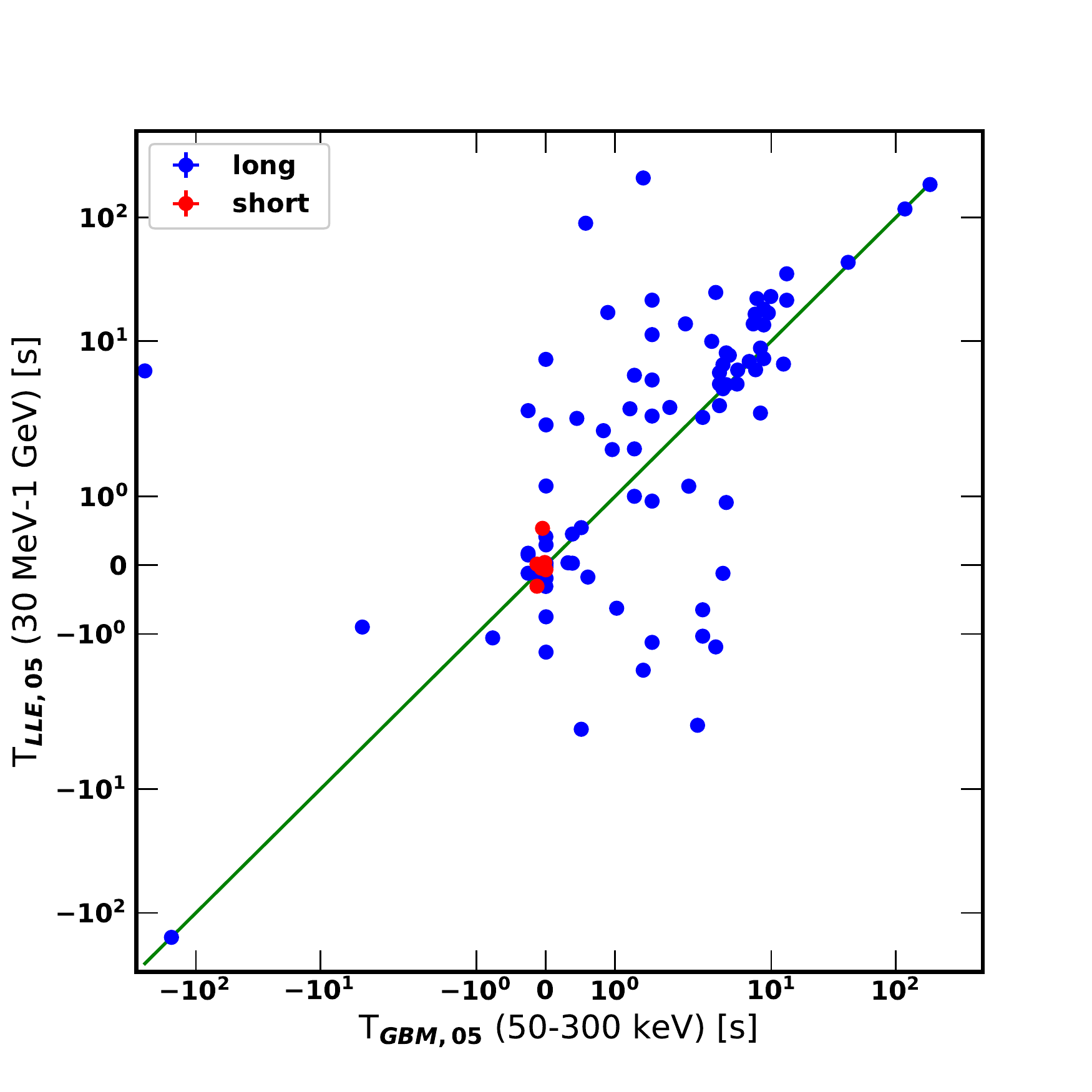} & 
\includegraphics[width=0.45\columnwidth,trim=5 5 40 40,clip=true]{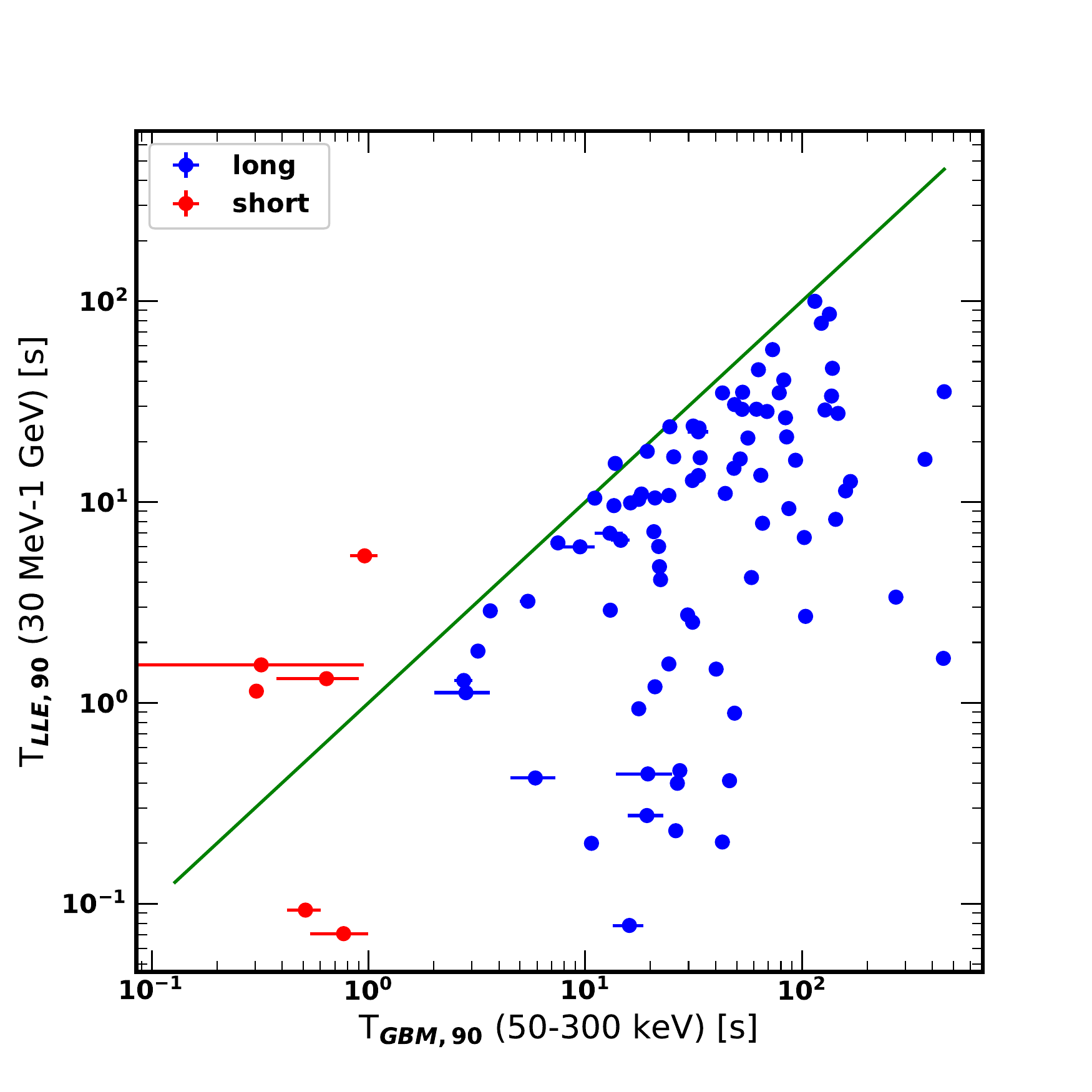}
\end{tabular}
\caption{Onset times (\tllf; left panel) and durations (\tlln; right panel) calculated using LLE data in the 30 MeV--1 GeV energy range vs. the corresponding quantities (\tf and \tn) calculated using GBM data in the 50--300 keV energy range. The solid green line denotes where value are equal. Blue and red circles represent long and short GRBs, respectively.}
\label{fig_GBMT05_LLET05}
\end{figure}
\subsection{LLE onset and duration}
If we restrict our considerations to the LLE analysis, where the bulk of the emission is in the energy range from 30 MeV to 1 GeV, we see that the left panel of Figure~\ref{fig_GBMT05_LLET05} shows how the onset times are relatively similar to the onset times as measured by the GBM. Here, two GRBs are not shown: GRB\,120624B and GRB\,150513A. Both GRBs triggered \Swift\ before they triggered GBM, 257\,s \citep{2012GCN.13381....1B} and 157\,s \citep{2015GCN.17810....1K} before the GBM trigger time, respectively. 
As a result, since all our calculations are referred to GBM trigger times, \tllf is negative and omitted from the figure (see Table \ref{tab_durations}). 

In contrast to the emission above 100 MeV, the right panel of Figure \ref{fig_GBMT05_LLET05} indicates that the duration of the signal in LLE is systematically shorter than the duration of the signal in the GBM, as was seen also in the \fcat. If we assume that the LLE emission is dominated by the same emission episodes as that in the GBM, we can infer that the pulses which make up the time profile of the prompt emission are systematically shorter in the LLE range than at lower energies. This behavior has previously been reported by \citet{Norris:96,norris:2002} using BATSE data, as well as for several LAT-observed GRBs \citep[e.g.,][]{Axelsson+12, Bissaldi2017, 2018ApJ...864..163V}.
\subsection{Comparison to the GBM population}
Since the majority of our triggers come from the GBM, and the GBM has observed nearly all GRBs in our sample, we examine how the LAT--detected bursts are drawn from the general GBM population covering the same 10-year time period. For this comparison, we extracted the peak photon flux, as measured on a 1024 ms timescale, and energy fluence measured by the GBM in the 10--1000 keV energy range from the \gcat. Here, the GBM fluence is derived from the parameters of the best--fit spectral model applied to GBM data over a time interval where the signal-to-noise ($S/N$) ratio exceeds a predefined value \citep[$S/N > 3.5$; see][for more details]{Gruber2014_211}. This requirement ensures that there are enough counts to perform a spectral fit, but as a result the time interval does not always coincide with \tn.  Note that eight GRBs, two triggered by \Swift and six by the IPN, were not detected by the GBM and are omitted from this comparison and from the following figures.

\begin{figure}[t!]
\centering
\begin{tabular}{ccc}
\includegraphics[width=0.32\textwidth]{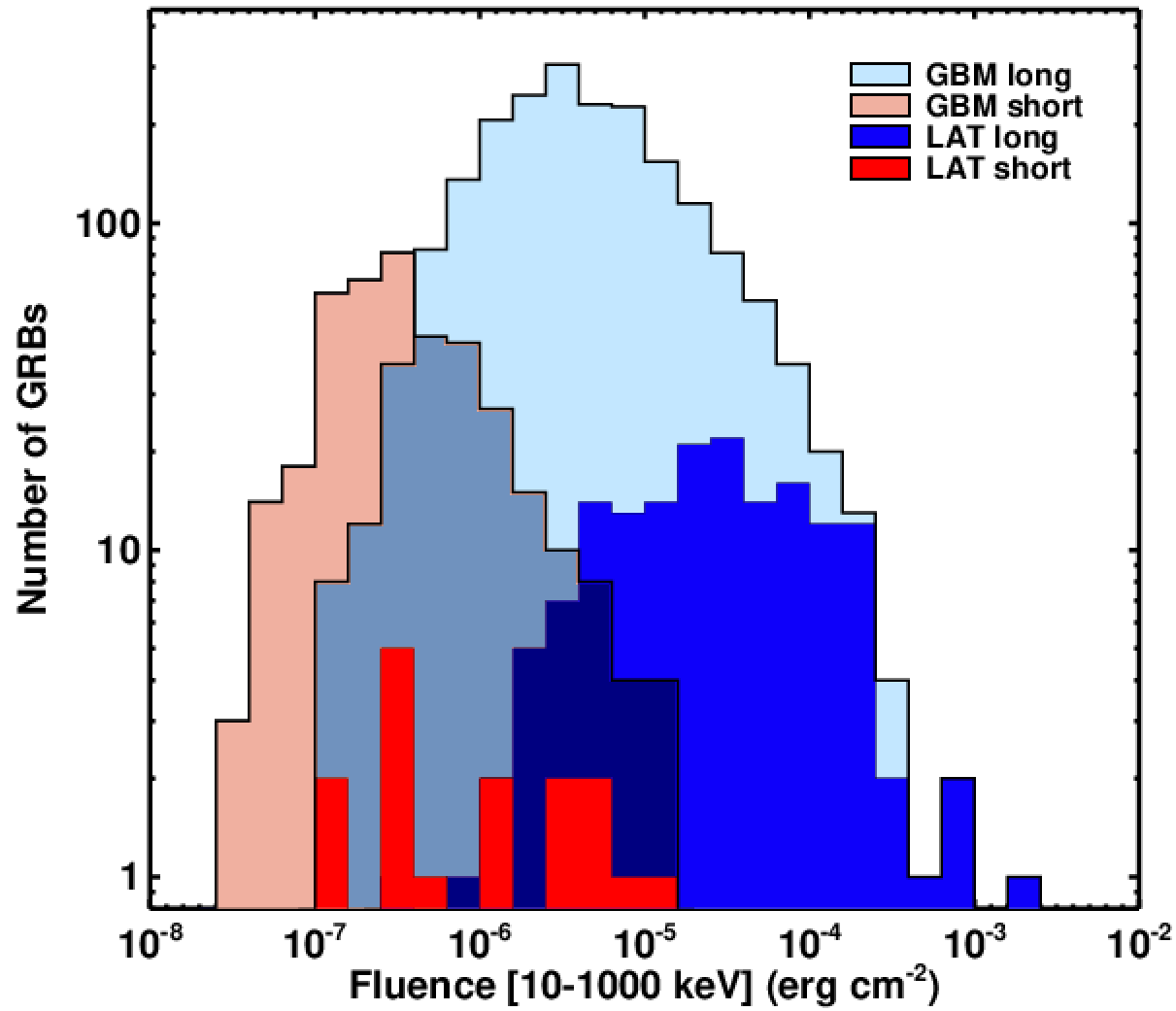} & 
\includegraphics[width=0.312\textwidth]{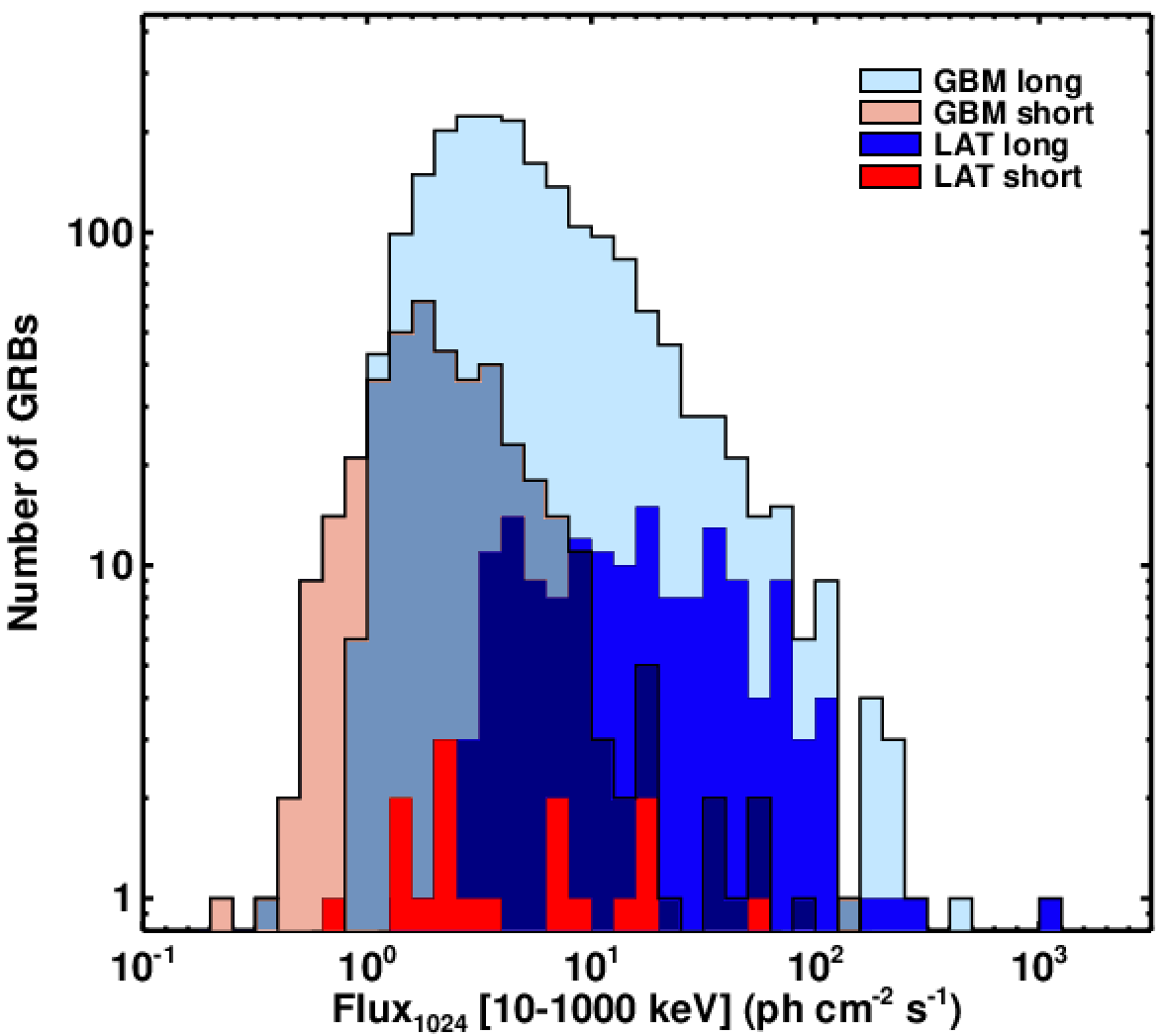} &
\includegraphics[width=0.322\textwidth]{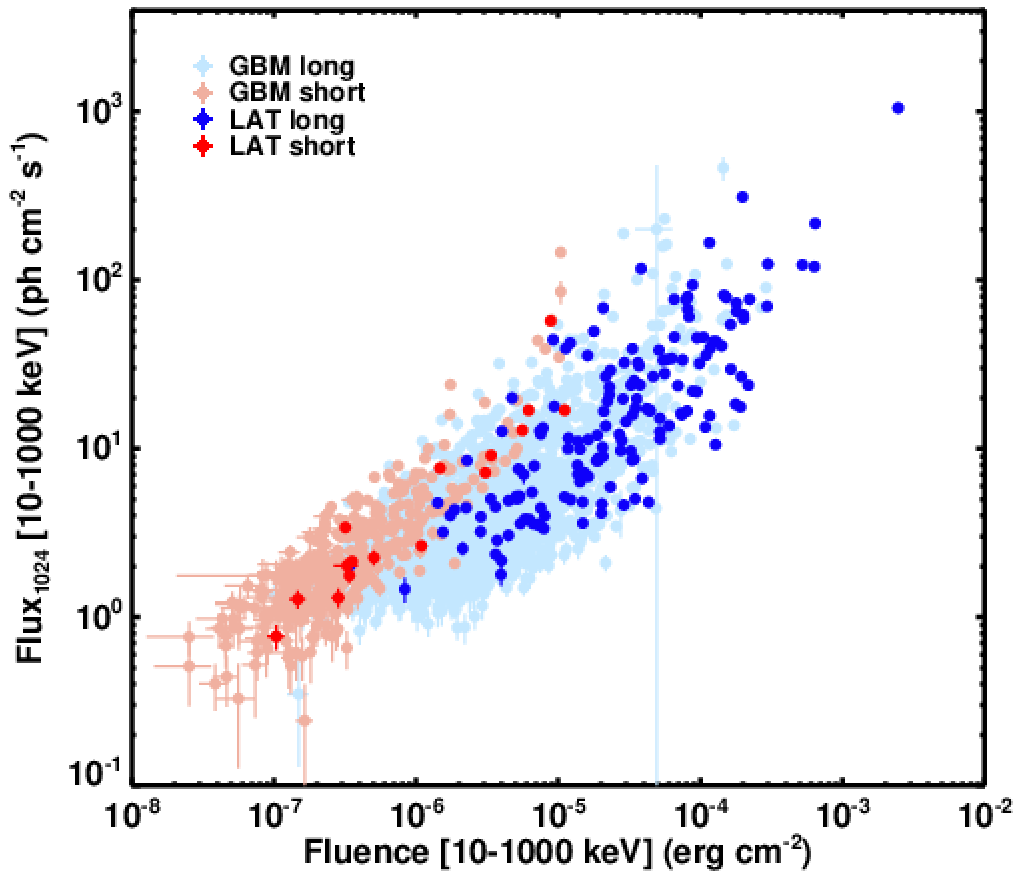}
\end{tabular}
\caption{Left panel: The distribution of energy fluence calculated in the 10--1000\,keV energy range for 178 bursts detected by the LAT compared to the entire sample of 2357 GRBs detected by GBM over the same time period. Middle panel: The distribution of peak photon flux in the 10--1000\,keV energy range for the same sample of LAT and GBM detected populations. Right panel: The GBM peak photon flux (10--1000 keV), as measured on a 1024\,ms timescale, versus the energy fluence (10--1000 keV) derived using a spectral model fit to a single spectrum over the entire duration of the burst.}
\label{fig_Fluence_Catalogs}
\end{figure}

Figure \ref{fig_Fluence_Catalogs} shows the distribution of the energy fluence (left panel) and of the peak photon flux (middle panel) for 178 bursts detected by the LAT compared to the entire sample of 2357 GRBs detected by GBM over the same time period. Here we have also made a distinction between short and long bursts for both the LAT (16 sGRBs and 162 lGRBs) and GBM (400 sGRBs and 1957 lGRBs) populations, showing a bifurcation in the range of flux and fluence values covered by these two classes of bursts. The right panel shows the peak photon flux plotted against the energy fluence for the LAT bursts compared again to the entire GBM burst catalog. Again, we separate short and long bursts for both the LAT and GBM populations. 

These comparisons show that although the majority of the LAT--detected GRBs come from the GBM--detected bursts with the highest peak flux and fluence, they cover a large range. LAT--detected short (long) bursts are present with a fluence $>10^{\,-7}$ erg/cm$^{\,2}$ ($>8\times 10^{\,-7}$ erg/cm$^{\,2}$) and with a peak flux $>0.8$ ph/cm$^{\,2}$/s ($>1.5$ ph/cm$^{\,2}$/s). The LAT--detected long GRBs cover more than two orders of magnitude in both distributions, and the  prominence of bright GRBs is even less pronounced in the short GRB sample. The spread is also evident from the right panel in Figure~\ref{fig_Fluence_Catalogs}, where the cluster of LAT events is only slightly shifted with respect to the GBM one. The burst with the highest fluence (and flux) is GRB\,130427A. It is worth noting that Figure~\ref{fig_Fluence_Catalogs} does not include any selection on the $\theta$ angle.

\FloatBarrier

\subsection{Flux, fluences and photon indexes from the time integrated analysis}

The results of the likelihood analysis are summarized in Table 4. For each time window, we report the number of detected and predicted LAT events in the ROI, the resulting test statistic, the spectral index obtained using a power-law fit, and the LAT flux and fluence calculated in the 100 MeV--100 GeV energy range. For \nredshift GRBs with known redshift we also report the total radiated energy ($E_{\rm iso}$).

\begin{figure}[t!]
\centering
\begin{tabular}{cc}
\includegraphics[width=0.45\columnwidth,trim=0 5 40 40,clip=true]{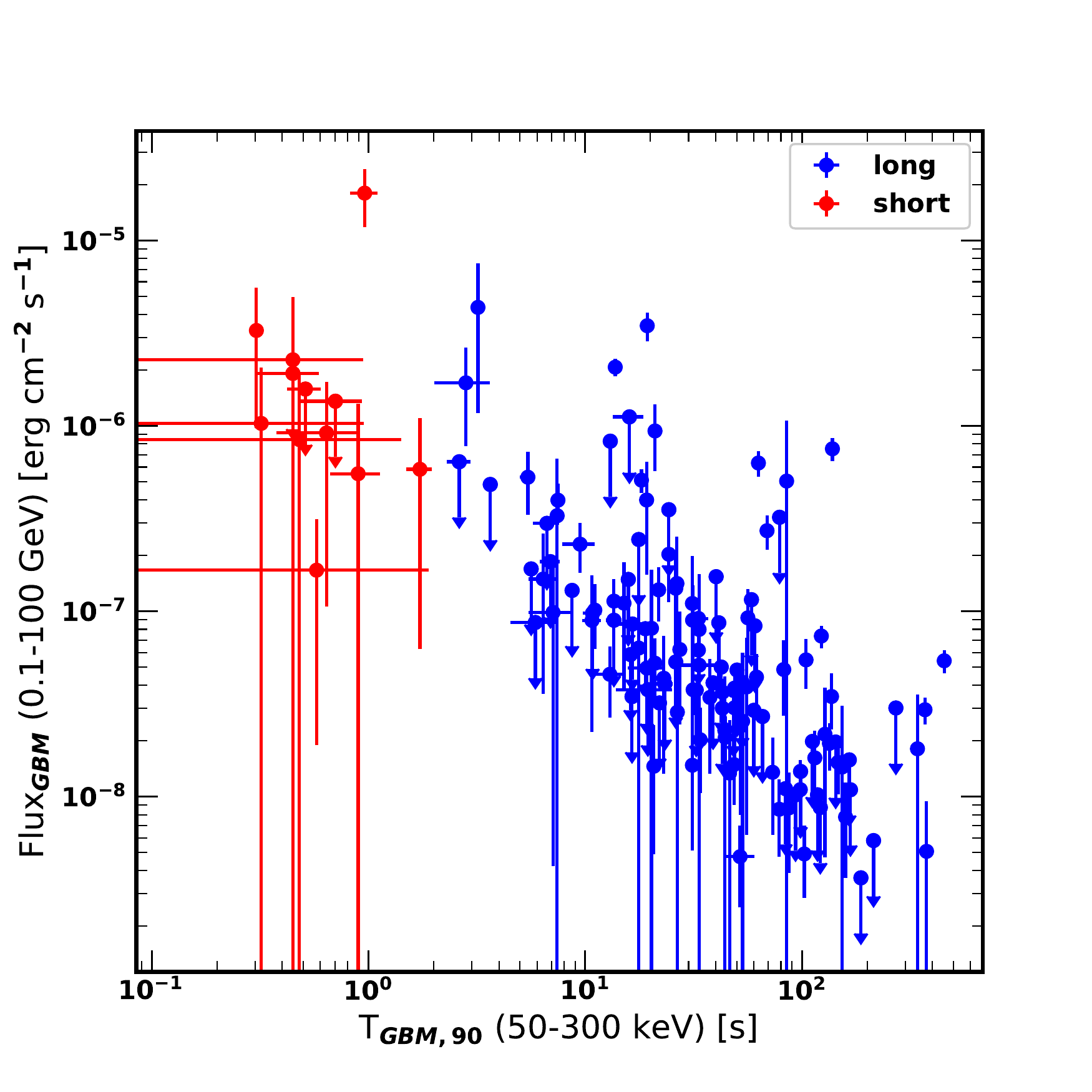} & 
\includegraphics[width=0.45\columnwidth,trim=0 5 40 40,clip=true]{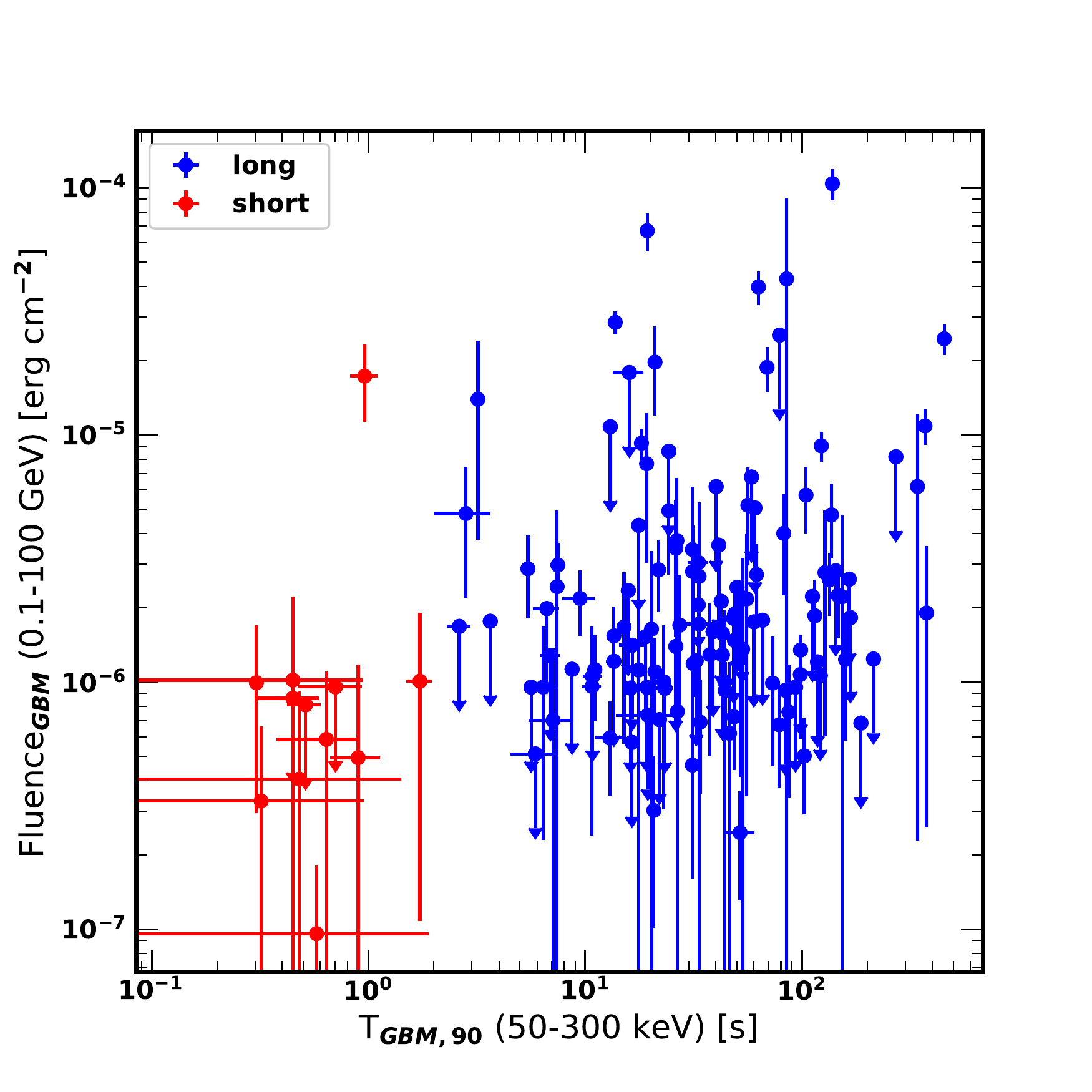} \\
\includegraphics[width=0.45\columnwidth,trim=0 5 40 40,clip=true]{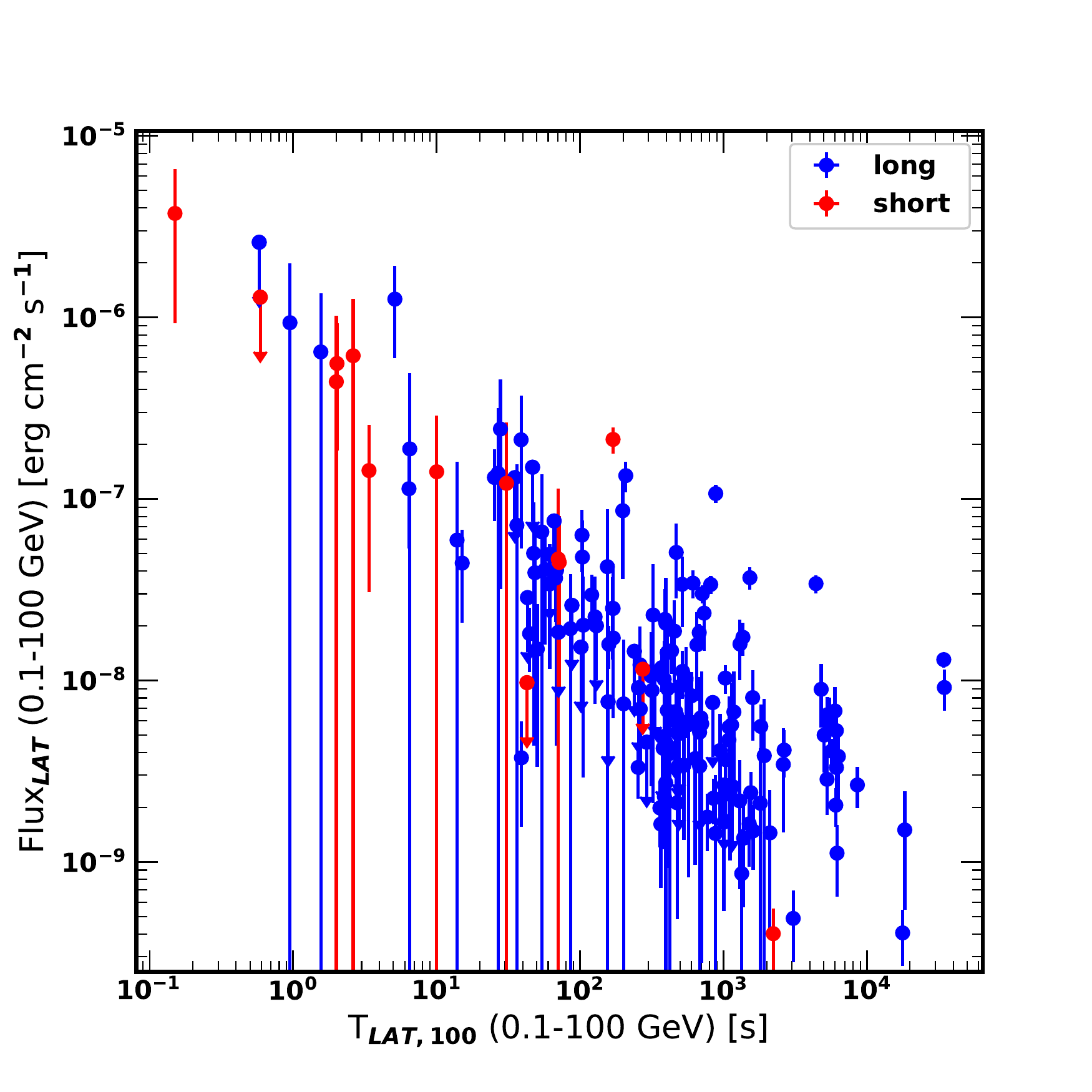} & 
\includegraphics[width=0.45\columnwidth,trim=0 5 40 40,clip=true]{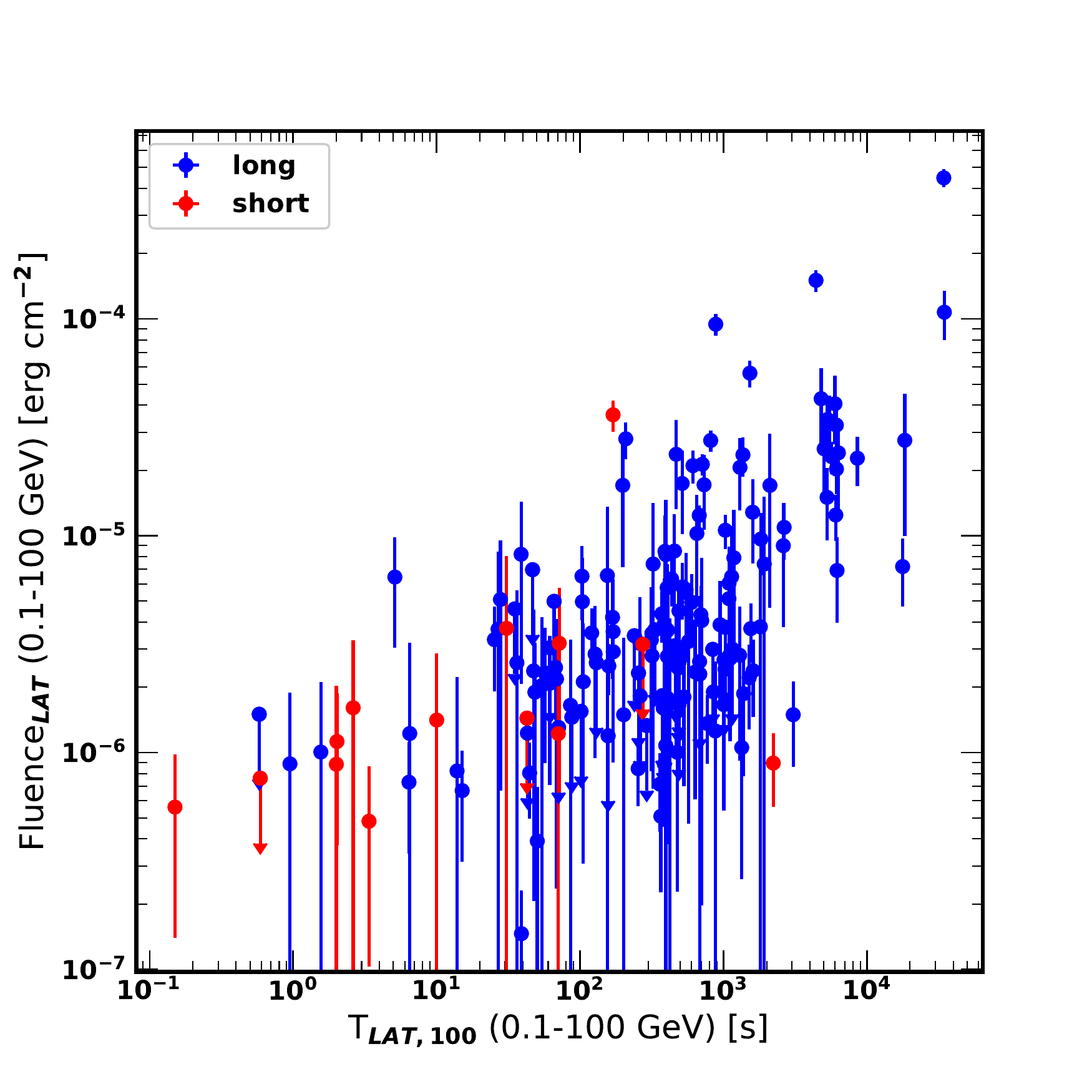}
\end{tabular}
\caption{Fluxes (left panels) and fluences (right panels) calculated in the 100 MeV--100 GeV energy range vs. GRB durations. Upper panels show fluxes and fluences evaluated in the ``GBM" time window vs. durations calculated in the 50--300 keV energy range (\tn). Bottom panels show fluxes and fluences evaluated in the ``LAT" time window vs. durations calculated in the 100 MeV--100 GeV energy range (\toz). Blue and red circles represent long and short GRBs, respectively.}
\label{fig_T90_FLUX_FLUENCE}
\end{figure}

Figure~\ref{fig_T90_FLUX_FLUENCE} shows the distributions of fluxes (left panels) and fluences (right panels) as a function of the measured duration of the signal in the ``GBM'' (top row) and ``LAT'' (bottom row) time windows. LAT fluxes decrease with increasing burst duration in both time windows, as expected. In the ``GBM'' time window, the LAT fluence seems to be clustered around a value of $10^{-6}$\,erg\,cm$^{-2}$ for the majority of lGRBs (regardless of duration), while sGRBs show slightly lower values. Both groups have bursts which are very much brighter than the average. At late times, there is instead a tendency for the fluence to increase with duration. The same conclusion can be drawn from the fluence values in the "LAT" time window, where most of the values are distributed around $\sim 5 \times 10^{-6}$\,erg\,cm$^{-2}$ and there is a less evident spread towards higher values.  

Comparing our results to figure~11 in the \fcat, we find that the four ``hyperfluent'' GRBs are no longer outliers. Instead, they are part of a continuous distribution. The range in both flux and fluence has also increased dramatically as compared to the sample in the \fcat.

\begin{figure}[b!]
\centering
\begin{tabular}{cc}
\includegraphics[width=0.45\columnwidth,trim=5 5 40 40,clip=true]{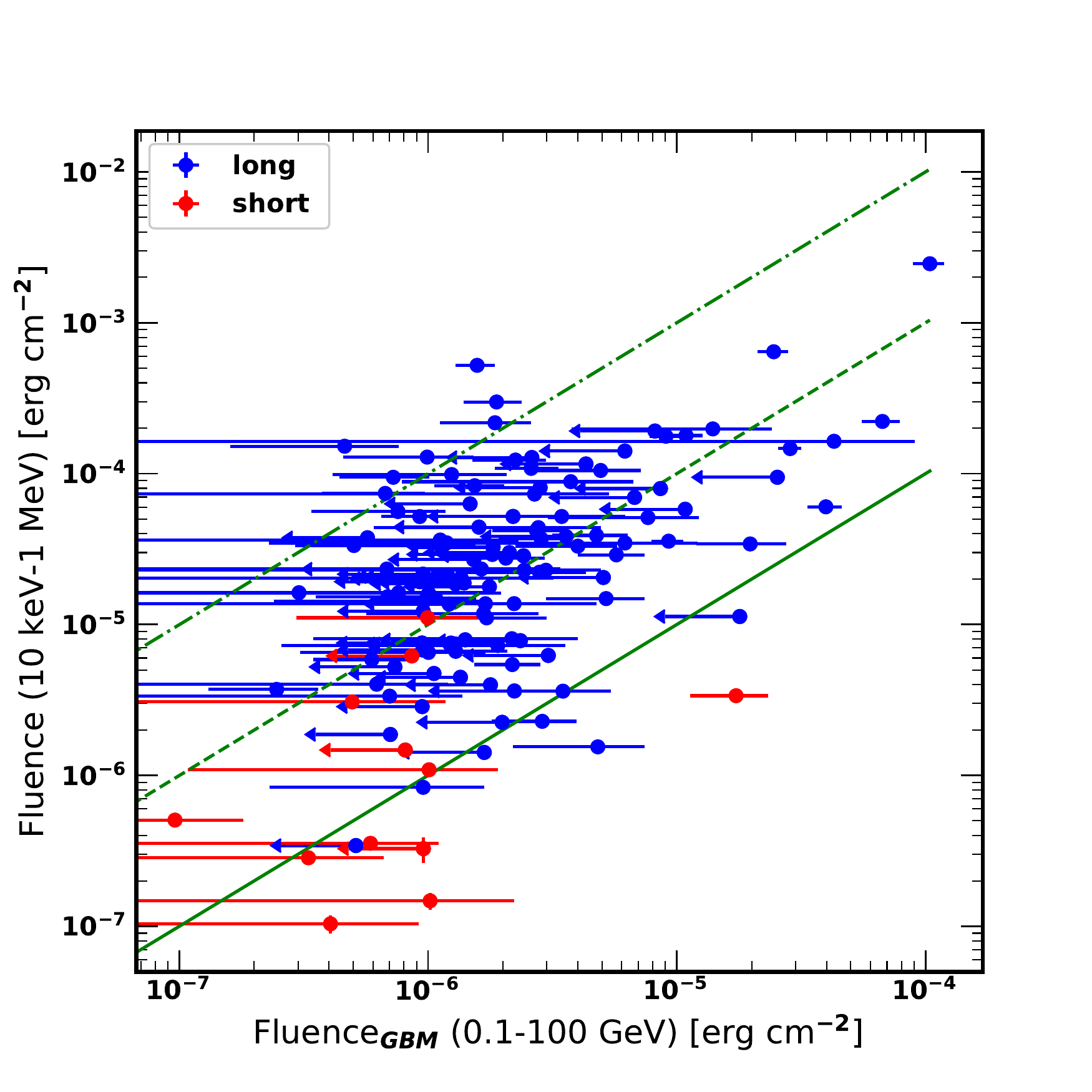} & 
\includegraphics[width=0.45\columnwidth,trim=5 5 40 40,clip=true]{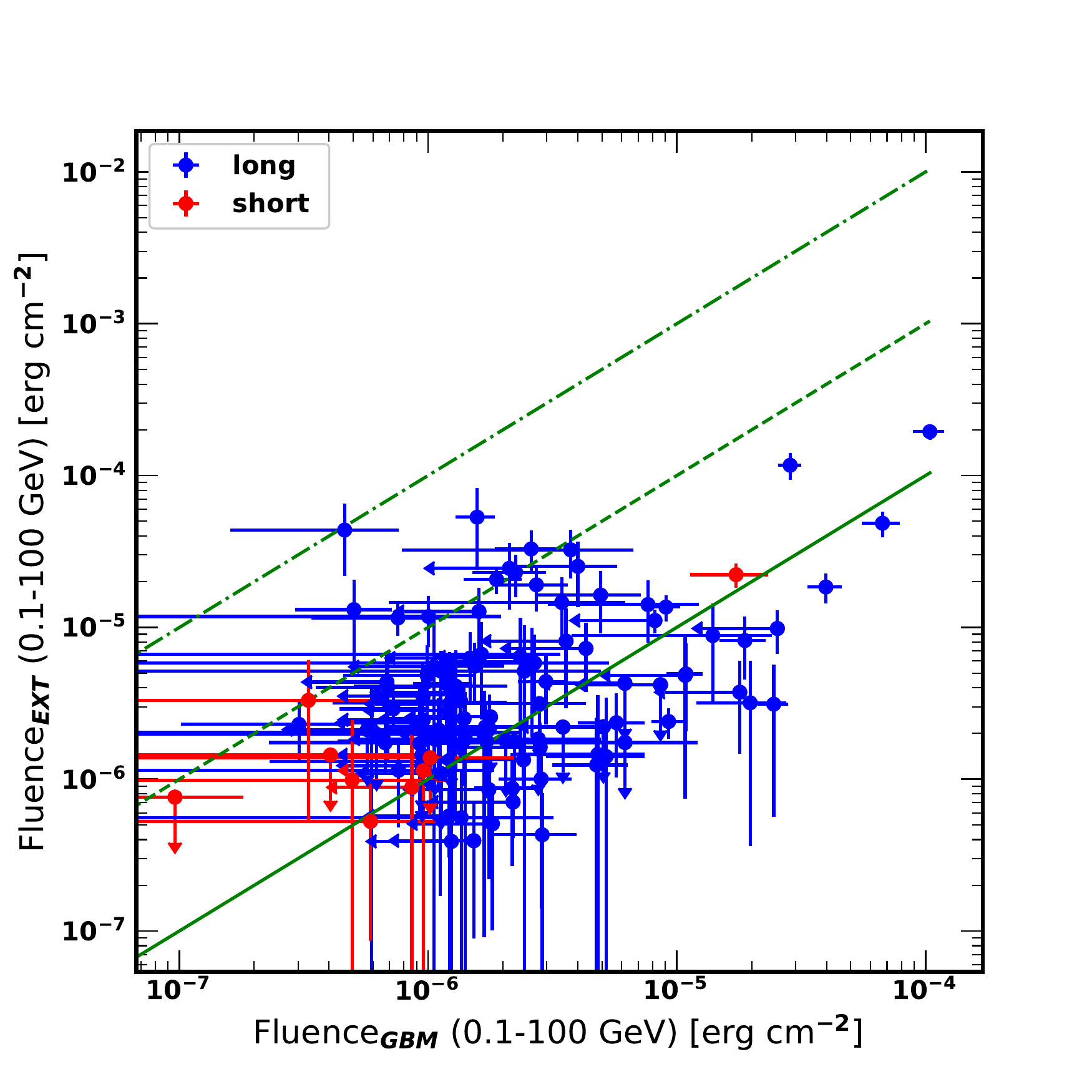}
\end{tabular}
\caption{Left panel: Fluences calculated in the 10--1000 keV energy range vs. fluences calculated in the 100 MeV--100 GeV energy range. All values are estimated in the ``GBM" time window.
Right panel: Fluences calculated in the 100 MeV--100 GeV energy range evaluated in the ``EXT" time window vs. the same quantities evaluated in the ``GBM" time window. The solid green lines denote where values are equal. The dashed and dashed-dotted green lines are shifted by factors of 10 and 100, respectively.}
\label{fig_LATFLUENCE_GBMFLUENCE}
\end{figure}

In Figure~\ref{fig_LATFLUENCE_GBMFLUENCE} we then compare the LAT fluence calculated in the 100\,MeV--100\,GeV energy range during the ``GBM'' time window with the GBM fluence calculated between 10 keV and 1 MeV (left panel) and with the LAT fluence calculated in the same energy range during the ``EXT'' time window (right panel).
In the left panel, it can be noted that the ``GBM'' time window is dominated by the low-energy emission, with the 100\,MeV--100\,GeV energy range contributing only a small fraction of the emission for the majority of long GRBs. Indeed, most events are clustered to the left of the solid and dashed lines, which indicate equality and a factor of 10 less, respectively. For short GRBs this difference seems less pronounced, and several lie close to the solid line of equality. Comparing the ``GBM'' and ``EXT'' time windows in the right panel, the points are instead much closer to the line of equality, suggesting that the high-energy emission in the two time windows is comparable. As in Figure~\ref{fig_T90_FLUX_FLUENCE}, the four ``hyperfluent'' GRBs of the \fcat (GRBs 080916C, 090510,
090902B, and 090926A) are no longer outliers.

\begin{figure}[ht!]
\centering
\begin{tabular}{cc}
\includegraphics[width=0.45\columnwidth,trim=5 5 40 40,clip=true]{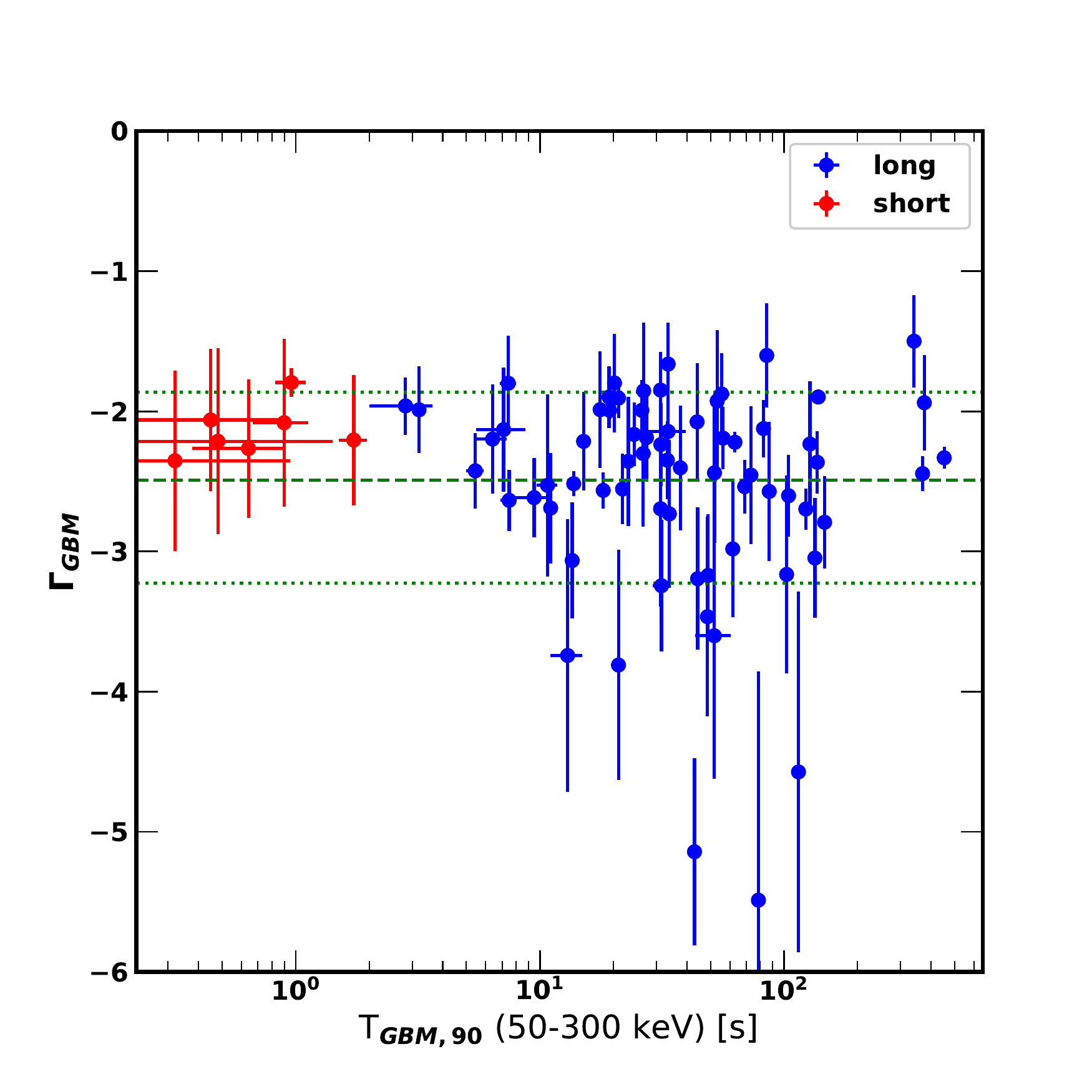} &
\includegraphics[width=0.45\columnwidth,trim=5 5 40 40,clip=true]{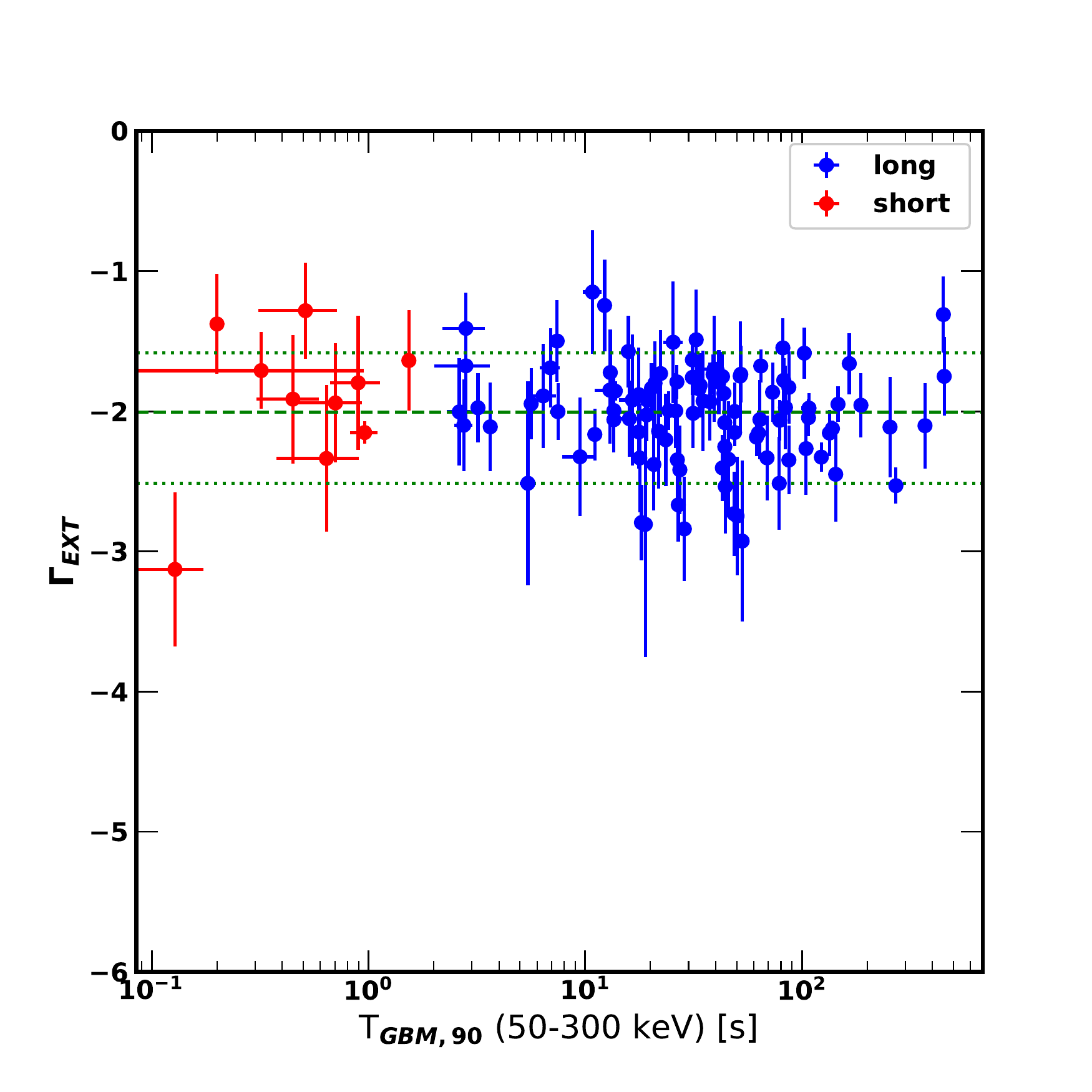} \\
\includegraphics[width=0.45\columnwidth,trim=5 5 40 40,clip=true]{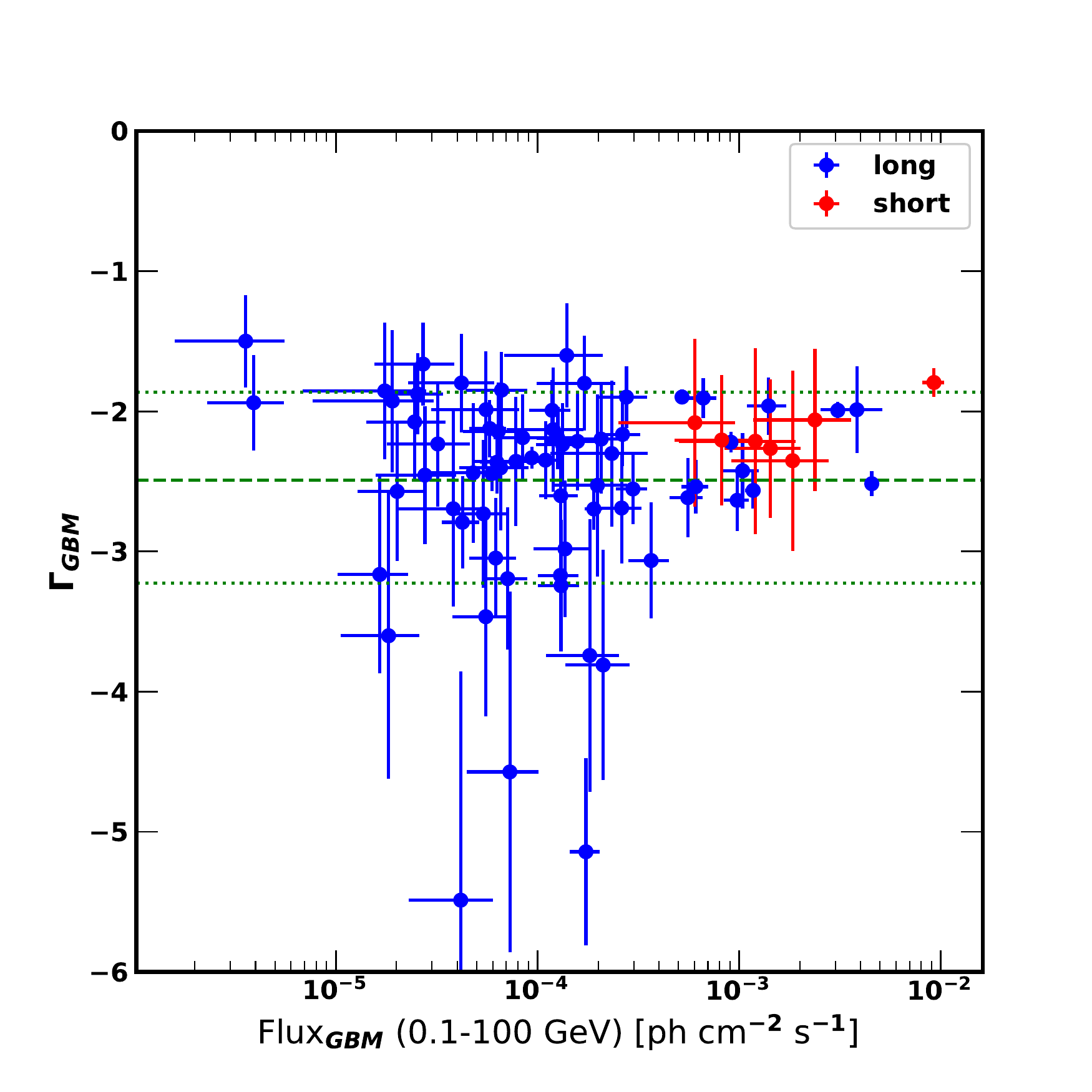} & 
\includegraphics[width=0.45\columnwidth,trim=5 5 40 40,clip=true]{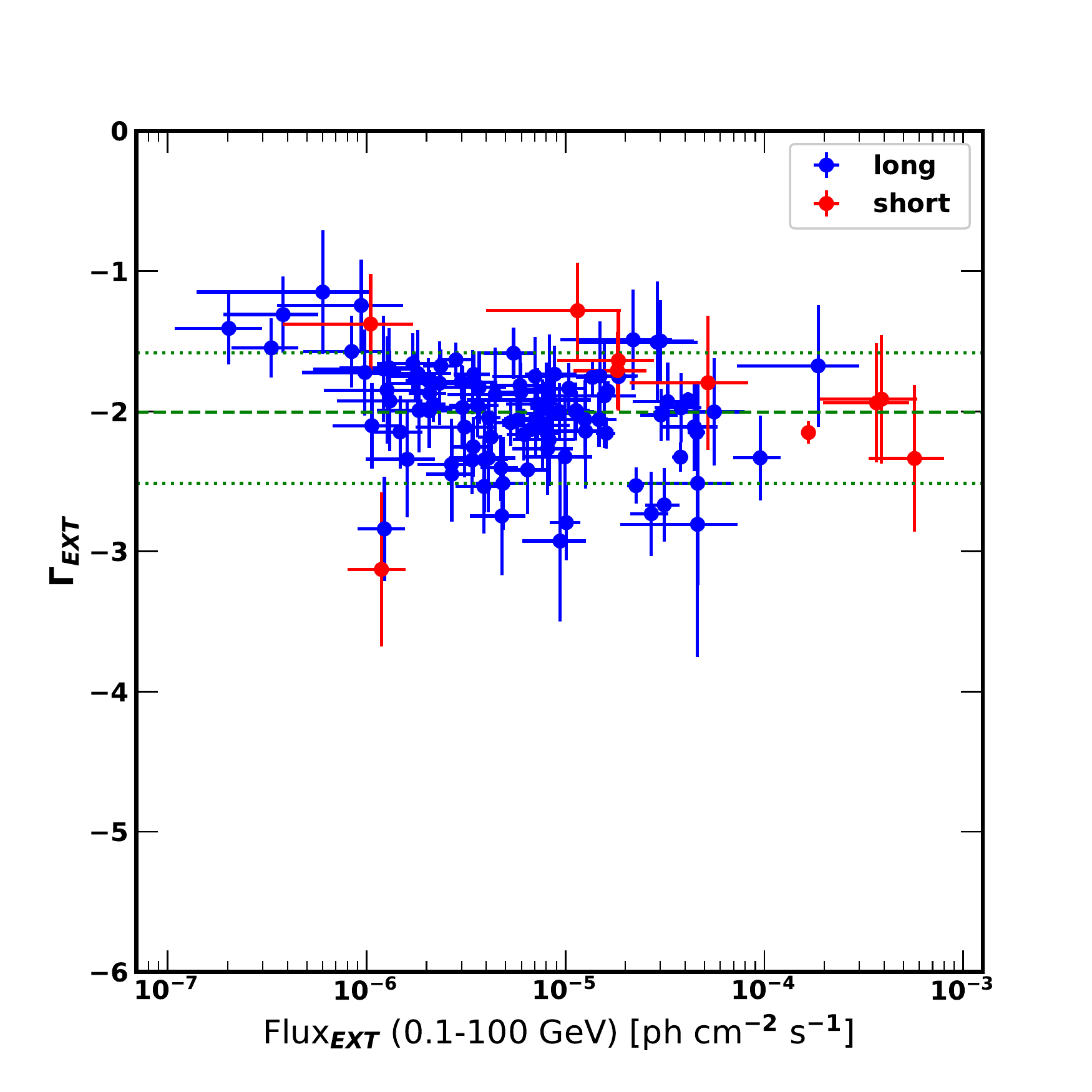}
\end{tabular}
\caption{Photon indexes $\Gamma_{\rm GBM}$ (left panels) and $\Gamma_{\rm EXT}$ (right panels) calculated in the 100 MeV--100 GeV energy range vs. duration calculated in the 50--300 keV energy range (top panels) and LAT 100 MeV--100 GeV fluxes (bottom panels). These are calculated in the ``GBM'' (bottom left) and ``EXT'' (bottom right) time windows, respectively. The green dashed lines denote the mean values and the dotted lines the 10\,$\%$ and 90\,$\%$ percentile of each distribution. Blue and red circles represent long and short GRBs, respectively.}
\label{fig_GRBindex_GBMT90}
\end{figure}

In Figure \ref{fig_GRBindex_GBMT90} we also compare the photon index measured by the LAT during the ``GBM'' time window ($\Gamma_{\rm GBM}$) and the ``EXT'' time window ($\Gamma_{\rm EXT}$). Both indexes are plotted as a function of the GRB duration as calculated in the 50--300 keV energy range (top panels) and of the LAT flux calculated in the ``GBM'' (bottom left panel) and ``EXT'' (bottom right panel) time windows, respectively.
The photon index shows no sign of being correlated neither with the GBM duration nor the flux in either time window, and is similar for long and short GRBs. The value is indeed similar between the two time windows, but is slightly harder in the ``EXT'' window. In the ``GBM'' time window, the values of the photon index are more scattered, with a mean value of $\Gamma_{\rm GBM} = -2.49$ and a 10$\%$ (90$\%$) percentile of $-3.22$ ($-1.86$). In the ``EXT'' time window, the values are more uniform, with a mean of $\Gamma_{\rm EXT} = -2.03$ and a 10$\%$ (90$\%$) percentile of $-2.45$ ($-1.6$). For comparison we recall the same values reported in the \fcat: $-2.08 \pm 0.04$ in the ``GBM'' time window and $-2.00 \pm 0.04$ in the ``EXT'' time window. While the latter is in agreement with the current value, the photon index during the ``GBM'' time window was much harder than the one we derive in the \tcat. Interestingly, it showed a weak inverse correlation with the duration of the burst (see figure~26 of the \fcat). This correlation is now less evident in the larger sample of bursts, but the $\Gamma_{\rm GBM}$ distribution still underlines the agreement with previous findings that the spectra of short-duration GRBs tend to be harder. No clear trend can be seen in the comparison with flux (lower panels), excepts a slight tendency for low-flux GRBs to show harder spectra when looking in the ``EXT'' window.
\FloatBarrier
\subsection{Energetics}
\begin{figure}[b!]
\begin{center}
\includegraphics[width=0.5\columnwidth,trim=0 0 0 0,clip=true]{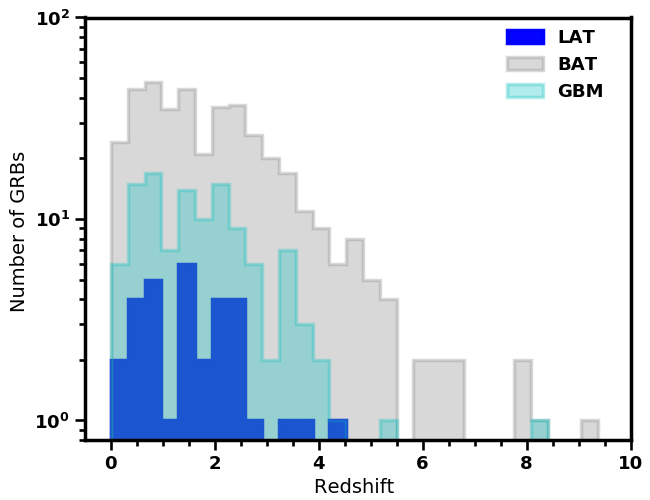}
\caption{Redshift distribution of 34 GRBs detected by LAT (blue histogram), 405 GRBs detected by \Swift-BAT (grey histogram) and 116 GRBs detected by GBM (cyan histogram).} 
\label{fig_redshift_dist}
\end{center}
\end{figure}

In order to more closely study the energetics of the bursts in this catalog, and to put the detections in a wider context, we focus on the GRBs with known redshift. We decided to compare our sample to other bursts with measured redshift detected by \Swift\ and GBM. As of the end of July 2018, \Swift-BAT has detected 1246 GRBs, of which $\sim 35\,\%$ have a measured redshift. In the case of the GBM-detected GRBs, only $\sim 5\,\%$ have measured redshift. The redshift distributions of 405 bursts detected by \Swift-BAT (grey histogram), 116 bursts detected by GBM (cyan histogram) and \nredshift bursts detected by LAT (blue histogram) are shown together in Figure \ref{fig_redshift_dist}. We see no obvious difference between the three distributions.

We next compare the isotropic radiated energy ($E_{\, iso}$) and the bolometric gamma-ray peak luminosity ($L_{\, iso}$) of LAT-detected GRBs to the same quantities in the \Swift\ and GBM samples. The values for $E_{\, iso}$ are computed according to Equation \ref{Eiso_sbolo} in the 1\,keV--10\,MeV energy range. In the case of GBM-detected GRBs, we adopt the fluence listed in the \gcat as computed from the best-fit spectral model, which is usually calculated on a slightly different time interval with respect to the burst \tn, according to the burst brightness.

In order to compute $E_{\, iso}$ of \Swift-detected events (with no GBM observation), we used the parameters of the best-fit spectral models obtained in the 15--350 keV energy range reported in the \Swift-BAT online catalog\footnote{\url{https://swift.gsfc.nasa.gov/results/batgrbcat/}} \cite[see][for more details]{Lien2016_829}. These are calculated over a time interval corresponding to a duration that contains 100$\%$ of the burst emission. For both the GBM and BAT $E_{\, iso}$ calculation, we only consider bursts for which the spectral parameters are globally well-constrained \citep[cf.][]{Gruber2014_211}.  Thus we find 116 (405) GBM (BAT) GRBs which satisfy these criteria, out of which 108 (376) are lGRBs and 8 (29) are sGRBs. The LAT sample comprises 32 lGRBs (2 of the 34 were not detected by the GBM, as previously discussed) and only one sGRB (090510).

We also calculate the isotropic luminosity $L_{\, iso}$, which takes into account the GRB prompt emission spectrum and is defined in a one second time interval centered around the time of the peak flux. It can be expressed as
\begin{equation}
L_{\, iso} =4\pi \,d_{{\,L}}^2 \,\,P(E_1, E_2, z),  
\label{Liso_sbolo}
\end{equation}
where $P(E_1, E_2, z)$ represents the bolometric peak flux, defined as
\begin{equation}
P(E_1, E_2, z) =  \displaystyle\int_{{E_{\,1}/{(1+z)}}}^{{E_{\,2}}/{(1+z)}}E\,N(E)\,dE.
\label{Pbolo}
\end{equation} 
As with $E_{\,iso}$, $L_{\,iso}$ is computed in the 1 keV--10 MeV energy range, for GBM-detected GRBs, using the 1-s peak flux of the best-fit model as reported in the \gcat. We again consider only GRBs whose time-integrated spectra are well-defined, as reported in the GBM and \Swift-BAT GRB catalogs. This leaves us with 394 BAT GRBs and with the same number (116) of GBM GRBs. The slightly lower number of BAT GRBs is expected, as the time interval (and thereby the number of photon counts) is smaller.

\begin{figure}[t!]
\centering
\begin{tabular}{cc}
\includegraphics[width=0.48\columnwidth,trim=0 0 0 0,clip=true]{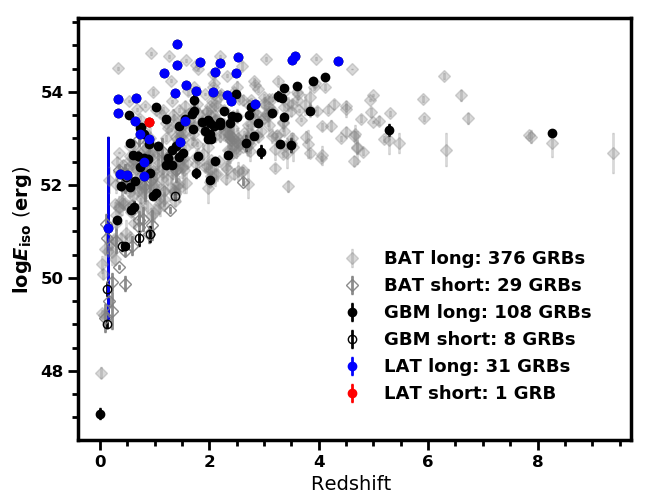} &
\includegraphics[width=0.48\columnwidth,trim=0 0 0 0,clip=true]{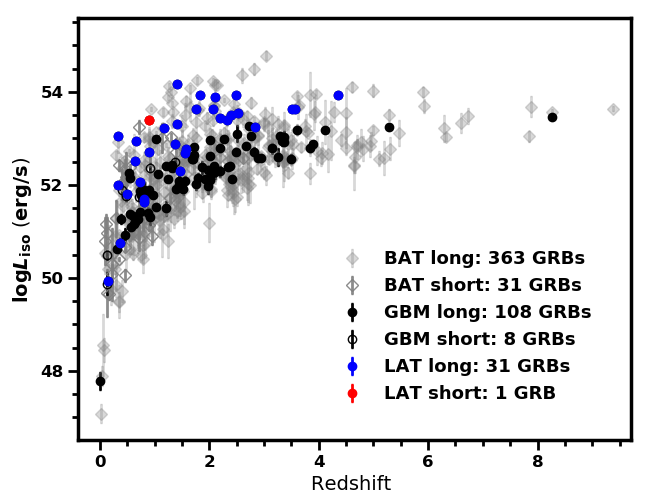}
\end{tabular}
\caption{Left panel: Isotropic radiated energy vs. redshift. Right panel: Isotropic luminosity vs. redshift.
Blue/red circles indicate the LAT long/short GRBs, black and grey circles indicate GBM/\Swift-BAT GRBs, with short bursts always marked by empty symbols.} 
\label{Eiso_redshift}
\end{figure}

Figure \ref{Eiso_redshift} shows the distribution of $E_{\,iso}$ (left panel) and $L_{\,iso}$ (right panel) as a function of redshift. \Swift-BAT and GBM bursts are indicated by gray and black points, respectively, with long (short) bursts marked with full (empty) symbols, respectively. LAT long and short bursts are marked with the standard blue and red circles used in this paper.
LAT-detected GRBs populate the top portion of both distributions, as was previously seen in the \fcat. At that time, this figure only contained 9 LAT-detected GRBs with redshift. It is worth noting that quite a few bursts have a moderate 1 keV--10 MeV $E_{\,iso}$ ($\lesssim 10^{\,-53}$ erg), yet have nevertheless been detected by the LAT.
\begin{figure}[t!]
\centering
\begin{tabular}{cc}
\includegraphics[width=0.45\columnwidth,trim=5 5 180 10,clip=true]{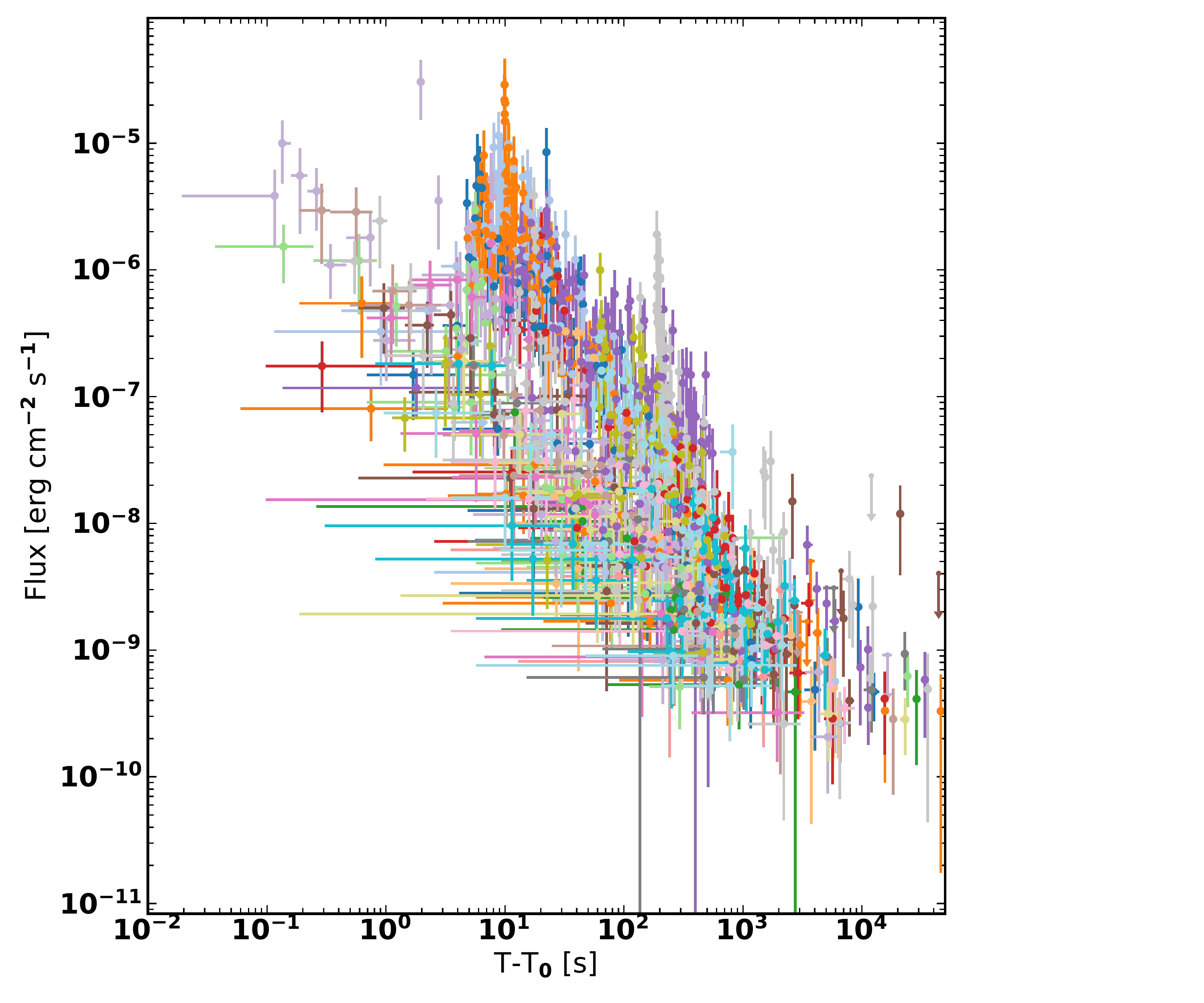} & 
\includegraphics[width=0.45\columnwidth,trim=5 5 180 10,clip=true]{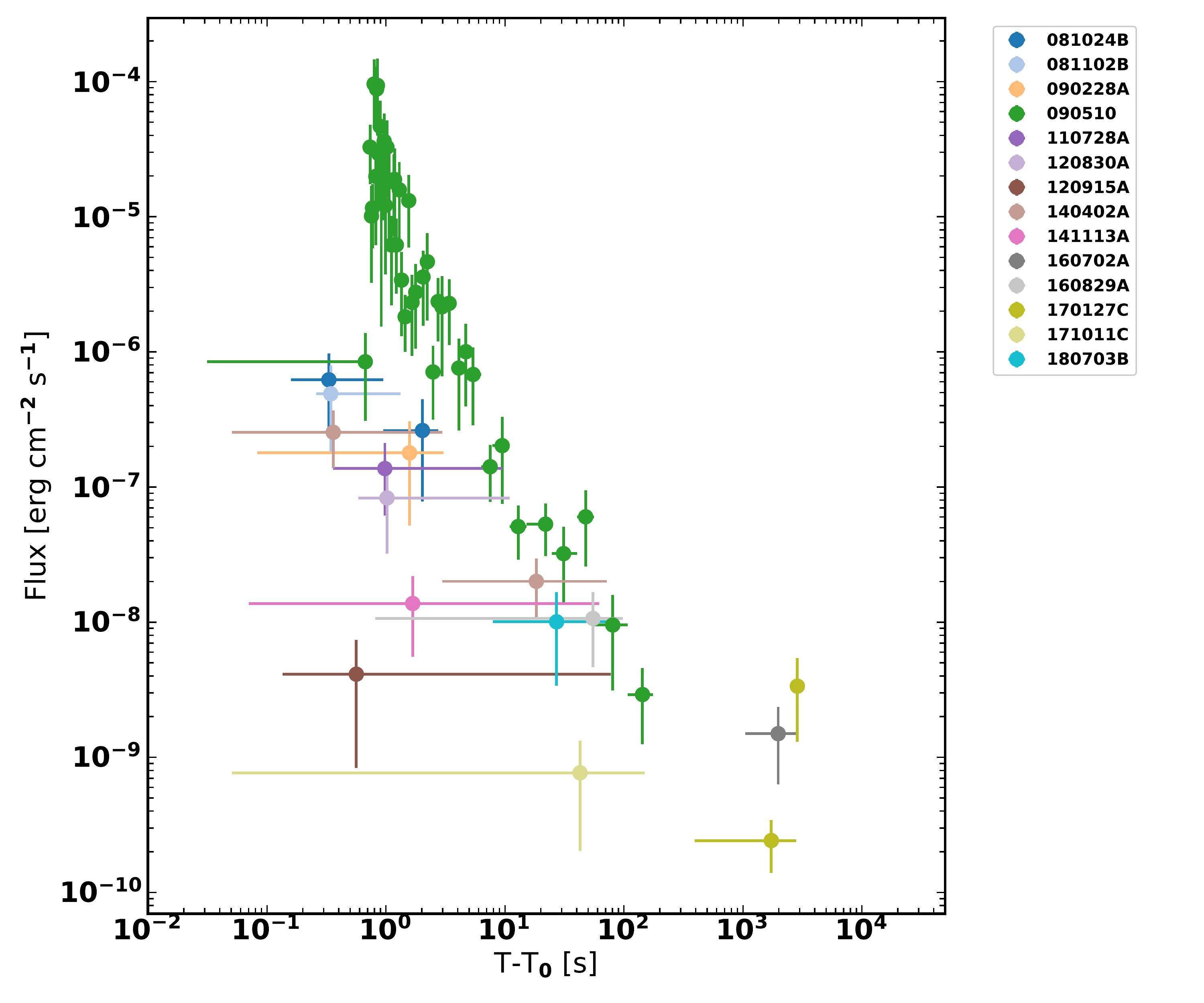}
\end{tabular}
\caption{Flux calculated in the 100 MeV--100 GeV energy range vs. elapsed time after trigger for \nagl long (left panel) and \nags short (right panel) bursts. Each color represents a separate GRB.}
\label{Extended_flux}
\end{figure}
\subsection{Time-resolved light curves}
\label{sec:res:temporal_decay}
We now turn to the temporal decay of the high-energy extended emission. Using the analysis described in Section \ref{sec:time-resolved}, we were able to determine the evolution of the flux as a function of time for \nagl long and \nags short GRBs in our sample. This is shown in the two panels of Figure \ref{Extended_flux},
displaying the temporal decay of long (left panel) and short (right panel) bursts separately. Each event is marked with a different color. The light curves of both sGRBs and lGRBs show a fairly large spread in the observer frame.  

In order to determine the corresponding temporal decay index, we perform a fit of all the light curves maximizing the $\chi^2$, with two different spectral models, namely (1) a simple power law (PL):
\begin{equation}
F(t) = F_0 \,\left(\frac{t}{T_0}\right) ^{-\alpha}, \label{eq_Flux_PL}
\end{equation}
where ${\alpha}$ is the temporal decay index, $T_0$ is the GRB trigger time and $F_0$ the normalization flux; and (2) a broken power law (BPL):
\begin{equation}
F(t) \propto \,t ^{-p} \;\;\left\{
        \begin{array}{ll}
           p=\alpha_1\;\; {\rm for}\; t<T_{b}      \\
           p=\alpha_2\;\; {\rm for}\; t\geq T_{b}, \\
        \end{array}
        \right.
\end{equation}
with index $\alpha_1$ for times before the break time $T_b$, and index $\alpha_2$ afterwards.
If there are at least three flux points (with TS$>$10) in the light curve after the \tnf, we fit a PL, and if there are at least four flux points we also try a BPL. 

\begin{figure}[t!]
\begin{tabular}{ccc}
\includegraphics[width=0.32\columnwidth,trim=0 0 0 0,clip=true]{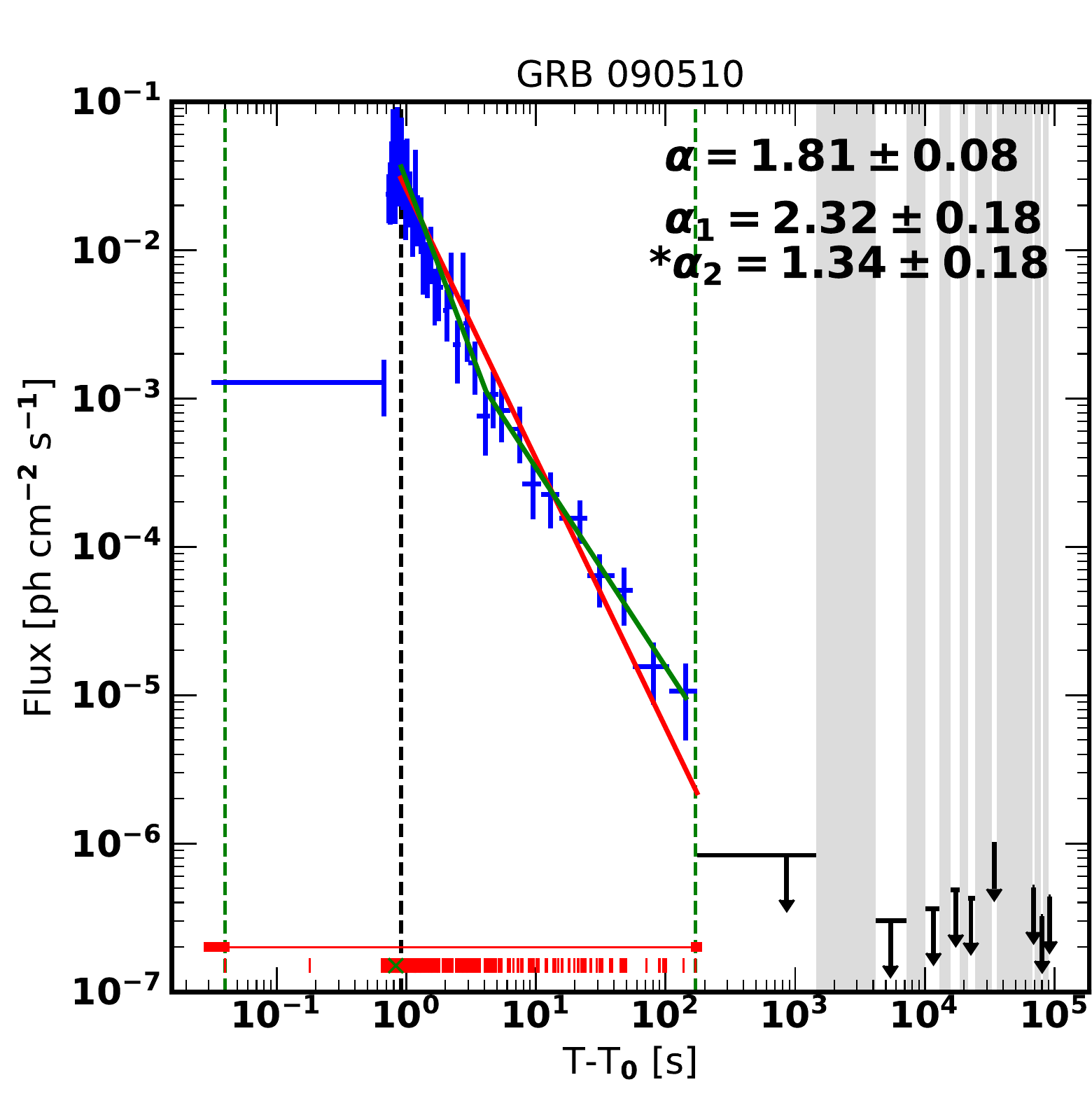} &
\includegraphics[width=0.32\columnwidth,trim=0 0 0 0,clip=true]{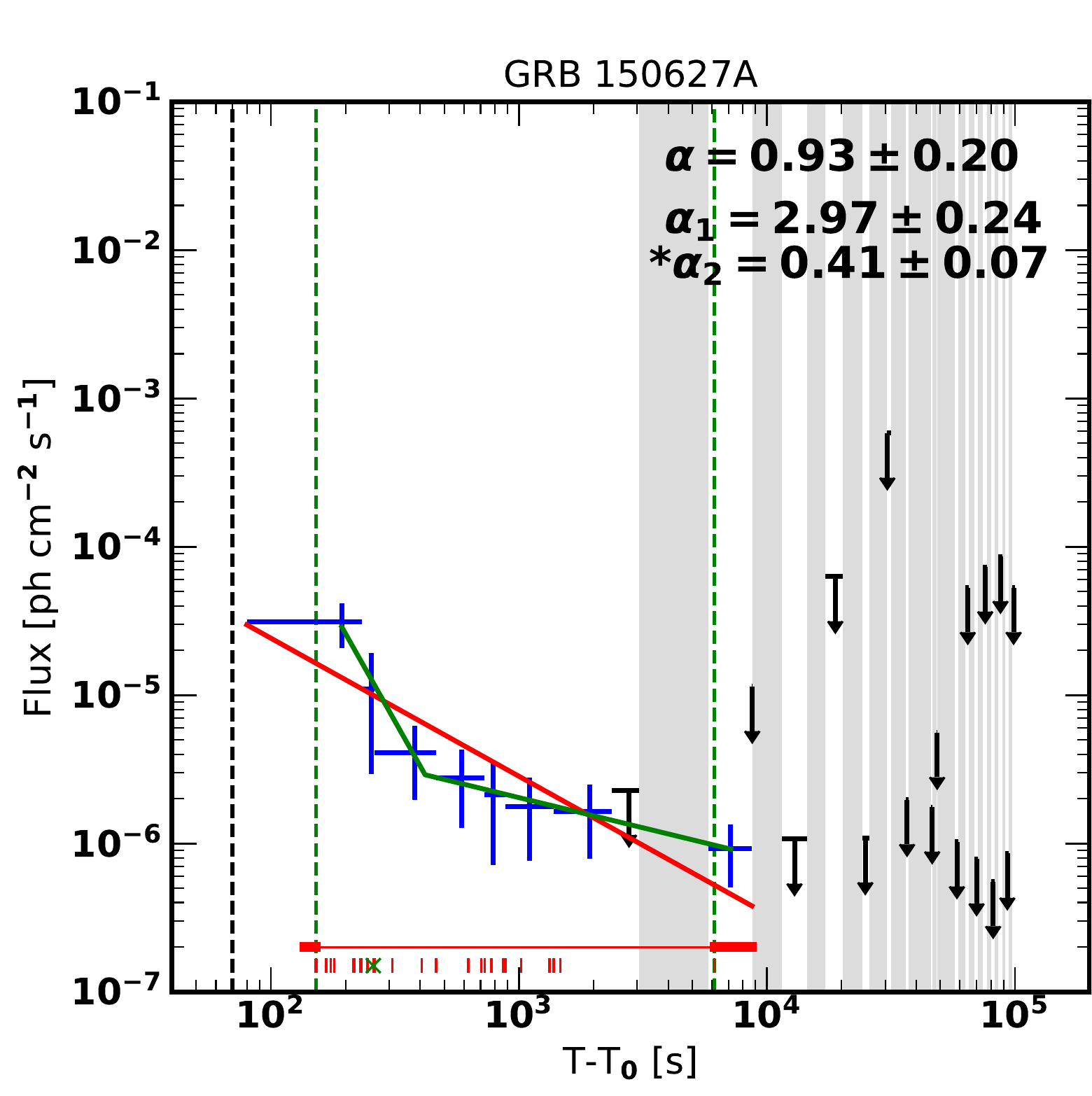} &
\includegraphics[width=0.32\columnwidth,trim=0 0 0 0,clip=true]{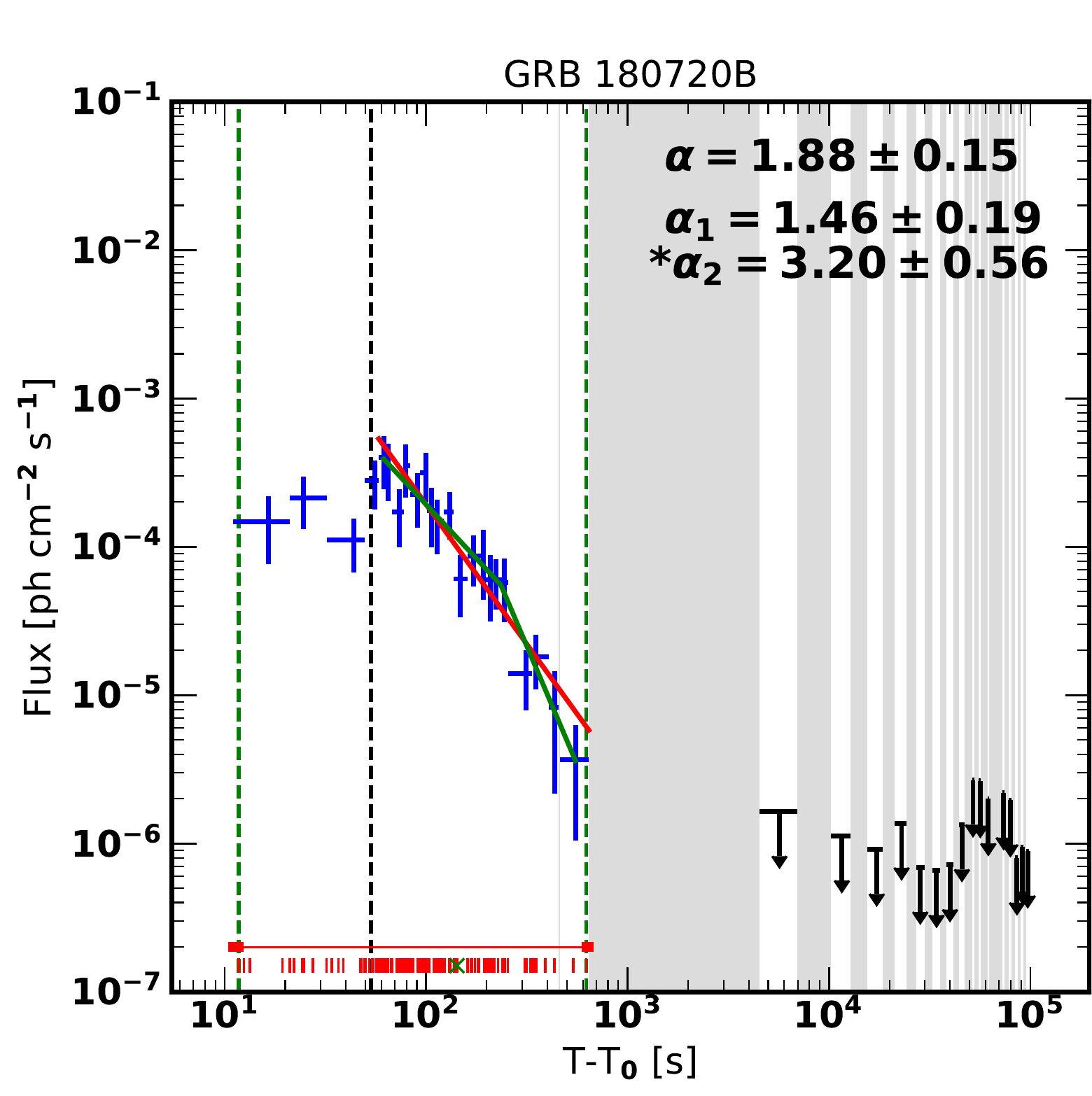} 
\end{tabular}
\caption{
Temporally extended emission for three GRBs that are best fit by a BPL model (a fourth one, GRB 130427A, is shown in the left panel of Figure\ref{fig_Longest_T100}). GRB names are shown in the title in each panel. The red vertical markers indicate the arrival times of the events with probability $>$0.9 to be associated with the GRB, with the green cross being the event with maximum energy. The green vertical dashed lines indicate the first and the last event. The vertical dashed black line marks the \tnf. The PL and BPL fits are indicated with solid red and green lines, respectively. Fit parameters are given in the top right corner of each panel. The asterisk highlights the value selected as temporal decay index in the analysis (which, in these cases, always corresponds to $\alpha_2$ of the BPL model).}
\label{fig_breaks}
\end{figure}

The results of the fits are presented in Table~\ref{tab_extended}.
By fitting the flux temporal decay with a broken power law we find a significant improvement in \agbpl cases. We show three examples in Figure~\ref{fig_breaks} (GRBs 090510, 150627A and 180720B; a fourth, GRB\,130427A, has already been shown in the left panel of Figure~\ref{fig_Longest_T100}). The BPL fit is indicated with a solid green line and the corresponding fit values are given in the top right corner of each panel. The PL fit is shown with a red solid line for comparison. In all but two cases the light curves manifest a steep-to-shallow decay, while for GRB\,171120A and GRB\,180720B (right panel in Figure~\ref{fig_breaks}) the decay steepens after the break.

\begin{figure}[t!]
\begin{center}
\includegraphics[width=0.5\columnwidth]{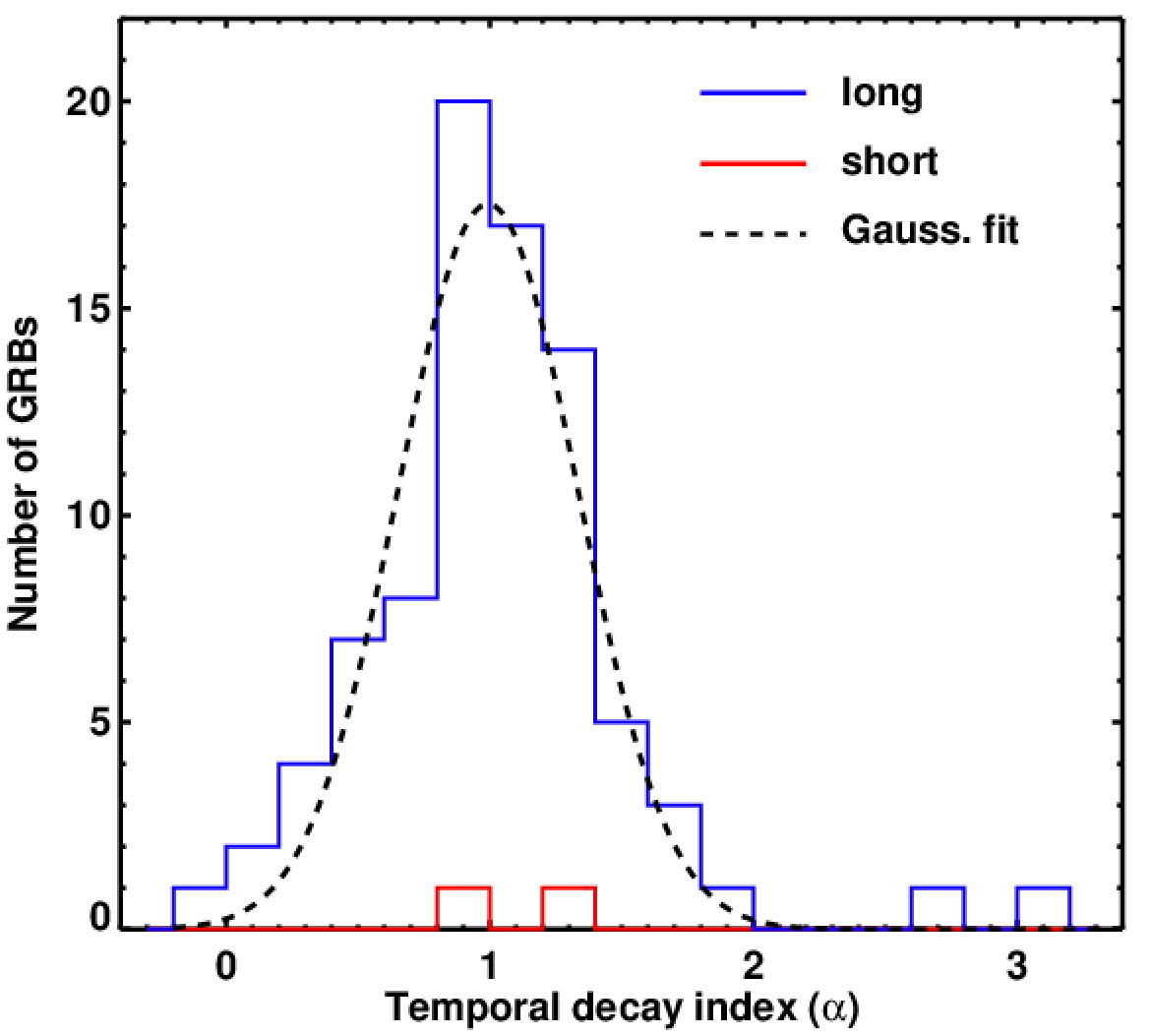}
\caption{Distribution of the temporal decay indexes for \agfitl long (blue histogram) and \agfits short (red histogram) bursts, calculated in the ``ETX" time window. A Gaussian fit of the lGRB indexes (mean $0.99 \pm 0.04$; standard deviation $0.80 \pm 0.07$) is superimposed on the distributions (dashed black line).}
\label{Temp_Decay_Index}
\end{center}
\end{figure}

The distribution of late-time temporal decay indexes ($\alpha$ or $\alpha_2$) is displayed in Figure \ref{Temp_Decay_Index}, together with a Gaussian fit (black dashed line) to the distribution. The distribution comprises 88 GRBs, \agfitl long and \agfits short bursts.
Among the long (short) GRBs, \agspll (\agspls) are best fit with the PL model, while \agbpll (\agbpls) prefer a BPL model.
This is a large increase compared to the \fcat, where only 9 GRBs had enough data to allow the decay index to be determined, ranging from 0.8 (for GRB\,090902B) to 1.8 (for GRB\,080916C) and with a mean value of $\sim$ 1.1. We now find a mean value of $0.99 \pm 0.04$ with a standard deviation of $0.80 \pm 0.07$, still in agreement with the results presented in the \fcat. 

\begin{figure}[t!]
\centering
\begin{tabular}{ccc}
\includegraphics[width=0.45\columnwidth,trim=0 0 0 0,clip=true]{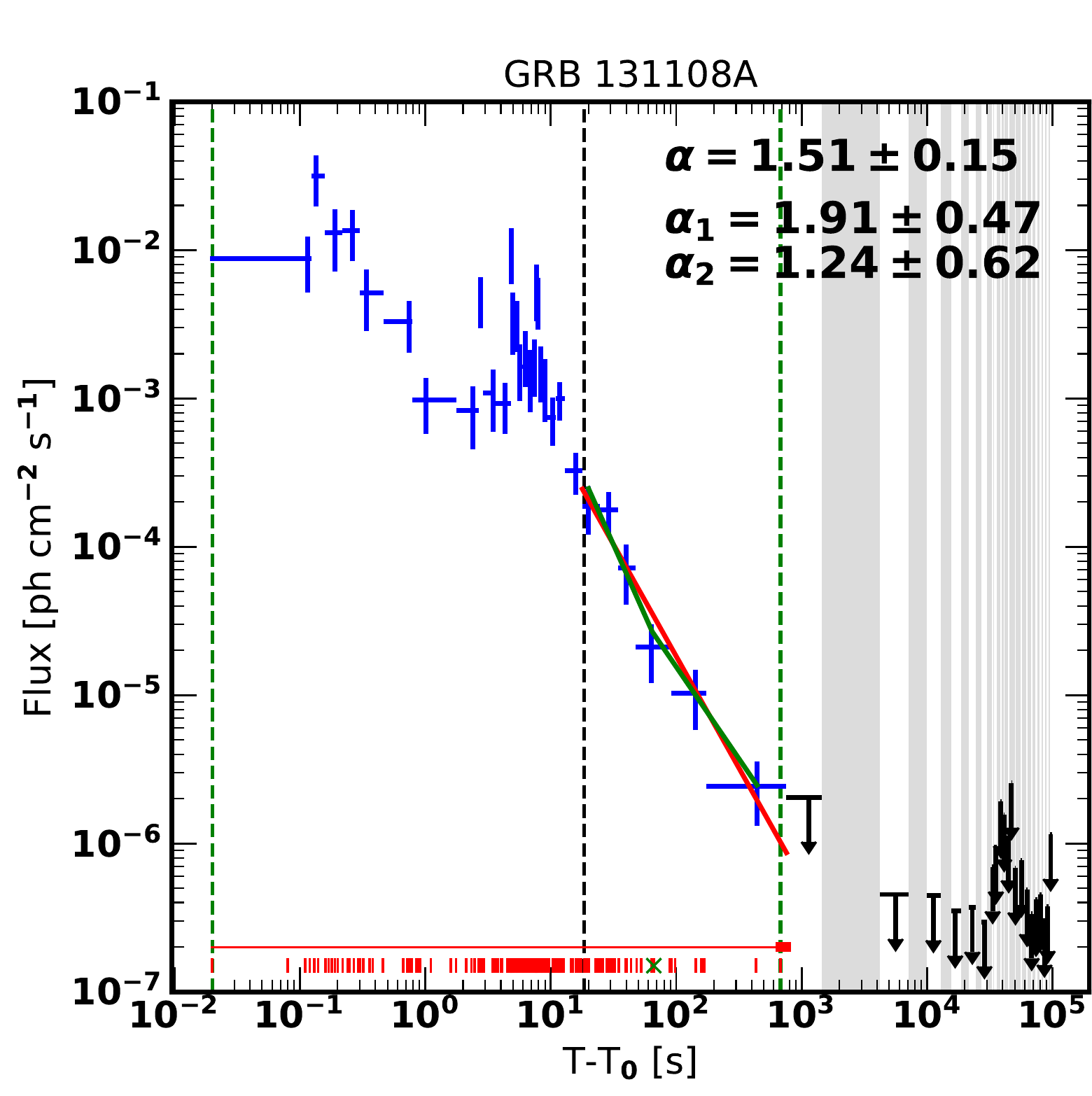}&
\includegraphics[width=0.45\columnwidth,trim=0 0 0 0,clip=true]{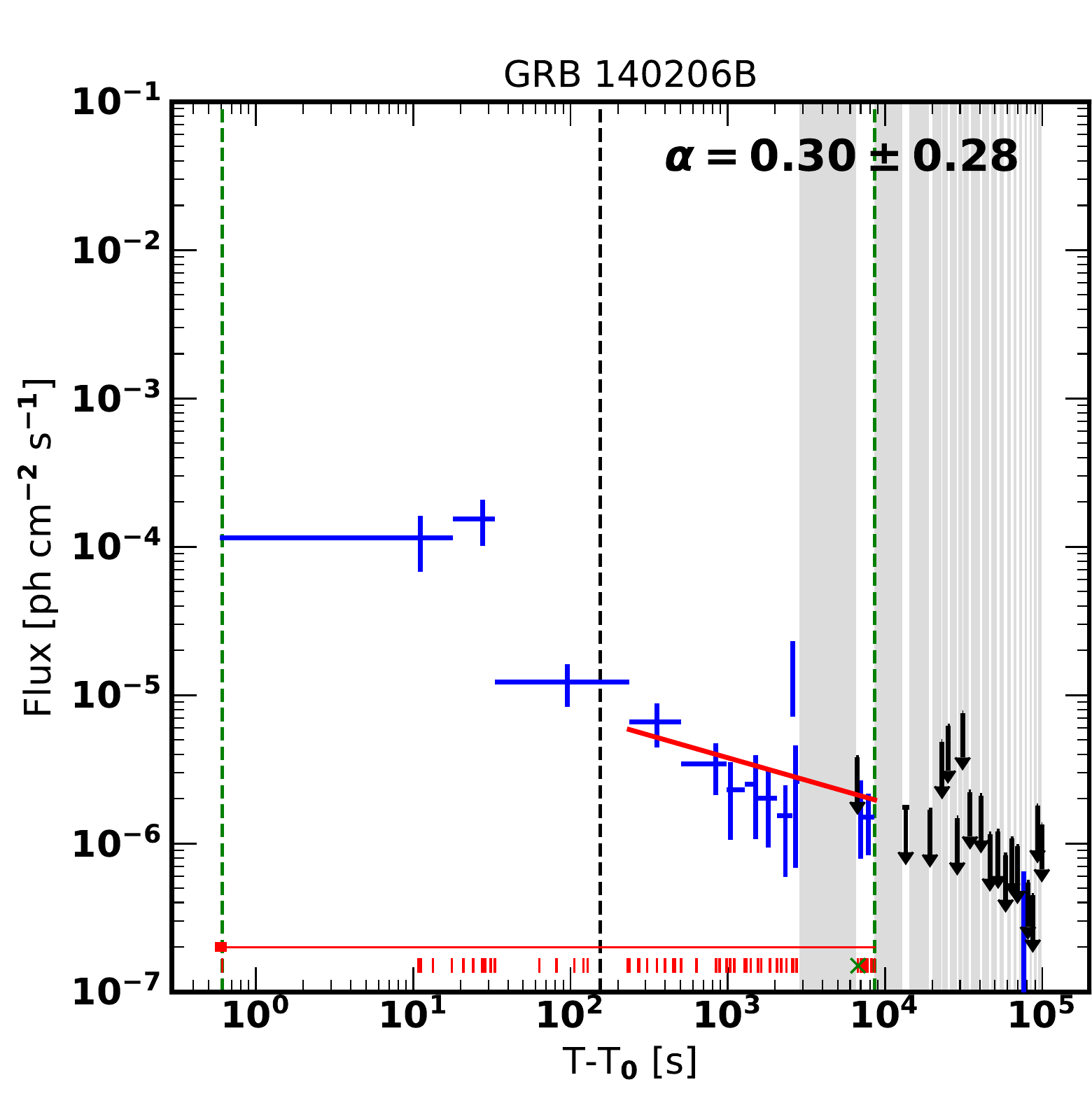} \\
\includegraphics[width=0.45\columnwidth,trim=0 0 0 0,clip=true]{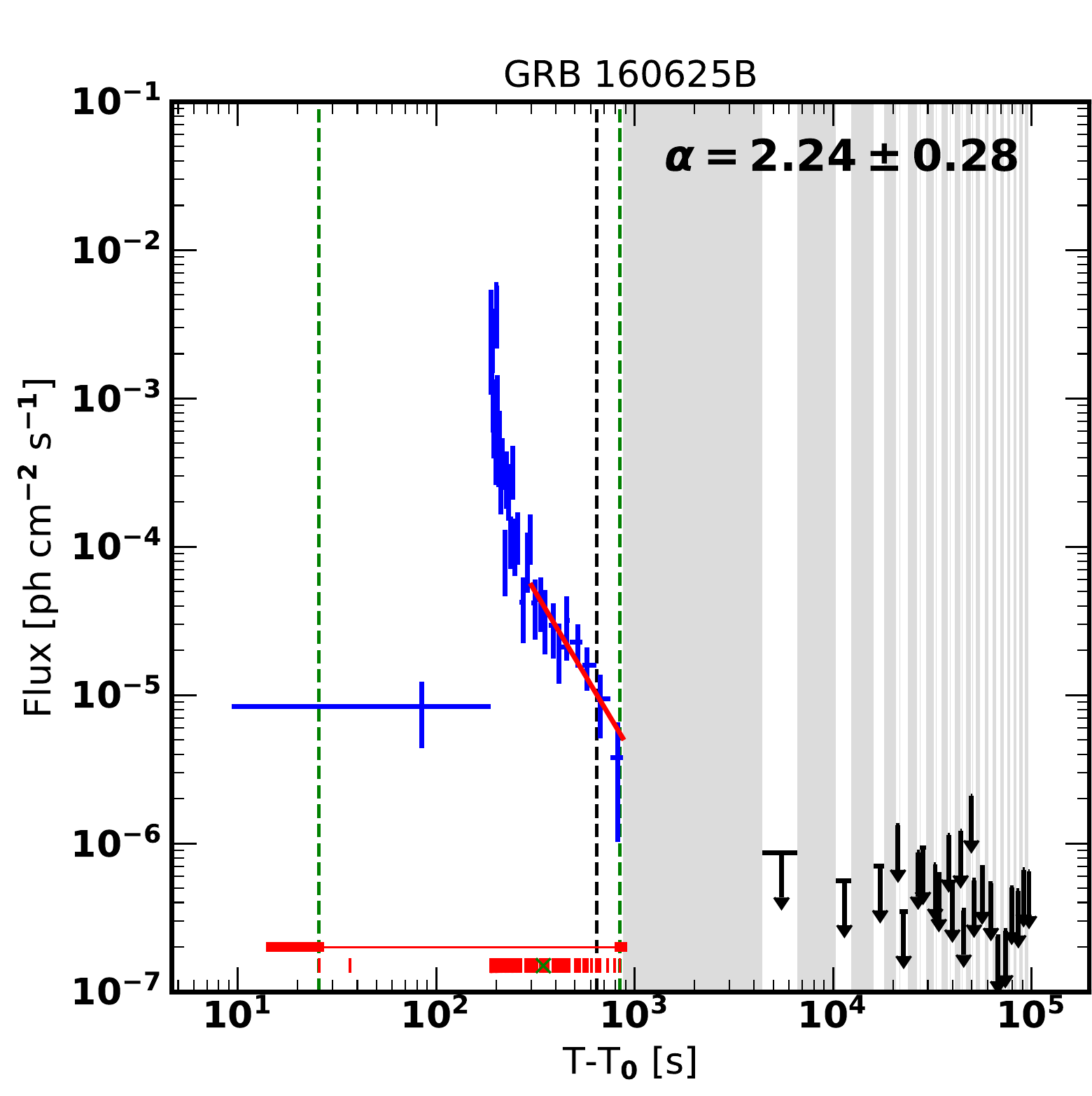} & 
\includegraphics[width=0.45\columnwidth,trim=0 0 0 0,clip=true]{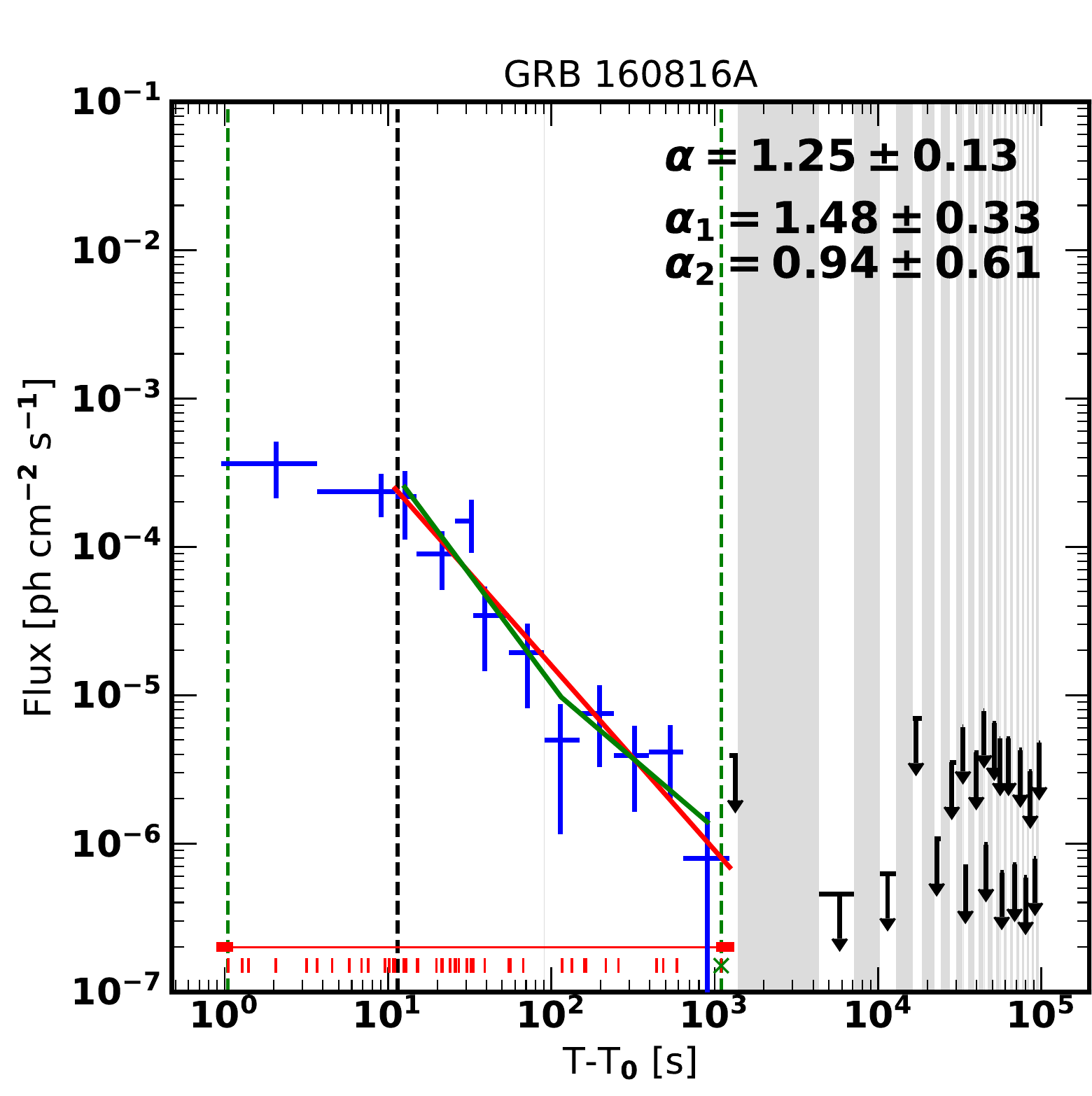} 
\end{tabular}
\caption{Temporal extended emission for four GRBs showing a peculiar behavior (the name of each GRB is shown in the title of each panel). Markers are the same as Figure~\ref{fig_breaks}.}
\label{fig_strangeLC}
\end{figure}

These results and their interpretation will be discussed in more detail in Sections \ref{sec_HE_em_prof} and \ref{sec_clos_rel}, but it is worth noting that there are several cases in which a BPL would likely be required if the data in the ``GBM'' time window had been included. Furthermore, several GRBs show features in the light curve which deviate from both a PL or BPL. Four light curves exemplifying both these cases are displayed in Figure~\ref{fig_strangeLC}. 

In the figure, GRB\,131108A (top left) exemplifies how some bursts display strong variability in the LAT $>$100~MeV light curve during the GBM emission. GRB\,140206B (top right) instead shows the possible presence of a late-time high-energy pulse. GRB\,160625B (bottom left) shows a strong pulse with a very sharp decay.  GRB\,160816A (bottom right panel) would clearly require a break to accommodate the data in the ``GBM'' time window. However, in the ``EXT'' time window used for the fits, a PL model is statistically preferred. These peculiar features are all observed for the first time in the \tcat.
Some light curves, like GRB 140206B (top right panel in Figure~\ref{fig_strangeLC}), show a sharp step between the end of the ``GBM'' and the beginning of the ``EXT'' time window (marked by the vertical black dashed line). In these cases, a fit of the complete dataset would again favor a BPL model rather than the PL one currently used in the ``EXT'' time window. In a few even more extreme cases, the light curve could not be fit at all in the ``EXT'' time window, since there is only a single point in the light curve after the ``GBM'' time window.

\begin{figure}[ht!]
\centering
\begin{tabular}{cc}
\includegraphics[width=1.0\columnwidth,trim=105 25 100 20,clip=true]{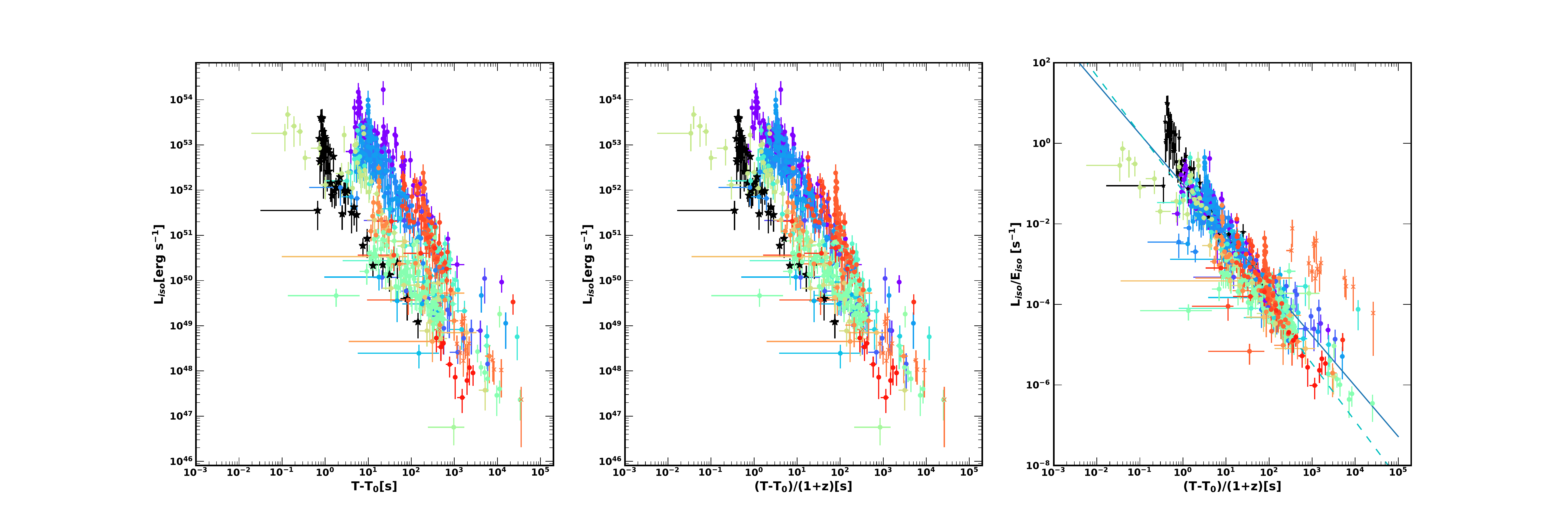}
\end{tabular}
\caption{Isotropic luminosity $L_{iso}$ calculated in the 100 MeV-100 GeV rest frame energy range for the \nredshift GRBs in our sample with measured redshift. The {\it left panel} shows $L_{iso}$ vs. the time elapsed since the trigger. 
GRB~090510 is the only short GRB with known redshift and it is marked with black stars. 
In the {\it middle panel} $L_{iso}$ is plotted against the vs. the time elapsed since the trigger in the rest frame. In the {\it right panel}, we show $L_{iso}$ divided by the isotropic energy (calculated in the 1 keV -- 10 MeV rest frame energy range) vs. the time elapsed since the trigger in the rest frame. We mark GRB\,160623A, which represents an outlier of the distribution (see text for explanation), with {\it orange crosses}. The solid line shows a linear fit, giving a decay index of $1.25 \pm 0.03$. For comparison, we also show a dashed line with decay index $10/7$.}
\label{Extended_Luminosity_all}
\end{figure}

In Figure \ref{Extended_Luminosity_all}, we show the 100\,MeV--100\,GeV luminosity evolution for the \nredshift GRBs in our sample with measured redshift. Among those, there is only one short burst, namely GRB 090510. The three panels of the figure show first the light curves in the observer frame (left), the evolution as a function of time in the source frame (center), and finally the luminosity divided by isotropic energy ($E_{\, iso}$; right), calculated in the 1\,keV--10\,MeV energy range. Each correction brings the light curves closer together, and in the right panel there is a remarkable alignment of all the GRBs. 
This analysis was done following the one presented by \citet{2014MNRAS.443.3578N}, where a similar result was found. It is worth noting that in  the right panel of Figure~\ref{Extended_Luminosity_all} one of the GRBs, GRB\,160623A, does not align with the others indicating a possible outlier. This burst was occulted by the Earth for a large part of its duration, and the GBM trigger occurred $\sim$50 s after the start of the GRB based on the Konus-Wind light curve. 
This likely leads us to underestimate the total energy release $E_{\rm \, iso}$ of the burst and thus to overestimate the normalization of the light curve. In the right panel we also include a linear fit to all \nredshift GRBs, indicated by the solid line. The decay index is $1.25 \pm 0.03$. For comparison, we also show a dashed line with decay index $10/7$ (see further Sect.~\ref{sec_HE_em_prof}).
\subsection{High energy events}
\label{Res_HE_Events}
The highest energy GRB photon ever recorded by \Fermi thus far is a 94.1 GeV event connected with GRB 130427A \citep{2014Sci...343...42A}. While displaying photon energies of a few hundred MeV is a common feature among the LAT-detected GRBs, higher energies are relatively rare. Table \ref{tab_energymax_gbm} summarizes the highest-energy photon characteristics for each burst in our sample. It lists the total number of photons detected with probability $>90\%$ of belonging to the burst, as well as the energy, arrival time and probability of the highest energy photon detected in the ``GBM'' time window. We also list the same quantities calculated in the time resolved analysis. 

\begin{figure}[t!]
\begin{center}
\includegraphics[width=0.45\columnwidth]{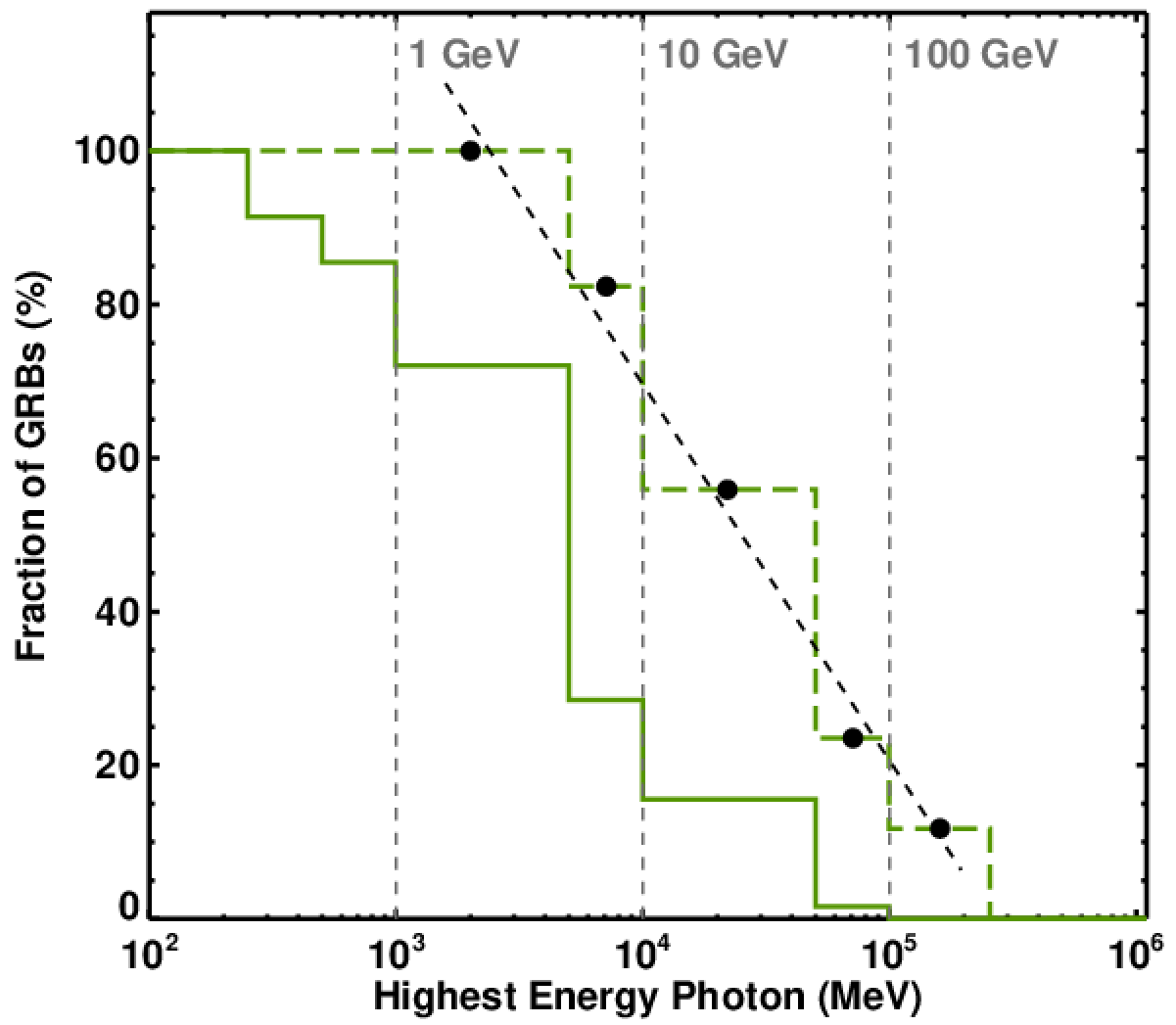}
\caption{Fraction of GRBs with the highest energy photon detected above selected threshold energies (250 MeV, 500 MeV, 1 GeV, 5 GeV, 10 GeV, 50 GeV) ({\it green solid line}). The distribution of the source-frame-corrected energies for the redshift sample is indicated with the {\it dashed green line}. The {\it dashed black line} denotes a linear fit to the values corresponding to the center of each bin.}
\label{HEP_distr}
\end{center}
\end{figure}

Figure \ref{HEP_distr} shows the fraction of GRBs detected above selected energy thresholds (250 MeV, 500 MeV, 1 GeV, 5 GeV, 10 GeV, 50 GeV). A sharp drop from $\sim$70\,\%\ to $\sim$30\,\%\ is seen at 5 GeV. There are three GRBs with emission above 50 GeV (2\,\%), namely GRB 130427A (95 GeV), GRB 140928A (52 GeV) and GRB 160509A  (52 GeV). 

Our sample of \nredshift GRBs with measured redshift also allows us to study the source-frame-corrected energies. This is shown as the dashed line in Figure \ref{HEP_distr}. This distribution shows a more gradual decrease with energy: almost 80\,\% of the included GRBs have a maximum source-frame photon energy above 5\,GeV, and $\sim$12\,\% (4 GRBs) above 100\,GeV. The highest source-frame energy is a 147\,GeV photon from GRB 080916C, at $z=4.35$ \citep{2013ApJ...774...76A}. 
The figure also includes a linear fit to the bin centers of the source-frame distribution. The fit is remarkably good, showing that the fraction of GRBs decreases as $A\times \log(E/1\,{\rm MeV})+B$, where $A = -49 \pm 4$ and $B = 266 \pm 21$.

\begin{figure}[t!]
\begin{center}
\includegraphics[width=0.45\columnwidth,trim=5 5 40 40,clip=true]{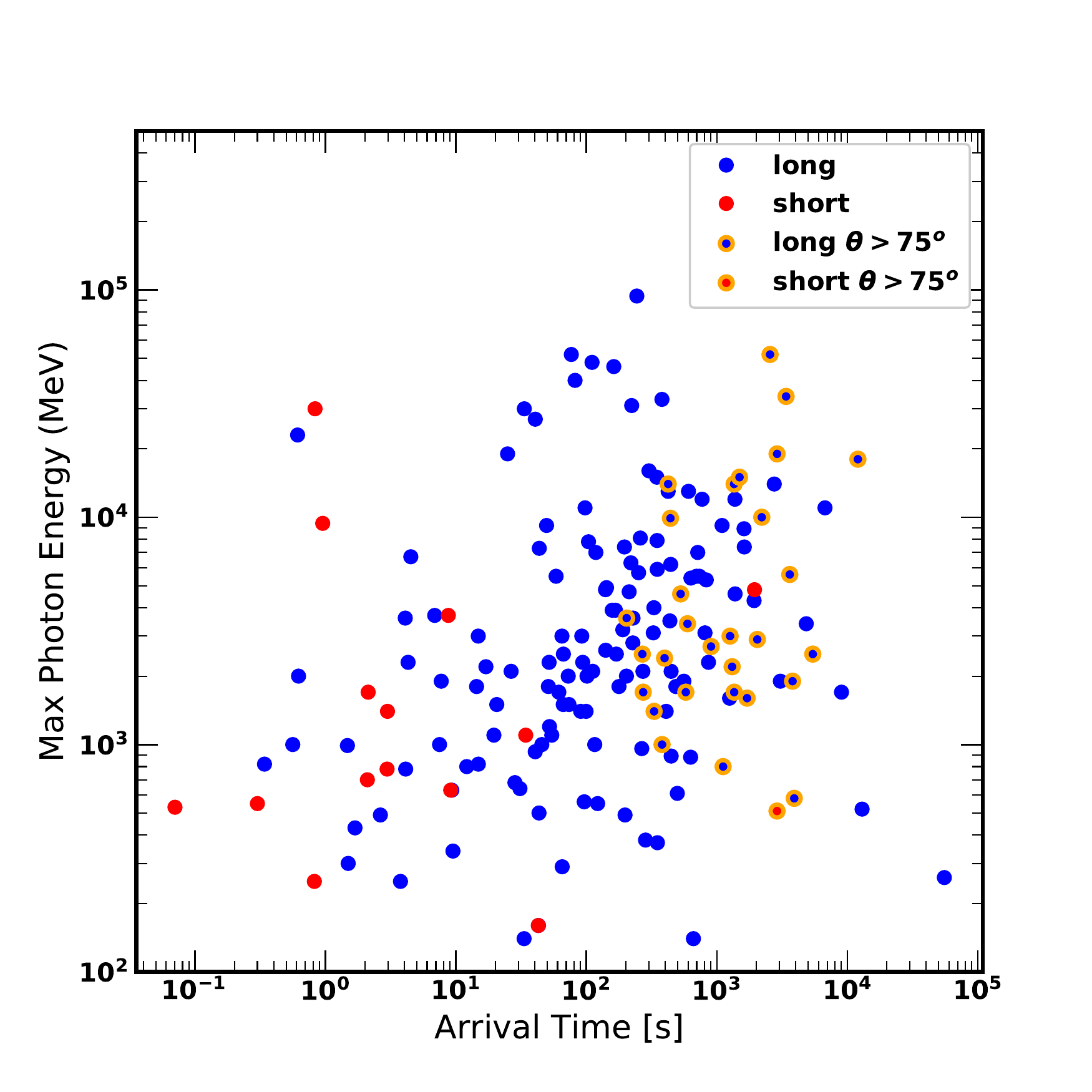}
\includegraphics[width=0.45\columnwidth,trim=5 5 40 40,clip=true]{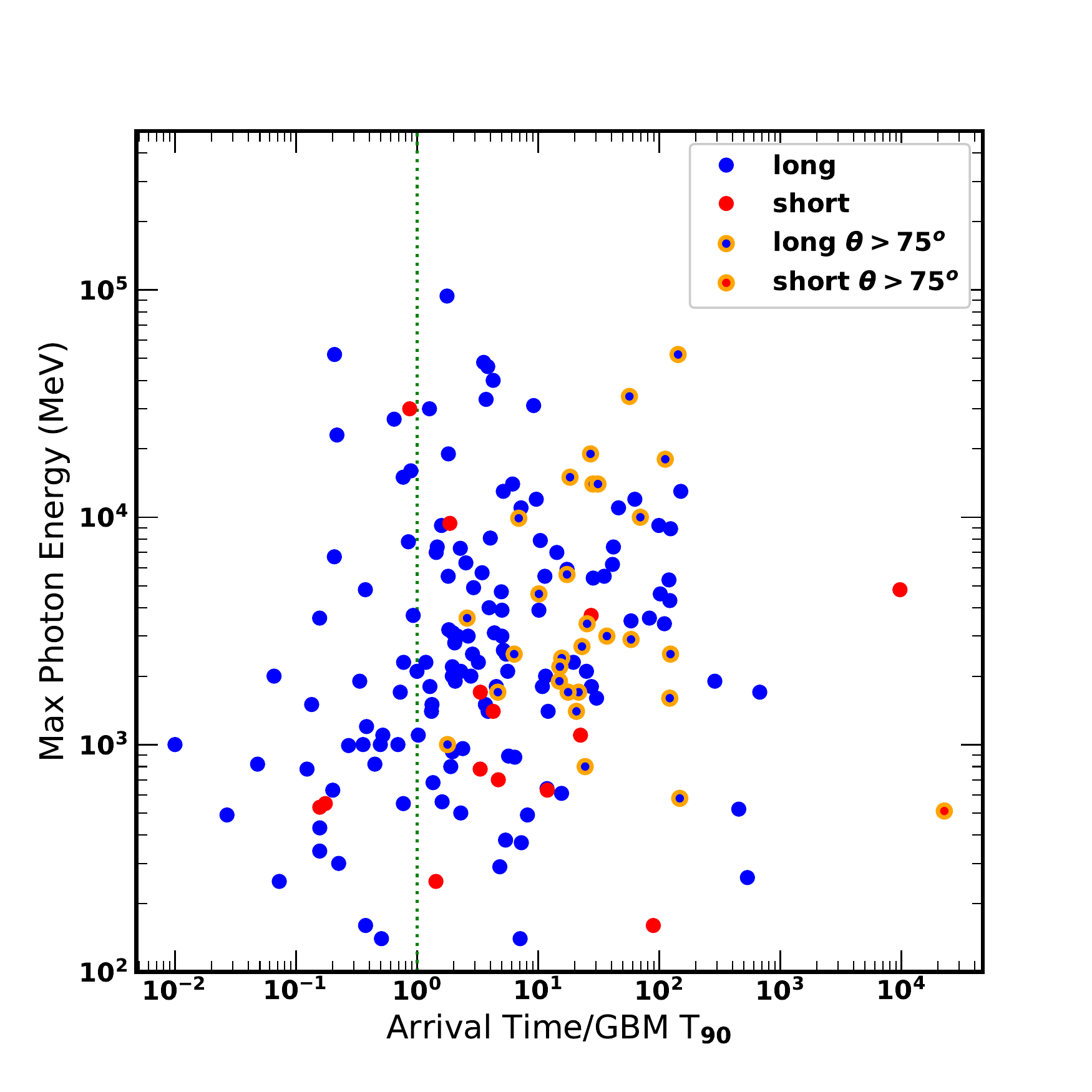}
\caption{Energy vs. arrival time for the highest energy photon of each GRB. 
In the right panel, the arrival time is normalized to the duration (T$_{90}$) calculated in the 50--300 keV energy range (indicated by the {\it dashed vertical line}). {\it Blue} and {\it red circles} represent long and short GRBs, respectively.}
\label{HEP_En_Time}
\end{center}
\end{figure}

Figure~\ref{HEP_En_Time} shows the energy of the highest-energy photon in each GRB as a function of arrival time (left panel). In the right panel of the figure, the arrival time is shown as a fraction of \tn, calculated in the 50--300\,keV range. No clear pattern can be distinguished, with long and short bursts overlapping in the right panel. The one clear outlier in the right panel is the short GRB 170127C, where the highest energy photon (500\,MeV) was detected almost 3\,ks after the trigger. This GRB was outside the LAT FoV at T$_0$, and the data therefore only cover the time from around 300\,s after the trigger (see also Sect.~\ref{sec:onset_duration} and Figs.~\ref{fig_THETA} and \ref{fig_170127067}).

In order to check for more high-energy photons to be detected beyond the standard energy range ($> 100$ GeV) we performed an additional analysis up to 300 GeV. No LAT detection was made above 100 GeV. Table~\ref{tab_energymax_Michele} presents the 29 GRBs in the sample detected above $> 10$ GeV. All high-energy photons with a probability $> 90$\%\ and with an observed energy $> 10$ GeV are reported for each burst. As in Table \ref{tab_energymax_gbm}, we specify the photon energy (listed in decreasing order), its arrival time, the GRB redshift and the source-frame-corrected energy ($E_{\,sf}$). Values of $E_{\,sf}>100$ GeV are marked in bold. The only short burst listed in this table is GRB 090510: in this case, a $\sim$30 GeV photon was detected 830 ms after the GBM trigger time. GRB 130427A holds the record with 17 photons detected above 10 GeV, with the highest event ever detected from a GRB (the 94.1 GeV photon) observed 243 s post trigger. The second burst with the most HE photons is GRB 090902B, with seven photons detected above 10 GeV. There are two bursts where a high-energy photon is detected at very late times ($>$10 ks): GRB\,130427A ($\sim$34 ks) and GRB\,160623A ($\sim$12 ks).
\FloatBarrier
\section{Discussion}
\label{sec:discuss}
We will now discuss our results and compare them to previous results, in particular what was seen in the \fcat. The discussion generally follows the outline of Sect.~\ref{sec:results}, starting with the LAT sample as a whole, then continuing with a comparison to the GBM sample. Finally we will consider the energetics, temporal decay and possibilities for detections at very high energy (VHE). The aim of this section is to put our results in a wider context; broader implications in the framework of theoretical models will be discussed in Sect.~\ref{sec:interpret}.
\subsection{Detectability of LAT bursts and LAT detection rate}
\label{Disc_DetLAT}
Before launch it was estimated that the LAT would detect 10--12 GRBs per year above 100\,MeV \citep{2009ApJ...701.1673B}. The results from the first GRB catalog showed 28 GRBs detected in the first three years of the mission, slightly below expectations. The current work instead shows that the LAT has exceeded expectations, with \nlike GRBs detected above 100\,MeV in 10 years. This is in large part due to the continuous improvements in event analysis and detection algorithms. 

It is interesting also to look specifically at the highest energies of the LAT-detected GRBs. As can be seen from Figure~\ref{HEP_distr}, only a small fraction of GRBs are seen in the upper energy range of the LAT. For example, 20\% of the GRBs have detected emission above 10\,GeV, corresponding to $\sim3$ GRBs per year. We will further discuss the occurrence of these highest-energy events in Sect.~\ref{Disc_HE_phot}.

Another interesting aspect to discuss is that while the highest-energy photon most often arrives after the \tn (right panel of Figure~\ref{HEP_En_Time}), there is also a large fraction where it arrives during this interval. Interestingly, this fraction seems to vary slightly with energy: below 1\,GeV more than half of the events arrive during \tn, whereas 70\% of events arrive later at energies above 10\,GeV. We also note that there is little difference between the short and long GRB populations regarding normalized arrival time, although the maximum energy of sGRBs appears on average a bit lower. This could link the highest-energy photons to a process common in the two scenarios, and independent from the progenitor, such as external shocks in the circumburst medium. This will be further discussed in Sect.~\ref{sec:interpret}.

As of writing, the observing pattern adopted after the solar array malfunction (see Sect.~\ref{sec:instruments}) could potentially lead to a loss in GRB detections, as events occurring in the unobserved hemisphere of the sky may not be monitored until long after the event. However, studying the detections reported in this catalog, no such event was detected even during the previous observing strategy. The exception to this are the ARRs, as they could bring bright events into the FoV. It is thus likely that the detection rate will be slightly impacted while ARRs remain disabled. Even if the LAT detects late-time emission, without ARRs there is the risk of losing high-energy prompt emission.
\subsection{Onset and duration of the high-energy emission}
As in the \fcat, we can firmly establish the general trend that high-energy emission from GRBs tends to have delayed onset and longer duration as compared to emission at lower energies. However, the panel (c) of Figure~\ref{fig_GBMT05_LATT05} also shows that when high-energy emission is detected, it starts during the prompt phase in $>$60\%\ of the cases. The majority of the other GRBs were outside the LAT FoV at trigger time, meaning the fraction is likely even higher. This is an interesting result which provides valuable input to models of the emission mechanisms and will be discussed further in Sect.~\ref{sec:interpret}.

Figure~\ref{fig_GBMT05_LATT05} shows how varied the difference between \tf (calculated in the 50--300\,keV range) and \tz (100\,MeV--10\,GeV) can be. In some cases, the LAT emission is completely contemporaneous with the GBM. In other cases, the LAT emission starts hundreds or even thousands of seconds later. Considering only bursts which were in the LAT FoV at the time of trigger, there are a number of cases where the high-energy emission came much later than the one at lower energies. For example, GRB 160503A shown in the right panel of Figure~\ref{fig_Like_T0} was at $\theta=25.1$\deg\ at \tz, and remained in the FoV for over 2\,ks without any high-energy emission being seen. The first detection instead came much later, at 5\,ks. These extreme delays are much longer than the ones seen in the \fcat, and represent a new result in the \tcat.

In addition to greater delays, we now report much longer durations. In the \fcat, the longest duration reported was $>800$\,s for GRB 090902B. In the \tcat, many GRBs have durations of order $10^3$\,s, with the longest duration being 35\,ks (GRB\,160623A). In general, the durations have increased also for most bursts contained in the \fcat, likely due to better sensitivity as a result of Pass 8 (we note, however, that the duration estimates were made using a different technique as described in Sect.~\ref{sec:time-resolved}, so the numbers are not directly comparable). 
\subsection{Comparison with GBM flux and fluence distribution}
As already presented in Sect.~\ref{sec:results}, the LAT-detected GRBs tend to sample the upper range of the GBM flux and fluence distribution (Figure~\ref{fig_Fluence_Catalogs}). At the high end of the GBM fluence distribution, the LAT detects a high fraction of the GRBs; above $\sim10^{-4}$\,erg\,cm$^{-2}$ the two distributions practically overlap (left panel). The few additional bursts seen by the GBM could be explained by some GRBs being outside the LAT FoV or at high zenith angles. The effect of the $\theta$ angle is further investigated in \citet{2018ApJ...861...85A}, who show that this is the main factor determining LAT detectability. The bias towards high GBM flux and fluence for LAT-detected bursts is therefore not related to a difference in sensitivity between the instruments.

Looking at the high end of the flux distribution (middle panel of Figure~\ref{fig_Fluence_Catalogs}), the fraction of GRBs not seen by the LAT is higher. This shows that the flux might not be such a good indicator for high-energy emission. While the fluence is a measure of the total energy output, the flux simply shows the ``strength'' of the peak. The light curves also show clear differences, and peaks at low energies are not necessarily mirrored at high energies. An example of this is GRB 180728A, where the GBM peak flux was about 230 photons\,cm$^{-2}$\,s$^{-1}$ \citep{2018GCN.23053....1V}, but no detection was made at high energy even though the burst was inside the LAT FoV ($\theta \sim 35^{\rm o}$). This is likely due to the low value of $E_{\rm \, peak}$, 80\,keV; indeed, the flux was dominated by energies below 50\,keV.

However, the LAT has also detected GRBs which have relatively low fluence in the GBM. These outliers are predominantly short GRBs, where the low fluence is naturally explained by the short duration. The fluence distribution of sGRBs in the GBM is also shifted to lower values overall. Also for sGRBs there is a tendency for the LAT to sample the higher fluence end of the GBM population, but this is much less marked than for the lGRBs. 

The result that the LAT-detections are biased towards the brighter GRBs was clear already in the \fcat. However, with the larger sample, the picture presented here becomes more nuanced. The sGRBs show that high-energy emission can be produced even at lower fluence, raising the question of why not more lGRBs are detected. There may be differences in the emission mechanisms or environments between the classes which explain why low-fluence sGRBs are more likely to be detected in high-energy emission than lGRBs of similar fluence. For instance, lGRBs are expected to have a denser circumburst medium, as they are coupled to massive stars which have strong stellar winds.

The possibility of GeV emission from sGRBs is particularly interesting in the light of GW170817 and the associated sGRB 170817A \citep{2017ApJ...848L..12A,2017ApJ...848L..13A,2017ApJ...848L..14G}. This event was outside the LAT FoV, and had a fairly low fluence of $2.8\times10^{-7}$\,erg\,cm$^{-2}$. However, Figure~\ref{fig_Fluence_Catalogs} shows that the LAT has detected GRBs with a similar fluence. This is very promising in view of the upcoming observation period scheduled to start in spring 2019, showing a strong potential for LAT detections of similar events \citep[see also][]{2018ApJ...861...85A}.
\subsection{Origin of emission below 100 MeV}
In our sample almost twice as many GRBs are detected above 100\,MeV (\nlike) as in the $30-100$\,MeV LLE range (\nlle). Several studies have found evidence for a separate spectral component behind the emission above 100\,MeV, and high-energy cut-offs between the LLE and LAT energy ranges have also been seen \citep[e.g.,][]{2018ApJ...864..163V}. The behavior of the emission below 100\,MeV further shows more similar temporal behavior to the GBM range than do the data above 100\,MeV (cf. Figures \ref{fig_GBMT05_LATT05} and \ref{fig_GBMT05_LLET05}). The LLE-only GRBs in our sample would then be the result of the low-energy emission being strong enough (and/or $E_{\rm \, peak}$ at high enough energy) to extend into the LLE range. As the burst evolves, $E_{\rm \, peak}$ moves to lower energies and the emission in the LLE range will therefore appear to fade before that at lower energies. This explains the fact that the duration at $30-100$\,MeV is almost always shorter than the one measured by the GBM. This picture also explains the fact that the LLE emission appears to start earlier than the GBM emission in a few cases (left panel in Figure~\ref{fig_GBMT05_LLET05}). As the duration is shorter in the LLE range, this will naturally make the value of \tllf shorter as well. Again it is just a sign that the emission in the GBM range lasts much longer than the one in the LLE range.

For GRBs detected in both LAT and LLE energy ranges, the duration is generally shorter in the $30-100$\,MeV band. While direct comparison of the \tlln and \toz\ should be done with caution, the large differences in duration here clearly point to an intrinsic origin rather than observational bias. For example, in the LLE range the effective area is up to a factor of 2-3 lower than that above 1 GeV; however, the \toz can be more than an order of magnitude longer than the \tlln.

Although there is much to suggest that the LAT emission is often the result of a separate component, we caution that there are also cases where the GRB spectrum is seen in the full range from keV to GeV, and well fit by a single component extending also to energies above 100\,MeV \citep[e.g.,][]{GRB080916C:Science,Axelsson+12}. So while the dominance of LAT-only detections supports a separate emission process $>100$\,MeV in general, proper spectral analysis must be made to draw conclusions about an individual burst.
\subsection{Energetics in the prompt and afterglow phases}
\label{sec:enpromptag}
We first investigate the energy output in different time windows. Figure~\ref{fig_T90_FLUX_FLUENCE} allows us to qualitatively compare the energy output during the total duration of each GRB as measured in the $50-300$\,keV (GBM) and 100\,MeV--10\,GeV (LAT) ranges, respectively. In the ``GBM'' duration window, the flux (top left panel) is inversely proportional to the duration. A possible interpretation is that a limited amount of total energy is available for any given GRB. A longer duration then means that the average flux will decrease. Comparing the fluences (top right panel), long and short GRBs are clearly separated, with all but one of the sGRBs having fluence below $10^{-6}$\,erg\,cm$^{-2}$. Within each group, fluence is independent of duration suggesting that the separation arises due to different energy budgets, as might be expected if the groups arise from two different progenitor scenarios. 

In contrast, long and short GRBs show no clear separation in the ``LAT'' duration window (bottom panels). While the populations occupy slightly different regions of the plots, there is a smooth transition between them. This may indicate that the emission in this window is dominated by the afterglow, and that this process is similar regardless of progenitor. In the right panel, the fluence shows a tendency to increase with duration. This could have many explanations: a variable energy budget, meaning more energetic afterglows last longer; an effect of differences in the circumburst medium, leading to varying radiative efficiency; or variations in the viewing angle with respect to the jet. We caution that there is likely also an observational bias: GRBs with favorable conditions can be studied longer; these will then both have a higher number of detected photons (i.e., greater fluence) and a longer measured duration.

Looking closer at the energetics, we can compare the LAT 100\,MeV--10\,GeV fluence to that of the 10\,keV--1\,MeV range in the GBM. We make this comparison in the ``GBM'' time window (left panel of Figure~\ref{fig_LATFLUENCE_GBMFLUENCE}), which can be seen as a proxy for the prompt phase of the burst. We again stress that no joint spectral fits have been made, and that the fluences are calculated based on different spectral fits in each range: The GBM fits are taken from the \gcat while a power-law fit is used for the LAT range (see Sect.~\ref{sec:catanalysis}). In the \fcat, it was found that the high energy fluences from the joint GBM-LAT spectral fit agreed with those from LAT-only analysis, with a small discrepancy seen only for the brightest GRBs. Therefore, we do not expect any significant bias in our results, but a more thorough study needs to be performed before detailed conclusions can be drawn. Such analyses will be presented in future publications.

As found in the \fcat, the bulk of the population emits most of its radiation in the GBM energy range, with the high-energy emission reaching $\lta$20\% of that in the lower band. However, given the larger sample in the \tcat, it is clear that this mainly applies to long GRBs. The short GRBs are more clustered around the line of equality, and several have higher fluence in the 100\,MeV--10\,GeV range than in the 10\,keV--1\,MeV one. We therefore conclude that short GRBs in the prompt phase tend to emit a relatively larger fraction of their energy at higher energies, as compared to long GRBs. 

In the next step we compare the relative energy output in the 100\,MeV--10\,GeV range during the ``GBM'' (prompt) and ``EXT'' (afterglow) time windows of each burst (right panel of Figure~\ref{fig_LATFLUENCE_GBMFLUENCE}). Similar to the results found in the \fcat, most GRBs are clustered around the line of equality, meaning that in the LAT energy range, comparable amounts of energy are released in the prompt and afterglow phases. For long GRBs there is a large fraction where the fluence during the afterglow phase is greater than that during the prompt phase. The short GRBs instead tend to have equal fluence in the two time windows.

In the \fcat, there appeared to be a correlation between the photon index and GBM duration, with shorter bursts tending to have harder spectra. This was coupled to the general property that short GRBs on average have harder spectra. Figure~\ref{fig_GRBindex_GBMT90} shows that this overall trend is not held up in the current larger sample. Furthermore, although the short GRBs on average have slightly harder spectra in the ``GBM'' time window (top left panel), this difference is not seen in the ``EXT'' window (top right panel). This could merely be an indication that the GRB spectra are sometimes ``contaminated'' by emission from lower energies, i.e. the $\beta$ value of the Band function. This value is typically just below $-2$ and the distribution has a long tail to lower values \citep{Gruber2014_211}. The larger number of lGRBs means that we are more likely to see the effect in this population. Comparing the spectral index between the ``GBM'' and ``EXT'' windows, the small difference ($\Delta\Gamma\sim0.5$) might indicate that we are seeing the early afterglow already in the ``GBM'' window. This possibility will be discussed further in Sect.~\ref{sec:interpret}.

The lower left panel of Figure~\ref{fig_GRBindex_GBMT90} shows that the spectral index is not dependent on flux in the ``GBM'' window. Instead, the lower right panel would appear to suggest a correlation between flux and photon index in the ``EXT'' time window, such that stronger emission is coupled to softer spectra. This is especially obvious in bursts with lower flux. However, we caution that this is likely an observational bias. Weaker bursts will only be detected if they have harder spectra.
\subsubsection{Intrinsic energetics}
In the \fcat, there were 10 LAT-detected GRBs with measured redshift. In the \tcat, the number has increased to \nredshift, allowing us to better study their intrinsic properties. Already Figure~\ref{fig_Fluence_Catalogs} suggests that the LAT GRBs are among the brightest. Comparing with the {\it Swift} and GBM samples in Figure~\ref{Eiso_redshift}, it is clear that this is an intrinsic property, and that the LAT preferentially detects brighter GRBs, regardless of redshift. The two panels in Figure~\ref{Eiso_redshift} show that in order for the LAT to detect a GRB at high redshift, such a burst must be intrinsically very bright. For example, all detected GRBs at $z>1$ have $E_{\,iso}>10^{52}$ erg while detecting a GRB at $z>4$ requires $E_{\,iso}>10^{54}$ erg. After 10 years, GRB080916C is still the most distant GRB observed by the LAT, at $z=4.35$.

\subsubsection{The highest energy photons}
\label{Disc_HE_phot}
We now turn to the highest-energy photons detected by the LAT, as presented in Figure~\ref{HEP_distr}. Although less than 80\% of GRBs do not reach above 5\,GeV in the observer frame, the figure indicates that higher energies are more common in the source frame, with $\sim$15\% of GRBs reaching source-frame-corrected energies above 100\,GeV. We note, however, that as there is only one sGRB with measured redshift, the rest-frame distribution is dominated by lGRBs. Interestingly, the fraction of GRBs detected drops smoothly as the threshold energy in the rest frame increases. It is tempting to connect this behavior to the underlying spectral distribution, such that we are seeing the limit determined by the intrinsic spectral shape, which seems to be similar for all GRBs (see Figure~\ref{fig_GRBindex_GBMT90}). In bright bursts more high-energy emission is produced, allowing GeV emission to be observed. Faint bursts will produce too little GeV emission and the LAT will only detect MeV photons. 

\begin{figure}[t!]
\centering
\begin{tabular}{cc}
\includegraphics[width=0.45\columnwidth]{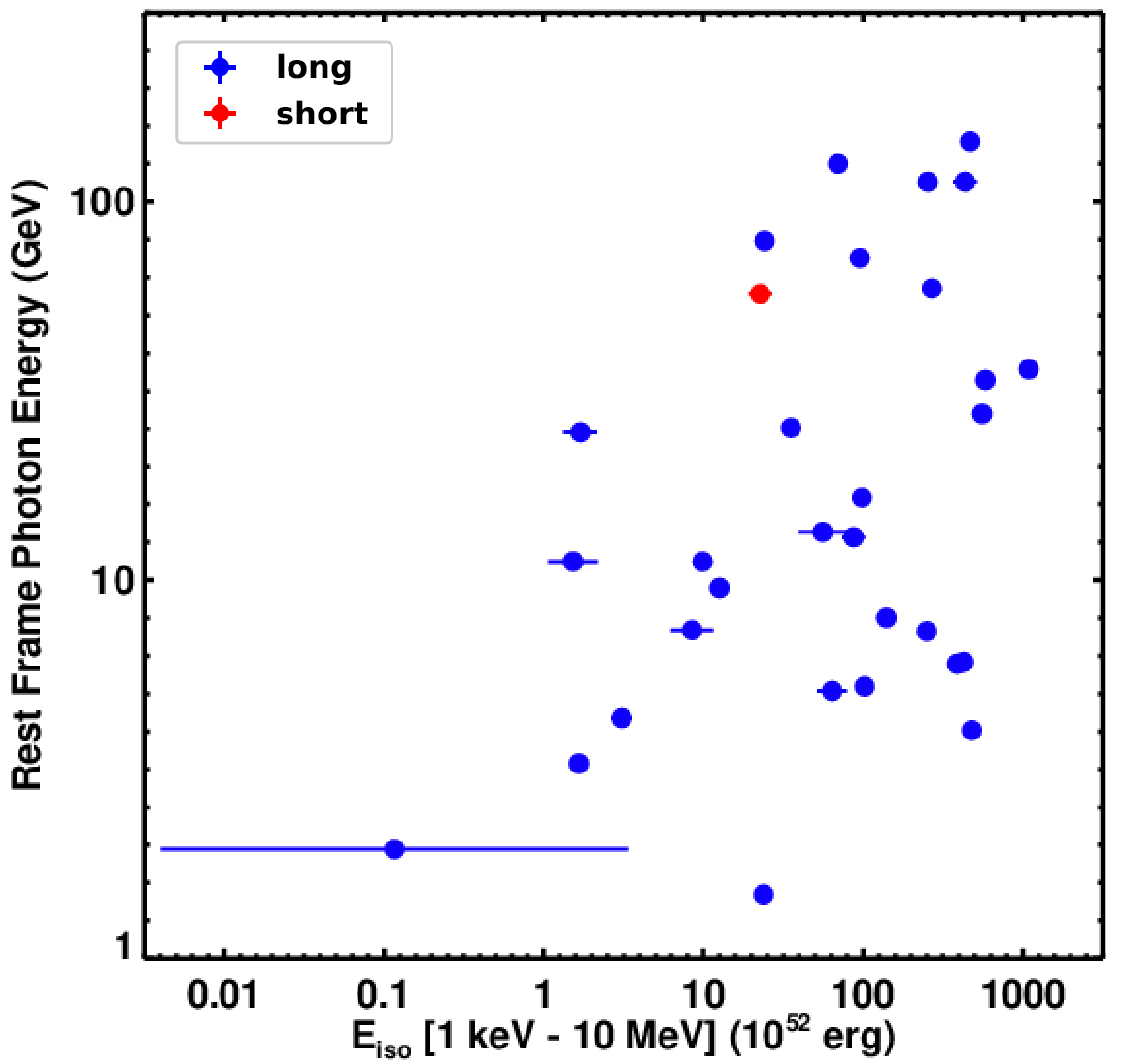} &
\includegraphics[width=0.45\columnwidth]{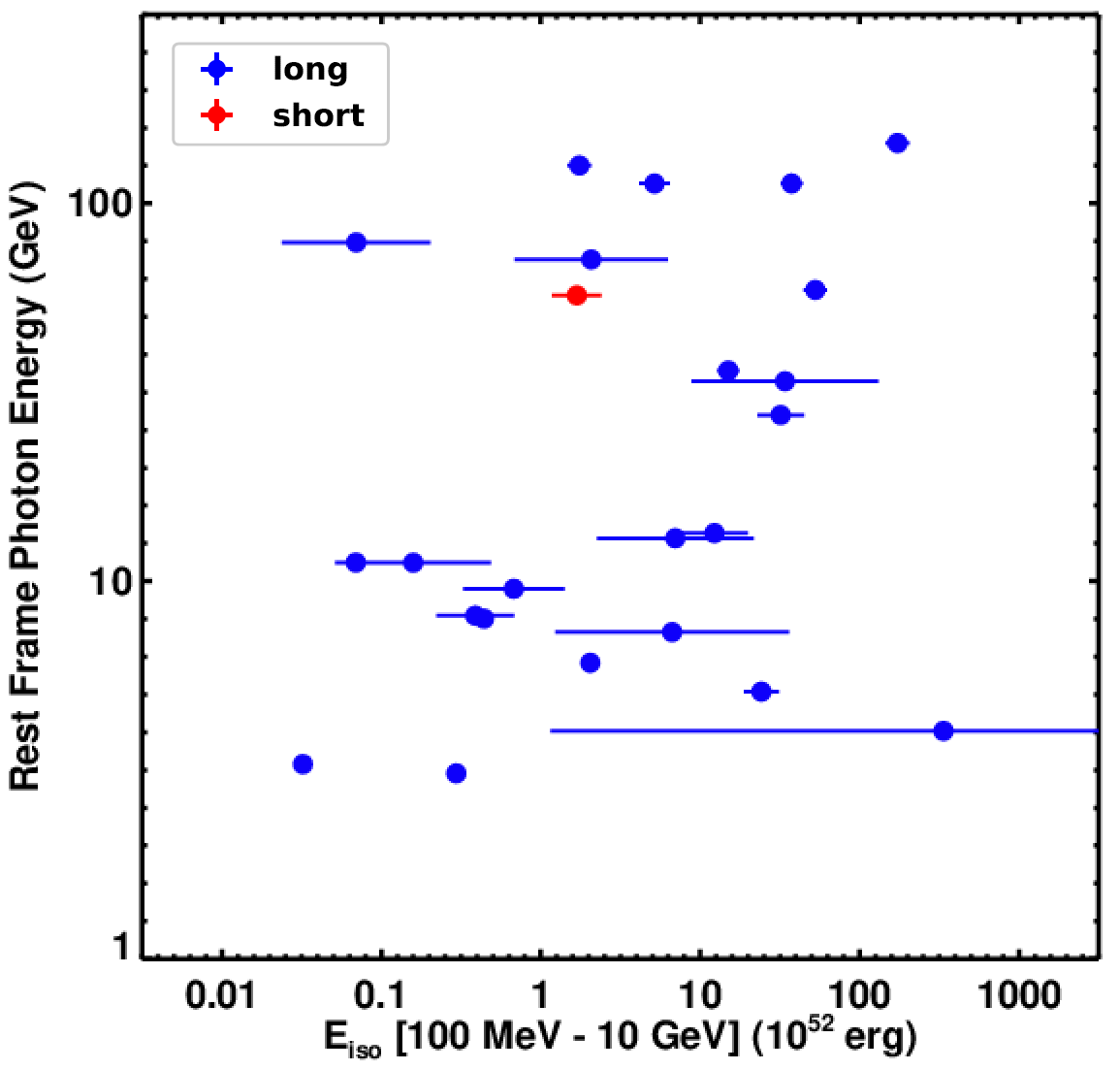}
\end{tabular}
\caption{Energy of the highest-energy photon calculated in the rest frame vs. $E_{\rm iso}$ in the 1 keV--10 MeV (left panel) and in the 100 MeV--10 GeV (right panel) energy ranges for each GRB with redshift estimation. {\it Blue} and {\it red circles} represent long and short GRBs, respectively.}
\label{HEP_En_EIso}
\end{figure}

Although it is clear that the high-energy emission in general can last a long time, the highest-energy photon in a given GRB in some cases arrives relatively soon after the trigger. For example, this photon arrives within $<2$\,s from the trigger time for $\sim50$\%\ of the sGRBs in the left panel of Figure~\ref{HEP_En_Time}. The right panel of Figure~\ref{HEP_En_Time} also shows that also among the GRBs with a maximum photon energy $>10$\,GeV, the highest energy photon is in some cases detected before the emission in the 50--300 keV energy range is over. Particle acceleration in GRBs must be efficient in order to produce such high-energy gamma rays within such a short time. Considering internal opacity of the jet outflow also leads us to conclude that a high bulk Lorentz factor is necessary in order for us to be able to detect these photons. Indeed, these considerations have been used to estimate a Lorentz factor well above 500 in several LAT-detected bursts \citep[see, e.g.,][]{GRB080916C:Science,GRB090902B:Fermi}.

In the \fcat, there was a trend for the photons with highest source-frame energy to appear in the GRBs with highest 1 keV--10 MeV $E_{\rm iso}$ (figure 21 in the \fcat), and the one short GRB with redshift (GRB 090510) did not follow the pattern seen in long GRBs. However, these results were based on only 9 GRBs. With the greater statistics in the \tcat, the trend not only persists but is also extended over a greater range, as shown in the left panel of Figure \ref{HEP_En_EIso}. This indicates that the maximum rest-frame energy produced is simply a function of the total energy output. The right panel instead shows the relation between the highest energy and $E_{\rm iso}$ calculated during the ``GBM'' time window in the LAT energy range (100\,MeV--100\,GeV). In this range, the correlation is less obvious and the points are more scattered. As the energy output in the LAT range is much smaller than at lower energies (see the left panel in Figure~\ref{fig_LATFLUENCE_GBMFLUENCE}), it is perhaps not surprising that the low-energy $E_{\rm iso}$ is a more reliable estimator of the energy budget. However, there is a small number of GRBs that deviate from the general trend, showing low rest-frame energies despite a large $E_{\rm iso}$. More detailed studies are required in order to understand this behavior. In both panels the region of GRB\,090510, which is still the only short GRB with a redshift measurement, is populated also by long GRBs. This again points to similar conditions for long and short GRBs. 
Finally, the high-energy gamma rays from GRBs offer a valuable way to probe more physics than the burst itself. For example, the interaction of $>10$\,GeV gamma rays from sources at cosmological distance with optical and UV photons of the extragalactic background light (EBL) causes absorption of gamma rays, modifying the high-energy part of the spectrum. High-redshift GRBs can thereby serve as a probe of EBL opacity \citep{Desai2017}. For example, EBL attenuation could contribute to the differences between the observer and rest-frame distributions in Figure~\ref{HEP_distr}.
\subsection{High energy emission temporal profile}
\label{sec_HE_em_prof}
In the {\fcat} the study of the temporal decay of the extended emission led to the hypothesis that this is part of an early afterglow component very similar to the afterglow component seen in the X--ray band. The stable decay index with a value around $\alpha=1$ for the small sample of bursts in the \fcat was taken as support for adiabatic expansion within the fireball scenario.
The index reported here remains centered around $\alpha=0.99$. If the extended component is fast cooling afterglow emission from either a radiative or an adiabatic fireball in a constant density environment, the decay indexes (given a photon index of $-2$) would be $10/7$ or $1$, respectively \citep{sari97,Katz+97,2010MNRAS.403..926G}. The peak in the decay index distribution would thus primarily suggest the adiabatic scenario. Nevertheless, the range of measured indexes allows also for a radiative fireball in at least a few individual bursts. 

The luminosity distribution seen in Figure~\ref{Extended_Luminosity_all} shows that the GRBs with known redshift all have similar decay indexes. Yet although the slope of the luminosity decay is similar among the GRBs its distribution versus the elapsed time since the trigger time is scattered (left panel). In the center panel, the time axis for each light curve has been corrected to the time in the rest frame, which reduces the spread. If the luminosity is divided by the isotropic energy of each burst, measured in the 1\,keV--10\,MeV range, the resulting light curves cluster even more (right panel in Figure~\ref{Extended_Luminosity_all}), as reported by \citet{2010MNRAS.403..926G} and \citet{2014MNRAS.443.3578N}. With higher statistics with respect to these studies (which were limited to a few LAT GRBs observed up until 2013), the clustering points to a strong correlation between the isotropic luminosity of the high-energy emission and the prompt isotropic energy calculated in the range 1\,keV--10\,MeV. 

In the analysis performed in Sections~\ref{sec:time-resolved} and \ref{sec:res:temporal_decay}, some light curves are found to be best fit with a spectral break. Such features may be related to the X-ray afterglow plateaus, which are well established but their nature is still debated. To investigate this, we analyze all LAT GRBs with TS$ \ge 64$, with a sufficient number of photons. This choice of TS is dictated by the need of observing the $\gamma$-ray afterglow for a sufficiently long time to allow us to characterize the nature of the deviation from a power law. We then determine whether the light curves can be fitted within the phenomenological \citet{2007ApJ...662.1093W} model, used to fit the X-ray plateaus (the details of the analysis will be presented in Dainotti et al., in preparation).

\begin{figure}[t!]
\centering
\includegraphics[width=0.5\columnwidth]{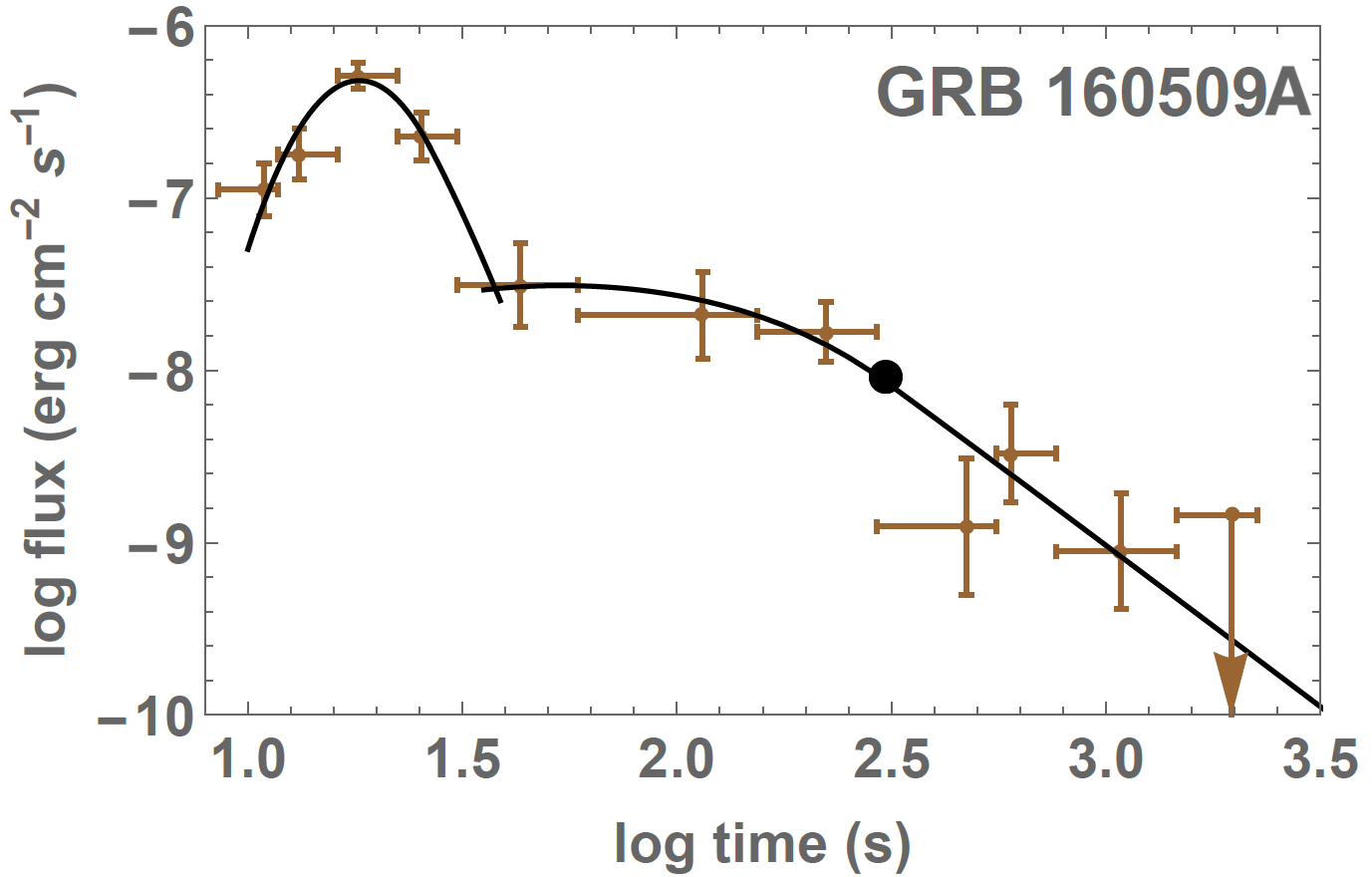}
\caption{Light curve of GRB\,160509A together with the fit using the \citet{2007ApJ...662.1093W} model model (solid line). The black dot indicates the end time of the plateau emission and its corresponding flux.}
\label{fig:160509_plateau}
\end{figure}

We find that 4 GRBs (160509A, 090902B, 150627A, 160816A) show a late-time flattening reminiscent of the X-ray plateaus (their light curves can be seen in Figure~\ref{fig_breaks} and \ref{fig_strangeLC}). The light curve of GRB 160509A, together with the \citet{2007ApJ...662.1093W} model, is presented in Figure~\ref{fig:160509_plateau}. From this resemblance we infer that there is indication of the existence of a plateau in a few {\lat} light curves. Our recovered values for $T_{\gamma}$, which is the time at the end of the plateau emission, are smaller than those typically found in the X-ray regime \citep[from hundreds to a few thousand seconds;][]{2007ApJ...662.1093W}. This suggests that more energetic bursts have smaller plateau durations. Although a one-to-one comparison is difficult due to the lack of corresponding X-ray observations for those bursts, it is nevertheless interesting to mention this trend for future reference and possible new observations.
\subsection{Prospects for GRB detections at VHE}
The detection of VHE emission from GRB 190114C reported by the MAGIC collaboration \citep{2019ATel12390....1M} opened up a new era in the observations of GRBs. The bright source was seen across the electromagnetic spectrum, from radio to VHE. While studies of this burst are certain to yield interesting results, more VHE detections will be necessary to build an understanding of the mechanisms behind such emission.

The Cherenkov Telescope Array \citep[CTA;][]{2011ExA....32..193A,2017arXiv170907997C} will mark an important milestone in the VHE range, covering the whole sky with unprecedented sensitivity. The CTA can steer to a GRB location in less than 30\,s after receiving a trigger notice from a space-based instrument \citep[see, e.g.,][]{2013APh....43..252I, 2013ExA....35..413G}. In an optimistic case, one can therefore assume that the CTA is on target within $\sim$50\,s after the onset of the burst. 

The GRB sample in the \tcat allows us to make a simple and phenomenological estimate of the detectability by VHE instruments. For this exercise, we use the observed LAT spectra as input for CTA simulations, considering only the \nredshiftl lGRBs with redshift measurements. We performed an ad hoc spectral analysis in a two-hour interval starting at \tz+50\,s in the energy range from 100 MeV to 100 GeV. The LAT data are fit using a power-law spectrum. The output spectral parameters (flux, $F_{0}$, and photon index, $\Gamma$), were then used as input for the CTA tools software package {\tt ctools}\footnote{\url{http://cta.irap.omp.eu/ctools/users/index.html}}, version 1.5.2, with the Instrument Response Functions (IRFs) 
Prod2\footnote{\url{https://www.cta-observatory.org/science/cta-performance/}} referred to the CTA configuration in the Southern hemisphere site. The CTA simulation requires as input the GRB location and spectral model, and gives as output a TS map showing whether the source is detected or not. The extrapolation into the CTA range assumes either a power law (PL) or a cut-off power law (CUTPL) spectral model. The background is already taken into account in the used IRFs. We performed simulations over an energy range of 50 GeV to 50 TeV, assuming that the CTA received the LAT detection notice with several different time delays: 0\,h, 0.5\,h, 1\,h, 2\,h, 4\,h, 6\,h, 10\,h. We considered a two-hour CTA observational interval following the notice time, assuming a flux temporal decay of $F = F_{0} \,t ^{-\alpha}$, where ${\alpha}$ is the temporal decay index taken from Table~\ref{tab_likelihoods}. For the 0\,h delay we used all \nredshiftl GRBs, while for all the other analyses the number of cases was limited to the 27 GRBs where $\alpha$ could be constrained.

\begin{figure}[!t]
  \centering
\includegraphics[width=0.5\columnwidth]{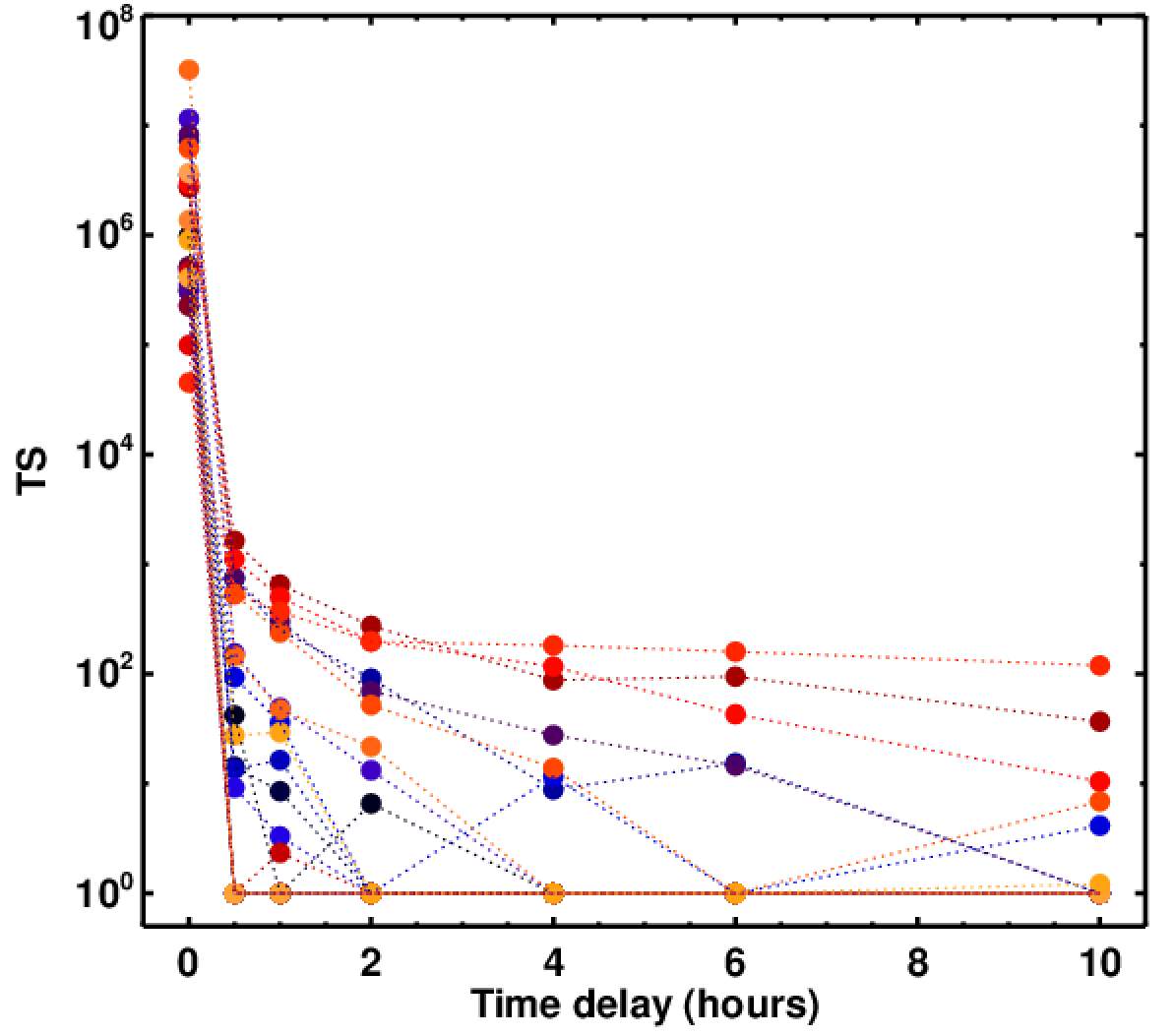}
  \caption{Test Statistic evolution with respect to the time delay (in hours) of the CTA alert reception, obtained from the analysis with a power-law model, with respect to the beginning of the LAT-detected extended emission. The figure shows the TS values for 27 GRBs. For clarity, TS values less than 1 are fixed to 1.}
\label{fig:TS}
\end{figure}

An important factor to consider in the very high energy range ($>50$\,GeV) is that the gamma-ray spectrum can be attenuated by $\gamma$$\gamma$ absorption due to the annihilation of high-energy photons with photons from the EBL. In order to include the EBL attenuation in our simulation, the cutoff energy of the CUTPL model was chosen as the critical energy defined by \citet{GRB090926A}. For the GRBs at redshift $>1$ (20 GRBs), we fixed the critical energy at 30 GeV \citep[following  figure 2 of][]{2012Sci...338.1190A}, while for the others at redshift $<1$ (13 GRBs) the critical energy was chosen following \citet{2012Sci...338.1190A}: $E_{\rm crit}(z) = 170\,(1+z)^{-2.38}$\,GeV. Spectral analyses of the CTA simulated data were then performed using both PL and CUTPL spectral models.

The analyses performed without any time delay (0\,h) reveal that the CTA is able to detect all \nredshiftl GRBs with high TS, as shown in  Figure~\ref{fig:TS} for the PL case. This shows that CTA telescopes would readily detect GRBs given ideal observational conditions during night time, i.e., assuming that the CTA was already on target or started observing immediately after a GRB trigger. However, the LAT data generally require more than six hours to be downlinked and analyzed on the ground. Figure \ref{fig:TS} also shows the TS evolution with respect to the time delay between the LAT detection and the CTA alert reception time in the case of a PL model. The TS values decrease rapidly with time and given a 10\,h delay from the LAT detection, the CTA can detect only two GRBs among 27 (with TS $>25$). In the case of the CUTPL model, all GRBs are detected only in the most optimistic observing condition, and the TS values fall below 25 for all 27 GRBs already with a 0.5\,h delay. These results show the strong need for rapid alerts and precise localizations to enable follow-up observations at very-high energies. 

We stress that the analysis presented here is the result of a very simplified scenario using a single spectral component. Extending this study to give more reliable estimates of the detectability with VHE instruments will require both more realistic models for the high-energy emission mechanisms in GRBs as well as using proper theoretical models of EBL. 
\subsection{LAT sGRBs and implications for GW observations}
The recent detection of GW\,170817 and the coincident GRB\,1701817A clearly showed a coupling between merging neutron stars, gravitational waves and short GRBs. This has made the rapid search for counterparts also in the MeV-GeV range a high priority for the LAT. The LAT collaboration has previously reported on follow-up studies of several GW triggers: GW\,150914 \citep{2016ApJ...823L...2A}, 
LVT\,151012 and GW\,151226 \citep{2017ApJ...835...82R}, and GW\,170817 \citep{2018ApJ...861...85A}. The first three triggers were associated with mergers of two black holes, so no electromagnetic counterpart was expected. In the last case, the location could only be observed by the LAT starting 1100\,s after the GBM trigger time, and again no detection was made. Nevertheless, these events have shown the importance of a dedicated pipeline to search for high-energy counterparts for a larger class of transient events than just GRBs. The LTF pipeline described in Sect.~\ref{sec:ltft} has formed the baseline for these searches. In addition, dedicated searches have been developed to cover various time intervals (also very long ones) around the GW triggers, which are thoroughly described in \citet{2017ApJ...841L..16V}. A particular challenge is the large area given by the LIGO probability map (several hundred square degrees), much larger than for a typical GRB.

As already said, due to observational constraints the LAT could observe the location of GW\,170817/GRB\,170817A only after more than 1 ks. In the sample of \nlikes sGRBs presented here, most events are only detected up to a few seconds after the GBM trigger. However, GRB\,160702A was not seen until $\sim1900$\,s after the trigger and lasted for about 300\,s. Also GRB\,170127C has an unusually long duration: from $\sim650$\,s to $\sim2900$\,s after the trigger. Neither of these events were especially bright (flux $<10^{-5}$ erg/cm$^{-2}$/s), showing that the LAT has the potential to find counterparts even to GW events which occur outside the FoV. Overall, the LAT has found high-energy counterparts to 5\% of the sGRBs seen by the GBM. A critical discriminating factor which has been carefully addressed by \citet{2018ApJ...861...85A} is the fact that the LAT can detect fainter sGRBs the earlier the observation starts, ideally within 100-200 s post trigger. They calculated that the LAT can observe (either a detection or upper bound) $\sim 23$\% of the full-sky sGRB population within 100 s from their GBM trigger. Moreover, assuming that the LAT will have the same efficiency for GRB/GW triggers and a rate of joint GBM/GW events of 1 (2) per year, they obtained a $\sim 5$\% ($\sim 10$\%) probability of detecting one or more GRB/GW with the LAT in one year, respectively. This was derived assuming that GRB/GW events will be representative of the entire GBM-detected sGRB population when observed in gamma-rays. Indeed, two other important factors affecting the detectability of such events by the LAT might be the off-axis viewing angle and the GRB distance.

Recently, \citet{2019arXiv190106158V} presented a search for GRBs in the entire 10-year \Fermi-GBM sample that show similar characteristics to GRB 170817A, i.e., a non-thermal pulse followed by a thermal component. They identified 13 candidates, and concluded that the observed similarities likely arose as a result of the same processes that shaped the gamma-ray signal of GRB 170817A. We examined their sGRB sample, and found that GRB 120915A has been also detected by the LAT and is included in the \tcat\. The short duration in the LAT (\toz $=0.6\, \pm\,0.3$\,s) clearly ties the high-energy emission to the prompt phase. This again shows that the LAT is likely to detect such events as long as they are in the field of view at the time of the GW trigger, as in the case of GRB 120915A (with $\theta=11^\circ$).

\section{Theoretical implications}
\label{sec:interpret}
Observations of GRBs with {\lat} have significantly broadened our understanding of the nature of high-energy emission from these powerful transients, and have led to renewed theoretical activities to model this emission \citep[see, e.g.,][for a review]{2013FrPhy...8..661G}. Already the \fcat hinted at several new prominent characteristics of the high-energy emission: delayed onset compared to the GBM trigger; a hard spectrum, often requiring an additional component when modeling together with GBM data in the prompt phase; variable emission during the early prompt phase; and temporally-extended emission beyond the prompt phase without measurable short timescale variation. Although no joint GBM-LAT spectral fits have been performed for the \tcat, with its greatly enhanced sample the results presented here have firmly established at least the temporal properties as being general to the population as a whole. Modeling these characteristics gives us clues to the particle acceleration sites, environment and radiation processes, while measurements of fluence, redshift and energetics help us to classify the LAT-detected GRBs in the context of the broader GRB population. 

Simultaneous multiwavelength observations of a handful of LAT-detected GRBs, e.g., GRBs 090510 \citep{2010ApJ...716.1178A}; 110731A \citep{2013ApJ...763...71A} and 130427A \citep{2014Sci...343...42A, 2014Sci...343...48M} have further helped constrain the origin of a few high-energy emission features, such as the temporally-extended emission. Regardless of the origin of high-energy emission, merely the fact that we detect it implies that the bulk Lorentz factor of the GRB fireball \citep{2001ApJ...555..540L, 2008ApJ...677...92G, 2012MNRAS.421..525H} is large. In particular, the fireball can be completely transparent to high-energy emission for sufficiently large bulk Lorentz factors \citep{2004ApJ...613.1072R}.
Furthermore, detection of $\gtrsim 10$ GeV photons from GRBs have constrained models of the extragalactic background light \citep{2010ApJ...723.1082A} and the models leading to violation of Lorentz invariance \citep{2009Natur.462..331A}. Finally, the detection of the short GRB 170817A by {\Fermi}-GBM \citep{2017ApJ...848L..14G} in coincidence with the gravitational wave event GW170817 \citep{2017ApJ...848L..13A,2017PhRvL.119p1101A, 2017ApJ...848L..12A} has proven that at least some short GRBs originate from mergers of binary neutron stars. 

The search for the origin of GeV emission from GRBs has intensified since {\Fermi} was launched and the LAT has detected an increasing number of GRBs. The results presented here make these efforts even more promising, as the large sample of the \tcat shows that high-energy emission can be seen also from some fainter GRBs (see Figure~\ref{fig_Fluence_Catalogs}). In the following, we will examine how GeV emission from GRBs can be used to test theoretical predictions. These examples are not exhaustive - rather, they are meant to illustrate various ways in which LAT observations can inform theoretical models.

\subsection{Delayed onset and long duration: challenges for emission models}
\label{subsec:synchro}
Synchrotron-Self-Compton (SSC) models for GeV emission is a natural extension of the synchrotron model for keV-MeV emission from GRBs, in the context of the internal shock scenario \citep[see, e.g.,][]{2010ApJ...716.1178A, 2009ApJ...698L..98W, 2009A&A...498..677B}. However, in a simple SSC model the Compton emission can be delayed compared to synchrotron emission by at most the flux variability time scale ($\lesssim 1$~s for long GRBs). Such models therefore cannot account for the $\gg 1$~s delay observed in a large fraction of LAT GRBs (see Figure~\ref{fig_GBMT05_LATT05}). SSC emission from late internal shocks, with much longer variability time at GeV than in MeV, can explain delayed LAT onset in some cases \citep[see, e.g.,][]{2009ApJ...707.1404T, 2011A&A...526A.110D}. 

A possible alternative for GeV emission during the prompt phase is hadronic emission, either proton-synchrotron radiation and radiation from $e^\pm$ pair cascades \citep{2010OAJ.....3..150R} or photohadronic interactions and associated cascade radiation \citep{2009ApJ...705L.191A}. In contrast to many SSC models, hadronic models can account for delayed onset of GeV emission via a hard power-law component with its flux increasing over time due to proton acceleration and cooling of cascade particles. Hadronic models, however, require $\gtrsim 100$ times more energy in the hadrons than in leptons which is challenging to produce in GRBs. Furthermore, neither SSC nor hadronic emission mechanisms active during the prompt phase are able to explain the extended GeV emission beyond the keV-MeV \tn.

The delayed onset, smooth light curve and temporally extended emission $\gtrsim 100$\,MeV have led to the development of an early afterglow interpretation of high-energy radiation from the external forward shock \citep{2009MNRAS.400L..75K, 2010MNRAS.403..926G, 2010ApJ...716.1178A, 2010ApJ...724L.109R, 2010ApJ...720.1008C}. The power-law spectrum and temporally extended emission are then interpreted as synchrotron radiation from the external shock. Modeling of simultaneous multiwavelength (optical to GeV) data has shown this approach to generally fit observations for several GRBs, such as GRB 090510, 110731A and 130427A. However, detection of $\gtrsim 10$ GeV photons at late times from a number of GRBs (see Figure~\ref{HEP_En_Time}) is not easy to explain with the synchrotron interpretation (nor are the VHE photons from GRB 190114C seen by MAGIC). The maximum photon energy from electron-synchrotron radiation is $h\nu_{\rm max} \approx 0.24~\phi^{-1} (1+z)^{-1}\Gamma (t)$ GeV, where $\phi^{-1} \lesssim 1$ is the acceleration efficiency for electrons and $\Gamma (t)$ is the bulk Lorentz factor \citep[see, e.g.,][]{2010ApJ...724L.109R}. 

In light of the difficulties to simultaneously match both the delayed onset and long duration of GeV emission, alternative models have also been proposed. For instance, several authors have pointed out that the prompt MeV emission should cause the creation of $e^\pm$ pairs in the medium ahead of the blast wave \citep[e.g.,][]{2000ApJ...538..105T, 2001ApJ...554..660M, 2002ApJ...565..808B}. This pair loading led \citet{2005ApJ...618L..13B} to suggest that GeV emission arises due to inverse Compton scattering of the prompt MeV radiation streaming through the external medium.

In \citet{2014ApJ...788...36B}, this scenario was shown to predict a delayed onset of $\sim 1-10$\,s, in agreement with the majority of LAT GRBs. However, it is unclear if also the much longer ($>100$\,s) delays seen in a few GRBs can be explained. This would require either a very low Lorentz factor of the blast wave (not easy to reconcile with GeV emission) or a suppression of the blast-wave luminosity until it has reached a very large radius. Likewise, while the model can explain a GeV duration several times greater than that at lower energies, it may face difficulties for the LAT GRBs where the GeV duration is more than an order of magnitude longer (see panel (c) of Figure~\ref{fig_GBMT05_LATT05}). Observational tests so far have also predicted rather large densities of the circumburst medium (comparable to stellar winds of $\sim 10^{-5}$\,\msun\,yr$^{-1}$), which may prove challenging as more sGRBs are detected, as binary neutron stars are not expected to have a dense surrounding medium.
\subsection{Fluence and energetics of LAT GRBs: probing the central engine}
LAT GRBs constitute a subsample of the GBM GRBs with a flux distribution peaking at a higher value than the complete GBM sample (see Figure~\ref{fig_Fluence_Catalogs}). This is also true for the fluence as evident from the same figure. On the other hand, the redshift distribution of the LAT GRBs is rather typical of Swift-detected GRBs (see Figure~\ref{fig_redshift_dist}). As a result, the isotropic energy release and peak luminosity (Figure~\ref{Eiso_redshift}) are also typically higher for the LAT GRBs across the redshifts. These results have implications for the GRB central engines.

The nature of the GRB central engine is a fundamental question in astrophysics. A newly-born magnetar is one of the two leading hypotheses (the other is a newly-born black hole). The vacuum/force-free electromagnetic spin-down power of a neutron star with surface (equatorial) dipole magnetic field $B_{\rm dip}$ and spin period $P$ is given by \citep{2006ApJ...648L..51S}:
\begin{equation}
{\dot E}_{\rm FF} \approx \frac{4\pi^4 B_{\rm dip}^2 R^6}{P^4c^3} \sim 10^{49} \left( \frac{B_{\rm dip}}{10^{15}~{\rm G}} \right)^2 \left( \frac{P}{1~{\rm ms}} \right)^{-4} \left( \frac{R}{10~{\rm km}} \right)^6 ~{\rm erg/s},
\end{equation}
where $R$ is the neutron star radius. The maximum isotropic $\gamma$-ray power is then $\sim 10^{49}$-$10^{51}$~erg/s, after taking into account a jet-beaming factor of up to $\sim 200$ for a $\sim 6^\circ$ jet opening angle. In comparison, the measured peak luminosity for a number of LAT-detected GRBs exceeds these values, e.g., GRB 080916C ($\approx 6\times 10^{53}$~erg/s) and GRBs 090510 and 160625B ($\approx 4\times 10^{53}$~erg/s). The observations of LAT GRBs therefore significantly constrain the parameter space of the models with a magnetar as the GRB central engine \citep[see, e.g.,][]{1992ApJ...392L...9D, 1992Natur.357..472U, 2011MNRAS.413.2031M}.
\FloatBarrier
\subsection{Closure relations: test of the standard external forward shock model}
\label{sec_clos_rel}

As described in Sect.~\ref{subsec:synchro}, a possible explanation for the temporally extended emission is synchrotron emission from an external forward shock developed by collisions between outgoing ejecta and the circumburst medium.
According to the model, a power-law distribution of electrons radiates synchrotron emission, and the spectrum and corresponding light curve are described as a series of broken power laws \citep[e.g.][]{Sari1998,Chevalier2000,Zhang2006,Gao2013}. The indexes of each segment of both spectrum and corresponding light curve are either constant or a function of the electron spectral index, and the relation between the spectral and temporal indexes is called the ``closure relation''. The closure relations have different forms depending on the physical conditions, such as the electron cooling regime, surrounding environment profile, electron spectral index, and jet geometry. 

A systematic analysis of 59 GRBs testing a set of closure relations against temporal and spectral indexes computed from LAT extended emission was performed in Tak et al. (2019, submitted). Among the analyzed GRBs, 81\% of them can be classified by a set of closure relations. A third of these allow both interstellar medium (ISM) and wind environments, while the remainder favor an ISM environment. The observed spectra and light curves of the extended emission in LAT GRBs are thus generally consistent with the standard external forward shock model in most cases, but there is still a significant fraction which cannot be explained by this model. The GRBs which do not match any closure relation tend to show a slower temporal decay ($\alpha <  1$). This implies the need for continuous energy injection or other physical sources to sustain the fluxes.

As already mentioned in Sect.~\ref{sec_HE_em_prof}, some light curves show a break or even hints of a plateau at late times. Such features can cause problems for the models used above, and instead a more phenomenological approach needs to be adopted. Using the parameters derived from fitting the light curves with the \citet{2007ApJ...662.1093W} model, we tested also these GRBs against the closure relations. Unfortunately, due to the large uncertainties no conclusive results could be reached at least at this stage. Thus, further analysis and future observations are required to cast light on this topic. Additional tests in other theoretical frameworks \citep[e.g., the models proposed by][]{2014ApJ...788...36B,2014ApJ...789..159I} will be the object of a forthcoming paper.

\section{Summary and conclusions}
We have here presented the second {\lat} GRB catalog. Compared to the \fcat there are several changes and improvements to the analysis, beyond the three times longer timescale:

\begin{itemize}
\item Pass 8 data have replaced Pass 6 data.
\item Over the past two years, a new detection algorithm has been developed at energies above 100\,MeV.
\item A Bayesian blocks algorithm has been employed to look for detections at 30-100\,MeV.
\item All available triggers (both GBM and other space-based missions) were used in the search for LAT counterparts.
\end{itemize}

All potential detections have been processed by a dedicated and standardized analysis. This differs both from the analysis performed for the \fcat as well as the standard LAT real-time analysis. In total we find \ngrb detections, a much larger sample than the 35 GRBs of the \fcat. 

The results presented in the \tcat have strengthened some of the conclusions from the \fcat, most notably the temporal characteristics. A major difference from the \fcat is a new definition of the duration, where we consider the first and last detected photon. Nevertheless, the properties of later onset and longer duration for the high-energy emission remain unchanged. Other trends suggested by the \fcat, most notably regarding the energetics and differences between long and short GRBs, have been shown not to hold with the larger sample in the \tcat.

While this catalog describes the temporal properties of the detections at both keV and MeV energy ranges, the spectral analysis is limited to the LAT range (100\,MeV--100\,GeV). This has shown to be problematic in some cases, such as when testing for breaks in the spectrum. These results are not surprising, as the division between the energy ranges is purely artifical. Nevertheless, a joint analysis is beyond the scope of this paper, and will instead be the focus of a dedicated study presented in a separate publication.

The larger sample in this catalog is also reflected in the \nredshift GRBs with measured redshift. This has allowed us to better investigate the energetics also in the rest frame, probing both the highest-energy photons as well as the closure relations. The remarkable alignment seen in the decay of the luminosity further points to a strong correlation between isotropic luminosity in the rest frame MeV to GeV range, and isotropic energy as measured at lower energies.

The \tcat presents the results of the first 10 years of \lat; It represents an initial step in understanding the processes behind high-energy emission from gamma-ray bursts. It is our hope that the {\Fermi} data presented here will lead to a large number of follow-up analyses and continue to drive the evolution of theoretical models. Within the \lat\ collaboration, several such studies are already underway: focused studies on the closure relations, energetics and the connection to VHE emission. The large sample of the \tcat will also allow more complete analysis of the detectability of MeV emission from GRBs as well as detailed comparison between the energy output in the LAT range during the prompt vs afterglow phases. The collaboration is also in the process of setting up an online repository, similar to the analysis of the \tcat, which will automatically be updated with each new GRB detection. The past few years have truly shown the strength of \Fermi\ observations, and its importance will continue for many years to come. This is especially true as the era of multi-messenger studies of GRBs has just begun.
\begin{acknowledgments}
The {\lat} Collaboration acknowledges generous ongoing support
from a number of agencies and institutes that have supported both the
development and the operation of the LAT as well as scientific data analysis.
These include the National Aeronautics and Space Administration and the
Department of Energy in the United States, the Commissariat \`a l'Energie Atomique and the Centre National de la Recherche Scientifique / Institut National de Physique Nucl\'eaire et de Physique des Particules in France, the Agenzia Spaziale Italiana and the Istituto Nazionale di Fisica Nucleare in Italy, the Ministry of Education, Culture, Sports, Science and Technology (MEXT), High Energy Accelerator Research Organization (KEK) and Japan Aerospace Exploration Agency (JAXA) in Japan, and the K.~A.~Wallenberg Foundation, the Swedish Research Council and the Swedish National Space Agency in Sweden.

Additional support for science analysis during the operations phase is gratefully acknowledged from the Istituto Nazionale di Astrofisica in Italy and the Centre National d'\'Etudes Spatiales in France. This work performed in part under DOE Contract DE-AC02-76SF00515. M. Axelsson gratefully acknowledges funding from the European Union's Horizon 2020 research and innovation programme under the Marie Sk{\l}odowska-Curie grant agreement No 734303 (NEWS).

\end{acknowledgments}
\appendix
\section{Tables}
\subsection{Sample of LAT bursts}
The complete list of the \lat\ 10-year catalog GRBs is presented in Table~\ref{tab_GRBs}. The first column gives the GRB name as it appeared in the corresponding GCN. The following two columns display the GRB trigger date (in UTC) and time (in Mission Elapsed Time, or MET, in seconds from 2001 January 1 UT 00:00:00). Columns 4 through 6 report the best reconstructed position using only \lat\ data. Coordinates are expressed in the equatorial frame of epoch J2000.  The localization error (column 6) is the error obtained from the LTF analysis. The boresight angle $\theta$ and the zenith angle $\zeta$ at trigger time are given in columns 7 and 8. Column 9 indicates the presence or absence of an ARR.  The detection significance of the LTF and LLE analysis are listed in columns 10 and 11. The LTF TS value corresponds to the highest TS obtained during one of the four time window listed in Tab.~\ref{tab_intervals_fit}. LLE-only events, for which no position calculation was possible, report the best location available for that event. The reference  to the burst location is then given in column 12. When available, we report the redshift of the host galaxy in column 13, together with its corresponding reference (column 14).

\begin{longrotatetable}
\startlongtable
\begin{deluxetable}{lrrrrrrrrrrrrr}
\tabletypesize{\tiny}
\tablecolumns{14}
\tablewidth{0pt}
\tablecaption{Sample of \textit{Fermi}-LAT GRBs, from August 2008 to August 2018. For LLE only GRBs, the localizations are from:\\ GCN13845: \citep{2012GCN.13845....1R}, GCN14257: \citep{2013GCN.14257....1C}, GCN17810:\citep{2015GCN.17810....1K}, GCN18403:\citep{2015GCN.18403....1G},GCN19927:\citep{2016GCN.19927....1E}.\label{tab_GRBs}}
\tablehead{
\colhead{ GRB NAME} & \colhead{DATE} & \colhead{Trigger Time} & \colhead{R.A.} & \colhead{Dec.} & \colhead{Loc. Err.} & \colhead{$\theta$} & \colhead{$\zeta$} & \colhead{ARR} & \colhead{LTF TS} & \colhead{LLE Sig.} & \colhead{Loc. GCN} & \colhead{Redshift} & \colhead{GCN} \\ 
\colhead{ } & \colhead{(UT)} & \colhead{(MET)} & \colhead{Deg., J2000} & \colhead{Deg., J2000} & \colhead{Deg.} & \colhead{Deg.} & \colhead{\de} & \colhead{} & \colhead{} & \colhead{$\sigma$ } & \colhead{} & \colhead{} & \colhead{}}
\startdata
 080818B  &    2008-08-18 22:40:49  & 240792050.080 &     109.46  &       15.0  & 0.47 &  56.4  &  75.0  & 0 & 26 & - &  & $-$ &  \\
 080825C  &    2008-08-25 14:13:48  & 241366429.110 &     234.83  &      -3.84  & 0.65 &  60.2  &  26.3  & 0 & 56 & 8 &  & $-$ &  \\
 080916C  &    2008-09-16 00:12:45  & 243216766.620 &     119.89  &      -56.7  & 0.07 &  49.0  &  83.8  & 0 & 2200 & 46 &  & 4.35 & \citep{2009AA...498...89G} \\
  081006  &    2008-10-06 14:29:34  & 244996175.170 &      135.0  &     -61.95  & 0.38 &  10.8  &  40.3  & 0 & 97 & - &  & $-$ &  \\
  081009  &    2008-10-09 03:20:58  & 245215259.060 &     251.57  &      18.92  & 0.28 &  69.6  &  55.4  & 0 & 50 & - &  & $-$ &  \\
 081024B  &    2008-10-24 21:22:40  & 246576161.860 &     323.01  &      20.84  & 0.31 &  19.2  &  30.6  & 0 & 83 & 8 &  & $-$ &  \\
 081102B  &    2008-11-02 08:45:00  & 247308301.510 &     212.95  &      30.33  & 0.58 &  48.2  &  47.6  & 0 & 26 & - &  & $-$ &  \\
 081122A  &    2008-11-22 12:28:12  & 249049693.210 &     338.82  &      37.57  & 0.27 &  21.5  &  25.9  & 1 & 38 & - &  & $-$ &  \\
\enddata
\tablenotetext{a}{File used: LAT2CATALOG-v14-LTF }
\end{deluxetable}

\end{longrotatetable}
\subsection{Comparison of duration estimation}
Table \ref{tab_durations} presents the different burst duration estimations. The GRB class (listed in column 2; L = long, S = short) is derived from the GBM duration in the 50--300 keV energy range (long bursts have \tn$>$2 s and short bursts have \tn$<$2 s, respectively). All GBM values (\tf, \tnf and \tn, columns 3 to 5) are taken from the \gcat. The LLE onset times and durations (\tllf, \tllnf and \tlln), listed in columns 6 to 8, are calculated in the 30 MeV--1 GeV energy range. The LAT onset time \tz, end time \tone and duration \toz, calculated in the 100 MeV--100 GeV energy range, are given in columns 9 to 11).
%
\startlongtable
\begin{deluxetable}{lrrrrrrrrrr}
\tabletypesize{\tiny}
\tablecolumns{11}
\tablewidth{0pt}
\tablecaption{Comparison between duration estimators for the detected \Fermi-LAT GRBs.\label{tab_durations}}
\tablehead{
\colhead{ GRB NAME} & \colhead{CLASS} & \colhead{T$_{\rm GBM,05}$} & \colhead{T$_{\rm GBM,95}$} & \colhead{T$_{\rm GBM,90}$} & \colhead{T$_{\rm LLE,05}$} & \colhead{T$_{\rm LLE,95}$} & \colhead{T$_{\rm LLE,90}$} & \colhead{T$_{\rm LAT,0}$} & \colhead{T$_{\rm LAT,1}$} & \colhead{T$_{\rm LAT,100}$} \\ 
\colhead{ } & \colhead{} & \colhead{s } & \colhead{s } & \colhead{s } & \colhead{s } & \colhead{s } & \colhead{s } & \colhead{s } & \colhead{s } & \colhead{s }}
\startdata
080818B & (L) & $-$0.51  & 12.86 & 13.38 & $-$ & $-$ & $-$ & 9018.9 & 9482.8 & 500 $\pm$ 200 \\
080825C & (L) & 1.2  & 22.21 & 20.99 & 2.84 & 4.05 & 1.20 & 3.1 & 173.5 & 170.4 $\pm$ 0.7 \\
080916C & (L) & 1.3  & 64.26 & 62.98 & 1.00 & 46.55 & 45.55 & 3.0 & 1531.8 & 1500 $\pm$ 200 \\
081006 & (L) & $-$0.26  & 6.14 & 6.40 & $-$ & $-$ & $-$ & 0.8 & 62.3 & 60 $\pm$ 20 \\
081009 & (L) & 1.3  & 42.69 & 41.34 & $-$ & $-$ & $-$ & 67.8 & 1250.1 & 1200 $\pm$ 200 \\
081024B & (S) & $-$0.064  & 0.58 & 0.64 & $-$0.04 & 1.28 & 1.32 & 0.2 & 2.2 & 2.02 $\pm$ 0.04 \\
081102B & (S) & $-$0.064  & 1.66 & 1.73 & $-$ & $-$ & $-$ & 0.3 & 0.4 & 0.15 $\pm$ 0.03 \\
081122A & (L) & $-$0.26  & 23.04 & 23.30 & $-$ & $-$ & $-$ & 23.5 & 66.5 & 40 $\pm$ 20 \\
081203A & (L) & * & 214.00 & 214.00 & $-$ & $-$ & $-$ & 291.9 & 379.7 & 90 $\pm$ 10 \\
081224 & (L) & 0.74  & 17.18 & 16.45 & $-$ & $-$ & $-$ & 76.5 & 339.2 & 260 $\pm$ 40 \\
090102 & (L) & 1.5  & 28.16 & 26.62 & $-$ & $-$ & $-$ & 3915.9 & 4404.8 & 500 $\pm$ 300 \\
090217 & (L) & 0.83  & 34.11 & 33.28 & 1.96 & 15.50 & 13.54 & 4.1 & 71.6 & 68 $\pm$ 10 \\
090227A & (L) & 0.003  & 16.19 & 16.19 & 1.15 & 11.05 & 9.90 & 2.9 & 204.3 & 201.4 $\pm$ 0.2 \\
090227B & (S) & $-$0.016  & 0.29 & 0.30 & 0.04 & 1.18 & 1.14 & $-$ & $-$ & - \\
090228A & (S) & * & 0.45 & 0.45 & $-$ & $-$ & $-$ & 0.1 & 2.1 & 2.0 $\pm$ 0.1 \\
090323 & (L) & 8.7  & 142.59 & 133.89 & 7.24 & 93.41 & 86.18 & 9.3 & 5321.6 & 5312 $\pm$ 8 \\
090328 & (L) & 4.4  & 66.05 & 61.70 & 8.05 & 37.02 & 28.97 & 14.4 & 6150.6 & 6140 $\pm$ 70 \\
090427A & (L) & * & 15.00 & 15.00 & $-$ & $-$ & $-$ & 422.9 & 435.6 & 10 $\pm$ 10 \\
090510 & (S) & $-$0.048  & 0.91 & 0.96 & 0.54 & 5.93 & 5.40 & 0.0 & 170.0 & 170 $\pm$ 2 \\
090531B & (S) & * & 0.77 & 0.77 & $-$0.07 & 0.00 & 0.07 & $-$ & $-$ & - \\
090626 & (L) & 1.5  & 50.43 & 48.90 & 21.48 & 22.37 & 0.89 & 9.0 & 557.3 & 550 $\pm$ 90 \\
090720B & (L) & $-$0.26  & 10.50 & 10.75 & $-$ & $-$ & $-$ & 0.1 & 1.7 & 1.6 $\pm$ 0.4 \\
090902B & (L) & 2.8  & 22.14 & 19.33 & $-$0.65 & 17.23 & 17.88 & 0.5 & 884.2 & 880 $\pm$ 60 \\
090926A & (L) & 2.2  & 15.94 & 13.76 & 1.15 & 16.71 & 15.56 & 2.2 & 4419.5 & 4420 $\pm$ 50 \\
091003 & (L) & 0.83  & 21.06 & 20.22 & $-$ & $-$ & $-$ & 1.0 & 392.0 & 390 $\pm$ 10 \\
091031 & (L) & 1.4  & 35.33 & 33.92 & $-$1.53 & 15.09 & 16.62 & 0.1 & 408.2 & 410 $\pm$ 30 \\
091120 & (L) & 1.0  & 51.20 & 50.18 & $-$ & $-$ & $-$ & 31.8 & 803.9 & 770 $\pm$ 20 \\
091127 & (L) & 0.003  & 8.70 & 8.70 & $-$ & $-$ & $-$ & 8.6 & 35.4 & 30 $\pm$ 10 \\
100116A & (L) & 0.58  & 103.11 & 102.53 & 89.94 & 96.59 & 6.66 & 77.5 & 730.4 & 650 $\pm$ 50 \\
100213C & (L) & * & 60.00 & 60.00 & $-$ & $-$ & $-$ & 2707.2 & 3389.0 & 700 $\pm$ 300 \\
100225A & (L) & $-$0.26  & 12.74 & 12.99 & 2.75 & 9.74 & 6.99 & 6.6 & 1012.4 & 1010 $\pm$ 20 \\
100325A & (L) & $-$0.38  & 6.72 & 7.10 & $-$ & $-$ & $-$ & 0.2 & 1.2 & 0.9 $\pm$ 0.2 \\
100414A & (L) & 1.9  & 28.35 & 26.50 & $-$ & $-$ & $-$ & 18.7 & 5506.1 & 5490 $\pm$ 30 \\
100423B & (L) & 1.6  & 18.11 & 16.51 & $-$ & $-$ & $-$ & 166.7 & 180.6 & 14 $\pm$ 7 \\
100511A & (L) & 0.83  & 43.27 & 42.43 & $-$ & $-$ & $-$ & 11.6 & 6338.2 & 6300 $\pm$ 700 \\
100620A & (L) & 0.19  & 52.03 & 51.84 & $-$ & $-$ & $-$ & 3.8 & 42.8 & 39 $\pm$ 6 \\
100724B & (L) & 8.2  & 122.88 & 114.69 & 2.63 & 102.44 & 99.81 & 9.1 & 53.6 & 45 $\pm$ 6 \\
100728A & (L) & 13  & 178.69 & 165.38 & $-$ & $-$ & $-$ & 248.6 & 1340.5 & 1090 $\pm$ 40 \\
100826A & (L) & 8.7  & 93.70 & 84.99 & 18.20 & 39.28 & 21.09 & 33.5 & 61.4 & 28 $\pm$ 6 \\
101014A & (L) & 1.4  & 450.82 & 449.42 & 208.72 & 210.38 & 1.67 & 2270.7 & 4196.0 & 1930 $\pm$ 50 \\
101107A & (L) & 2.3  & 378.12 & 375.81 & $-$ & $-$ & $-$ & 134.0 & 239.4 & 105 $\pm$ 4 \\
101123A & (L) & 41  & 145.41 & 103.94 & 43.43 & 46.13 & 2.70 & $-$ & $-$ & - \\
101227B & (L) & 0.77  & 154.12 & 153.35 & $-$ & $-$ & $-$ & 17.3 & 23.8 & 7 $\pm$ 3 \\
110120A & (L) & 0.003  & 26.18 & 26.17 & 0.29 & 0.53 & 0.23 & 0.6 & 1112.8 & 1100 $\pm$ 200 \\
110123A & (L) & 0.70  & 18.56 & 17.86 & $-$ & $-$ & $-$ & 51.6 & 527.0 & 480 $\pm$ 20 \\
110213A & (L) & $-$0.77  & 33.54 & 34.30 & $-$ & $-$ & $-$ & 1261.4 & 1944.2 & 700 $\pm$ 400 \\
110328B & (L) & 2.6  & 86.53 & 83.97 & $-$3.05 & 23.23 & 26.27 & 102.5 & 737.2 & 630 $\pm$ 60 \\
110428A & (L) & 2.7  & 8.32 & 5.63 & $-$ & $-$ & $-$ & 7.3 & 393.5 & 386.2 $\pm$ 0.3 \\
110518A & (L) & * & 35.00 & 35.00 & $-$ & $-$ & $-$ & 1968.2 & 2364.2 & 400 $\pm$ 200 \\
110529A & (S) & $-$0.13  & 0.38 & 0.51 & 0.01 & 0.11 & 0.09 & $-$ & $-$ & - \\
110625A & (L) & 3.8  & 30.72 & 26.88 & $-$ & $-$ & $-$ & 205.9 & 577.2 & 400 $\pm$ 100 \\
110721A & (L) & 0.003  & 21.82 & 21.82 & $-$0.75 & 5.25 & 6.00 & 0.0 & 120.6 & 121 $\pm$ 3 \\
110728A & (S) & $-$0.13  & 0.58 & 0.70 & $-$ & $-$ & $-$ & 0.4 & 3.0 & 2.6 $\pm$ 0.9 \\
110731A & (L) & 0.003  & 7.49 & 7.49 & 2.11 & 8.37 & 6.26 & 1.1 & 436.0 & 430 $\pm$ 10 \\
110903A & (L) & $-$0.26  & 341.00 & 341.25 & $-$ & $-$ & $-$ & 46.7 & 370.1 & 320 $\pm$ 50 \\
110921B & (L) & 0.90  & 18.56 & 17.66 & 17.07 & 18.01 & 0.93 & 7.4 & 321.6 & 310 $\pm$ 50 \\
111210B & (L) & * & 60.00 & 60.00 & $-$ & $-$ & $-$ & 6.5 & 394.0 & 390 $\pm$ 70 \\
120107A & (L) & 0.064  & 23.10 & 23.04 & $-$ & $-$ & $-$ & 1.3 & 48.8 & 50 $\pm$ 10 \\
120226A & (L) & 4.4  & 57.34 & 52.99 & 4.46 & 33.38 & 28.93 & 29.4 & 283.6 & 250 $\pm$ 20 \\
120316A & (L) & 1.5  & 28.16 & 26.62 & 11.31 & 11.71 & 0.40 & 15.0 & 545.0 & 530 $\pm$ 30 \\
120328B & (L) & 3.8  & 33.54 & 29.70 & 4.51 & 7.25 & 2.74 & $-$ & $-$ & - \\
120420B & (L) & 0.003  & 254.92 & 254.91 & $-$ & $-$ & $-$ & 3501.7 & 3908.0 & 410 $\pm$ 10 \\
120526A & (L) & 3.1  & 46.72 & 43.65 & $-$ & $-$ & $-$ & 692.2 & 3306.3 & 2600 $\pm$ 200 \\
120624B & (L) & $-$260  & 14.34 & 271.36 & 5.76 & 9.12 & 3.36 & 73.7 & 1103.9 & 1030 $\pm$ 30 \\
120709A & (L) & $-$0.13  & 27.20 & 27.33 & $-$0.12 & 0.34 & 0.46 & 0.1 & 695.9 & 700 $\pm$ 10 \\
120711A & (L) & 62  & 106.50 & 44.03 & $-$ & $-$ & $-$ & 393.3 & 5431.6 & 5000 $\pm$ 800 \\
120729A & (L) & $-$1.0  & 24.45 & 25.47 & $-$ & $-$ & $-$ & 396.9 & 432.0 & 40 $\pm$ 10 \\
120830A & (S) & * & 0.90 & 0.90 & $-$ & $-$ & $-$ & 0.7 & 10.7 & 10.0 $\pm$ 0.2 \\
120911B & (L) & * & 69.00 & 69.00 & 7.14 & 35.37 & 28.23 & 9.2 & 217.8 & 209 $\pm$ 2 \\
120915A & (S) & $-$0.32  & 0.26 & 0.58 & $-$ & $-$ & $-$ & 0.2 & 0.8 & 0.6 $\pm$ 0.3 \\
120919B & (L) & 2.0  & 120.07 & 118.02 & $-$ & $-$ & $-$ & 121.2 & 605.3 & 500 $\pm$ 200 \\
121011A & (L) & 1.0  & 66.82 & 65.79 & $-$0.63 & 7.21 & 7.84 & $-$ & $-$ & - \\
121029A & (L) & $-$0.90  & 14.91 & 15.81 & $-$ & $-$ & $-$ & 119.0 & 1926.5 & 1810 $\pm$ 70 \\
121123B & (L) & 2.3  & 44.80 & 42.50 & $-$ & $-$ & $-$ & 268.0 & 1651.3 & 1400 $\pm$ 200 \\
121225B & (L) & 9.5  & 67.97 & 58.50 & 16.90 & 21.11 & 4.21 & $-$ & $-$ & - \\
130305A & (L) & 1.3  & 26.88 & 25.60 & 5.31 & 22.09 & 16.78 & $-$ & $-$ & - \\
130310A & (L) & 4.1  & 20.10 & 16.00 & 4.13 & 4.21 & 0.08 & 67.2 & 560.1 & 490 $\pm$ 30 \\
130325A & (L) & 0.77  & 7.68 & 6.91 & $-$ & $-$ & $-$ & 324.1 & 1029.7 & 700 $\pm$ 200 \\
130327B & (L) & 2.0  & 33.28 & 31.23 & $-$ & $-$ & $-$ & 8.0 & 523.2 & 520 $\pm$ 10 \\
130427A & (L) & 4.1  & 142.34 & 138.24 & $-$0.12 & 46.19 & 46.30 & 0.1 & 34366.2 & 34400 $\pm$ 300 \\
130502B & (L) & 7.2  & 31.49 & 24.32 & 13.80 & 15.37 & 1.56 & 12.4 & 1316.6 & 1300 $\pm$ 200 \\
130504C & (L) & 8.7  & 81.92 & 73.22 & 13.54 & 70.90 & 57.36 & 42.6 & 590.3 & 550 $\pm$ 20 \\
130518A & (L) & 9.9  & 58.50 & 48.58 & 23.00 & 37.71 & 14.72 & 26.8 & 343.6 & 320 $\pm$ 10 \\
130606B & (L) & 5.4  & 57.60 & 52.23 & $-$ & $-$ & $-$ & 130.5 & 527.3 & 400 $\pm$ 40 \\
130702A & (L) & 0.77  & 59.65 & 58.88 & $-$ & $-$ & $-$ & 272.3 & 1283.3 & 1000 $\pm$ 100 \\
130821A & (L) & 3.6  & 90.62 & 87.04 & 24.79 & 34.07 & 9.28 & 33.9 & 6103.7 & 6000 $\pm$ 2000 \\
130828A & (L) & 13  & 150.27 & 136.96 & 35.08 & 68.78 & 33.70 & 17.1 & 616.8 & 600 $\pm$ 10 \\
130907A & (L) & * & 210.00 & 210.00 & $-$ & $-$ & $-$ & 3618.4 & 4010.9 & 400 $\pm$ 300 \\
131014A & (L) & 0.96  & 4.16 & 3.20 & 1.68 & 3.49 & 1.81 & 1.9 & 200.4 & 199 $\pm$ 5 \\
131029A & (L) & 1.0  & 105.47 & 104.45 & $-$ & $-$ & $-$ & 38.7 & 557.2 & 520 $\pm$ 50 \\
131108A & (L) & 0.32  & 18.50 & 18.18 & 0.04 & 10.99 & 10.96 & 0.0 & 678.1 & 680 $\pm$ 40 \\
131209A & (L) & 2.8  & 16.38 & 13.57 & $-$ & $-$ & $-$ & 14.3 & 374.8 & 360 $\pm$ 60 \\
131216A & (L) & 0.003  & 19.27 & 19.26 & $-$0.19 & 0.09 & 0.28 & $-$ & $-$ & - \\
131231A & (L) & 13  & 44.54 & 31.23 & 21.45 & 34.25 & 12.80 & 23.1 & 4824.2 & 4800 $\pm$ 700 \\
140102A & (L) & 0.45  & 4.10 & 3.65 & 2.38 & 5.25 & 2.87 & 3.1 & 60.2 & 60 $\pm$ 10 \\
140104B & (L) & 11  & 198.15 & 187.14 & $-$ & $-$ & $-$ & 227.3 & 1174.0 & 900 $\pm$ 200 \\
140110A & (L) & $-$0.26  & 9.22 & 9.47 & 0.15 & 6.13 & 5.98 & 0.6 & 159.4 & 159 $\pm$ 8 \\
140124A & (L) & $-$13  & 108.87 & 121.54 & $-$ & $-$ & $-$ & 93.6 & 468.2 & 370 $\pm$ 90 \\
140206B & (L) & 7.5  & 154.18 & 146.69 & 5.88 & 33.48 & 27.60 & 0.6 & 8585.0 & 8584.43 $\pm$ 0.01 \\
140219A & (L) & $-$2.6  & 74.50 & 77.06 & $-$ & $-$ & $-$ & 959.8 & 1802.0 & 840 $\pm$ 50 \\
140323A & (L) & 5.1  & 116.48 & 111.43 & $-$ & $-$ & $-$ & 226.8 & 715.6 & 500 $\pm$ 100 \\
140402A & (S) & $-$0.13  & 0.19 & 0.32 & $-$0.31 & 1.24 & 1.55 & 0.1 & 71.7 & 72 $\pm$ 5 \\
140416A & (L) & $-$2.8  & 28.96 & 31.74 & $-$ & $-$ & $-$ & 1174.8 & 2207.4 & 1000 $\pm$ 300 \\
140523A & (L) & 0.58  & 19.78 & 19.20 & $-$ & $-$ & $-$ & 1.6 & 470.2 & 470 $\pm$ 20 \\
140528A & (L) & 1.0  & 14.59 & 13.57 & $-$ & $-$ & $-$ & 81.1 & 1377.9 & 1300 $\pm$ 300 \\
140619B & (L) & $-$0.26  & 2.56 & 2.82 & 0.17 & 1.30 & 1.12 & 0.2 & 5.3 & 5 $\pm$ 1 \\
140723A & (L) & * & 56.32 & 56.32 & $-$0.31 & 20.51 & 20.82 & 0.6 & 103.8 & 103 $\pm$ 1 \\
140729A & (L) & 0.51  & 56.06 & 55.55 & $-$ & $-$ & $-$ & 4.8 & 73.4 & 68.6 $\pm$ 0.4 \\
140810A & (L) & 6.7  & 88.32 & 81.67 & $-$ & $-$ & $-$ & 460.8 & 18827.5 & 18400 $\pm$ 100 \\
140825A & (L) & * & 14.00 & 14.00 & $-$ & $-$ & $-$ & 1463.0 & 1702.3 & 200 $\pm$ 100 \\
140928A & (L) & $-$11  & 7.17 & 17.92 & $-$ & $-$ & $-$ & 1676.6 & 2554.7 & 880 $\pm$ 40 \\
141012A & (L) & $-$26  & 11.78 & 37.63 & $-$ & $-$ & $-$ & 3.1 & 107.0 & 100 $\pm$ 10 \\
141028A & (L) & 6.7  & 38.15 & 31.49 & 6.86 & 30.75 & 23.89 & 10.9 & 500.5 & 490 $\pm$ 80 \\
141102A & (L) & $-$0.064  & 2.56 & 2.62 & $-$ & $-$ & $-$ & 3.5 & 57.7 & 54 $\pm$ 1 \\
141113A & (S) & $-$0.064  & 0.38 & 0.45 & $-$ & $-$ & $-$ & 0.1 & 3.5 & 3.4 $\pm$ 0.1 \\
141207A & (L) & 1.3  & 22.27 & 20.99 & 1.69 & 12.17 & 10.48 & 1.8 & 734.3 & 730 $\pm$ 10 \\
141221B & (L) & $-$1.3  & 31.23 & 32.51 & $-$ & $-$ & $-$ & 22.2 & 58.5 & 36 $\pm$ 7 \\
141222A & (L) & * & 2.75 & 2.75 & $-$0.03 & 1.26 & 1.29 & 34.2 & 440.6 & 410 $\pm$ 10 \\
150118B & (L) & 7.7  & 47.87 & 40.19 & 22.12 & 23.59 & 1.47 & 4.3 & 51.0 & 50 $\pm$ 20 \\
150202B & (L) & 1.5  & 168.96 & 167.43 & 4.86 & 17.51 & 12.65 & 44.9 & 115.7 & 70 $\pm$ 50 \\
150210A & (L) & 0.003  & 31.30 & 31.29 & 0.03 & 2.55 & 2.52 & 0.8 & 169.4 & 170 $\pm$ 20 \\
150314A & (L) & 0.61  & 11.30 & 10.69 & $-$0.17 & 0.03 & 0.20 & 0.1 & 3064.3 & 3100 $\pm$ 100 \\
150403A & (L) & 3.3  & 25.60 & 22.27 & 9.97 & 14.08 & 4.11 & 399.5 & 970.5 & 600 $\pm$ 200 \\
150416A & (L) & 0.51  & 33.79 & 33.28 & $-$3.28 & 19.07 & 22.36 & $-$ & $-$ & - \\
150510A & (L) & 0.38  & 52.29 & 51.90 & 0.03 & 16.41 & 16.38 & 2.4 & 172.5 & 170 $\pm$ 10 \\
150513A & (L) & $-$160  & 1.79 & 158.98 & $-$157.75 & $-$146.39 & 11.36 & $-$ & $-$ & - \\
150514A & (L) & 0.003  & 10.82 & 10.81 & $-$ & $-$ & $-$ & 442.4 & 597.6 & 200 $\pm$ 200 \\
150523A & (L) & 1.8  & 84.22 & 82.43 & 2.92 & 43.39 & 40.48 & 3.9 & 6129.1 & 6130 $\pm$ 80 \\
150627A & (L) & 5.3  & 69.89 & 64.58 & 4.51 & 18.09 & 13.58 & 152.0 & 6143.0 & 6000 $\pm$ 1000 \\
150702A & (L) & 1.3  & 47.17 & 45.83 & $-$ & $-$ & $-$ & 512.6 & 1853.3 & 1340 $\pm$ 20 \\
150820A & (L) & $-$0.77  & 5.12 & 5.89 & $-$1.06 & $-$0.64 & 0.42 & $-$ & $-$ & - \\
150902A & (L) & 3.8  & 17.41 & 13.57 & 3.02 & 12.61 & 9.60 & 3.5 & 408.7 & 405 $\pm$ 2 \\
151006A & (L) & 1.5  & 94.98 & 93.44 & $-$1.12 & 15.02 & 16.14 & $-$ & $-$ & - \\
160101B & (L) & 0.003  & 22.02 & 22.01 & $-$1.26 & 3.50 & 4.76 & $-$ & $-$ & - \\
160310A & (L) & $-$0.26  & 25.60 & 25.86 & $-$ & $-$ & $-$ & 99.2 & 432.2 & 300 $\pm$ 200 \\
160314B & (L) & $-$3.8  & 94.72 & 98.56 & $-$ & $-$ & $-$ & 133.4 & 1285.4 & 1200 $\pm$ 200 \\
160325A & (L) & 2.0  & 44.99 & 42.95 & 13.82 & 14.02 & 0.20 & 5.0 & 1555.7 & 1600 $\pm$ 500 \\
160422A & (L) & 0.83  & 13.12 & 12.29 & $-$ & $-$ & $-$ & 769.6 & 1061.4 & 300 $\pm$ 300 \\
160503A & (L) & $-$2.6  & 26.11 & 28.67 & $-$ & $-$ & $-$ & 5324.7 & 23075.4 & 17750 $\pm$ 70 \\
160509A & (L) & 8.2  & 377.86 & 369.67 & 8.81 & 25.12 & 16.31 & 9.6 & 5687.4 & 6000 $\pm$ 1000 \\
160521B & (L) & 0.32  & 3.14 & 2.82 & $-$ & $-$ & $-$ & 90.2 & 2199.4 & 2100 $\pm$ 300 \\
160623A & (L) & $-$1.3  & 106.50 & 107.78 & $-$ & $-$ & $-$ & 401.5 & 35069.0 & 35000 $\pm$ 1000 \\
160625B & (L) & 190  & 641.84 & 453.38 & 184.86 & 220.27 & 35.41 & 25.6 & 840.5 & 810 $\pm$ 20 \\
160702A & (S) & * & 0.20 & 0.20 & $-$ & $-$ & $-$ & 1941.5 & 2215.6 & 300 $\pm$ 100 \\
160709A & (L) & * & 5.44 & 5.44 & 0.41 & 3.62 & 3.21 & 0.5 & 25.8 & 25 $\pm$ 2 \\
160816A & (L) & 0.38  & 11.46 & 11.07 & 0.45 & 10.91 & 10.46 & 1.1 & 1094.8 & 1090 $\pm$ 70 \\
160821A & (L) & 120  & 161.54 & 43.01 & 117.29 & 152.21 & 34.92 & 92.1 & 1459.2 & 1370 $\pm$ 40 \\
160829A & (S) & $-$0.064  & 0.45 & 0.51 & $-$ & $-$ & $-$ & 0.9 & 31.7 & 31 $\pm$ 6 \\
160905A & (L) & 3.8  & 37.38 & 33.54 & 5.57 & 28.92 & 23.34 & 7.8 & 463.3 & 460 $\pm$ 20 \\
160910A & (L) & 4.6  & 28.93 & 24.32 & 7.70 & 18.47 & 10.78 & 86.4 & 216.4 & 130 $\pm$ 50 \\
160917B & (L) & $-$4.6  & 9.98 & 14.59 & $-$0.90 & 5.55 & 6.45 & $-$ & $-$ & - \\
160917A & (L) & $-$0.26  & 19.20 & 19.46 & $-$0.12 & 0.33 & 0.44 & $-$ & $-$ & - \\
161015A & (L) & 0.26  & 15.36 & 15.10 & $-$ & $-$ & $-$ & 1.1 & 7.5 & 6.4 $\pm$ 0.2 \\
161109A & (L) & 3.8  & 27.39 & 23.55 & $-$ & $-$ & $-$ & 423.4 & 890.9 & 500 $\pm$ 200 \\
170115B & (L) & 0.51  & 44.80 & 44.29 & 0.55 & 11.58 & 11.03 & 1.3 & 1027.9 & 1027 $\pm$ 10 \\
170127C & (S) & * & 0.13 & 0.13 & $-$ & $-$ & $-$ & 664.8 & 2889.0 & 2220 $\pm$ 50 \\
170214A & (L) & 13  & 135.43 & 122.88 & 6.54 & 84.10 & 77.56 & 39.5 & 752.0 & 713 $\pm$ 8 \\
170228A & (L) & 1.6  & 62.24 & 60.67 & $-$ & $-$ & $-$ & 6.5 & 72.4 & 70 $\pm$ 10 \\
170306B & (L) & 4.6  & 23.55 & 18.94 & $-$ & $-$ & $-$ & 21.2 & 71.5 & 50 $\pm$ 10 \\
170329A & (L) & 0.26  & 33.79 & 33.54 & $-$ & $-$ & $-$ & 3.9 & 52.4 & 48 $\pm$ 7 \\
170405A & (L) & 7.4  & 86.02 & 78.59 & 16.56 & 51.51 & 34.95 & 17.8 & 868.0 & 850 $\pm$ 60 \\
170409A & (L) & 29  & 93.44 & 64.00 & $-$ & $-$ & $-$ & 178.4 & 440.3 & 260 $\pm$ 30 \\
170424A & (L) & 2.8  & 56.06 & 53.25 & 2.42 & 37.62 & 35.19 & 21.8 & 107.7 & 86 $\pm$ 2 \\
170510A & (L) & 2.8  & 130.56 & 127.75 & $-$1.03 & 27.68 & 28.71 & 26.8 & 154.2 & 127 $\pm$ 5 \\
170522A & (L) & 0.58  & 8.00 & 7.42 & $-$ & $-$ & $-$ & 2.8 & 41.6 & 40 $\pm$ 10 \\
170728B & (L) & * & 46.34 & 46.34 & $-$0.01 & 0.40 & 0.41 & 0.2 & 469.3 & 469.070 $\pm$ 0.010 \\
170808B & (L) & 4.1  & 21.76 & 17.66 & 6.47 & 16.77 & 10.31 & 13.7 & 6205.9 & 6190 $\pm$ 40 \\
170813A & (L) & $-$0.51  & 111.36 & 111.87 & $-$ & $-$ & $-$ & 9.5 & 265.3 & 260 $\pm$ 70 \\
170825B & (L) & $-$0.51  & 6.14 & 6.66 & $-$ & $-$ & $-$ & 0.9 & 1.5 & 0.6 $\pm$ 0.3 \\
170906A & (L) & 12  & 90.88 & 78.85 & $-$ & $-$ & $-$ & 162.1 & 1983.2 & 1821 $\pm$ 3 \\
170921B & (L) & 1.0  & 40.38 & 39.36 & $-$ & $-$ & $-$ & 901.7 & 1058.6 & 200 $\pm$ 200 \\
171010A & (L) & 17  & 123.91 & 107.27 & $-$ & $-$ & $-$ & 335.6 & 2984.8 & 2650 $\pm$ 50 \\
171011C & (S) & $-$0.45  & 0.03 & 0.48 & $-$ & $-$ & $-$ & 0.1 & 42.7 & 40 $\pm$ 20 \\
171102A & (L) & 7.7  & 56.06 & 48.38 & $-$ & $-$ & $-$ & 34.9 & 401.0 & 366 $\pm$ 10 \\
171120A & (L) & 0.003  & 44.06 & 44.06 & $-$ & $-$ & $-$ & 0.3 & 5276.0 & 5280 $\pm$ 10 \\
171124A & (L) & $-$0.70  & 25.47 & 26.18 & $-$ & $-$ & $-$ & 3.4 & 321.3 & 318 $\pm$ 7 \\
171210A & (L) & 3.6  & 146.69 & 143.11 & $-$1.19 & 7.01 & 8.20 & 5.9 & 1518.1 & 1510 $\pm$ 70 \\
171212B & (L) & $-$1.0  & 30.98 & 32.00 & $-$ & $-$ & $-$ & 116.6 & 497.0 & 400 $\pm$ 100 \\
180113C & (L) & 5.4  & 29.95 & 24.58 & 5.85 & 29.54 & 23.69 & $-$ & $-$ & - \\
180210A & (L) & 4.1  & 43.01 & 38.91 & $-$ & $-$ & $-$ & 23.1 & 1621.1 & 1600 $\pm$ 200 \\
180305A & (L) & 1.5  & 14.59 & 13.06 & 2.48 & 5.38 & 2.89 & 1613.8 & 2038.0 & 400 $\pm$ 400 \\
180526A & (L) & * & 87.00 & 87.00 & $-$ & $-$ & $-$ & 808.9 & 1948.6 & 1140 $\pm$ 50 \\
180703A & (L) & 1.5  & 22.27 & 20.74 & 0.93 & 8.05 & 7.12 & 3.8 & 1614.0 & 1610 $\pm$ 80 \\
180703B & (S) & 0.13  & 1.66 & 1.54 & $-$ & $-$ & $-$ & 8.0 & 78.5 & 70 $\pm$ 20 \\
180718B & (L) & 1.6  & 99.91 & 98.31 & $-$ & $-$ & $-$ & 1.1 & 16.2 & 15 $\pm$ 3 \\
180720B & (L) & 4.4  & 53.25 & 48.90 & 0.91 & 31.47 & 30.55 & 11.8 & 625.0 & 613 $\pm$ 9 \\
\enddata
\tablenotetext{a}{File used: LAT2CATALOG-v14-LTF }
\end{deluxetable}

\subsection{Likelihood analysis in different time window}
Table~\ref{tab_likelihoods} reports the likelihood analysis results in the different time windows. The definition of the time windows is explained in Table~\ref{tab_intervals_fit}.  
The time range of each interval in which we perform the analysis is given in column 2. Columns 3 and 4 display the number of selected events in the ROI and the number of events predicted by the best fit model obtained after likelihood maximization, respectively. The test statistic (TS) is given in the column 6. Note that only intervals with TS$>$20 are displayed in the table. The remaining columns contain the value of the spectral index (6), flux (7) and fluence (7) in each interval, and, when possible, the isotropic energy calculated in the 100 MeV--100 GeV rest frame energy band (column 8).
\begin{longrotatetable}
\startlongtable
\begin{deluxetable}{lrrrrrrrr}
\tabletypesize{\tiny}
\tablecolumns{9}
\tablewidth{0pt}
\tablecaption{Likelihood analysis\label{tab_likelihoods}}
\tablehead{
\colhead{ GRB NAME} & \colhead{Interval (t$_{0}$-t$_{1}$)} & \colhead{Trans. Ev.} & \colhead{Trans. Ev.} & \colhead{Test Statistic} & \colhead{Spectral Index} & \colhead{Flux} & \colhead{Fluence} & \colhead{E$\rm_{iso}$ (100 MeV--10 GeV)} \\ 
\colhead{ } & \colhead{s} & \colhead{in the ROI} & \colhead{Predicted} & \colhead{(TS)} & \colhead{ } & \colhead{cm$^{-2}$ s$^{-1}$ ($\times 10^{-5}$)} & \colhead{erg cm$^{-2}$ ($\times 10^{-5}$)} & \colhead{erg ($\times 10^{52}$)}}
\startdata
080818B &  LTF (0.0--10000.0) & 65 & 5.0 & 23 & $-$1.8 $\pm$ 0.4 & 0.08 $\pm$ 0.05 & 2 $\pm$ 2 & - \\
\hline
080825C &  GBM (1.2--22.2) & 11 & 9.6 & 51 & $-$3.8 $\pm$ 0.8 & 21 $\pm$ 7 & 0.11 $\pm$ 0.04 & - \\
 &  LTF (0.0--100.0) & 23 & 18.8 & 87 & $-$3.0 $\pm$ 0.4 & 8 $\pm$ 2 & 0.26 $\pm$ 0.08 & - \\
 &  EXT (22.2--173.5) & 14 & 10.1 & 23 & $-$2.4 $\pm$ 0.4 & 3 $\pm$ 1 & 0.2 $\pm$ 0.1 & - \\
 &  LAT (3.1--173.5) & 25 & 19.8 & 86 & $-$2.8 $\pm$ 0.4 & 5 $\pm$ 1 & 0.29 $\pm$ 0.10 & - \\
\hline
080916C &  GBM (1.3--64.3) & 255 & 246.8 & 2344 & $-$2.22 $\pm$ 0.07 & 91 $\pm$ 6 & 4.0 $\pm$ 0.6 & 170 $\pm$ 10 \\
 &  LTF (0.0--100.0) & 290 & 277.7 & 2546 & $-$2.21 $\pm$ 0.07 & 64 $\pm$ 4 & 4.5 $\pm$ 0.7 & 190 $\pm$ 10 \\
 &  EXT (64.3--1531.8) & 186 & 106.0 & 352 & $-$2.2 $\pm$ 0.1 & 1.6 $\pm$ 0.2 & 1.8 $\pm$ 0.4 & 68 $\pm$ 9 \\
\enddata
\tablenotetext{a}{File used: LAT2CATALOG-v14-LTF }
\end{deluxetable}

\end{longrotatetable}
\subsection{Temporally extended high-energy emission}
Table~\ref{tab_extended} displays the temporally extended high-energy emission analysis results. Column 2 and 3 give the peak flux and peak flux time. These are followed by the fit parameters of the simple power law ($\alpha$, column 4) and broken power law ($\alpha_1$, column 5, $\alpha_2$, column 6, and $T_b$, column 7). Column 8 reports the best fit decay index, indicating if the SPL or the BPL is the preferred model. The last column indicates the total duration of the extended emission.
\begin{longrotatetable}
\startlongtable
\begin{deluxetable}{lrrrrrrrr}
\tabletypesize{\tiny}
\tablecolumns{9}
\tablewidth{0pt}
\tablecaption{Temporally extended high-energy emission\label{tab_extended}}
\tablehead{
\colhead{ GRB NAME} & \colhead{Peak Flux} & \colhead{PeakFlux Time} & \colhead{Decay Index (SPL)} & \colhead{Decay Index 1 (BPL)} & \colhead{Decay Index 2 (BPL)} & \colhead{Break Time    (BPL)} & \colhead{Decay Index} & \colhead{Extended Emission Duration} \\ 
\colhead{ } & \colhead{cm$^{-2}$ s$^{-1}$ ($\times 10^{-5}$)} & \colhead{(sec)} & \colhead{ } & \colhead{ } & \colhead{ } & \colhead{(sec)} & \colhead{ } & \colhead{ s }}
\startdata
080825C & 21 $\pm$ 8 & 10 $\pm$ 20 & 1.45 $\pm$ 0.03 & - & - & - & 1.45 $\pm$ 0.03 & 170.4  \\
080916C & 1800 $\pm$ 700 & 5.87 $\pm$ 0.08 & 1.1 $\pm$ 0.1 & 1.6 $\pm$ 0.2 & 0.4 $\pm$ 0.3 & 3.1 $\pm$ 0.3 & 0.4 $\pm$ 0.3 & 1528.8  \\
081009 & 1.9 $\pm$ 1.0 & 100 $\pm$ 200 & 0.9 $\pm$ 0.2 & - & - & - & 0.9 $\pm$ 0.2 & 1182.4  \\
090323 & 13 $\pm$ 4 & 90 $\pm$ 30 & 1.1 $\pm$ 0.1 & 1 $\pm$ 2 & 1.1 $\pm$ 0.2 & 2.4 $\pm$ 0.4 & 1.1 $\pm$ 0.1 & 5312.3  \\
090328 & 16 $\pm$ 5 & 20 $\pm$ 50 & 0.99 $\pm$ 0.09 & 0.7 $\pm$ 0.4 & 1.1 $\pm$ 0.1 & 2.4 $\pm$ 0.5 & 0.99 $\pm$ 0.09 & 6136.3  \\
090510 & 6000 $\pm$ 3000 & 0.85 $\pm$ 0.01 & 1.81 $\pm$ 0.08 & 2.3 $\pm$ 0.2 & 1.3 $\pm$ 0.2 & 0.6 $\pm$ 0.2 & 1.3 $\pm$ 0.2 & 169.9  \\
090626 & 4 $\pm$ 1 & 100 $\pm$ 40 & 0.9 $\pm$ 0.2 & - & - & - & 0.9 $\pm$ 0.2 & 548.3  \\
090902B & 1300 $\pm$ 500 & 8.0 $\pm$ 0.1 & 1.63 $\pm$ 0.08 & 1.9 $\pm$ 0.2 & 1.2 $\pm$ 0.2 & 2.2 $\pm$ 0.3 & 1.2 $\pm$ 0.2 & 883.7  \\
090926A & 7000 $\pm$ 3000 & 9.91 $\pm$ 0.02 & 1.39 $\pm$ 0.08 & 1.8 $\pm$ 0.2 & 1.1 $\pm$ 0.2 & 2.0 $\pm$ 0.3 & 1.1 $\pm$ 0.2 & 4417.3  \\
091003 & 4 $\pm$ 2 & 20 $\pm$ 40 & 0.9 $\pm$ 0.2 & 0.7 $\pm$ 0.4 & 2 $\pm$ 1 & 2.5 $\pm$ 0.4 & 0.9 $\pm$ 0.2 & 391.0  \\
091031 & 30 $\pm$ 10 & 1 $\pm$ 3 & 1.3 $\pm$ 0.2 & - & - & - & 1.3 $\pm$ 0.2 & 408.1  \\
091120 & 0.8 $\pm$ 0.3 & 200 $\pm$ 300 & 0.54 $\pm$ 0.09 & - & - & - & 0.54 $\pm$ 0.09 & 772.1  \\
100116A & 7 $\pm$ 2 & 110 $\pm$ 20 & 2.7 $\pm$ 0.2 & - & - & - & 2.7 $\pm$ 0.2 & 652.9  \\
100414A & 40 $\pm$ 10 & 30 $\pm$ 30 & 1.3 $\pm$ 0.1 & 1.8 $\pm$ 0.2 & 0.3 $\pm$ 0.6 & 2.9 $\pm$ 0.3 & 0.3 $\pm$ 0.6 & 5487.5  \\
100423B & 0.3 $\pm$ 0.4 & 0 $\pm$ 200 & 0.2 $\pm$ 0.1 & - & - & - & 0.2 $\pm$ 0.1 & 13.9  \\
100511A & 1.1 $\pm$ 0.4 & 100 $\pm$ 200 & 0.58 $\pm$ 0.07 & - & - & - & 0.58 $\pm$ 0.07 & 6326.6  \\
100728A & 0.3 $\pm$ 0.1 & 300 $\pm$ 600 & 0.8 $\pm$ 0.5 & - & - & - & 0.8 $\pm$ 0.5 & 1091.8  \\
101014A & 0.05 $\pm$ 0.03 & 2500 $\pm$ 1000 & $-$0.2 $\pm$ 0.3 & - & - & - & $-$0.2 $\pm$ 0.3 & 1925.3  \\
110120A & 1.7 $\pm$ 0.7 & 50 $\pm$ 70 & 0.6 $\pm$ 0.4 & - & - & - & 0.6 $\pm$ 0.4 & 1112.3  \\
110428A & 9 $\pm$ 5 & 12 $\pm$ 10 & 1.0 $\pm$ 0.1 & - & - & - & 1.0 $\pm$ 0.1 & 386.2  \\
110518A & 0.11 $\pm$ 0.06 & 2000 $\pm$ 1000 & 0.7 $\pm$ 0.5 & - & - & - & 0.7 $\pm$ 0.5 & 396.0  \\
110625A & 8 $\pm$ 4 & 260 $\pm$ 30 & 0.6 $\pm$ 0.3 & - & - & - & 0.6 $\pm$ 0.3 & 371.4  \\
110721A & 600 $\pm$ 200 & 0.1 $\pm$ 0.2 & 1.0 $\pm$ 0.1 & - & - & - & 1.0 $\pm$ 0.1 & 120.6  \\
110731A & 1400 $\pm$ 500 & 5.63 $\pm$ 0.08 & 1.5 $\pm$ 0.1 & 1.8 $\pm$ 0.1 & 0 $\pm$ 1 & 2.4 $\pm$ 0.5 & 0 $\pm$ 1 & 434.8  \\
111210B & 1.0 $\pm$ 0.4 & 100 $\pm$ 200 & 0.5 $\pm$ 0.2 & - & - & - & 0.5 $\pm$ 0.2 & 387.5  \\
120226A & 1.8 $\pm$ 0.9 & 30 $\pm$ 60 & 1.21 $\pm$ 0.07 & - & - & - & 1.21 $\pm$ 0.07 & 254.3  \\
120316A & 2.0 $\pm$ 0.8 & 40 $\pm$ 60 & 0.9 $\pm$ 0.2 & - & - & - & 0.9 $\pm$ 0.2 & 530.0  \\
120526A & 0.4 $\pm$ 0.2 & 900 $\pm$ 500 & 0.7 $\pm$ 0.1 & - & - & - & 0.7 $\pm$ 0.1 & 2614.1  \\
120624B & 8 $\pm$ 3 & 380 $\pm$ 20 & 1.2 $\pm$ 0.3 & - & - & - & 1.2 $\pm$ 0.3 & 1030.2  \\
120709A & 60 $\pm$ 30 & 0 $\pm$ 2 & 0.7 $\pm$ 0.1 & - & - & - & 0.7 $\pm$ 0.1 & 695.8  \\
120711A & 2.1 $\pm$ 0.7 & 400 $\pm$ 200 & 1.6 $\pm$ 0.2 & - & - & - & 1.6 $\pm$ 0.2 & 5038.3  \\
120911B & 500 $\pm$ 200 & 20 $\pm$ 1 & 1.3 $\pm$ 0.2 & - & - & - & 1.3 $\pm$ 0.2 & 208.5  \\
130325A & 0.2 $\pm$ 0.1 & 200 $\pm$ 200 & 0.1 $\pm$ 0.3 & - & - & - & 0.1 $\pm$ 0.3 & 705.7  \\
130327B & 17 $\pm$ 6 & 35 $\pm$ 9 & 1.6 $\pm$ 0.2 & 1.3 $\pm$ 0.9 & 1.7 $\pm$ 0.3 & 1.9 $\pm$ 0.8 & 1.6 $\pm$ 0.2 & 515.2  \\
130427A & 300 $\pm$ 100 & 16.5 $\pm$ 0.6 & 1.24 $\pm$ 0.06 & 0.8 $\pm$ 0.2 & 1.4 $\pm$ 0.1 & 2.7 $\pm$ 0.2 & 1.4 $\pm$ 0.1 & 34366.0  \\
130502B & 40 $\pm$ 10 & 28 $\pm$ 8 & 1.44 $\pm$ 0.06 & - & - & - & 1.44 $\pm$ 0.06 & 1304.2  \\
130504C & 1.6 $\pm$ 0.6 & 100 $\pm$ 200 & 0.77 $\pm$ 0.06 & 0.7 $\pm$ 0.1 & 1.0 $\pm$ 0.5 & 2.4 $\pm$ 0.4 & 0.77 $\pm$ 0.06 & 547.7  \\
130518A & 5 $\pm$ 2 & 40 $\pm$ 60 & 1.1 $\pm$ 0.2 & - & - & - & 1.1 $\pm$ 0.2 & 316.9  \\
130606B & 1.3 $\pm$ 0.5 & 200 $\pm$ 200 & 0.7 $\pm$ 0.2 & - & - & - & 0.7 $\pm$ 0.2 & 396.7  \\
130821A & 2.2 $\pm$ 0.8 & 70 $\pm$ 80 & 1.0 $\pm$ 0.1 & - & - & - & 1.0 $\pm$ 0.1 & 6069.8  \\
130828A & 70 $\pm$ 30 & 35 $\pm$ 2 & 1.0 $\pm$ 0.4 & - & - & - & 1.0 $\pm$ 0.4 & 599.8  \\
131014A & 5000 $\pm$ 2000 & 2.0 $\pm$ 0.2 & 0.8 $\pm$ 0.2 & - & - & - & 0.8 $\pm$ 0.2 & 198.6  \\
131029A & 22 $\pm$ 8 & 50 $\pm$ 20 & 1.1 $\pm$ 0.2 & - & - & - & 1.1 $\pm$ 0.2 & 518.5  \\
131108A & 3000 $\pm$ 1000 & 0.13 $\pm$ 0.03 & 1.5 $\pm$ 0.2 & 1.9 $\pm$ 0.5 & 1.2 $\pm$ 0.6 & 1.8 $\pm$ 0.6 & 1.5 $\pm$ 0.2 & 678.1  \\
131209A & 2.6 $\pm$ 1.0 & 30 $\pm$ 60 & 0.8 $\pm$ 0.2 & - & - & - & 0.8 $\pm$ 0.2 & 360.5  \\
131231A & 5 $\pm$ 2 & 30 $\pm$ 40 & 1.0 $\pm$ 0.2 & - & - & - & 1.0 $\pm$ 0.2 & 4801.1  \\
140102A & 50 $\pm$ 20 & 4 $\pm$ 2 & 1.2 $\pm$ 0.4 & - & - & - & 1.2 $\pm$ 0.4 & 57.1  \\
140104B & 0.5 $\pm$ 0.3 & 900 $\pm$ 200 & 0.3 $\pm$ 0.6 & - & - & - & 0.3 $\pm$ 0.6 & 946.8  \\
140110A & 130 $\pm$ 50 & 1.0 $\pm$ 0.9 & 0.97 $\pm$ 0.02 & - & - & - & 0.97 $\pm$ 0.02 & 158.8  \\
140206B & 15 $\pm$ 5 & 30 $\pm$ 20 & 0.3 $\pm$ 0.3 & - & - & - & 0.3 $\pm$ 0.3 & 8584.4  \\
140402A & 50 $\pm$ 20 & 0 $\pm$ 3 & 0.87 $\pm$ 0.06 & - & - & - & 0.87 $\pm$ 0.06 & 71.5  \\
140523A & 30 $\pm$ 10 & 8 $\pm$ 8 & 1.0 $\pm$ 0.1 & - & - & - & 1.0 $\pm$ 0.1 & 468.6  \\
140810A & 0.4 $\pm$ 0.2 & 500 $\pm$ 200 & 0.8 $\pm$ 0.2 & - & - & - & 0.8 $\pm$ 0.2 & 18366.7  \\
141028A & 25 $\pm$ 10 & 20 $\pm$ 6 & 0.97 $\pm$ 0.03 & - & - & - & 0.97 $\pm$ 0.03 & 489.6  \\
141102A & 11 $\pm$ 6 & 10 $\pm$ 10 & 1.0 $\pm$ 0.2 & - & - & - & 1.0 $\pm$ 0.2 & 31.0  \\
141207A & 150 $\pm$ 60 & 8 $\pm$ 1 & 1.88 $\pm$ 0.03 & - & - & - & 1.88 $\pm$ 0.03 & 732.6  \\
141222A & 6 $\pm$ 3 & 50 $\pm$ 30 & 1.3 $\pm$ 0.4 & - & - & - & 1.3 $\pm$ 0.4 & 406.3  \\
150314A & 4 $\pm$ 1 & 30 $\pm$ 80 & 0.9 $\pm$ 0.1 & - & - & - & 0.9 $\pm$ 0.1 & 3064.2  \\
150523A & 20 $\pm$ 7 & 15 $\pm$ 7 & 1.0 $\pm$ 0.3 & - & - & - & 1.0 $\pm$ 0.3 & 6125.2  \\
150627A & 3 $\pm$ 1 & 200 $\pm$ 200 & 0.9 $\pm$ 0.2 & 3.0 $\pm$ 0.2 & 0.41 $\pm$ 0.07 & 2.62 $\pm$ 0.03 & 0.41 $\pm$ 0.07 & 5991.0  \\
150702A & 0.2 $\pm$ 0.1 & 800 $\pm$ 600 & 0.4 $\pm$ 0.3 & - & - & - & 0.4 $\pm$ 0.3 & 1224.5  \\
150902A & 60 $\pm$ 20 & 5 $\pm$ 3 & 1.0 $\pm$ 0.2 & - & - & - & 1.0 $\pm$ 0.2 & 405.2  \\
160325A & 1.7 $\pm$ 0.6 & 100 $\pm$ 100 & 0.74 $\pm$ 0.10 & - & - & - & 0.74 $\pm$ 0.10 & 1550.7  \\
160509A & 1500 $\pm$ 500 & 17.5 $\pm$ 0.1 & 1.1 $\pm$ 0.1 & 0.9 $\pm$ 0.3 & 1.3 $\pm$ 0.3 & 2.8 $\pm$ 0.5 & 1.1 $\pm$ 0.1 & 5677.8  \\
160521B & 1.5 $\pm$ 0.8 & 90 $\pm$ 70 & 1.3 $\pm$ 0.2 & - & - & - & 1.3 $\pm$ 0.2 & 11821.5  \\
160623A & 8 $\pm$ 4 & 500 $\pm$ 100 & 1.25 $\pm$ 0.09 & - & - & - & 1.25 $\pm$ 0.09 & 11758.4  \\
160625B & 400 $\pm$ 200 & 200.6 $\pm$ 0.3 & 2.2 $\pm$ 0.3 & - & - & - & 2.2 $\pm$ 0.3 & 814.9  \\
160816A & 40 $\pm$ 10 & 2 $\pm$ 3 & 1.2 $\pm$ 0.1 & 1.5 $\pm$ 0.3 & 0.9 $\pm$ 0.6 & 2.1 $\pm$ 0.8 & 1.2 $\pm$ 0.1 & 1093.7  \\
160821A & 300 $\pm$ 100 & 137.3 $\pm$ 0.7 & 1.15 $\pm$ 0.10 & - & - & - & 1.15 $\pm$ 0.10 & 1367.1  \\
160905A & 4 $\pm$ 2 & 40 $\pm$ 30 & 1.2 $\pm$ 0.3 & - & - & - & 1.2 $\pm$ 0.3 & 455.5  \\
161109A & 1.9 $\pm$ 0.8 & 500 $\pm$ 200 & 1.3 $\pm$ 0.5 & - & - & - & 1.3 $\pm$ 0.5 & 467.5  \\
170115B & 80 $\pm$ 30 & 7 $\pm$ 1 & 1.2 $\pm$ 0.2 & 2 $\pm$ 2 & 0.9 $\pm$ 0.7 & 2.3 $\pm$ 0.4 & 1.2 $\pm$ 0.2 & 1026.6  \\
170214A & 400 $\pm$ 100 & 63.0 $\pm$ 0.4 & 1.7 $\pm$ 0.3 & 2 $\pm$ 2 & 1.6 $\pm$ 0.5 & 2.3 $\pm$ 0.5 & 1.7 $\pm$ 0.3 & 712.5  \\
170405A & 8 $\pm$ 3 & 30 $\pm$ 30 & 1.27 $\pm$ 0.01 & - & - & - & 1.27 $\pm$ 0.01 & 850.1  \\
170409A & 1.8 $\pm$ 0.7 & 300 $\pm$ 200 & 1.3 $\pm$ 0.1 & - & - & - & 1.3 $\pm$ 0.1 & 261.9  \\
170808B & 4 $\pm$ 1 & 100 $\pm$ 100 & 1.0 $\pm$ 0.2 & - & - & - & 1.0 $\pm$ 0.2 & 6192.2  \\
170906A & 3 $\pm$ 1 & 200 $\pm$ 100 & 0.8 $\pm$ 0.1 & - & - & - & 0.8 $\pm$ 0.1 & 1732.2  \\
171010A & 3 $\pm$ 1 & 380 $\pm$ 70 & 1.3 $\pm$ 0.2 & 2.2 $\pm$ 0.7 & 1.0 $\pm$ 0.3 & 2.9 $\pm$ 0.2 & 1.0 $\pm$ 0.3 & 2649.2  \\
171102A & 1.4 $\pm$ 0.7 & 60 $\pm$ 90 & 1.02 $\pm$ 0.07 & - & - & - & 1.02 $\pm$ 0.07 & 315.0  \\
171120A & 2 $\pm$ 1 & 10 $\pm$ 40 & 0.6 $\pm$ 0.3 & 0.2 $\pm$ 0.6 & 1.1 $\pm$ 0.6 & 2.8 $\pm$ 0.5 & 1.1 $\pm$ 0.6 & 5275.7  \\
171124A & 40 $\pm$ 10 & 8 $\pm$ 4 & 0.89 $\pm$ 0.09 & - & - & - & 0.89 $\pm$ 0.09 & 317.9  \\
171210A & 0.3 $\pm$ 0.1 & 200 $\pm$ 500 & 0.7 $\pm$ 0.3 & - & - & - & 0.7 $\pm$ 0.3 & 1368.6  \\
180210A & 1.8 $\pm$ 0.6 & 140 $\pm$ 80 & 1.0 $\pm$ 0.2 & - & - & - & 1.0 $\pm$ 0.2 & 1598.1  \\
180526A & 3 $\pm$ 2 & 800 $\pm$ 200 & 1.3 $\pm$ 0.7 & - & - & - & 1.3 $\pm$ 0.7 & 1139.7  \\
180703A & 6 $\pm$ 3 & 10 $\pm$ 20 & 0.8 $\pm$ 0.2 & - & - & - & 0.8 $\pm$ 0.2 & 1610.1  \\
180720B & 40 $\pm$ 20 & 62 $\pm$ 5 & 1.9 $\pm$ 0.1 & 1.5 $\pm$ 0.2 & 3.2 $\pm$ 0.6 & 2.37 $\pm$ 0.08 & 3.2 $\pm$ 0.6 & 613.3  \\
\enddata
\tablenotetext{a}{File used: LAT2CATALOG-v14-LTF }
\end{deluxetable}

\end{longrotatetable}
\subsection{Highest energy events}
For each GRB, Table~\ref{tab_energymax_gbm} shows the highest energy photon detected in the GBM time window (columns 2-5), as well as the highest energy photon found overall in the time-resolved analysis (columns 6-9). For each time span the values given are: the total number of photons found with probability $>90\%$ to be associated to the GRB, the energy of the most energetic photon, its arrival time, and the probability to be associated with the GRB. 

Table~\ref{tab_energymax_Michele} lists the 29 GRBs from which at least one photon with energy greater than 10 GeV has been detected. For each such photon, the energy and arrival time are given. For GRBs with a measured redshift, this is listed along with the corresponding source-frame energy (E$_{\rm sf}$) of each photon.

\startlongtable
\begin{deluxetable}{@{\extracolsep{4pt}}lrrrrrrrr}
\tabletypesize{\tiny}
\tablecolumns{9}
\tablewidth{0pt}
\tablecaption{Highest energy events of \textit{Fermi}-LAT GRBs\label{tab_energymax_gbm}}
\tablehead{
\\ \colhead{} & \multicolumn{4}{c}{GBM Time window}& \multicolumn{4}{c}{Time resolved} \\ \cline{2-5}\cline{6-9}\colhead{ GRB NAME} & \colhead{Events} & \colhead{Energy} & \colhead{Arrival time} & \colhead{Probability } & \colhead{Events} & \colhead{Energy} & \colhead{Arrival time} & \colhead{Probability } \\ 
\colhead{ } & \colhead{(P$>$0.9)} & \colhead{(GeV)} & \colhead{(sec)} & \colhead{} & \colhead{(P$>$0.9)} & \colhead{(GeV)} & \colhead{(sec)} & \colhead{}}
\startdata
080818B & - & $-$ & $-$ & $-$ & 3 & 1.70 & 9018.86 & 1.00 \\
080825C & 9 & 0.30 & 21.38 & 0.93 & 13 & 0.68 & 28.28 & 1.00 \\
080916C & 237 & 27.00 & 40.50 & 1.00 & 320 & 27.00 & 40.50 & 1.00 \\
081006 & 8 & 0.71 & 1.80 & 1.00 & 12 & 0.80 & 12.08 & 0.99 \\
081009 & - & $-$ & $-$ & $-$ & 9 & 1.60 & 1250.13 & 0.99 \\
081024B & 6 & 0.45 & 0.26 & 1.00 & 12 & 1.70 & 2.12 & 1.00 \\
081102B & 6 & 0.74 & 1.33 & 0.97 & 5 & 0.55 & 0.30 & 1.00 \\
081122A & 1 & 0.15 & 5.05 & 0.93 & 4 & 2.50 & 66.53 & 1.00 \\
081203A & - & $-$ & $-$ & $-$ & 5 & 1.00 & 379.66 & 1.00 \\
081224 & - & $-$ & $-$ & $-$ & 4 & 1.80 & 177.49 & 1.00 \\
090102 & - & $-$ & $-$ & $-$ & 2 & 0.58 & 3915.89 & 0.99 \\
090217 & 19 & 0.82 & 14.83 & 1.00 & 23 & 0.82 & 14.83 & 1.00 \\
090227A & 1 & 0.43 & 2.90 & 1.00 & 4 & 2.30 & 51.68 & 1.00 \\
090227B & 2 & 0.13 & 0.21 & 1.00 & - & $-$ & $-$ & $-$ \\
090228A & 1 & 0.14 & 0.09 & 1.00 & 5 & 0.70 & 2.09 & 0.92 \\
090323 & 16 & 1.20 & 22.57 & 0.99 & 59 & 7.40 & 195.42 & 1.00 \\
090328 & 12 & 2.30 & 53.30 & 1.00 & 67 & 5.50 & 697.80 & 1.00 \\
090427A & - & $-$ & $-$ & $-$ & 2 & 14.00 & 422.87 & 1.00 \\
090510 & 64 & 30.00 & 0.83 & 1.00 & 260 & 30.00 & 0.83 & 1.00 \\
090626 & 3 & 0.12 & 9.00 & 0.99 & 20 & 2.10 & 111.63 & 1.00 \\
090720B & 6 & 0.43 & 1.68 & 0.99 & 5 & 0.43 & 1.68 & 1.00 \\
090902B & 245 & 14.00 & 14.16 & 1.00 & 480 & 40.00 & 81.74 & 1.00 \\
090926A & 252 & 3.30 & 9.48 & 1.00 & 410 & 19.00 & 24.83 & 1.00 \\
091003 & 5 & 2.70 & 6.47 & 1.00 & 22 & 5.90 & 348.63 & 1.00 \\
091031 & 9 & 0.60 & 6.44 & 1.00 & 25 & 1.40 & 408.18 & 1.00 \\
091120 & 3 & 0.20 & 31.79 & 0.99 & 10 & 7.00 & 712.58 & 1.00 \\
091127 & 1 & 1.60 & 8.61 & 1.00 & 3 & 2.20 & 16.94 & 1.00 \\
100116A & 8 & 0.84 & 101.30 & 0.99 & 20 & 33.00 & 378.98 & 1.00 \\
100213C & - & $-$ & $-$ & $-$ & 3 & 34.00 & 3389.03 & 1.00 \\
100225A & 6 & 0.34 & 6.62 & 0.97 & 8 & 3.00 & 64.85 & 1.00 \\
100325A & 6 & 0.82 & 0.34 & 1.00 & 6 & 0.82 & 0.34 & 1.00 \\
100414A & 4 & 0.47 & 24.50 & 1.00 & 48 & 30.00 & 33.36 & 1.00 \\
100423B & - & $-$ & $-$ & $-$ & 3 & 3.90 & 166.69 & 1.00 \\
100511A & 3 & 0.60 & 11.61 & 0.92 & 19 & 46.00 & 161.90 & 1.00 \\
100620A & 5 & 0.25 & 3.75 & 1.00 & 5 & 0.25 & 3.75 & 1.00 \\
100724B & 8 & 0.16 & 42.83 & 0.96 & 8 & 0.16 & 42.83 & 0.98 \\
100728A & - & $-$ & $-$ & $-$ & 7 & 3.10 & 325.06 & 1.00 \\
100826A & 5 & 1.70 & 61.39 & 1.00 & 4 & 1.70 & 61.39 & 1.00 \\
101014A & - & $-$ & $-$ & $-$ & 6 & 14.00 & 2750.71 & 1.00 \\
101107A & 6 & 4.80 & 139.85 & 1.00 & 7 & 4.80 & 139.85 & 1.00 \\
101227B & 3 & 1.50 & 20.53 & 1.00 & 3 & 1.50 & 20.53 & 1.00 \\
110120A & 2 & 0.47 & 0.87 & 0.99 & 10 & 2.00 & 72.46 & 1.00 \\
110123A & - & $-$ & $-$ & $-$ & 7 & 2.10 & 445.55 & 1.00 \\
110213A & - & $-$ & $-$ & $-$ & 2 & 3.00 & 1261.39 & 1.00 \\
110328B & 1 & 0.16 & 9.84 & 0.90 & 6 & 4.00 & 328.67 & 1.00 \\
110428A & 1 & 0.62 & 7.28 & 0.94 & 9 & 3.00 & 14.80 & 0.94 \\
110518A & - & $-$ & $-$ & $-$ & 4 & 2.90 & 2039.24 & 1.00 \\
110529A & 1 & 0.10 & 0.07 & 1.00 & - & $-$ & $-$ & $-$ \\
110625A & - & $-$ & $-$ & $-$ & 7 & 1.70 & 577.24 & 0.95 \\
110721A & 33 & 6.70 & 4.50 & 1.00 & 36 & 6.70 & 4.50 & 1.00 \\
110728A & 1 & 0.51 & 0.36 & 1.00 & 6 & 1.40 & 2.98 & 1.00 \\
110731A & 48 & 0.97 & 5.52 & 1.00 & 75 & 3.50 & 435.96 & 1.00 \\
110903A & 3 & 16.00 & 301.35 & 1.00 & 4 & 16.00 & 301.35 & 1.00 \\
110921B & 4 & 0.65 & 13.08 & 0.99 & 10 & 2.00 & 202.48 & 1.00 \\
111210B & 3 & 0.15 & 47.63 & 0.99 & 7 & 0.56 & 96.10 & 1.00 \\
120107A & 7 & 1.90 & 7.70 & 1.00 & 8 & 1.90 & 7.70 & 1.00 \\
120226A & 3 & 0.27 & 30.71 & 0.96 & 10 & 0.38 & 283.65 & 0.96 \\
120316A & 3 & 2.10 & 26.44 & 1.00 & 10 & 2.10 & 26.44 & 1.00 \\
120420B & - & $-$ & $-$ & $-$ & 4 & 1.90 & 3800.54 & 1.00 \\
120526A & - & $-$ & $-$ & $-$ & 20 & 14.00 & 1354.30 & 1.00 \\
120624B & 4 & 0.36 & $-$201.87 & 0.91 & 110 & 1.90 & 557.66 & 1.00 \\
120709A & 12 & 2.20 & 1.77 & 1.00 & 22 & 2.60 & 140.33 & 0.99 \\
120711A & - & $-$ & $-$ & $-$ & 29 & 2.50 & 5431.65 & 1.00 \\
120729A & - & $-$ & $-$ & $-$ & 3 & 2.40 & 396.87 & 1.00 \\
120830A & 3 & 0.48 & 0.75 & 1.00 & 6 & 0.78 & 2.96 & 1.00 \\
120911B & 50 & 1.20 & 15.51 & 0.96 & 66 & 1.40 & 90.29 & 1.00 \\
120915A & 2 & 0.18 & 0.14 & 1.00 & 2 & 0.25 & 0.82 & 0.99 \\
120919B & - & $-$ & $-$ & $-$ & 3 & 13.00 & 605.27 & 1.00 \\
121029A & 1 & 0.12 & 13.80 & 0.94 & 5 & 4.30 & 1926.46 & 1.00 \\
121123B & - & $-$ & $-$ & $-$ & 7 & 2.50 & 268.02 & 1.00 \\
121225B & 1 & 0.12 & 11.04 & 0.94 & - & $-$ & $-$ & $-$ \\
130310A & 2 & 0.84 & 4.12 & 1.00 & 11 & 1.40 & 329.90 & 1.00 \\
130325A & 1 & 0.23 & 2.19 & 0.99 & 4 & 5.30 & 828.57 & 1.00 \\
130327B & 7 & 4.60 & 20.23 & 1.00 & 42 & 9.20 & 49.42 & 1.00 \\
130427A & 252 & 77.00 & 18.64 & 1.00 & 600 & 94.00 & 243.13 & 1.00 \\
130502B & 21 & 2.00 & 27.47 & 1.00 & 68 & 31.00 & 222.10 & 0.97 \\
130504C & 7 & 2.20 & 54.16 & 1.00 & 19 & 5.70 & 250.82 & 1.00 \\
130518A & 8 & 0.39 & 50.71 & 0.94 & 32 & 2.10 & 270.81 & 1.00 \\
130606B & - & $-$ & $-$ & $-$ & 11 & 4.60 & 527.27 & 1.00 \\
130702A & - & $-$ & $-$ & $-$ & 3 & 1.70 & 272.29 & 0.99 \\
130821A & 7 & 2.80 & 64.29 & 1.00 & 47 & 6.30 & 219.13 & 0.96 \\
130828A & 32 & 1.20 & 52.09 & 1.00 & 32 & 1.20 & 52.09 & 1.00 \\
130907A & - & $-$ & $-$ & $-$ & 2 & 5.60 & 3618.37 & 1.00 \\
131014A & 10 & 1.20 & 1.98 & 1.00 & 23 & 1.80 & 14.37 & 0.98 \\
131029A & 23 & 1.20 & 71.66 & 1.00 & 35 & 3.20 & 189.93 & 1.00 \\
131108A & 135 & 1.20 & 5.21 & 1.00 & 200 & 1.50 & 66.33 & 1.00 \\
131209A & 2 & 0.22 & 15.56 & 0.92 & 8 & 0.29 & 65.13 & 0.91 \\
131231A & 5 & 0.96 & 37.94 & 1.00 & 35 & 48.00 & 110.29 & 1.00 \\
140102A & 3 & 0.37 & 3.42 & 0.99 & 12 & 2.30 & 4.29 & 1.00 \\
140104B & - & $-$ & $-$ & $-$ & 11 & 3.10 & 809.91 & 1.00 \\
140110A & 29 & 2.00 & 0.62 & 1.00 & 38 & 2.00 & 0.62 & 1.00 \\
140124A & 1 & 2.30 & 93.56 & 0.96 & 3 & 2.30 & 93.56 & 0.93 \\
140206B & 24 & 0.75 & 24.00 & 1.00 & 62 & 11.00 & 6735.90 & 1.00 \\
140219A & - & $-$ & $-$ & $-$ & 4 & 1.70 & 1356.56 & 1.00 \\
140323A & - & $-$ & $-$ & $-$ & 4 & 2.80 & 226.80 & 1.00 \\
140402A & 4 & 0.66 & 0.05 & 1.00 & 16 & 3.70 & 8.73 & 1.00 \\
140416A & - & $-$ & $-$ & $-$ & 3 & 10.00 & 2207.37 & 1.00 \\
140523A & 14 & 2.60 & 18.87 & 1.00 & 35 & 7.30 & 43.46 & 1.00 \\
140528A & - & $-$ & $-$ & $-$ & 3 & 4.60 & 1377.90 & 1.00 \\
140619B & 21 & 23.00 & 0.61 & 1.00 & 23 & 23.00 & 0.61 & 1.00 \\
140723A & 18 & 1.00 & 0.56 & 0.99 & 31 & 1.00 & 0.56 & 1.00 \\
140729A & 9 & 1.30 & 44.14 & 1.00 & 11 & 1.50 & 73.40 & 1.00 \\
140810A & - & $-$ & $-$ & $-$ & 13 & 15.00 & 1490.23 & 1.00 \\
140825A & - & $-$ & $-$ & $-$ & 3 & 1.60 & 1702.32 & 0.91 \\
140928A & - & $-$ & $-$ & $-$ & 4 & 52.00 & 2554.67 & 1.00 \\
141012A & 7 & 1.00 & 9.64 & 1.00 & 14 & 1.10 & 19.51 & 1.00 \\
141028A & 19 & 0.68 & 32.21 & 1.00 & 34 & 3.90 & 157.52 & 0.94 \\
141102A & 1 & 0.15 & 2.54 & 0.92 & 7 & 0.64 & 30.91 & 0.97 \\
141113A & 4 & 0.53 & 0.07 & 1.00 & 5 & 0.53 & 0.07 & 1.00 \\
141207A & 39 & 3.40 & 4.80 & 1.00 & 48 & 5.50 & 734.33 & 1.00 \\
141221B & 1 & 0.11 & 22.17 & 0.98 & 3 & 5.50 & 58.47 & 1.00 \\
141222A & - & $-$ & $-$ & $-$ & 11 & 3.60 & 227.42 & 1.00 \\
150118B & - & $-$ & $-$ & $-$ & 3 & 1.80 & 51.03 & 1.00 \\
150202B & 2 & 1.00 & 115.70 & 1.00 & 2 & 1.00 & 115.70 & 0.99 \\
150210A & 15 & 1.00 & 2.02 & 1.00 & 18 & 2.50 & 169.37 & 1.00 \\
150314A & - & $-$ & $-$ & $-$ & 13 & 1.90 & 3064.28 & 0.97 \\
150403A & - & $-$ & $-$ & $-$ & 5 & 5.40 & 631.76 & 1.00 \\
150510A & 5 & 0.58 & 2.43 & 1.00 & 9 & 2.00 & 100.91 & 1.00 \\
150513A & 8 & 2.20 & $-$56.40 & 1.00 & - & $-$ & $-$ & $-$ \\
150514A & 1 & 0.18 & 3.58 & 0.99 & 2 & 6.20 & 442.44 & 1.00 \\
150523A & 23 & 1.90 & 42.33 & 1.00 & 44 & 7.00 & 118.00 & 1.00 \\
150627A & - & $-$ & $-$ & $-$ & 26 & 8.10 & 258.66 & 1.00 \\
150702A & - & $-$ & $-$ & $-$ & 6 & 0.80 & 1115.33 & 0.99 \\
150902A & 18 & 0.41 & 6.18 & 0.99 & 40 & 11.00 & 97.49 & 1.00 \\
160310A & - & $-$ & $-$ & $-$ & 2 & 1.40 & 99.25 & 0.95 \\
160314B & - & $-$ & $-$ & $-$ & 4 & 0.88 & 628.72 & 0.94 \\
160325A & 2 & 0.31 & 4.98 & 0.96 & 21 & 3.00 & 92.02 & 1.00 \\
160422A & - & $-$ & $-$ & $-$ & 2 & 12.00 & 769.62 & 1.00 \\
160503A & - & $-$ & $-$ & $-$ & 11 & 0.52 & 12969.55 & 0.91 \\
160509A & 103 & 52.00 & 76.51 & 1.00 & 140 & 52.00 & 76.51 & 1.00 \\
160521B & - & $-$ & $-$ & $-$ & 8 & 13.00 & 422.62 & 1.00 \\
160623A & - & $-$ & $-$ & $-$ & 43 & 18.00 & 12038.53 & 1.00 \\
160625B & 250 & 15.00 & 346.18 & 1.00 & 260 & 15.00 & 346.18 & 1.00 \\
160702A & - & $-$ & $-$ & $-$ & 3 & 4.80 & 1941.52 & 1.00 \\
160709A & 24 & 0.99 & 1.47 & 1.00 & 28 & 0.99 & 1.47 & 1.00 \\
160816A & 15 & 1.10 & 1.40 & 1.00 & 46 & 9.20 & 1094.77 & 1.00 \\
160821A & 35 & 0.68 & 156.93 & 0.94 & 52 & 4.70 & 212.43 & 1.00 \\
160829A & - & $-$ & $-$ & $-$ & 3 & 9.40 & 0.95 & 1.00 \\
160905A & 5 & 2.20 & 22.03 & 1.00 & 22 & 7.90 & 347.76 & 1.00 \\
160910A & 3 & 0.15 & 23.57 & 0.91 & 4 & 0.49 & 197.48 & 0.99 \\
161015A & 9 & 1.00 & 7.47 & 1.00 & 10 & 1.00 & 7.47 & 1.00 \\
161109A & - & $-$ & $-$ & $-$ & 8 & 3.40 & 594.81 & 1.00 \\
170115B & 13 & 0.64 & 1.34 & 0.93 & 28 & 2.30 & 861.41 & 0.99 \\
170127C & - & $-$ & $-$ & $-$ & 10 & 0.51 & 2889.00 & 1.00 \\
170214A & 103 & 7.80 & 103.62 & 1.00 & 220 & 7.80 & 103.62 & 1.00 \\
170228A & 5 & 0.34 & 9.47 & 0.97 & 6 & 0.34 & 9.47 & 0.97 \\
170306B & 1 & 0.23 & 21.16 & 0.99 & 6 & 0.50 & 43.25 & 1.00 \\
170329A & 9 & 0.78 & 4.11 & 1.00 & 10 & 0.78 & 4.11 & 1.00 \\
170405A & 7 & 0.16 & 41.82 & 0.93 & 23 & 0.89 & 445.97 & 0.98 \\
170409A & - & $-$ & $-$ & $-$ & 10 & 9.90 & 440.27 & 1.00 \\
170424A & 3 & 1.10 & 54.20 & 1.00 & 7 & 1.10 & 54.20 & 1.00 \\
170510A & 6 & 1.00 & 45.47 & 1.00 & 6 & 1.00 & 45.47 & 1.00 \\
170522A & 6 & 3.70 & 6.86 & 1.00 & 10 & 3.70 & 6.86 & 1.00 \\
170728B & 4 & 0.63 & 9.27 & 0.99 & 7 & 0.63 & 9.27 & 0.92 \\
170808B & 2 & 0.14 & 13.73 & 1.00 & 18 & 1.80 & 484.36 & 1.00 \\
170813A & 3 & 0.58 & 9.52 & 1.00 & 4 & 0.96 & 265.30 & 0.96 \\
170825B & 2 & 0.30 & 1.49 & 1.00 & 2 & 0.30 & 1.49 & 0.99 \\
170906A & - & $-$ & $-$ & $-$ & 29 & 3.60 & 203.25 & 1.00 \\
170921B & - & $-$ & $-$ & $-$ & 2 & 2.70 & 901.67 & 1.00 \\
171010A & - & $-$ & $-$ & $-$ & 47 & 19.00 & 2890.98 & 0.99 \\
171011C & 3 & 0.51 & $-$0.09 & 1.00 & 2 & 0.16 & 42.72 & 0.99 \\
171102A & 2 & 0.14 & 34.91 & 0.98 & 8 & 0.37 & 349.89 & 0.99 \\
171120A & 6 & 1.60 & 5.63 & 1.00 & 29 & 3.40 & 4840.92 & 1.00 \\
171124A & 20 & 3.60 & 4.08 & 1.00 & 21 & 3.60 & 4.08 & 1.00 \\
171210A & 2 & 0.65 & 74.83 & 1.00 & 12 & 12.00 & 1374.49 & 1.00 \\
171212B & - & $-$ & $-$ & $-$ & 3 & 0.61 & 496.97 & 0.97 \\
180210A & 3 & 0.47 & 31.78 & 1.00 & 32 & 7.40 & 1621.12 & 1.00 \\
180305A & 1 & 0.20 & 6.18 & 0.99 & 2 & 8.90 & 1613.84 & 1.00 \\
180526A & - & $-$ & $-$ & $-$ & 8 & 2.20 & 1308.09 & 1.00 \\
180703A & 2 & 0.21 & 5.06 & 0.99 & 13 & 0.93 & 40.46 & 1.00 \\
180703B & - & $-$ & $-$ & $-$ & 5 & 1.10 & 34.22 & 1.00 \\
180718B & 11 & 0.49 & 2.63 & 0.96 & 9 & 0.49 & 2.63 & 0.99 \\
180720B & 19 & 0.63 & 38.77 & 1.00 & 130 & 4.90 & 142.43 & 1.00 \\
\enddata
\tablenotetext{a}{File used: LAT2CATALOG-v14-LTF }
\end{deluxetable}

\startlongtable
\begin{deluxetable}{lcccc}
\tablecolumns{5}
\tablewidth{0pt}
\tablecaption{{\it Fermi}-LAT GRBs with photon energies $E>10$ GeV.
\label{tab_energymax_Michele}}
\tablehead{
\colhead{GRB NAME}	&	\colhead{Energy}	&	\colhead{Arrival Time}	&	\colhead{z}	&	\colhead{E$_{\rm sf}$} \\
(CLASS) &	\colhead{(GeV)}	&	\colhead{(s)}	&	\colhead{}	&	\colhead{(GeV)}} 
\startdata
080916C (L)	&	27.4	&	40.5	&	4.35	&	{\bf 146.6}  	\\ 
 		 	&	12.4	& 	16.5	&			&	66.3	\\ 
090427A (L)	&	14.1	& 	422.9	&	$-$		&	$-$  	\\ 
090510 (S)	&	30.0  	&  	0.83	& 	0.90	&	56.8	\\
090902B (L)	&	40.0	&	81.7	&	1.82	&	{\bf 112.5}	\\
 &              21.7 	& 	331.9 	&			& 	61.2 	\\
 &              18.1 	& 	26.2  	& 			&	51.0 	\\
 &              15.4 	& 	45.6	&  			& 	43.4	\\
 &              14.2 	& 	14.2 	& 			& 	40.0 	\\
 &              12.7 	& 	42.4 	&  			& 	35.8	\\
 &				11.9 	& 	11.7 	&  			& 	33.6 	\\
090926A (L)	&	19.4	&	24.8	&	2.11	&	60.3 	\\
			&	10.4	&	3785.0	&			&	32.3	\\
100116A (L)	&	32.6	& 	379.0	& 	$-$		&	$-$		\\
			&	13.3	& 	296.0	&	$-$		&	$-$		\\
100213C (L)	&  	34.0	& 	3389.0  & 	$-$		&	$-$		\\
100414A (L)	& 	30.0 	& 	33.4 	& 	1.37 	&	70.6	\\
			& 	25.1 	& 	358.5 	& 			&	59.5	\\
100511A (L)	& 	46.0	&	161.9 	& 	$-$		&	$-$		\\	
			& 	18.4	& 	179.8 	& 	$-$		&	$-$		\\
101014A (L)	&	13.6	&	2750.7	& 	$-$		&	$-$		\\	
			&	11.2	&	2962.0	& 	$-$		&	$-$		\\	
110903A (L) & 	15.6 	& 	301.0	& 	$-$		&	$-$		\\
120526A (L) &   14.3 	&	1354.3	&   $-$		&	$-$		\\
120919B	(L) & 	12.7	& 	605.3	&   $-$		&	$-$		\\
130427A (L)	&	94.1	&	243.1	&	0.34	&	{\bf 126.1}		\\
			&	77.1	&	18.6	&			&	{\bf 103.3}		\\
			&	57.4 	& 	256.0	& 			& 	76.9 	\\
			&	38.7 	&	78.4 	& 			& 	51.9 	\\
			&	38.2 	& 	3409.0 	& 	 		& 	51.2 	\\
			&	33.6 	& 	34366.0 & 			& 	45.0 	\\
			&	28.4	& 	47.6 	& 			& 	38.0 	\\
			&	26.9	& 	84.7 	& 			& 	36.0 	\\
			&	25.4 	& 	141.0 	& 	 		& 	34.0 	\\
			&	19.3 	& 	6062.0 	& 		 	& 	25.9 	\\
			&	17.1	& 	217.0 	& 			& 	22.9 	\\
			&	14.9 	& 	119.3 	& 			& 	20.0 	\\
			&	12.9 	& 	80.5	& 			& 	17.3 	\\
            & 	12.2	& 	64.5 	&  			&  	16.3	\\
			&	12.0 	& 	23.5 	&			& 	16.1 	\\
			&	11.7 	& 	214.0	& 	 		& 	15.7 	\\
			&	10.8 	& 	23.2 	& 			& 	14.5 	\\
130502B (L) &   31.1 	& 	222.1 	&  	$-$		&	$-$		\\
            &   17.3 	& 	48.0 	&  	$-$		&	$-$		\\
131231A (L)	&	48.3	&	110.3	&	0.64	&	79.2	\\
			&	17.1	&	844.2	&			&	28.0	\\
140206B (L)	&	11.0	&	6735.9	&	$-$		&	$-$		\\
140416A	(L) & 	10.1	& 	2207.4	&   $-$		&	$-$		\\
140619B (L) &	22.7	& 	0.6 	&   $-$		&	$-$		\\
140810A	(L) & 	15.4	& 	1490.0	&  	$-$		&	$-$		\\
140928A (L)	&	51.7	&	2554.7	&	$-$		&	$-$		\\		
150902A (L) &   11.3    &   97.5    &   $-$		&	$-$		\\
160422A (L)	&	12.3	&	769.6	&	$-$		&	$-$		\\
160509A (L)	&	51.9	&	76.5	&	1.17	&	{\bf 112.6}	\\
			&	41.5	&	242.0	&			&	90.1	\\
160521B (L)	&	12.7	&	422.6	&	$-$		&	$-$		\\
160623A (L)	&	18.2	&	12038.5	&	0.37	&	24.9	\\
160625B (L)	&	15.3	&	346.2	&	1.41	&	36.9	\\
171010A (L) & 	19.0 	& 	2891.0	&	0.33	&	25.3	\\
171210A (L)	&	12.4	&	1374.5	&	$-$		&	$-$		\\
\enddata
\end{deluxetable}
\section{Description of the content of the fits file}
All the information used to produce plots and tables in this paper are saved into a \texttt{fits} file described in the following appendix.
\startlongtable
\begin{deluxetable}{l|c|l}
\tablecolumns{3}
\tablewidth{0pt}
\tablecaption{Definition of the column in the Fermi GRB LAT fits file
\label{tab_fits_file}}
\tablehead{\colhead{Name}&\colhead{Units}&\colhead{Description}}
\startdata\texttt{GCNNAME} &  & \makecell[tl]{Name as appear in the GCN distribution list} \\
\texttt{GRBNAME} &  & \makecell[tl]{Name of the GRB in YYMMDDFFF} \\
\texttt{GRBDATE} &  & \makecell[tl]{Date of the trigger } \\
\texttt{GRBMET} & s & \makecell[tl]{Mission Elapsed Time since 2001-01-01UT00:00:00} \\
\texttt{RA} & deg & \makecell[tl]{Right Ascension (J200)} \\
\texttt{DEC} & deg & \makecell[tl]{Declination (J200)} \\
\texttt{ERR} & deg & \makecell[tl]{Localization error from LTF analysis } \\
\texttt{REDSHIFT} &  & \makecell[tl]{Redshift of the GRB} \\
\texttt{LUMINOSITY\_DISTANCE} & cm & \makecell[tl]{Luminosity distance calculated using the redshift of the GRB (when available)} \\
\texttt{THETA} & deg & \makecell[tl]{Off-axis angle at the time of the trigger} \\
\texttt{ZENITH} & deg & \makecell[tl]{Zenith angle at the time of the trigger } \\
\texttt{ARR} &  & \makecell[tl]{If the GRB triggered an Autonomous Repoint Request, this flag is set to 1, otherwise is set to 0} \\
\texttt{DISTANCE2CLOSEST} & deg & \makecell[tl]{Distance to the closest 3FGL source} \\
\texttt{IRFS} &  & \makecell[tl]{Instrument response function used in the analysis} \\
\texttt{GBMT05} & s & \makecell[tl]{GBM T$_{05}$, when available. Otherwise this number is set to 0. } \\
\texttt{GBMT90} & s & \makecell[tl]{GBM T$_{90}$, when available. Otherwise this number correspond \\ to the estimated duration of the prompt emission.} \\
\texttt{GBMT95} & s & \makecell[tl]{Calclated as \texttt{GBMT05}+\texttt{GBMT05}} \\
\texttt{LLEBBBD\_SIG} & sigma & \makecell[tl]{Significance of the signal in the LLE data.}\\
\texttt{LLEBBBD\_SIG\_DETECTED} &  & \makecell[tl]{Wether the GRB is detected in LLE data.}\\
\texttt{LLET05} & s & \makecell[tl]{ LLE onset time (\tllf). }\\
\texttt{LLET90} & s & \makecell[tl]{ LLE duration time (\tlln).            }\\
\texttt{LLET95} & s & \makecell[tl]{ LLE end time (\tllnf).            }\\
\texttt{TL0} & s & \makecell[tl]{ LAT emission estimated onset time \tz.           }\\
\texttt{TL100} & s & \makecell[tl]{ LAT emission estimated duration \toz.             }\\
\texttt{TL1} & s & \makecell[tl]{ LAT emission estimated end \tone.                   }\\
\texttt{TL0\_L} & s & \makecell[tl]{Estimated lower limit on the LAT emission onset time \tz.}\\
\texttt{TL100\_ERR} & s & \makecell[tl]{Estimated error on the LAT emission duration \toz.}\\
\texttt{TL1\_U} & s & \makecell[tl]{Estimated upper limit on LAT emission end \tone.}\\
\texttt{LIKE\_BEST\_T0} & s & \makecell[tl]{Start of the time window for the likelihood analysis that returned the highest value of Test Statistics.}\\
\texttt{LIKE\_BEST\_T1} & s & \makecell[tl]{ End of the time window for the likelihood analysis that returned the highest value of Test Statistics.}\\
\texttt{LIKE\_BEST\_TS\_GRB} &  & \makecell[tl]{Value of the Test Statistics obtained by likelihood analysis that returned the highest value of Test Statistics.}\\
\texttt{LIKE\_BEST\_FLUX} & ph/cm$^{2}$/s & \makecell[tl]{Flux obtained by likelihood analysis that returned the highest value of Test Statistics.             }\\
\texttt{LIKE\_BEST\_FLUX\_ERR} & ph/cm$^{2}$/s & \makecell[tl]{Estimated error on the flux obtained by likelihood analysis that returned the highest \\ value of Test Statistics.           }\\
\texttt{LIKE\_BEST\_FLUX\_ENE} & erg/cm$^{2}$/s & \makecell[tl]{Energy flux obtained by likelihood analysis that returned the highest value of Test \\ Statistics.             }\\
\texttt{LIKE\_BEST\_FLUX\_ENE\_ERR} & erg/cm$^{2}$/s & \makecell[tl]{Estimated error on the energy flux obtained by likelihood analysis that returned \\ the highest value of Test Statistics. }\\
\texttt{LIKE\_BEST\_FLUENCE\_ENE} & erg/cm$^{2}$ & \makecell[tl]{  Energy fluence obtained by likelihood analysis that returned the highest value of \\ Test Statistics.              }\\
\texttt{LIKE\_BEST\_FLUENCE\_ENE\_ERR} & erg/cm$^{2}$/s & \makecell[tl]{Estimated error on the energy fluence obtained by likelihood analysis that \\ returned the highest value of Test Statistics.             }\\
\texttt{LIKE\_BEST\_GRBindex} &  & \makecell[tl]{Photon index of the power law that models the GRB in the likelihood analysis that returned the highest \\ value of Test Statistics. }\\
\texttt{LIKE\_BEST\_GRBindex\_ERR} &  & \makecell[tl]{Estimated error on the photon index of the power law that models the GRB in the likelihood \\ analysis that returned the highest value of Test Statistics.              }\\
\texttt{LIKE\_BEST\_EISO52\_RF} & 10$^{52}$ erg & \makecell[tl]{ Rest frame isotropic energy from 100\,MeV -- 10\,GeV in the likelihood \\ analysis that returned the highest value of Test Statistics.              }\\
\texttt{LIKE\_BEST\_EISO52\_RF\_ERR} &  10$^{52}$ erg & \makecell[tl]{Estimated error on the rest frame isotropic energy from 100\,MeV -- 10\,GeV \\  in the likelihood analysis that returned the highest value of Test Statistics.              }\\
\texttt{LIKE\_LAT\_T0} & s & \makecell[tl]{Start of the time window for the likelihood analysis in the LAT time window.}\\
\texttt{LIKE\_LAT\_T1} & s & \makecell[tl]{ End of the time window for the likelihood analysis in the LAT time window.}\\
\texttt{LIKE\_LAT\_TS\_GRB} &  & \makecell[tl]{Value of the Test Statistics obtained by likelihood analysis in the LAT time window.}\\
\texttt{LIKE\_LAT\_FLUX} & ph/cm$^{2}$/s & \makecell[tl]{Flux obtained by likelihood analysis in the LAT time window.             }\\
\texttt{LIKE\_LAT\_FLUX\_ERR} & ph/cm$^{2}$/s & \makecell[tl]{Estimated error on the flux obtained by likelihood analysis in the LAT time window.           }\\
\texttt{LIKE\_LAT\_FLUX\_ENE} & erg/cm$^{2}$/s & \makecell[tl]{Energy flux obtained by likelihood analysis in the LAT time window.             }\\
\texttt{LIKE\_LAT\_FLUX\_ENE\_ERR} & erg/cm$^{2}$/s & \makecell[tl]{Estimated error on the energy flux obtained by likelihood analysis in the LAT time \\ window. }\\
\texttt{LIKE\_LAT\_FLUENCE\_ENE} & erg/cm$^{2}$ & \makecell[tl]{  Energy fluence obtained by likelihood analysis in the LAT time window.              }\\
\texttt{LIKE\_LAT\_FLUENCE\_ENE\_ERR} & erg/cm$^{2}$/s & \makecell[tl]{Estimated error on the energy fluence obtained by likelihood analysis in the \\ LAT time window.             }\\
\texttt{LIKE\_LAT\_GRBindex} &  & \makecell[tl]{Photon index of the power law that models the GRB in the likelihood analysis in the LAT time window. }\\
\texttt{LIKE\_LAT\_GRBindex\_ERR} &  & \makecell[tl]{Estimated error on the photon index of the power law that models the GRB in the likelihood analysis \\ in the LAT time window.              }\\
\texttt{LIKE\_LAT\_EISO52\_RF} & 10$^{52}$ erg & \makecell[tl]{ Rest frame isotropic energy  from 100\,MeV -- 10\,GeV in the likelihood analysis \\  in the LAT time window. }\\
\texttt{LIKE\_LAT\_EISO52\_RF\_ERR} &  10$^{52}$ erg & \makecell[tl]{Estimated error on the rest frame isotropic energy  from 100\,MeV -- 10\,GeV \\ in the likelihood analysis in the LAT time window.}\\
\texttt{LIKE\_GBM\_T0} & s & \makecell[tl]{Start of the time window for the likelihood analysis in the GBM time window.}\\
\texttt{LIKE\_GBM\_T1} & s & \makecell[tl]{ End of the time window for the likelihood analysis in the GBM time window.}\\
\texttt{LIKE\_GBM\_TS\_GRB} &  & \makecell[tl]{Value of the Test Statistics obtained by likelihood analysis in the GBM time window.}\\
\texttt{LIKE\_GBM\_FLUX} & ph/cm$^{2}$/s & \makecell[tl]{Flux obtained by likelihood analysis in the GBM time window.             }\\
\texttt{LIKE\_GBM\_FLUX\_ERR} & ph/cm$^{2}$/s & \makecell[tl]{Estimated error on the flux obtained by likelihood analysis in the GBM time window.           }\\
\texttt{LIKE\_GBM\_FLUX\_ENE} & erg/cm$^{2}$/s & \makecell[tl]{Energy flux obtained by likelihood analysis in the GBM time window.             }\\
\texttt{LIKE\_GBM\_FLUX\_ENE\_ERR} & erg/cm$^{2}$/s & \makecell[tl]{Estimated error on the energy flux obtained by likelihood analysis in the GBM time window. }\\
\texttt{LIKE\_GBM\_FLUENCE\_ENE} & erg/cm$^{2}$ & \makecell[tl]{  Energy fluence obtained by likelihood analysis in the GBM time window.              }\\
\texttt{LIKE\_GBM\_FLUENCE\_ENE\_ERR} & erg/cm$^{2}$/s & \makecell[tl]{Estimated error on the energy fluence obtained by likelihood analysis in \\ the GBM time window.             }\\
\texttt{LIKE\_GBM\_GRBindex} &  & \makecell[tl]{Photon index of the power law that models the GRB in the likelihood analysis in the GBM time window. }\\
\texttt{LIKE\_GBM\_GRBindex\_ERR} &  & \makecell[tl]{Estimated error on the photon index of the power law that models the GRB in the likelihood \\ analysis in the GBM time window.              }\\
\texttt{LIKE\_GBM\_EISO52\_RF} & 10$^{52}$ erg & \makecell[tl]{ Rest frame isotropic energy  from 100\,MeV -- 10\,GeV in the likelihood analysis \\  in the GBM time window. }\\
\texttt{LIKE\_GBM\_EISO52\_RF\_ERR} & 10$^{52}$ erg  & \makecell[tl]{Estimated error on the rest frame isotropic energy  from 100\,MeV -- 10\,GeV \\  in the likelihood analysis in the GBM time window.}\\
\texttt{LIKE\_EXT\_T0} & s & \makecell[tl]{Start of the time window for the likelihood analysis in the EXT time window.}\\
\texttt{LIKE\_EXT\_T1} & s & \makecell[tl]{ End of the time window for the likelihood analysis in the EXT time window.}\\
\texttt{LIKE\_EXT\_TS\_GRB} &  & \makecell[tl]{Value of the Test Statistics obtained by likelihood analysis in the EXT time window.}\\
\texttt{LIKE\_EXT\_FLUX} & ph/cm$^{2}$/s & \makecell[tl]{Flux obtained by likelihood analysis in the EXT time window.             }\\
\texttt{LIKE\_EXT\_FLUX\_ERR} & ph/cm$^{2}$/s & \makecell[tl]{Estimated error on the flux obtained by likelihood analysis in the EXT time window.           }\\
\texttt{LIKE\_EXT\_FLUX\_ENE} & erg/cm$^{2}$/s & \makecell[tl]{Energy flux obtained by likelihood analysis in the EXT time window.             }\\
\texttt{LIKE\_EXT\_FLUX\_ENE\_ERR} & erg/cm$^{2}$/s & \makecell[tl]{Estimated error on the energy flux obtained by likelihood analysis in the EXT time window. }\\
\texttt{LIKE\_EXT\_FLUENCE\_ENE} & erg/cm$^{2}$ & \makecell[tl]{  Energy fluence obtained by likelihood analysis in the EXT time window.              }\\
\texttt{LIKE\_EXT\_FLUENCE\_ENE\_ERR} & erg/cm$^{2}$/s & \makecell[tl]{Estimated error on the energy fluence obtained by likelihood analysis in the EXT time window.             }\\
\texttt{LIKE\_EXT\_GRBindex} &  & \makecell[tl]{Photon index of the power law that models the GRB in the likelihood analysis in the EXT time window. }\\
\texttt{LIKE\_EXT\_GRBindex\_ERR} &  & \makecell[tl]{Estimated error on the photon index of the power law that models the GRB in the likelihood analysis \\  in the EXT time window.              }\\
\texttt{LIKE\_EXT\_EISO52\_RF} & 10$^{52}$ erg & \makecell[tl]{ Rest frame isotropic energy  from 100\,MeV -- 10\,GeV in the likelihood analysis \\ in the EXT time window. }\\
\texttt{LIKE\_EXT\_EISO52\_RF\_ERR} &  10$^{52}$ erg & \makecell[tl]{Estimated error on the rest frame isotropic energy  from 100\,MeV -- 10\,GeV \\ in the likelihood analysis in the EXT time window.}\\
\texttt{gtsrcprob\_ExtendedEmission\_MAXE} & MeV & \makecell[tl]{Maximum energy of the event with $>$90\% probability to be associated with the GRB,  \\ calculated during the time resolved analysis.}\\
\texttt{gtsrcprob\_ExtendedEmission\_MAXE\_P} &  & \makecell[tl]{Probability of the event with maximum energy calculated during the time resolved \\  analysis.     }\\
\texttt{gtsrcprob\_ExtendedEmission\_MAXE\_T} & s & \makecell[tl]{Arrival time of the event with maximum energy calculated during the time resolved \\ analysis. }\\
\texttt{gtsrcprob\_ExtendedEmission\_NTH} &  & \makecell[tl]{Number of events with probability $>$90\% to be associated with the GRB, calculated during \\  the time resolved analysis.           }\\
\texttt{LAT\_BPL\_CHI2} &  & \makecell[tl]{Value of the $\chi^{2}$ obtained by fitting the LAT light curve with a broken  power law model.            }\\
\texttt{LAT\_BPL\_F0} &  & \makecell[tl]{ Normalization of the broken  power law model.            }\\
\texttt{LAT\_BPL\_F0\_ERR} &  & \makecell[tl]{  Error on the normalization of the broken  power law model.             }\\
\texttt{LAT\_BPL\_IN1} &  & \makecell[tl]{ First index of the broken  power law model.              }\\
\texttt{LAT\_BPL\_IN1\_ERR} &  & \makecell[tl]{ Error on the first index of the broken  power law model.              }\\
\texttt{LAT\_BPL\_IN2} &  & \makecell[tl]{Second index of the broken  power law model.               }\\
\texttt{LAT\_BPL\_IN2\_ERR} &  & \makecell[tl]{ Error on the second index of the broken  power law model.              }\\
\texttt{LAT\_BPL\_TB} &  & \makecell[tl]{ Time of the break.            }\\
\texttt{LAT\_BPL\_TB\_ERR} &  & \makecell[tl]{ Error on the time of the break.            }\\
\texttt{LAT\_SPL\_CHI2} &  & \makecell[tl]{Value of the $\chi^{2}$ obtained by fitting the LAT light curve with a simple  power law model.               }\\
\texttt{LAT\_SPL\_IN1} &  & \makecell[tl]{ Index of the simple power law model            }\\
\texttt{LAT\_SPL\_IN1\_ERR} &  & \makecell[tl]{Error on the index of the simple power law model             }\\
\texttt{LAT\_F0} &  & \makecell[tl]{ Normalization of the simple power law model            }\\
\texttt{LAT\_F0\_ERR} &  & \makecell[tl]{ Error on the normalization of the simple power law model            }\\
\texttt{LAT\_IN} &  & \makecell[tl]{ Index that best describe the behavior at late time  (either  \texttt{LAT\_BPL\_IN2} or  \texttt{LAT\_SPL\_IN1}), depending \\ on the value of the respective $\chi^{2}$.         }\\
\texttt{LAT\_IN\_ERR} &  & \makecell[tl]{Error on the  index that best describe the behavior at late time.}\\
\texttt{SU\_TSINPUT} &  & \makecell[tl]{Test Statistic resulting from the LTF detection algorithm             }\\
\texttt{GBM\_assoc\_key} &  & \makecell[tl]{Name of the GRB in the Fermi GBM GRB catalog.}\\
\texttt{T90\_ERROR} &  s & \makecell[tl]{Error on the GBM \tn from the Fermi GBM GRB catalog.}\\
\texttt{FLUENCE} & erg/cm$^{2}$/s   & \makecell[tl]{Fluence in the  10\,keV--1\,MeV energy band, from the Fermi GBM GRB catalog.           }\\
\texttt{FLUENCE\_ERROR} & erg/cm$^{2}$/s   & \makecell[tl]{Error on the fluence in the  10\,keV--1\,MeV energy band, from the Fermi GBM GRB \\ catalog.             }\\
\texttt{LC\_START} & s & \makecell[tl]{Variable length array containing the starting points of the time interval in the time-resolved analysis.}\\
\texttt{LC\_MEDIAN} & s & \makecell[tl]{Variable length array containing the median points of the time interval in the time-resolved analysis.             }\\
\texttt{LC\_END} & s & \makecell[tl]{Variable length array containing the end points of the time interval in the time-resolved analysis.            }\\
\texttt{LC\_ENE\_FLUX} & erg/cm$^{2}$/s & \makecell[tl]{Variable length array containing the energy flux values in the time-resolved analysis.             }\\
\texttt{LC\_ENE\_FLUX\_ERR} & erg/cm$^{2}$/s & \makecell[tl]{ Variable length array containing the energy flux value errors in the time-resolved \\ analysis.                  }\\
\texttt{LC\_FLUENCE} & erg/cm$^{2}$ & \makecell[tl]{Variable length array containing the energy fluence values in the time-resolved analysis.                   }\\
\texttt{LC\_FLUX} & ph/cm$^{2}$/s & \makecell[tl]{Variable length array containing the photon flux values in the time-resolved analysis.                   }\\
\texttt{LC\_FLUX\_ERR} & ph/cm$^{2}$/s & \makecell[tl]{Variable length array containing the photon flux value errors in the time-resolved \\ analysis.             }\\
\texttt{LC\_INDEX} &  & \makecell[tl]{Variable length array containing the values of the photon index  in the time-resolved analysis.             }\\
\texttt{LC\_INDEX\_ERR} &  & \makecell[tl]{Variable length array containing the errors on the values of the photon index  in the time-resolved \\ analysis.                }\\
\texttt{LC\_TS} &  & \makecell[tl]{Variable length array containing the values of the Test Statistic in the time-resolved analysis.                }\\
\enddata
\end{deluxetable}

\bibliography{Fermi_GRB_2}
\bibliographystyle{yahapj}

\end{document}